%% file: main.tex
\documentclass[%
 reprint,
 superscriptaddress,
 amsmath,amssymb,amsthm,
 aps,prx,
]{revtex4-2}

\usepackage{graphicx}
\usepackage{dcolumn}
\usepackage{bm}
\usepackage{braket}
\usepackage[table]{xcolor}
\usepackage{xcolor}
\usepackage{amsthm}
\usepackage{upgreek}
\usepackage[normalem]{ulem}
\DeclareMathOperator{\trace}{Tr}
\DeclareMathOperator{\sign}{Sign}
\DeclareMathOperator{\hc}{h.c.}
\newcommand{\qop}{{\bf q}}
\newcommand{\Qop}{{\bf Q}}

\newcommand{\phiop}{{\bf \Phi}}

\newcommand{\hop}{{\bf h}}

\newcommand{\Pop}{{\bf p}}
\newcommand{\SSS}{{\bf S}}

\newcommand{\X}{{\bf X}}
\newcommand{\Z}{{\bf Z}}
\newcommand{\Y}{{\bf Y}}
\newcommand{\E}{{\bf E}}

\newcommand{\gateLengthFactor}{\lambda}

\let\normalrho\rho
\renewcommand{\rho}{\boldsymbol{\normalrho}}
\newcommand{\rhoo}{\boldsymbol{\normalrho}}

\newcommand{\nth}{n_{\mathrm{th}}}

\newcommand{\D}{{\bf D}}

\newcommand{\HHop}{{\bf H}}

\newcommand{\HH}{{\bf H}}
\newcommand{\GG}{{\bf G}}

\newcommand{\UU}{{\bf U}}
\newcommand{\VV}{{\bf V}}

\newcommand{\II}{{\bf 1}}

\newcommand{\N}{{\bf N}}

\newcommand{\LL}{{\bf L}}
\newcommand{\MM}{{\bf M}}
\newcommand{\NN}{{\bf N}}

\newcommand{\Rop}{{\bf R}}
\newcommand{\rop}{{\bf r}}
\newcommand{\sop}{{\bf s}}

\usepackage{soul}
\usepackage{siunitx}

\newcommand{\aop}{{\bf a}}
\newcommand{\Aop}{{\bf A}}
\newcommand{\adag}{{\bf a^{\dag}}}

\newcommand{\bop}{{\bf b}}
\newcommand{\Bop}{{\bf B}}
\newcommand{\bdag}{{\bf b^{\dag}}}

\newcommand{\bigparskip}{\\[16pt]}
\newcommand{\bigpar}[1]{\paragraph*{\bf\emph{#1.}}}

\usepackage{titletoc}

\usepackage[titles]{tocloft}
\makeatletter
\renewcommand{\@tocrmarg}{1.5em plus1fil}
\makeatother

\usepackage[colorlinks=false, linktocpage=true, hypertexnames=false]{hyperref}

\usepackage[all]{hypcap}

\usepackage[OT2,T1]{fontenc}
\DeclareSymbolFont{cyrletters}{OT2}{wncyr}{m}{n}
\DeclareMathSymbol{\Sha}{\mathalpha}{cyrletters}{"58}

\newcommand{\cA}{\mathcal A}
\newcommand{\cD}{\mathcal D}
\newcommand{\oa}{{\bf a}}
\renewcommand{\dag}{\dagger}

\usepackage[capitalize]{cleveref}

\newtheorem{estimate}{Estimate}

\usepackage{svg}

\binoppenalty=\maxdimen
\relpenalty=\maxdimen

\begin{document}

\preprint{APS/123-QED}

\title{Dissipative protection of a GKP qubit in a high-impedance superconducting circuit driven by a microwave frequency comb}

 \author{L.-A. Sellem}
 \affiliation{Laboratoire de Physique de l’Ecole Normale Supérieure, Mines Paris-PSL, Inria, ENS-PSL, Université PSL, CNRS, Sorbonne Université, Paris, France}
 \author{A. Sarlette}
 \affiliation{Laboratoire de Physique de l’Ecole Normale Supérieure, Mines Paris-PSL, Inria, ENS-PSL, Université PSL, CNRS, Sorbonne Université, Paris, France}
 \affiliation{%
 Department of Electronics and Information Systems, Ghent University, Belgium}
     \author{Z. Leghtas}
     \affiliation{Laboratoire de Physique de l’Ecole Normale Supérieure, Mines Paris-PSL, Inria, ENS-PSL, Université PSL, CNRS, Sorbonne Université, Paris, France}
 \author{M. Mirrahimi}
 \affiliation{Laboratoire de Physique de l’Ecole Normale Supérieure, Mines Paris-PSL, Inria, ENS-PSL, Université PSL, CNRS, Sorbonne Université, Paris, France}
\author{P. Rouchon}
\affiliation{Laboratoire de Physique de l’Ecole Normale Supérieure, Mines Paris-PSL, Inria, ENS-PSL, Université PSL, CNRS, Sorbonne Université, Paris, France}
\author{P. Campagne-Ibarcq}%
 \email{philippe.campagne-ibarcq@inria.fr}
\affiliation{Laboratoire de Physique de l’Ecole Normale Supérieure, Mines Paris-PSL, Inria, ENS-PSL, Université PSL, CNRS, Sorbonne Université, Paris, France}

\date{\today}

\begin{abstract}
We propose a novel approach to generate, protect and control GKP qubits. It employs a microwave frequency comb parametrically modulating a Josephson circuit  to enforce a dissipative dynamics of a high impedance circuit mode,  autonomously stabilizing the finite-energy GKP code. The encoded GKP qubit is robustly protected against all dominant  decoherence channels plaguing superconducting circuits but quasi-particle poisoning. In particular, noise from  ancillary modes leveraged for  dissipation engineering does not  propagate at  the logical level. In a state-of-the-art experimental setup,  we estimate that the encoded qubit lifetime could extend  two orders of magnitude  beyond the break-even point, with substantial margin for improvement through progress in fabrication and control electronics.  Qubit initialization, readout and control via Clifford gates can be performed while maintaining the code stabilization, paving the way toward the assembly of GKP qubits in a fault-tolerant quantum computing architecture.
\end{abstract}

\maketitle

\startcontents[maintext]
\addtocontents{toc}{\protect\setcounter{tocdepth}{0}}
\begingroup
\hyphenchar\font=-1 
\printcontents[maintext]{l}{1}{\section*{Contents}}
\hyphenchar\font=`\- 
\endgroup
\addtocontents{toc}{\protect\setcounter{tocdepth}{3}}

\input{article_body.tex}

\clearpage

\phantomsection
\addcontentsline{toc}{section}{\protect\numberline{}Appendix}
\stopcontents[maintext]

\input{appendix}

\stopcontents[supmat]

\resumecontents[maintext]
\bibliography{biblio.bib}

\end{document}

%% file: article_body.tex
\section{Introduction}

Despite considerable progress realized over the past decades in better isolating quantum systems from their fluctuating environment, noise levels in all explored physical platforms remain far too high to run useful quantum algorithms. Quantum error correction (QEC) would  overcome this roadblock by encoding a logical qubit in a high-dimensional physical system and correcting noise-induced evolutions before they accumulate and lead to logical flips. In stabilizer codes, such errors are unambiguously revealed by  measuring \textit{stabilizer operators}~\cite{gottesman1997stabilizer}, which commute with the logical Pauli operators and thus do not perturb the encoded qubit.   A central assumption behind QEC is that a physical system  only interacts with its noisy environment  via  low-weight operators. For instance, in discrete variable codes such as the toric code~\cite{kitaev2003fault}, the surface code~\cite{freedman2001projective,bravyi1998quantum} or the color code~\cite{bombin2006topological}, the logical qubit is encoded in a collection of physical two-level systems devoid of many-body interactions. In bosonic codes such as the GKP code~\cite{gottesman2001encoding,grimsmo2021quantum},
the Schrödinger cat code~\cite{Cochrane1999cats,mirrahimi2014dynamically}
and the binomial code~\cite{michael2016new,hu2019quantum},
the qubit is encoded in a quantum oscillator
whose interactions, denoted here as \textit{ low-weight interactions},
involve a small number of photons. More precisely, these interactions are mediated by a  coupling Hamiltonian which is a low-order polynomial of the oscillator annihilation and creation operators $\aop$ and $\adag$. Under these assumptions, noise does not directly induce logical flips between well-chosen code states. Specifically, codes are constructed such that several two-level systems should flip in order to induce a logical flip in the former case, and that a multi-photonic transition should occur in the latter case. Admittedly, logical flips may occur indirectly as    low-weight interactions can generate a high-weight  evolution operator,  but this evolution takes time and is correctable provided that QEC is performed sufficiently fast. \\

\begin{figure}[htbp]
		\centering
		\includegraphics[width=1\columnwidth]{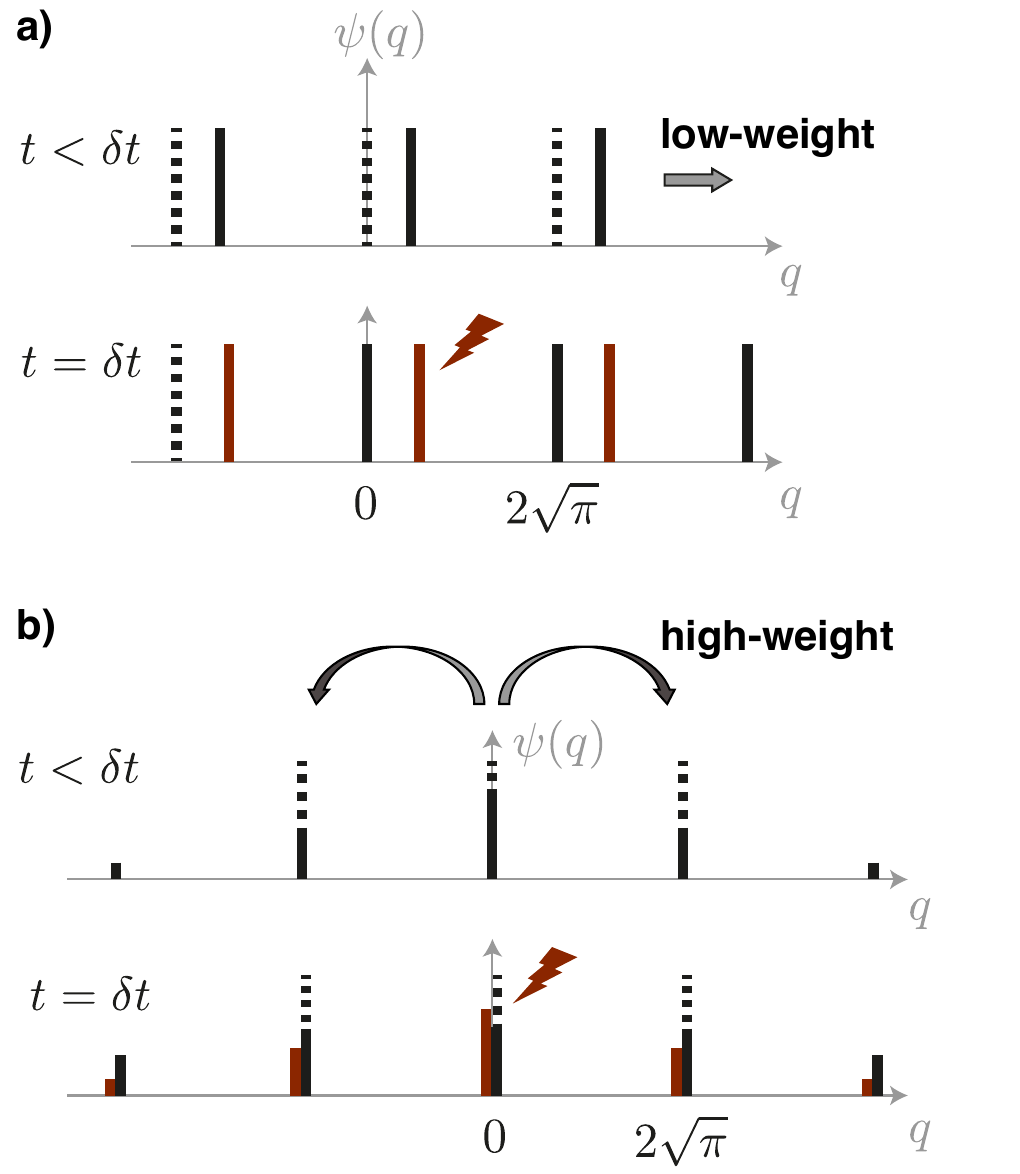}
		\caption{{\bf a) Low-weight interactions.}  $\HH=-g\Pop {\bf B}$ is an example of low-weight Hamiltonian employed in recent experiments stabilizing the GKP code. It entails a continuous displacement of a GKP state along the $q$ quadrature of an oscillator (plain black lines, initial state represented by dashed black  lines), conditioned on an ancillary mode observable ${\bf B}$. The interaction duration $\delta t$ is chosen  such that the state is displaced by one period of the square GKP lattice after the evolution. However, if noise modifies the value of ${\bf B}$ during the interaction (red lightning), the final target state is shifted (red lines) and the GKP qubit may be flipped (see Sec.~\ref{sec:gkp}). {\bf b) High-weight interactions.}  $\HH=-g \mathrm{cos}(2\sqrt{\pi}\Pop) {\bf B}$  is a high-weight  (modular) Hamiltonian that entails a hopping dynamics along the GKP lattice.  If noise modifies the value of ${\bf B}$ during the interaction, the relative weights of the final state peaks may be affected but not their positions, so that no logical flip may occur.   }
		\label{fig:weight}
	\end{figure}

 The aforementioned bosonic codes are appealing for their moderate hardware overhead, but a paradox emerges in their operation: some of their stabilizers are  high-weight operators that do not appear naturally in the system interactions. A common strategy to measure these stabilizers is to map their  value to an ancilla system via an evolution operator  generated from a low-weight interaction. It was successfully employed to stabilize cat codes~\cite{leghtas2015confining}, binomial codes~\cite{hu2019quantum} and the GKP code~\cite{fluhmann2018encoding}, but results in the opening of uncorrectable error channels. As illustrated in Fig.~\ref{fig:weight}a in the case of the GKP code, while the interaction is   carefully timed so that the overall  evolution operator leaves code states unaffected in the absence of noise, ancilla errors during the interaction propagate as uncontrolled long shifts of the target system,  triggering logical flips. Partial QEC of the ancilla~\cite{puri2018stabilized} or error mitigation~\cite{ma2020path,shi2019fault,siegele2023robust} was proposed to suppress this advert effect, but the robust implementation of these ideas is a major experimental challenge~\cite{rosenblum2018fault}. An alternative strategy, more robust but experimentally more demanding, consists in engineering
  high-weight interactions so that the target system  interacts at all time  with the ancilla via its stabilizer operators only. In this configuration, ancilla noise propagates to the target system as an evolution operator generated by the stabilizers only, which leaves the logical qubit unaffected (see Fig.~\ref{fig:weight}b).  \\

Focusing on the GKP code, the two stabilizers are  commuting trigonometric functions of the oscillator position and momentum  (high-weight operators), which generate discrete translations along a grid in phase-space. The phase of these so-called \textit{modular operators}~\cite{Neumann1955,Aharonov1969,Popescu2010,fluhmann2018sequential} reveals spurious small shifts of the oscillator state in phase-space while supporting no information on the encoded qubit state. Most proposals~\cite{travaglione2002preparing,Pirandola2006,terhal2016encoding,motes2017encoding,weigand2020realizing,royer2020stabilization} and all experimental demonstrations~\cite{fluhmann2018encoding,campagne2020quantum,de2022error,sivak2022real,nord_quantique_exp} of GKP state preparation and error-correction are based on variants of phase-estimation~\cite{kitaev1995quantum,svore2013faster} of the stabilizers. Phase-estimation falls into the first category of stabilizer measurement strategies described above, and therefore leaves the target system open to uncorrectable error channels. In this paper, we consider the second, more  robust strategy and aim at engineering high-weight interactions involving only the  two modular stabilizers. The state of the oscillator would then only hop  along the GKP code lattice in phase-space (see Fig.~\ref{fig:weight}b for schematic hopping along one phase-space quadrature). But how can we engineer a coupling Hamiltonian involving two modular operators?\\

An isolated Josephson junction behaves as an inductive element whose dynamics is governed by a modular flux operator. However, in most circuitQED experiments~\cite{blais2021circuit}, the junction is shunted by a low-impedance circuitry, so that it effectively acts on the circuit modes as a weakly non-linear, low-weight, operator. In contrast, connecting the junction to a circuit whose impedance exceeds the quantum of resistance---a regime recently attained in circuitQED---reveals its truly modular nature~\cite{cohen2017degeneracy}. Unfortunately, experimental implementations of the dual \textit{coherent phase-slip element}, whose dynamics is governed by a modular charge operator~\cite{mooij2006superconducting} are not yet coherent enough for practical use~\cite{astafiev2012coherent}. Moreover, the doubly modular Hamiltonian implemented by the association of these two elements would only stabilize a single GKP state and not a two-dimensional code manifold~\cite{le2019doubly}.  The  $0-\pi$ qubit~\cite{brooks2013protected,groszkowski2018coherence} is an elementary protected circuit that would circumvent these two  pitfalls.  In this circuit, an effective coherent phase-slip behavior emerges in the low energy dynamics of an ultra-high impedance \textit{fluxonium} mode~\cite{manucharyan2009fluxonium,pechenezhskiy2020superconducting}. When appropriately coupled to a \textit{transmon} mode~\cite{koch2007charge}, the quasi-degenerate ground manifold is spanned by a pair of two-mode GKP states~\cite{conrad2022gottesman}. However, fully fledged GKP states are only obtained in an extreme parameter regime currently out of reach~\cite{groszkowski2018coherence}. Recently, Rymarz \textit{et al.}~\cite{rymarz2021hardware} proposed an alternative approach to offset the lack of a phase-slip element. Building on an idea suggested in the original GKP proposal~\cite{gottesman2001encoding}, they realized that two Josephson junctions bridged by a high-impedance gyrator would implement a doubly modular Hamiltonian stabilizing quasi-degenerate GKP states. However, existing gyrators are either far too limited in impedance and bandwidth~\cite{chapman2017widely,lecocq2017nonreciprocal,barzanjeh2017mechanical} or rely on strong magnetic fields incompatible with superconducting circuits~\cite{mahoney2017chip}.  \\

\begin{figure*}[htbp]
		\centering
		\includegraphics[width=1.7\columnwidth]{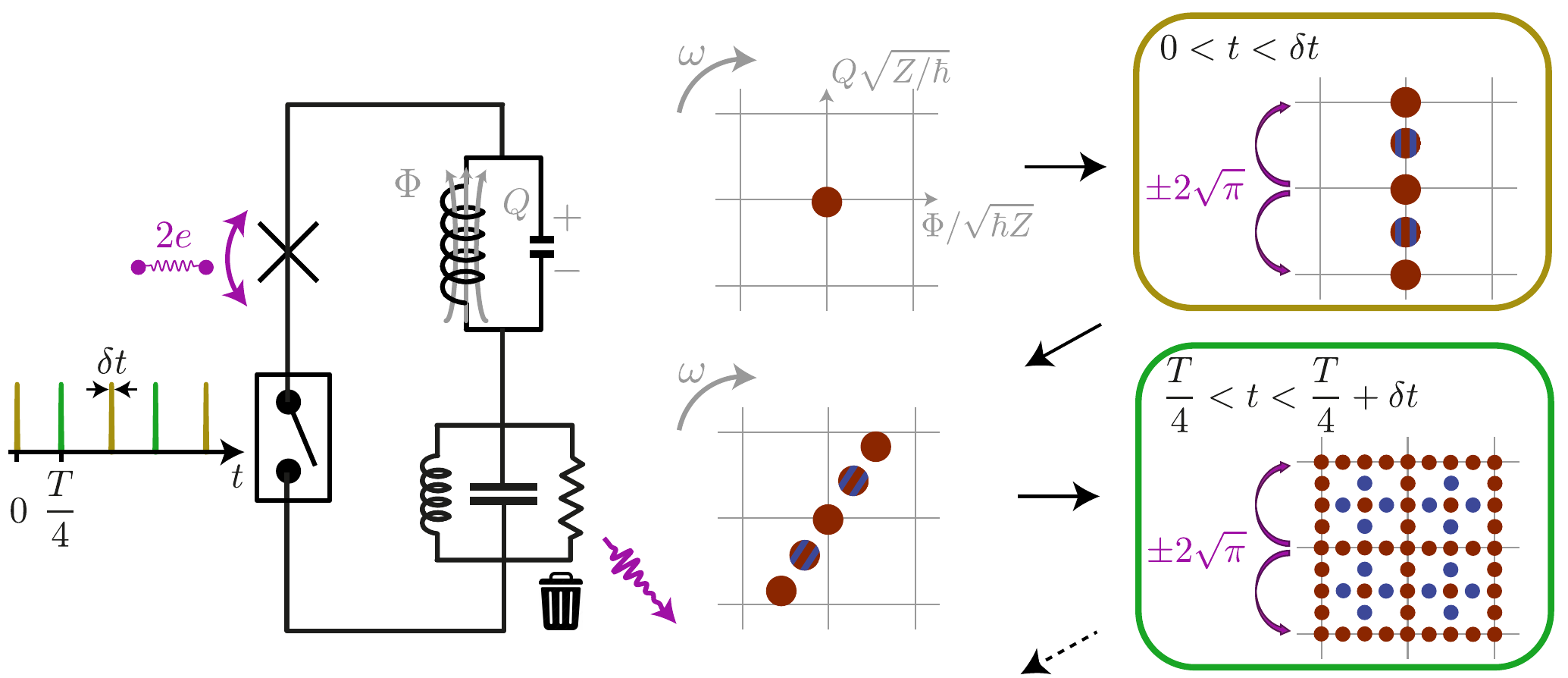}
		\caption{ {\bf  Schematic representation of modular dissipation engineering}. A switch controls the coherent tunneling of Cooper pairs (charge $2e$) across a Josephson junction placed in parallel  with a two-mode circuit. The target mode (top) has a high impedance $Z$ such that, in normalized phase-space coordinates, tunneling events translate its state by $\pm 2\sqrt{\pi}$ along the charge axis. The switch is controlled with a train of sharp pulses (duration $\delta t$) activating tunneling every quarter of a period $T$ of the target oscillator. In between pulses, the oscillator state rotates freely in phase-space at $\omega=2\pi/T$. Overall, the target mode  dynamics is generated by discrete shifts along a square grid matching the GKP lattice (gray grid with period $2\sqrt{\pi}$ overlaid with Wigner diagrams of the  oscillator state). A lower impedance ancillary mode (bottom), also driven by Cooper pair tunneling, dissipates excitations into a cold load (purple wriggled arrow) to ensure that the target mode dynamics is irreversible,  autonomously stabilizing the GKP code.  }
		\label{fig:schematic}
	\end{figure*}

In this paper, we propose to engineer a true doubly modular Hamiltonian in the rotating frame of a state-of-the-art Josephson circuit. The method, similar to the twirling-based engineering introduced in Ref.~\cite{conrad2021twirling},  is schematically represented in Fig.~\ref{fig:schematic}. A Josephson junction allows the coherent tunneling of Cooper pairs across a high-impedance circuit mode, translating its state by $\pm 2e$ along the charge axis of phase-space. Modulating the tunneling rate with fast pulses, we ensure that such translations occur every quarter period of the target mode only, and let the state rotate freely in phase-space in-between pulses. As a result, the state evolves in discrete steps on a square grid, which matches the GKP code lattice for the proper choice of target mode impedance.  We combine this novel approach with dissipation-engineering techniques successfully employed to stabilize Schrödinger cat states~\cite{leghtas2015confining,lescanne2020exponential}, so that the target oscillator autonomously stabilizes in the GKP code manifold. Mathematical analysis and numerical simulations show that this strategy can enhance the logical qubit coherence far beyond  that of the underlying circuit. Moreover, we describe how to control encoded qubits with fault-tolerant Clifford gates, paving the way toward  a high-fidelity quantum computing architecture based on GKP qubits.\\

The paper is organized as follows. In Sec.~\ref{sec:gkp}, we review the properties of  idealized GKP states and their realistic, finite-energy counterparts. In Sec.~\ref{sec:modular dissip}, we propose a dissipative dynamics based on four modular Lindblad operators stabilizing the finite-energy GKP code, and benchmark its error-correction performances against the dominant decoherence channels plaguing superconducting resonators.  In Sec.~\ref{sec:hamiltonian}, we show how to engineer a doubly modular  Hamiltonian in a high-impedance, parametrically driven Josephson circuit. In Sec.~\ref{sec:dissipationengineering}, we combine this method with reservoir engineering techniques to obtain the target modular dissipation. In Sec.~\ref{sec:noise}, we briefly discuss the impact of various noise processes and that of circuit fabrication constraints and disorder. We refer the reader to the Appendices for a more detailed analysis. Finally, in Sec.~\ref{sec:gates} we sketch how to  control encoded GKP qubits with protected Clifford gates and how to measure their Pauli operators.

\section{The GKP code}
\label{sec:gkp}
GKP introduced coding \textit{grid states} as superpositions of periodically spaced position states of a quantum oscillator. For simplicity’s sake, we consider throughout this paper square grid states---see \cref{smsec:hexacode} for generalization to hexagonal grid states---defined as
 \begin{equation}
 \begin{split}
	 \label{infiniteStates}
        &|+Z_{\infty}\rangle=\sum_{n\in \mathbb{Z}}|n \eta\rangle_q =\sum_{n\in \mathbb{Z}}|\frac{2\pi n}{\eta} \rangle_p \\
         &|-Z_{\infty}\rangle=\sum_{n\in \mathbb{Z}}|(n+\tfrac{1}{2}) \eta\rangle_q =\sum_{n\in \mathbb{Z}}(-1)^n|\frac{2\pi n}{\eta} \rangle_p
 \end{split}
 \end{equation}
where $\eta=2\sqrt{\pi}$ and $|r\rangle_q$ (respectively $|r\rangle_p$) denotes an eigenstate with eigenvalue $r$  of the oscillator normalized position $\qop=(\aop +\adag)/\sqrt{2}$ (respectively momentum $\Pop=(\aop -\adag)/(i\sqrt{2})$). One can show that any pair of orthogonal logical states have distant support in phase-space, providing the code robustness against position and momentum shift errors. Since the evolution of an oscillator quasi-probability distribution in phase-space is  local  under the action of noise coupling  via low-weight operators~\cite{Cahill1969,gottesman2001encoding,cohen2017autonomous}, this robustness extends to all dominant error channels in superconducting resonators.\\

Error-syndromes are extracted by measuring the phase  of the code stabilizers $\SSS_q=e^{ i\eta \qop}$ and $\SSS_p=e^{-i\eta \Pop}$, which is 0 inside the code manifold. Given that the logical qubit can be perfectly decoded as long as the oscillator is not shifted by more  than $\sqrt{\pi}/2$, we define \textit{generalized Pauli} operators $\Z=\mathrm{Sgn}\big(\mathrm{cos}(\frac{\eta}{2}\qop)\big)$, $\X=\mathrm{Sgn}\big(\mathrm{cos}(\frac{\eta}{2}\Pop)\big)$
and $\Y=i \X \Z$.
Here, the superoperator $\mathrm{Sgn}(\cdot)$ denotes the sign of a real-valued operator and is applied to the logical operators introduced by GKP. With our definition, $\X$, $\Y$ and $\Z$ respect the Pauli algebra composition rules throughout the oscillator Hilbert space and coincide with the logical qubit Pauli operators inside the code manifold.   The qubit they define  can remain pure whilst the oscillator state is not. Moreover, we verify that they commute with the  stabilizers, which can thus be measured without  perturbing the encoded qubit. More generally, a noisy environment coupling to the oscillator via the stabilizer operators does not induce logical errors: this is the core idea guiding our approach. Finally, we note that $\X$ and $\Z$ are directly measurable, for instance by trivially decoding the outcome of a homodyne detection respectively along $\Pop$ or $\qop$.\\

Even though infinitely squeezed grid states are physically unrealistic, GKP suggested that these desirable features would be retained for the normalized, finitely squeezed states $|\pm Z_{\Delta}\rangle= \E_{\Delta} |\pm Z_{\infty}\rangle$ where $\E_{\Delta}=e^{-\Delta \adag \aop}$
with $\Delta\ll 1$~\cite{menicucciRefFiniteGKP,royer2020stabilization,matsuura2020equivalence}.  Analogously to the infinitely squeezed case, these two states are $+1$-eigenstates of the commuting, normalized, stabilizers $\SSS_q^{\Delta}=\E_{\Delta} \SSS_q \E_{\Delta}^{-1}$ and $\SSS_p^{\Delta}=\E_{\Delta} \SSS_p \E_{\Delta}^{-1}$. However, they are not orthogonal since their wavefunction peaks are Gaussian with a non-zero standard deviation $\sigma=(\mathrm{tanh}(\Delta))^{\frac{1}{2}}$. Orthogonal, finite-energy logical states can be rigorously defined as their symmetric and antisymmetric superpositions, and  Pauli operators for the finite-energy code can be defined therefrom. Nevertheless, in the following, we retain the encoded qubit as defined by the  $\X$, $\Y$ and $\Z$  operators. Even though this definition does not allow the preparation of a pure logical state at finite energy, it is operationally relevant as these observables can be measured experimentally (either by homodyne detection or following the method described in Sec.~\ref{sec:readout}). Moreover, the qubit maximum purity is exponentially close to 1 as $\Delta$ approaches $0$, so that the encoded qubit is well suited for quantum information processing applications for only modest average photon number in the grid states: we find
$1-(\langle \X \rangle ^2 + \langle \Y \rangle ^2 + \langle \Z \rangle ^2)^{1/2}\simeq2\times10^{-8}$
for a pure finite-energy code state containing $\overline{n}=10 $~photons.

\section{Protection of GKP qubits by modular dissipation}
\label{sec:modular dissip}
\subsection{Convergence toward the code manifold and  errors induced by modular dissipation}

\begin{figure}[htbp]
		\centering
		\includegraphics[width=1.0\columnwidth]{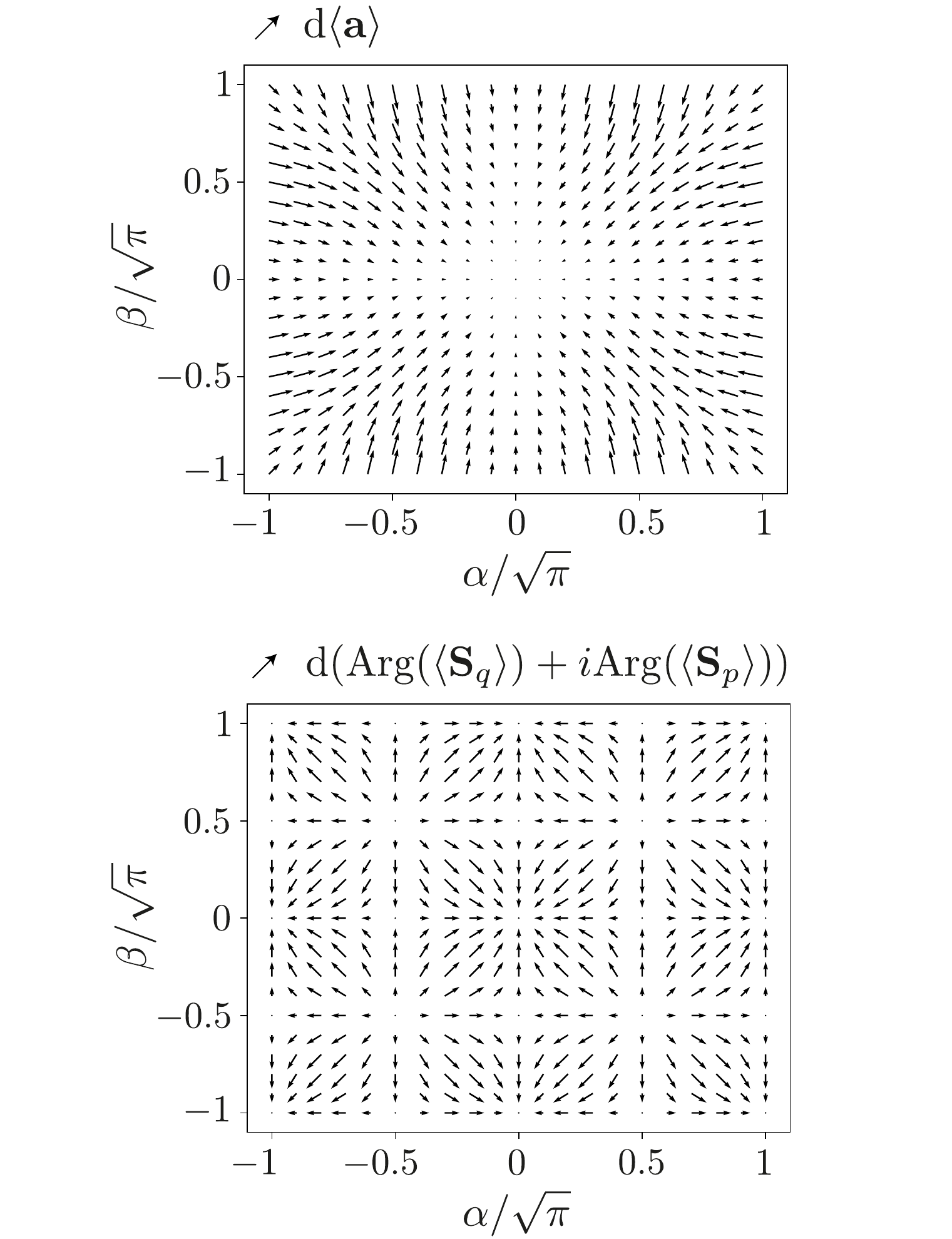}
		\caption{{\bf Modular dissipation phase portraits}. For a finite-energy code state ($ \mathrm{sinh}(\Delta)=0.2/\eta$) displaced by $\alpha+i\beta$ in phase-space, arrows encode the evolution of the state center of mass (top panel) and modular coordinates (bottom panel)  entailed by the Lindblad operators~\eqref{eq:lindissip} over a short time step $\mathrm{d}t\ll 1/\Gamma$. Arrows length are rescaled to arbitrary units. }
		\label{fig:phaseportrait}
	\end{figure}

In Ref.~\cite{sellem2022exponential}, it was shown that a dissipative dynamics based on four Lindblad operators  derived from the two finite-energy code stabilizers and their images by a $\pi$ rotation in phase space stabilizes the code manifold. More precisely, denoting $\mathcal{D}[\LL]$ the  dissipator formed from an arbitrary operator $\LL$ and defined by its action on the density matrix  $\mathcal{D}[\LL](\rhoo)= \LL \rhoo \LL^{\dag}-\frac{1}{2}(\LL^{\dag}\LL \rhoo +  \rhoo \LL^{\dag}\LL)$, the finite-energy code states are fixed points of the Lindblad equation
 \begin{equation}
 \frac{\mathrm{d}\rhoo}{\mathrm{d}t}=\Gamma \sum^3_{k=0} \mathcal{D}[\MM_k](\rhoo),
\label{eq:lindblad}
 \end{equation}
where $\MM_k =  \Rop_{\frac{k\pi}{2}}(\SSS_q^{\Delta}-\II)\Rop^{\dag}_{\frac{k\pi}{2}}$, $\Rop_{\theta}=e^{i\theta \adag \aop}$ performs a rotation by $\theta$ in phase-space and $\Gamma$ is the dissipation rate. Indeed, the Lindblad operators $\MM_k$ are stabilizers of the GKP code, offset by  $-\II$ to ensure that they cancel on the code manifold. Moreover, any initial state of the oscillator converges exponentially toward the code manifold at a rate set by $\Gamma$ and $\Delta$~\cite{sellem2022exponential}. \\

Unfortunately, the $\MM_k$ operators are products of trigonometric \textit{and} hyperbolic functions of $\qop$ and $\Pop$, which would prove formidably challenging to engineer in an experimental system.
Here, we propose to approximate them to first order in $\Delta$ by products of trigonometric and linear functions of $\qop$ and $\Pop$ with the operators
\begin{equation}
    \LL_k= \mathcal{A} \Rop_{\frac{k\pi}{2}}  e^{i\eta \qop}(\II- \epsilon \Pop)\Rop^{\dag}_{\frac{k\pi}{2}}-\II,
    \label{eq:lindissip}
\end{equation} where $\epsilon=\eta~\mathrm{sinh}(\Delta)$ is a small parameter and the scalar factor $\mathcal{A}=e^{-\eta \epsilon/2 }$ originates from the non commutativity of $\qop$ and $\Pop$ in the Baker-Campbell-Hausdorff formula.\\

In order to qualitatively apprehend the dynamics entailed by these modular Lindblad operators, we represent in   Fig.~\ref{fig:phaseportrait} the evolution of a displaced  code state $\rhoo_{\alpha + i \beta }=e^{-i\alpha \Pop + i \beta \qop  }|+ Z_{\Delta} \rangle \langle + Z_{\Delta} | e^{+i\alpha \Pop - i \beta \qop  }$
 over an infinitesimal time step $\mathrm{d}t\ll 1/\Gamma$. On the top panel, arrows represent the variation of the state center of mass (vector complex coordinates proportional to $\mathrm{d}\mathrm{Tr}(\aop~\rhoo_{\alpha + i \beta})$). A single attractor at the origin of phase-space  pins the grid state normalizing envelope. On the bottom panel, arrows represent the variation of the state position and momentum modulo $2\pi/\eta$ (vector complex coordinates proportional to $\mathrm{d}\mathrm{Tr}(\mathrm{Arg}[\SSS_q \rhoo_{\alpha + i \beta}]+i\mathrm{Arg}[\SSS_p \rhoo_{\alpha + i \beta}])$). Multiple attractors appear for $\alpha,~\beta=0~\mathrm{mod}~2\pi/\eta$ pinning the grid peaks onto the GKP code lattice. Note that here, we employ the displaced grid state $\rhoo_{\alpha + i \beta}$  as a sensitive position and momentum shift detector~\cite{Duivenvoorden2017}, but initializing the oscillator in a less exotic state such as a coherent state centered in $\alpha, \beta$ yields similar phase portraits, albeit smoothed by the state quadrature fluctuations. These observations hint at a convergent dynamics toward the finite-energy code manifold, irrespective of the oscillator initial state. This contrasts with the Lindblad dynamics based on only two modular dissipators introduced in Ref.~\cite{royer2020stabilization}, for which we observe dynamical instabilities (see \cref{sm_ssec_instability_lindblad2dissip}).\\

 Quantitatively, we show that, under this four-dissipator dynamics, the  expectation values of the infinite-energy code stabilizers  converge to their steady state value at a rate $\Gamma_c \gtrsim \mathcal{A}\epsilon \eta \Gamma $
 and that the oscillator energy remains bounded (see \cref{ssec:energy_estimates}), proving that the dynamics is indeed stable.
 Note that, due to the linear approximation of hyperbolic functions we made to obtain the operators~\eqref{eq:lindissip}, the state reached by the oscillator after a few $1/\Gamma_c$ does not strictly belong to the code manifold, but consists in a statistical mixture of shifted code states. In terms of phase-space quasiprobability distribution, this results in broader peaks for the stabilized grid states. Yet, the overlap of a peak with its neighbors remains exponentially small as $\epsilon$ decreases, so that high-purity encoded states can still be prepared, and population leakage between two orthogonal logical states occurs on a timescale much longer than $1/\Gamma_c$. Quantitatively, we  show that when $\epsilon \ll 1$, the generalized Pauli operators $\X$ and $\Z$ decay at a rate  $\Gamma^0_L=\frac{4}{\pi}\mathcal{A}\epsilon \eta \Gamma e^{-\frac{4}{ \mathcal{A}\epsilon \eta}}$, while $\Y$ decays twice faster, as expected for the square GKP code. These residual logical errors induced by the engineered modular dissipation itself  vanish when $\epsilon \rightarrow 0$. However, the confinement rate $\Gamma_c$ onto the code manifold---loosely understood as the rate at which stochastic shifts from additional noise channels are corrected---also vanishes in this limit. Therefore, when correcting against intrinsic noise of the target oscillator, the value of $\epsilon$ should be optimized to balance errors induced by the modular dissipation itself with those resulting from excursions outside the code manifold induced by intrinsic noise. We quantitatively analyze this trade-off in the next section.\\

 \subsection{Error-correction of low-weight noise channels by modular dissipation}
\label{sec:errorcorrec}

\begin{figure}[htbp]
		\centering
		\includegraphics[width=1.0\columnwidth]{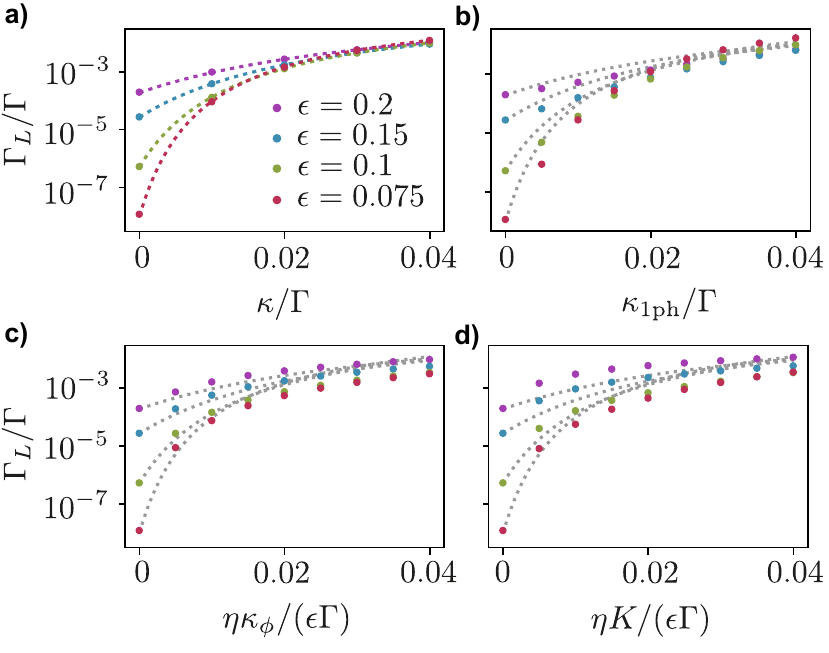}
		\caption{{\bf GKP qubit protection by modular dissipation}. The decay  rate $\Gamma_L$ of the Pauli operators $\Z$ and $\X$  is extracted from numerical simulations (dots) when varying the strength of some intrinsic noise channel relative to  the modular dissipation rate $\Gamma$. For all low-weight noise channels considered, errors appear to be exponentially suppressed in the weak noise  limit. {\bf a)} Quadrature noise modeled by two Lindblad operators $\sqrt{\kappa}\qop$ and $\sqrt{\kappa}\Pop$. Dashed lines are predictions by spectral analysis of the Lindblad superoperator (see \cref{sm__ssec__eigenvalues}). {\bf b)} Single-photon dissipation modeled by a Lindblad operator $\sqrt{\kappa_{1\mathrm{ph}}}\aop$.  {\bf c)} Pure dephasing modeled by a  Lindblad operator $\sqrt{\kappa_{\phi}}\aop^{\dag}\aop$. {\bf d)} Kerr Hamiltonian perturbation of the form $\frac{K}{2}(\aop^{\dag}\aop)^2$. For (c-d), note the rescaling of the x-axis by $ \eta/\epsilon=2\overline{n}$. For (b-d), dashed gray lines  reproduce the dashed colored lines in (a), un-rescaled, for comparison.   }
		\label{fig:gammavskappa}
	\end{figure}

We first analyze the simple case  of a Gaussian white noise channel---also known as quadrature noise--- entering the Lindblad dynamics as two spurious dissipators
$\mathcal{D}[\sqrt{\kappa}\qop]$ and $\mathcal{D}[\sqrt{\kappa}\Pop]$. We show that, in the limit of weak intrinsic dissipation $\kappa \ll \Gamma_c$, the decay rate of the generalized Pauli operators $\X$ and $\Z$ reads $\Gamma_L=\frac{4}{\pi}\mathcal{A}\epsilon \eta \Gamma e^{-4/( \mathcal{A}\tilde{\epsilon}\eta)}$, where $\tilde{\epsilon}=\epsilon+\frac{ \kappa}{2\mathcal{A}^2\epsilon \Gamma}$ (see \cref{sm__ssec__eigenvalues}).
The minimum flip rate is obtained for $\epsilon \simeq (\frac{\kappa}{2\mathcal{A}^2\Gamma})^{\frac{1}{2}}$ and  reads $\Gamma_L \simeq \frac{4\eta}{\pi}(\frac{\kappa \Gamma}{2})^{\frac{1}{2}} e^{-(\frac{8\Gamma}{\eta^2\kappa})^{\frac{1}{2}} }$.
This exponential scaling ensures that logical errors can be heavily suppressed for a modest ratio  $\Gamma/\kappa$, as illustrated by Fig.~\ref{fig:gammavskappa}a. There, we represent the decay rate of the generalized Pauli operators $\X$ and $\Z$ extracted  by spectral analysis of the Lindblad superoperator (dashed lines), in quantitative agreement with a full Lindblad master equation simulation (dots). The latter is  computationally  much more costly but proves necessary to investigate  more realistic noise models for which no simulation shortcut was found.
In particular, we verify numerically that errors entailed by single-photon dissipation
\footnote{%
	The relatively low mode frequency $\omega_a = 2\pi \times 150$~MHz proposed later in \cref{tableparam}
	can lead to non-negligible thermal population at typical cryogenic temperatures.
	In order to take this effect into account, one could consider a one-photon gain process
	on top of one-photon loss dissipation.
	In that case, two Lindblad operators should be included in the simulations:
	$\LL_{\mathrm{loss}} = \sqrt{\kappa (1+n_{\mathrm{th}})} \, \aop$
	and
	$\LL_{\mathrm{gain}} = \sqrt{\kappa n_{\mathrm{th}}} \, \aop^\dagger$,
	with $n_{\mathrm{th}} = \left( \exp(\frac{\hbar \omega_a}{k_B T}) -1 \right)^{-1}$
	the average thermal photon number, $T$ the temperature
	and $k_B$ the Boltzmann constant.
				This setting can be intuitively understood
	as interpolating between quadrature noise and pure photon loss.
	Indeed, at the level of dissipators, we have
	$\cD[\qop] + \cD[\Pop] = \cD[\aop] + \cD[\aop^\dag]$
	so that the dissipators associated to one-photon loss and gain satisfy:
	$\cD[\LL_\mathrm{loss}] + \cD[\LL_\mathrm{gain}] = \kappa_- \,
		\left( s ( \cD[\qop] + \cD[\Pop]) + (1-s) \cD[\aop]\right)$
	for $\kappa_- = \kappa (1+n_{\mathrm{th}})$ and $s = n_{\mathrm{th}}/ (1+n_{\mathrm{th}})$.
},
pure dephasing and a Kerr Hamiltonian perturbation all appear  to be exponentially suppressed when increasing the modular dissipation rate (see Fig.~\ref{fig:gammavskappa}b-d). The logical error rates induced by the two latter processes---entering the Lindblad equation via fourth order polynomials in $\qop$ and $\Pop$---are qualitatively captured by a  mean-field approximation which boils down to quadrature noise scaled up by the grid states mean photon number $ \overline{n}=\eta/(2\epsilon)$ (dashed gray lines  in Fig.~\ref{fig:gammavskappa}c-d).  These numerical considerations support the intuition that modular dissipation  can suppress errors induced by  arbitrary finite-weight noise channels, albeit with degraded performances when considering higher-weight processes. In the limit of infinite-weight noise processes, \textit{i.e.} modular noise channels, errors are not corrected.

\section{Modular Hamiltonian engineering in a Josephson circuit}
\label{sec:hamiltonian}

\begin{figure}[htbp]
		\centering
		\includegraphics[width=1.0\columnwidth]{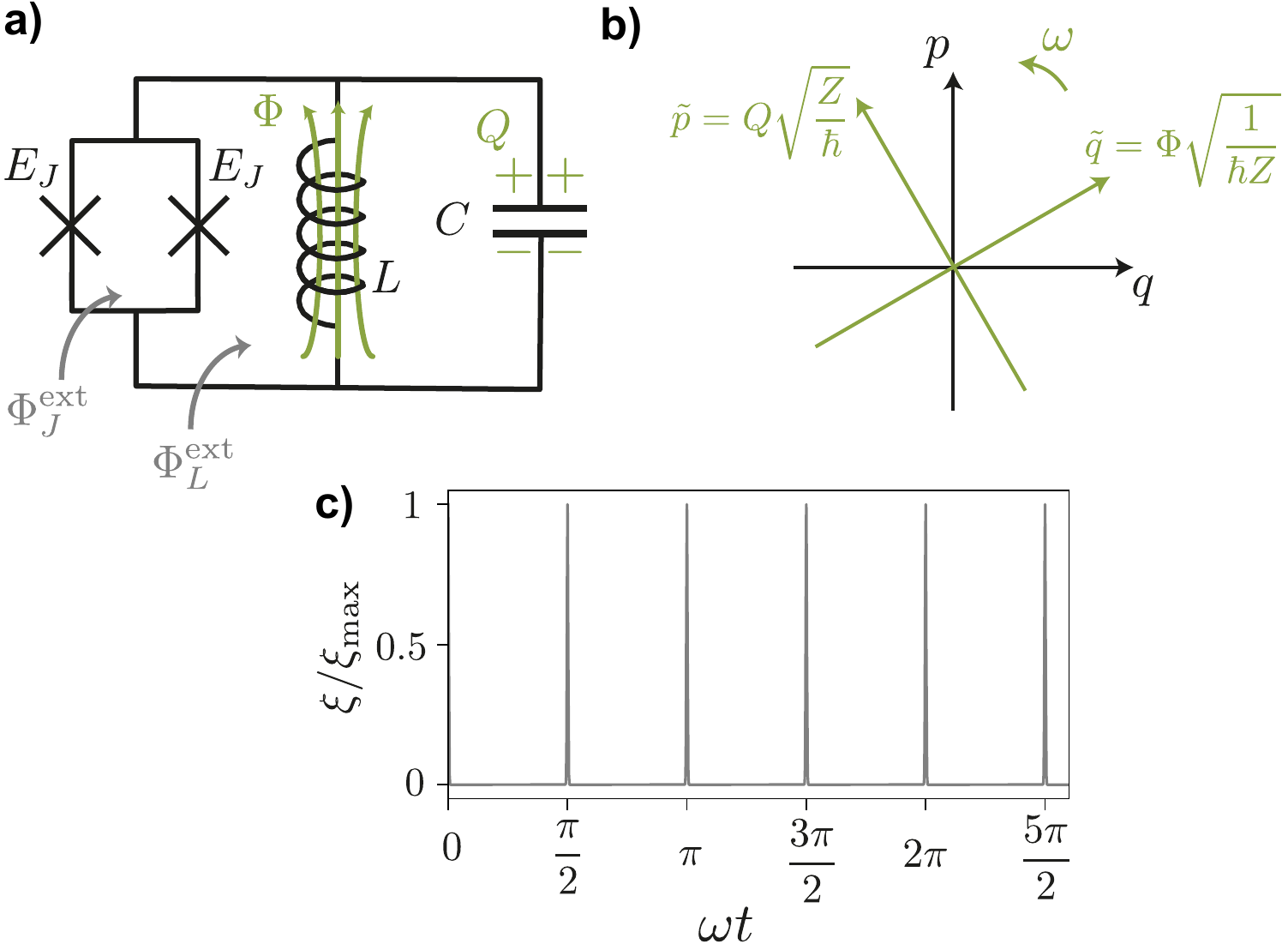}
		\caption{{\bf Engineering of the GKP Hamiltonian.} {\bf a) } A Josephson ring is placed in parallel  with an  $LC$ resonator of large impedance $Z=\sqrt{\frac{L}{C}}=2R_Q$. Magnetic fluxes  threading the circuit loops  $\Phi_{J}^{\mathrm{ext}}(\xi)$ and $\Phi_{L}^{\mathrm{ext}}(\xi)$ are both functions of a control signal $\xi$. They are used to   tune the circuit effective Josephson energy  $2E_J\xi$ and the phase of Josephson tunneling.  {\bf b) } The laboratory frame  flux coordinate $\Phi$ periodically aligns with the coordinates $q$ and $p$ of the frame rotating at $\omega=1/\sqrt{LC}$. {\bf c) } The  signal $\xi(t)$ controlling the circuit effective Josephson energy consists in a train of short bias pulses, so that the  Josephson energy  takes non-zero values  at these instants only. In the RWA, the effective Hamiltonian contains modular functions  of  $\qop$ \textit{and} $\Pop$, with spatial frequencies $2\sqrt{\pi}$ for the chosen circuit impedance, both stemming from the Josephson modular flux operator.}
		\label{fig:simpleCircuit}
	\end{figure}

For the sake of pedagogy, we now describe a control method to engineer a Hamiltonian involving the two modular stabilizers of the infinite-energy GKP code in a simple superconducting circuit. The method is similar to that introduced in \cite{kolesnikow2023} and the key ideas of the protocol for modular dissipation engineering described in Sec.~\ref{sec:dissipationengineering} are already present in this toy example. The goal here is to synthesize the GKP Hamiltonian
\begin{equation}
\HH_{\mathrm{GKP}}=-E \big(\mathrm{cos}(\eta \qop)+ \mathrm{cos}(\eta \Pop) \big),
\label{eq:HGKP}
\end{equation}
in the rotating frame of a superconducting resonator. This Hamiltonian has a degenerate ground state corresponding to the two infinite-energy GKP states $|\pm Z_{\infty}\rangle$.\\

We consider the circuit pictured in Fig.~\ref{fig:simpleCircuit}a. The inductor and capacitor form a quantum oscillator whose conjugate variables are the flux threading the inductor $\Phi$ and the charge on the capacitor $Q$. The corresponding operators can be reduced as $\tilde{\qop}=\frac{1}{\sqrt{\hbar Z}} \phiop$ and $\tilde{\Pop}=\sqrt{ \frac{Z}{\hbar }} \Qop$, where $Z=\sqrt{L/C}$ is the circuit impedance, so  as to verify $[\tilde{\qop}, \tilde{\Pop}]=i$ and to display equal fluctuations in the vacuum state. The $LC$ oscillator is placed in parallel  with a ring made of two Josephson junctions with equal energy $E_J$. We apply two magnetic fluxes  $\Phi_J^{\mathrm{ext}}=\varphi_0(\pi-2\mathrm{Arcsin}(\xi(t)))$ and $\Phi_L^{\mathrm{ext}}=-\Phi_J^{\mathrm{ext}}/2$, where $\varphi_0=\hbar/(2e)$ is the reduced flux quantum and $\xi$ is an AC bias signal, respectively through the Josephson ring loop and the loop formed with the inductor. In presence of these flux biases, the Josephson ring behaves as a single junction with time-varying energy and null tunneling phase~\cite{lescanne2020exponential}, acting on the $LC$ resonator via the Hamiltonian
\begin{equation}
\begin{aligned}
    \HH_J(t)
       &= -2  E_J \xi(t) \mathrm{cos}(\phiop/\varphi_0).
\end{aligned}
\end{equation}
Designing the circuit to have an impedance  $Z=2R_Q$, where $R_Q=\frac{h}{4 e^2}\simeq 6.5~\mathrm{k}\Omega$ is the resistance quantum, the  circuit Hamiltonian in reduced coordinates reads
\begin{equation}
    \HH_0(t)= \frac{\hbar \omega}{2}(\tilde{\qop}^2 + \tilde{\Pop}^2)-2  E_J \xi(t) \mathrm{cos}(\eta \tilde{\qop}),
\end{equation}
where $\omega=1/\sqrt{LC}$. We now place ourselves in the interaction picture to cancel out the dynamics of the linear part of the circuit. In the $(q,p)$ frame rotating at $\omega$, the  sole remaining dynamics is governed by the Josephson term, a modular function of the now rotating quadrature operator  $\tilde{\qop}(t)=\mathrm{cos}(\omega t) \qop +\mathrm{sin}(\omega t) \Pop$. This operator aligns with $\qop$ or $\Pop$ every quarter period of the oscillator (see Fig.~\ref{fig:simpleCircuit}b). The idea is to bias the Josephson ring with a train of short flux pulses in order to activate Josephson tunneling at these precise instants only (see Fig.~\ref{fig:simpleCircuit}c). Letting $\xi(t) \simeq  \xi_1  \Sha_{\frac{\pi}{2\omega}}(t)$ where  $\xi_1$ is the integrated amplitude of each pulse  and $\Sha_T$ denotes a Dirac comb of period $T$, in the Rotating Wave Approximation (RWA), we  obtain the effective Hamiltonian
\begin{equation}
\begin{split}
\HH_{\mathrm{RWA}}&=-2 E_J \overline{\xi(t) \mathrm{cos}(\eta \tilde{\qop}(t) } )\\
&=-E \big(\mathrm{cos}(\eta \qop)+ \mathrm{cos}(\eta \Pop) \big)\\
&=\HH_{\mathrm{GKP}}
\end{split}
\end{equation}
 with $E=\frac{2 E_J\omega}{\pi} \xi_1$. It is straightforward to combine this doubly modular Hamiltonian with a small  quadratic potential $\frac{\hbar \delta}{2}(\qop^2 + \Pop^2)$ with $\hbar\delta \ll E$ in order to get
finite-energy GKP states as quasi-degenerate ground states~\cite{rymarz2021hardware}. Indeed, such a weakly confining potential is simply obtained by increasing the duration between the  pulses of the  bias train to $\frac{\pi}{2(\omega-\delta)}$.\\

Here, we stress that  we described this method as an example of modular dynamics engineering only. It does not provide  a protected qubit \textit{per se} as would a circuit implementing the same Hamiltonian in the laboratory frame~\cite{gottesman2001encoding,rymarz2021hardware}. Indeed, the GKP code states are not stable upon loss of a photon. For a system directly governed by the static Hamiltonian $\HH_{\mathrm{GKP}}$ and prepared in the ground manifold, photon emission into a cold bath would violate energy conservation and photon loss thus does not occur. This argument does not hold when $\HH_{\mathrm{GKP}}$ is engineered in the rotating frame from a time-dependent Hamiltonian. In that case, photon emission into the environment can occur even at zero temperature, pulling the oscillator state out of the ground manifold of $\HH_{\mathrm{GKP}}$. Stabilization of the GKP code manifold could still be achieved by coupling the circuit to a \textit{colored} bath engineered to enforce energy relaxation in the rotating frame~\cite{putterman2022stabilizing}\footnote{The idea of colored bath engineering is to induce relaxation between two energy levels $i$ and $j$ of $\HH_{\mathrm{GKP}}$ verifying $|\langle j |\aop| i\rangle |>0$ by coupling parametrically   the target mode  to an ancillary dissipative mode via an interaction of the form $g_{ij} e^{i(\omega_{ij}+\omega_a-\omega_b)t} \aop \bop^{\dagger}+h.c.$. The coupling strength should be chosen such that  $g_{ij}\ll |\omega_{ij}-\omega_{ik}|$ for any other level $k$ in order not to induce spurious transitions to $|k\rangle$. Other types of couplings, in particular of the form $g_{ij} e^{i(\omega_{ij}-\omega_a-\omega_b)t} \aop^{\dagger} \bop^{\dagger}+h.c.$ may be needed to induce transitions for which $\aop$ has a negligible matrix element \cite{sivak2022real}}.

\section{Modular dissipation engineering in a Josephson circuit}
\label{sec:dissipationengineering}
\subsection{Modular dissipators from modular interactions}
Armed with the previous example, we now turn to engineering the modular dissipative dynamics described in Sec.~\ref{sec:modular dissip}. We here stress that the dissipation entailed by the four modular Lindblad operators~\eqref{eq:lindissip} is sufficient to stabilize and perform error correction of the GKP qubit. No further Hamiltonian dynamics is needed, and in particular the Hamiltonian~\eqref{eq:HGKP} does not appear in the system master equation. In order to engineer the target dissipative dynamics, we first note  that the Lindblad operators~\eqref{eq:lindissip} can be substituted with the following linear combinations
 \begin{equation}
 \begin{split}
 \LL_{q,s}&=(\LL_0+\LL_2)/\sqrt{2}\\
  \LL_{q,d}&=(\LL_0-\LL_2)/(\sqrt{2}i)\\
 \LL_{p,s}&=(\LL_1+\LL_3)/\sqrt{2}\\
  \LL_{p,d}&=(\LL_1-\LL_3)/(\sqrt{2}i)
\label{eq:lindblad2}
\end{split}
 \end{equation}
Second, following a standard procedure (see \cref{ssec:general_strategy}), each Lindblad operator $\LL_{r,l}$ with $r=q$ or $p$, $l=s$ or $d$  is obtained by coupling the target mode $a$ to an ancillary mode $b$, damped at rate $\kappa_b$, via an interaction Hamiltonian
\begin{equation}
    \HH^{\mathrm{int}}_{r,l}=\hbar g \LL_{r,l} \bdag + h.c.
    \label{eq:hint0}
\end{equation}
Indeed, adiabatically eliminating the mode $b$ in the limit $g \ll \kappa_b$, the two-mode dynamics reduces to a single-mode dissipative dynamics with the desired Lindblad operator $\LL_{r,l}$, at a rate $\Gamma=4g^2/\kappa_b$. Third, we define rotated quadrature operators of the target and ancillary modes  $\qop_a^{\Theta_a}=e^{ i \Theta_a\aop^{\dag}\aop} ~ \qop_a ~ e^{- i\Theta_a \aop^{\dag}\aop} $ and $\qop_b^{\Theta_b}=e^ { i  \Theta_b   \bop^{\dag}\bop} ~ \qop_b ~ e^{ -i \Theta_b \bop^{\dag}\bop}$, and we remark that the  Hamiltonian \eqref{eq:hint0} is approximated at first order in  $\epsilon$~\footnote{An arbitrarily accurate approximation is obtained by considering quadrature operators rotated by arbitrarily small  angles $\theta_a=\frac{ \epsilon \epsilon' }{2\eta}$ and $\theta_b=\frac{\pi}{2}-\frac{\eta\epsilon \epsilon'}{4}$  with $\epsilon' \rightarrow 0$, and scaling the terms associated to $j=\pm1$ in the sum by $2/\epsilon'$
\label{footnote:finite_differences}
} by
\begin{equation}
\begin{split}
       \HH^{\mathrm{int}}_{r,l} \simeq  \quad 2 \hbar g ~ \Big(& ~ (\delta_l-1)~\qop^0_b~+ \\
       &\mathcal{A}\sum_{j =0, +1, -1} ~ \mathrm{cos}(\eta \qop^{\phi_r+j\theta_a}_a - \delta_l \frac{\pi}{2}   )~\qop^{j\theta_b}_b  ~\Big)
\end{split}
\label{eq:hint}
\end{equation}
with $\phi_r=0$ for $r=q$, $\phi_r=\pi/2$ for $r=p$, $\theta_a=\frac{\epsilon }{2\eta}$, $\theta_b=\frac{\pi}{2} - \frac{\eta\epsilon }{4}$,  $\delta_l=0$ for $l=s$ and $\delta_l=1$ for $l=d$. The modular interactions  in this Hamiltonian (second line) all have the same form and can be activated in the rotating frame of a two-mode Josephson circuit as described in the next section. The linear term (first line) is trivially implemented by driving the ancillary mode resonantly. \\

Note that activating simultaneously four Lindblad operators  necessitates to activate four  interaction Hamiltonians with four distinct ancillary modes, which would all appear in series with the target mode in Fig.~\ref{fig:dissipCircuit}a. A  hardware-efficient alternative consists in activating them sequentially, leveraging a single ancillary mode as pictured in Fig.~\ref{fig:dissipCircuit}a, and switching from one operator to the next at a rate slower than $\kappa_b$---giving the ancillary mode sufficient time to reach its steady state and justifying its adiabatic elimination---but faster than $\Gamma$---accurately reproducing the target four-dissipator dynamics by Trotter decomposition. This strategy  drastically reduces the experimental  complexity, at the cost of a fourfold reduction of the modular dissipation rate $\Gamma$. With these considerations in mind, we now focus on the activation of a single Lindblad operator $\LL_{r,l}$ and assume that the full target dynamics  is easily derived thereof. \\

\subsection{Activating modular interactions in the rotating frame}
\label{sec:modularengineering}

\begin{figure}[htbp]
		\centering
		\includegraphics[width=1.0\columnwidth]{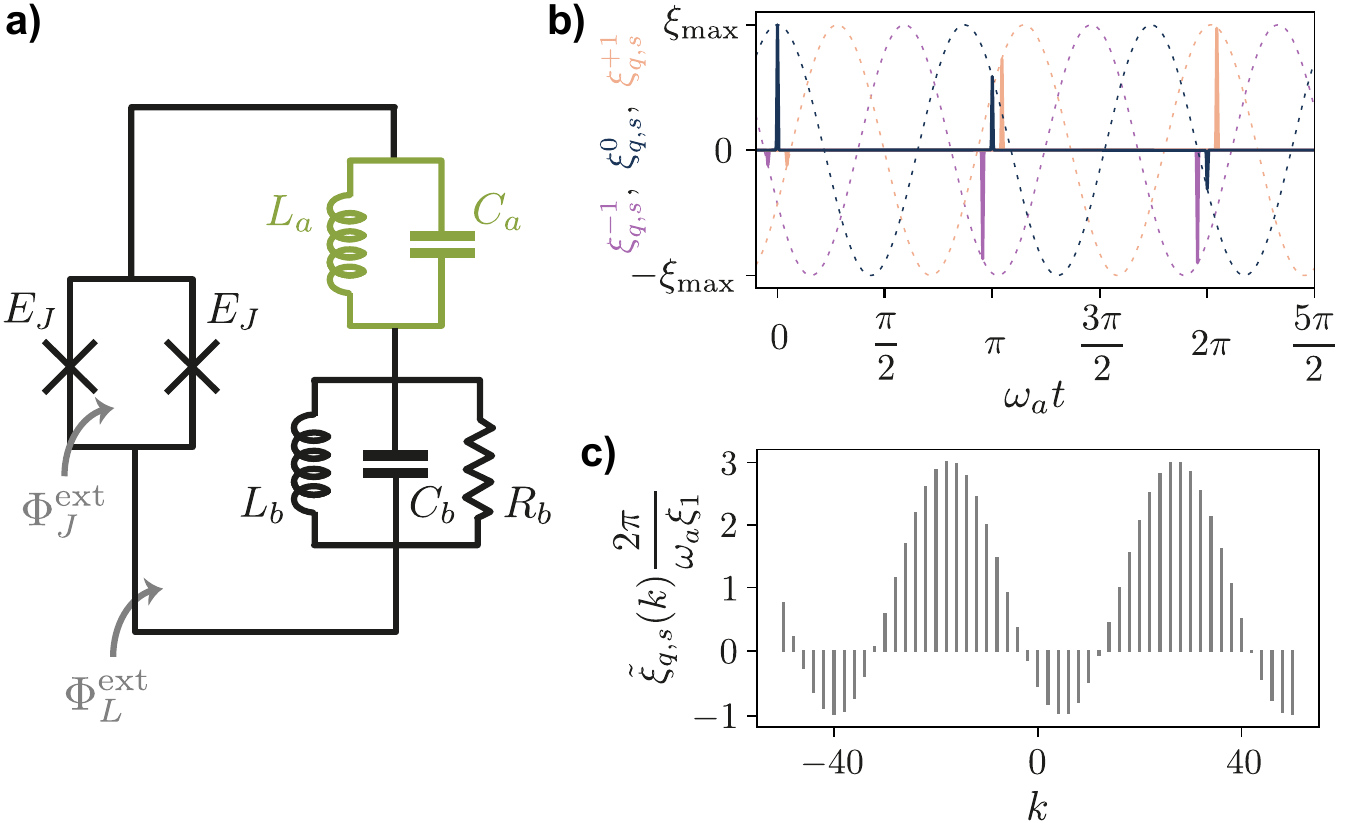}
		\caption{{\bf Engineering modular interactions.} {\bf a) } A Josephson ring is placed in parallel  with a high-impedance target resonator (green) and a low-impedance, dissipative, ancillary resonator (black). The  Josephson tunneling amplitude  $2E_J \xi(t)$ and  phase are adjusted with the control fluxes $\Phi_{J,L}^{\mathrm{ext}}$ biasing the circuit.  {\bf b) } Each modular Lindblad operator $\LL_{r,l}$ in~\eqref{eq:lindblad2} is activated with an  AC bias signal  consisting of  three pulse trains  $\xi_{r,l}=\xi_{r,l}^{-1}+\xi_{r,l}^0+\xi_{r,l}^{+1}$ (trains respectively colored in purple, black, and orange), pulses within each train  being separated by half a period of the target resonator. Each train  modulates a carrier at $\omega_{b}$, the three carriers being phase-shifted by $\sim \pm \frac{\pi}{2}$ from one another (dashed lines with same colors as the pulse trains). {\bf c) }  In frequency domain, the bias signal $\mathcal{F}[\xi_{r,l}](\omega)=\sum_{k\in \mathbb{Z}}\tilde{\xi}_{r,l}(k)(\delta(\omega-\omega_b-k\omega_a)+\delta(\omega+\omega_b+k\omega_a))$ is a real-valued frequency comb  centered at $\pm \omega_b$ and whose  amplitude $\tilde{\xi}_{r,l}(k)$ oscillates with a period $\frac{4\pi \eta}{\epsilon}$. The signal represented in \textbf{b-c} corresponds to the activation of $\LL_{q,s}$ and contains only even harmonics $k\in2\mathbb{Z}$.  For readability, we set $\omega_{b} =2.3~ \omega_a$ and $\epsilon=1$  in \textbf{b}, which increases the phase-shift between the carriers at $\omega_b$ beyond $\pi/2$  (respectively $\omega_{b}/ \omega_a \rightarrow \infty$ in \textbf{c} so that the comb does not overlap with its mirror image centered at $-\omega_b$), which is not the regime in Table~\ref{table}.		 }
		\label{fig:dissipCircuit}
	\end{figure}

 The method and circuit to activate  the modular  interactions in the Hamiltonian~\eqref{eq:hint}  is analogous to the GKP Hamiltonian engineering technique described in Sec.~\ref{sec:hamiltonian}. Here, we consider the  multimode circuit pictured in Fig.~\ref{fig:dissipCircuit}a. The Josephson ring  is shunted by the target resonator with impedance  $Z_a=2 R_Q$ placed in series with a low-impedance  dissipative ancillary mode $b$ ($Z_b\ll R_Q$, $\kappa_b \sim \omega_b Z_b/R_b$). Note that this circuit should not necessarily represent a physical device: it suffices  to represent the Foster decomposition~\cite{foster1924reactance,nigg2012black,smith2016quantization} of a linear environment connected to the two ports of the Josephson ring. \\

Compared to Sec.~\ref{sec:hamiltonian}, the DC flux bias point is modified following
\begin{equation}
\begin{split}
	\Phi_J^{\mathrm{ext}}&=\varphi_0\big(\pi+2\mathrm{Arcsin}(\xi(t))\big)\\
\Phi_L^{\mathrm{ext}}&=-\frac{\Phi_J^{\mathrm{ext}}}{2}+\varphi_0\frac{\pi}{4}
\label{eq:fluxbias}
\end{split}
\end{equation}
 in order to give a non-trivial phase to the Josephson tunneling~\cite{lescanne2020exponential}. The circuit Hamiltonian then reads
\begin{equation}
    \HH_{0}(t)= \hbar \omega_a \adag \aop + \hbar \omega_b \bdag \bop +
        2 E_J \xi(t) ~\mathrm{cos}\big(\frac{\phiop}{\varphi_0} - \frac{\pi}{4}\big)
    \label{eq:ho}
\end{equation}
 where the generalized phase operator across the series of resonators reads $\phiop=\varphi_0 (\eta_a \tilde{\qop}_a+\eta_b \tilde{\qop}_b)$, and the vacuum phase fluctuations of each mode across the Josephson ring are given by $\eta_a=\sqrt{2\pi Z_a/R_Q}=2\sqrt{\pi}$ and  $\eta_b=\sqrt{2\pi Z_b/R_Q}\ll 1$. Importantly, these values do not need to be fine-tuned in circuit fabrication as one can adapt the system controls to accommodate a value of $\eta_a$ exceeding $2\sqrt{\pi}$ (see Sec.~\ref{sec:noise} and \cref{sec:disorder_miscab}).   \\

Placing ourselves in the rotating frame of both $a$ and $b$, the Hamiltonian becomes
\begin{equation}
        \HH(t)=2 E_J \xi(t) ~\mathrm{cos}\big(\eta_a \tilde{\qop}_a(t) + \eta_b \tilde{\qop}_b(t) - \pi/4 \big)
\label{eq:H2}
\end{equation}
where the  quadrature operators $\tilde{\qop}_a(t)$ and $\tilde{\qop}_b(t)$ respectively rotate at $\omega_a$ and $\omega_b$ in phase-space. \\

Reminding the reader that $r=q~\text{or}~p$ and $l=s~\text{or}~d$ label the   Hamiltonian~\eqref{eq:hint} employed to engineer one of the Lindblad dissipators~\eqref{eq:lindblad2}, we now consider the AC bias signal
\begin{equation}
\label{eq:ideal_controls}
\begin{split}
&\xi_{r,l}(t) = \sum_{j =0, +1, -1} \xi_{r,l}^j(t)\\
&\quad \quad ~ =\sum_{j =0, +1, -1} \xi_1 \mathrm{cos}(\omega_b t ~-~j\theta_{b}) ~\times \\
&\Big(\Sha_{\frac{2\pi}{\omega_a}}(t-\frac{j\theta_a+\phi_r}{\omega_a}) ~+~ (-1)^{\delta_l} \Sha_{\frac{2\pi}{\omega_a}}(t-\frac{j\theta_a+\phi_r+\pi}{\omega_a} ) \Big)
 \end{split}
 \end{equation}
consisting of three  trains of  Dirac pulses---pulse integrated amplitude $\xi_1$---modulating carriers at frequency $\omega_b$, the pulses within each train being separated by half a period of the target resonator and having either constant  or alternating signs.  Each train activates one of the three modular interactions in the target Hamiltonian \eqref{eq:hint} with the same label $j$ and the same definition for $\theta_a$, $\theta_b$, $\phi_r$ and $\delta_l$. Indeed, a pulse train with phase $\Theta_a$ allows Josephson tunneling when the  operator $\tilde{\qop}_a(t)$ aligns or anti-aligns with the rotated quadrature  $\qop_a^{\Theta_a}$. Together with a carrier with phase $\Theta_b$, it selects out, in the RWA, terms of the form $\mathrm{cos}(\pm \eta_a \qop^{\Theta_a}_a) \qop_b^{\Theta_b}$ and $\mathrm{sin}(\pm \eta_a \qop^{\Theta_a}_a) \qop_b^{\Theta_b}$~\footnote{Here, we have assumed that $\omega_a$ and $\omega_b$ are not commensurable and neglected terms in $(\bdag \bop)^k \rop_b$ with $k>0$, whose only impact is to renormalize the modular interaction strength  $g\rightarrow e^{-\eta_b^2/4}g$ as detailed in \cref{sm:sec__rwa}.}  
Finally, choosing  pulses with constant  or alternating  signs ensures that only cosine \textit{or} sine operators survive the RWA---depending on which Lindblad operator is targeted.  In Fig.~\ref{fig:dissipCircuit}b, we represent the  bias signal  when activating $\LL_{q,s}$. In  frequency domain, it is a frequency comb  centered at $\pm \omega_b$ (see Fig.~\ref{fig:dissipCircuit}c, mirror image around $-\omega_b$ not shown) and whose amplitude oscillates with a period $\frac{4\pi\eta_a}{\epsilon}\omega_a$. The signals activating other Lindblad operators are obtained by alternating the pulses sign in time domain and/or alternating the harmonics sign in frequency domain.\\

Overall,  the target Hamiltonian \eqref{eq:hint} is activated at a rate
$g=E_J \eta_b \omega_a \xi_1 /(2\sqrt2 \pi \hbar\mathcal{A})$.
Note that,
to engineer modular dissipation operators from this effective Hamiltonian, we performed an adiabatic elimination of the ancillary mode---requiring $g\ll \kappa_b$.
This adiabatic elimination is valid only if it takes places on a much slower
timescale than the RWA producing the effective Hamiltonian in the first place---requiring in turn $\kappa_b \ll \omega_a$ (see \cref{ssec:approx_formulae}).
Moreover, we choose $\omega_a \ll \omega_b$ to avoid frequency collisions that would enable high-order processes involving multiple photons of the ancilla in the RWA.  Given that protection of the logical qubit requires the modular dissipation rate $\Gamma \sim \frac{g^2}{\kappa_b}$ to be larger than the target resonator photon loss rate $\kappa_a$, the system parameters should respect
 \begin{equation}
 \kappa_a  \ll g^2/\kappa_b \ll \kappa_b \ll \omega_a \ll \omega_b.
 \label{eq:hierarchy}
 \end{equation}
 This regime is attainable in a state-of-the-art circuit (see Tab.~\ref{table}) comprising a high-impedance mode resonating in the 100~MHz range. This unusually low resonance frequency is needed to respect the above hierarchy, and to ensure that flux bias pulses are sufficiently short with respect to the target oscillator period, as detailed in the next section.  \\

\section{Implementation with state-of-the-art circuits and control electronics}

\label{sec:noise}
The goal of this section is to propose realistic  experimental parameters for the stabilization of GKP qubits and to estimate the impact of various experimental imperfections.
We first remind the reader that the impact of intrinsic, low-weight noise processes affecting the target resonator was analyzed in \cref{sec:modular dissip} and shown to be robustly suppressed by the modular dissipation.
Here, we consider the noise sources induced by the dissipation engineering itself
in realistic experimental conditions.
In \cref{sec:hot_ancilla}, we explore the propagation of ancilla noise
by extending our analysis of modular dissipation engineering to the case of an ancilla mode
suffering from thermal excitation and dephasing.
In \cref{sec:harmonics}, we focus on errors induced by the finite-bandwidth of the bias signal,
while in \cref{sec:jjasym}, we focus on the impact of circuit fabrication disorder. 
Mitigating the former prompts the use of a target mode resonating at low frequency $\omega_a$
to embed GKP qubits,
while our mitigation strategy for the latter relies on a RWA only valid
if $\hbar\omega_a$ dominates over specific energy scales of the circuit.
Therefore, the circuit parameters we propose in \cref{tableparam} results from a trade-off
and entail spurious logical errors.
Nevertheless, we estimate that the decay rate of the GKP qubit Pauli operators could still be
two orders of magnitude lower than the intrinsic dissipation rates of the circuit
for these realistic parameters. Finally, in \cref{sec:1overf} and ~\cref{sec:QP}, we describe qualitatively the impact of low-frequency drifts in the  bias signal and of quasi-particle poisoning. We also lay out possible mitigation strategies that will be investigated in a future work.

\input{mt_ancilla_noise.tex}

\subsection{Limited bandwidth and accuracy of the flux bias signal}
\label{sec:harmonics}

\begin{figure}[htbp]
		\centering
		\includegraphics[width=0.75\columnwidth]{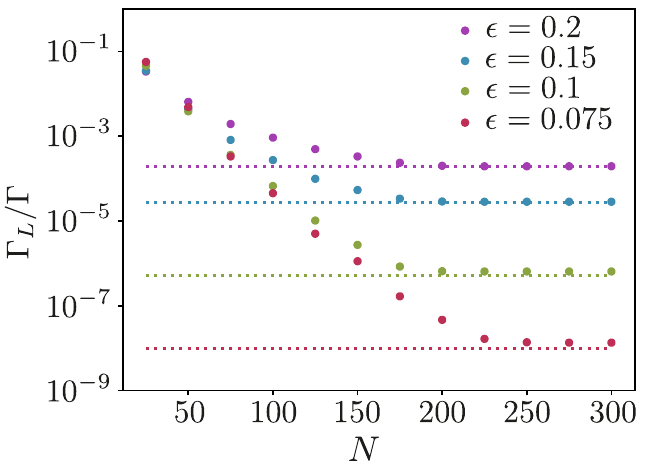}
		\caption{{\bf Truncated frequency comb.  }
		Decay rate of the generalized Pauli operators $\X$ and $\Z$ under modular dissipation engineered with a frequency comb containing a finite number   of harmonics $N$ (dots), in absence of intrinsic noise of the target resonator.  A finite bandwidth signal yields spurious errors at a rate decreasing exponentially with $N$.
		The mismatch between
		the decay rate found for $N\rightarrow \infty$
		and
		the rate found for the ideal dissipators \eqref{eq:lindissip} (dashed lines)
		is due to the additional approximation introduced when going
		from Eq.~\eqref{eq:hint0} to Eq.~\eqref{eq:hint}
		(see~\cite{Note3} and \cref{sm:sec__rwa} for details):
		even with an idealized infinite frequency comb,
		the engineered dissipators do not exactly coincide with
		the ideal dissipators in~\eqref{eq:lindissip}.
		The computation of the dissipation operators activated by a truncated bias signal
		$\tilde \xi_N$ relies on a RWA and is presented in \cref{smsec:numerics_rwa1}.}
		\label{fig:comb}
\end{figure}

A central hypothesis  to the dissipation engineering technique detailed in Sec.~\ref{sec:modular dissip} is that the width of the flux pulses that bias the circuit is negligible with respect to the target oscillator period. In  frequency domain, this figure of merit directly relates to the number of harmonics $N$ in the frequency comb forming the bias signal $\xi$ (see Fig.~\ref{fig:dissipCircuit}c, we drop the subscript ${r,l}$ for simplicity). This number should be quantitatively optimized: on the one hand, it should not be too small for the aforementioned hypothesis to hold, but  picking an unnecessarily large $N$ would place prohibitive constraints on the circuit design---for a fixed control signal bandwidth, one can only increase $N$ by decreasing the target mode resonance frequency---and limit the modular dissipation rate for a given maximum value of the bias signal $\xi_{\mathrm{max}}$~\footnote{The modular interaction strength $g$ is proportional to the bias pulses integrated amplitude $\xi_1$, which decreases with $N$ following $\xi_1=2\pi\xi_{\mathrm{max}}/((2N+1)\omega_a)$}. \\

To this end, we perform numerical simulations, in the RWA (see \cref{smsec:numerics_rwa1}), considering Lindblad operators activated by a bias signal  $\tilde{\xi}_{N}(k)$ obtained by truncating the Fourier series  $\tilde{\xi}(k)$ (setting $\tilde{\xi}(k)=0~\text{for}~|k|>N$, see Fig.~\ref{fig:dissipCircuit}c for a representation of $\tilde{\xi}(k)$). The evolution of the target oscillator state is computed for the corresponding imperfect modular dissipation in absence of any other decoherence channel. The decay rate of the generalized Pauli operators $\X$ and $\Z$ is extracted for each value of $N$, and represented in Fig.~\ref{fig:comb}. Truncation of the bias comb leads to spurious logical flips at a rate independent of $\epsilon$ and exponentially decreasing with $N$. In the long term, this scaling is encouraging  as one does not need to increase the  control signal bandwidth indefinitely to robustly protect the encoded information. In the short term, combs containing $N\sim100$ harmonics are needed to suppress the logical error rate significantly beyond the break-even point (see Tab.~\ref{tableparam}). Limiting microwave drives to the 0-20~GHz range, which corresponds  to the bandwidth of standard  laboratory equipment and is below the typical plasma frequency of Josephson junctions~\footnote{Josephson junctions feature an intrinsic capacitance omitted in Fig.~\ref{fig:dissipCircuit}a,  which, combined with their kinetic inductance, form an oscillator typically resonating around 10---50~GHz for standard microfabrication techniques.   }, this places the target mode resonance frequency in the sub-GHz range (see Tab.~\ref{tableparam}). \\

Delivering a precise, wideband, microwave signal  to a superconducting circuit cooled down in the quantum regime  is a major experimental challenge. If this signal is generated at room temperature, one needs to account for \textit{a priori} unknown dispersion of the feedlines. Therefore, the complex amplitudes of $2N+1$ phase-locked, monochromatic microwave signals  need to be individually calibrated (see \cref{sm:control_miscab} for quantitative estimates of the impact of miscalibration). Recent advances in  digital synthesis of microwave signals allows for the automation of these calibrations. An alternative strategy consists in generating the frequency comb  directly on-chip with a dedicated Josephson circuit~\cite{solinas2015radiation,solinas2015josephson} in order to deliver a precise, wideband comb with no need for complex calibrations.

\subsection{Fabrication constraints and disorder}
\label{sec:jjasym}

Inaccuracy on the energy of  Josephson junctions is the main source of disorder in superconducting circuits, with a typical mismatch of the order of a few percents from the targeted value to the one obtained in fabrication. In the circuit  depicted in Fig.~\ref{fig:dissipCircuit}a, this leads to uncertainty on the value of the \textit{superinductance} $L_a$, typically implemented by a chain of Josephson junctions~\cite{masluk2012microwave}, and to a small energy mismatch between the two  junctions forming the ring. Fortunately, these parameters do not need to be fine-tuned in our approach.\\

Indeed, an  inductance $L_a$ differing from its nominal value only results in a modified  target mode impedance $Z_a$, and  therefore in modified phase fluctuations  across the Josephson ring $\eta_a=\sqrt{2\pi Z_a /R_Q}$. Here we remind the reader that the target value $\eta_a=2\sqrt{\pi}$ was chosen to match the length of the square GKP  lattice unit cell. However, as detailed in Sec.~\ref{sec:gates}, there exists a continuous family of GKP codes obtained by symplectic transformation of the square code lattice. The diamond-shaped unit cells of these codes  still have an area of $4\pi$, but longer edges. As long as $\eta_a>2\sqrt{\pi}$, one simply adjusts the timing of flux bias pulses to stabilize such a non-square code. We verify in simulation that the accuracy with which this  adjustment needs to be performed is well within  reach of current experimental setups (see \cref{ssec:excessive_impedance}).\\

We now  consider the effect of a small asymmetry of the circuit Josephson ring. We remind the reader that in our dissipation engineering scheme, the effective Josephson energy of the ring is cancelled by threading the ring with half a quantum of magnetic flux---corresponding to the DC contribution in $\Phi_J^{\mathrm{ext}}$---except at precise instants when it is activated with sharp flux pulses---corresponding to the AC contribution in $\Phi_J^{\mathrm{ext}}$. Mismatch between the two junction energies lead to imperfect cancellation in-between pulses, potentially generating  shifts  of the target oscillator state by $\eta_a$ along a random axis in phase-space (see Fig.~\ref{fig:schematic}). As detailed in \cref{sec:disorder_miscab}, this adverse effect can be mitigated by slightly adjusting the circuit DC  bias point so that the imperfectly cancelled Josephson
Hamiltonian becomes non-resonant and drops out in the RWA. This RWA is only valid if the energy mismatch between junctions is much smaller than the target mode frequency $\omega_a$, placing a new constraint on the circuit parameters. In Tab.~\ref{tableparam}, we choose a Josephson energy as low as $E_J=h\times 500~$MHz---which we still consider experimentally realistic while keeping the junctions plasma frequency above 20~GHz (see \cref{sec:disorder_miscab})---such that a $2\%$ mismatch should be tolerable. We leave quantitative analysis  of the robustness of this strategy for future work and note that it may be combined with the method sketched in the next section for a more robust  suppression of the impact of  imperfectly cancelled  Josephson energy.

\subsection{$1/f$ magnetic flux noise}
\label{sec:1overf}

While its microscopic origin is still debated, low-frequency magnetic flux noise (referred to as \textit{1/f} noise) is ubiquitous in superconducting circuits~\cite{paladino20141}. In practice, such noise will induce slow drifts in the DC bias point of our proposed circuit, which cannot be detected and compensated on short ($\sim 1~$ms) timescales. A small offset to the magnetic flux $\Phi_L^{\mathrm{ext}}$ threading the rightmost loop of the circuit (see Fig.~\ref{fig:dissipCircuit} and Eq.~\ref{eq:fluxbias}) is not expected to affect significantly the performances of our protocol. Indeed, it
only impacts the phase of the Josephson term in \eqref{eq:H2}, slightly unbalancing the rates of the engineered modular dissipators \eqref{eq:lindblad2}. On the other hand, an offset to the magnetic flux threading the Josephson ring $\Phi_J^{\mathrm{ext}}$ results in an imperfectly cancelled  Josephson energy in between fast bias pulses, similar to that induced by a mismatch on the energy of the two junctions. \\

In detail, a small offset $2 \varphi_0 e_J$   in the magnetic flux $\Phi_J^{\mathrm{ext}}$ threading the Josephson ring (with $e_J\ll1$) yields a spurious term 
\begin{equation}
2E_J e_J (1-\xi^2(t))^{1/2}\cos(\eta_a \tilde{\qop}_a(t) + \eta_b \tilde{\qop}_b(t) - \pi/4 +e_J \big)
\label{eq:spuriousHamil}
\end{equation}
in the circuit Hamiltonian \eqref{eq:H2} (expressed in the rotating frame, with $\tilde{\qop}_a(t)$ and $\tilde{\qop}_b(t)$ respectively rotating at $\omega_a$ and $\omega_b$ in phase-space). This time-dependent Hamiltonian may generate long shifts of the target oscillator along a random axis, triggering logical errors. Unfortunately,  here, adapting the circuit bias to make this spurious Hamiltonian non-resonant is not an option as the value of $e_J$ is unknown. \\

A possible strategy to mitigate the impact of  magnetic flux offsets is to dynamically vary the circuit parameters in order to  decrease the value of $\eta_a$ in between bias pulses. This adjustment may be realized using tunable inductors as detailed in Sec.~\ref{sec:gates}. If $\eta_a \simeq 2\sqrt{\pi}$ when $\tilde{\qop}_a$ is close to aligning with $\qop_a$ or $\Pop_a$ but takes a value  $\delta \eta_a \ll 2\sqrt{\pi}$ at any other time, the time-dependent Hamiltonian \eqref{eq:spuriousHamil} contains only terms of the form $e^{i (\alpha \qop_a + \beta \Pop_a) }$ with $\alpha, \beta$ in the neighborhood of either $2\sqrt{\pi}$ or $0$. It may then only trigger shifts which are approximately aligned with the GKP lattice, and are thus correctable.  Crucially, $\delta \eta_a$ does not need to strictly cancel for this strategy to be effective. We will investigate quantitatively the feasibility and performances of this scheme in a future work.\\

\subsection{Quasi-particle poisoning}
\label{sec:QP}

 Quasi-particles are excitations of the circuit electron fluid  above the superconducting gap~\cite{glazman2021bogoliubov}. The probability for such excitations should be negligible at the working temperature  of circuit QED experiments (10~mK), but normalized densities of quasi-particles in the range $x_{qp}\sim10^{-5}-10^{-7}$ are typically observed. A quasi-particle with charge $e$ tunneling through the Josephson ring is expected to translate the target mode by $\pm \sqrt{\pi}$ in normalized units, which can directly lead to a logical flip.  In the long term, this uncorrected error channel could limit the coherence time of the logical qubit. Quantitative estimates of the logical error rate induced by a given density of quasi-particles will be sought in a future work. Note that quasi-particle poisoning is detrimental to all circuitQED architectures, and is thus actively investigated. Recent progress in identifying and suppressing sources of out-of-equilibrium quasi-particles~\cite{cardani2021reducing,mannila2022superconductor,anthony2022stress,bertoldo2023cosmic}, as well as in trapping and annihilating them~\cite{nsanzineza2014trapping,wang2014measurement,gustavsson2016suppressing,patel2017phonon,henriques2019phonon,martinis2021saving,marchegiani2022quasiparticles} could conceivably lead to efficient suppression strategies in the near future.

\begin{table}
\begin{tabular} {|c|c|c|}
\hline
\rowcolor{lightgray}
Parameter & Symbol   &  Value \\
\hline
Target mode inductance& $L_a$   & 14~$\mu$H \\
 (inductive energy)&&($h\times$12~MHz)\\
  \hline
  Josephson junction energy&$E_J$ & $h\times$500~MHz\\
\hline
Target mode  capacitance&$C_a $  & 80~fF \\
 (charging energy)&&($h\times$240~MHz)\\
\hline
Target mode  frequency&$\omega_a$   & 2$\pi\times$150~MHz   \\
\hline
Target mode photon loss rate& $\kappa_a$ & 2$\pi\times$300~Hz\\
\hline
Ancillary  mode frequency& $\omega_b$   & 2$\pi\times$5~GHz   \\
 \hline
 Ancillary  mode phase  & $\eta_b$ & 0.3 \\
 fluctuations across the ring&&\\
  \hline
  Ancillary  mode photon loss rate & $\kappa_b$ & 2$\pi\times$0.5~MHz \\
  \hline
 Number of harmonics in bias comb& $N$& 100 \\
   \hline
 Maximum modulation signal &$\xi_{\mathrm{max}}$ & 0.2\\
\hline
  Modular interaction rate & $g$ & 2$\pi\times$100~kHz \\
 \hline
Modular dissipation rate&$ \Gamma$ & 2$\pi\times$20~kHz\\
\hline
Decay rate of $\X$ and $\Z$&$ \Gamma_L$ & 2$\pi\times$4~Hz\\
 Pauli operators&&\\
\hline
\end{tabular}
	\caption{ {\bf Proposed circuit parameters.}  The target mode has an impedance $Z_a=2R_Q$ and resonates in the radio-frequency range to allow biasing of the circuit with a frequency comb containing $N=100$ harmonics within a 20~GHz bandwidth.  This  requires to load the circuit with an ultra-high inductance, which can be implemented at the cost of only a small number of parasitic modes appearing in the operating band (see \cref{sec:array}) with state-of-the-art techniques~\cite{pechenezhskiy2020superconducting}. The energy of the two Josephson junctions is chosen low enough that a $2\%$ energy mismatch may be compensated \textit{in situ} (see Sec.~\ref{sec:jjasym}), and other parameters are chosen to respect the hierarchy \eqref{eq:hierarchy}. The estimated decay rate of the generalized Pauli operators only accounts for errors induced by photon loss of the target mode and by  truncation of the bias frequency comb. The latter error channel significantly dominates over the former, so that increasing further the number of harmonics in the bias comb would yield a much more robust GKP qubit.
}
\label{table}
\label{tableparam}
\end{table}

 \section{Protected Clifford gates and Pauli measurements}
 \label{sec:gates}

 In their seminal paper~\cite{gottesman2001encoding} GKP defined \textit{fault-tolerant} operations on GKP qubits as transformations of the embedding oscillators that do not amplify shift errors: small shifts should remain small throughout the operation. In particular, they proposed to perform fault-tolerant Clifford gates in the infinite-energy code through symplectic transformations of the oscillators quadratures. These transformations can be driven with simple low-weight  Hamiltonians whose  strength do not need to be calibrated with high precision as errors induced by slightly off evolutions are mapped to short displacements and are thus correctable.\\

  In this section, we extend these results and derive target evolutions implementing Clifford gates in the finite-energy  code. In contrast with the infinite-energy  case, these  evolutions  are not unitary and thus not trivially driven: a practical driving scheme  remains to be found. Fortunately, one can circumvent the problem by slowly varying the  parameters of the dissipation described in the previous sections such that its fixed points follow the desired code states trajectory in phase-space throughout the gate. In the limit where the gate duration $T_{\mathrm{gate}}$ is much longer than $1/\Gamma_c$ ($\Gamma_c$ is the confinement rate onto the code manifold, see Sec.~\ref{sec:modular dissip}), we expect dissipation to coral the target state with no additional drives, as was proposed for the control of cat qubits~\cite{guillaud2019repetition}. \\
 
  Since the modular dissipation is always on throughout the gate and that  symplectic transformations do not amplify errors---in the GKP sense---we expect the exponential scaling of the logical error rates found in Sec.~\ref{sec:errorcorrec} to hold when applying gates, and thus the GKP qubits to remain \textit{protected}.   Admittedly,  the GKP definition for error amplification is only qualitative, and the amplitude of shift errors along symplectically transformed quadratures is expected to quantitatively  vary during the gate. In particular, for quadrature noise, the effective noise rate scales quadratically with the  length of the transformed quadratures, thereby renormalizing the numerical prefactors in the expression of  the logical error rate found in Sec.~\ref{sec:errorcorrec}. Quantitative analysis of error rates during gates performed in finite time and for more realistic noise models will be the subject of a future work. Similarly, we  propose in Sec.~\ref{sec:readout} a method to measure the Pauli operators of a GKP qubit that does not require to drop the modular dissipation. We expect  the lifetime of the measured operator not to be impacted by the measurement, but leave  rigorous analysis of the performances of this \textit{protected} readout to future research.

 \begin{figure*}[htbp]
		\centering
		\includegraphics[width=1.7\columnwidth]{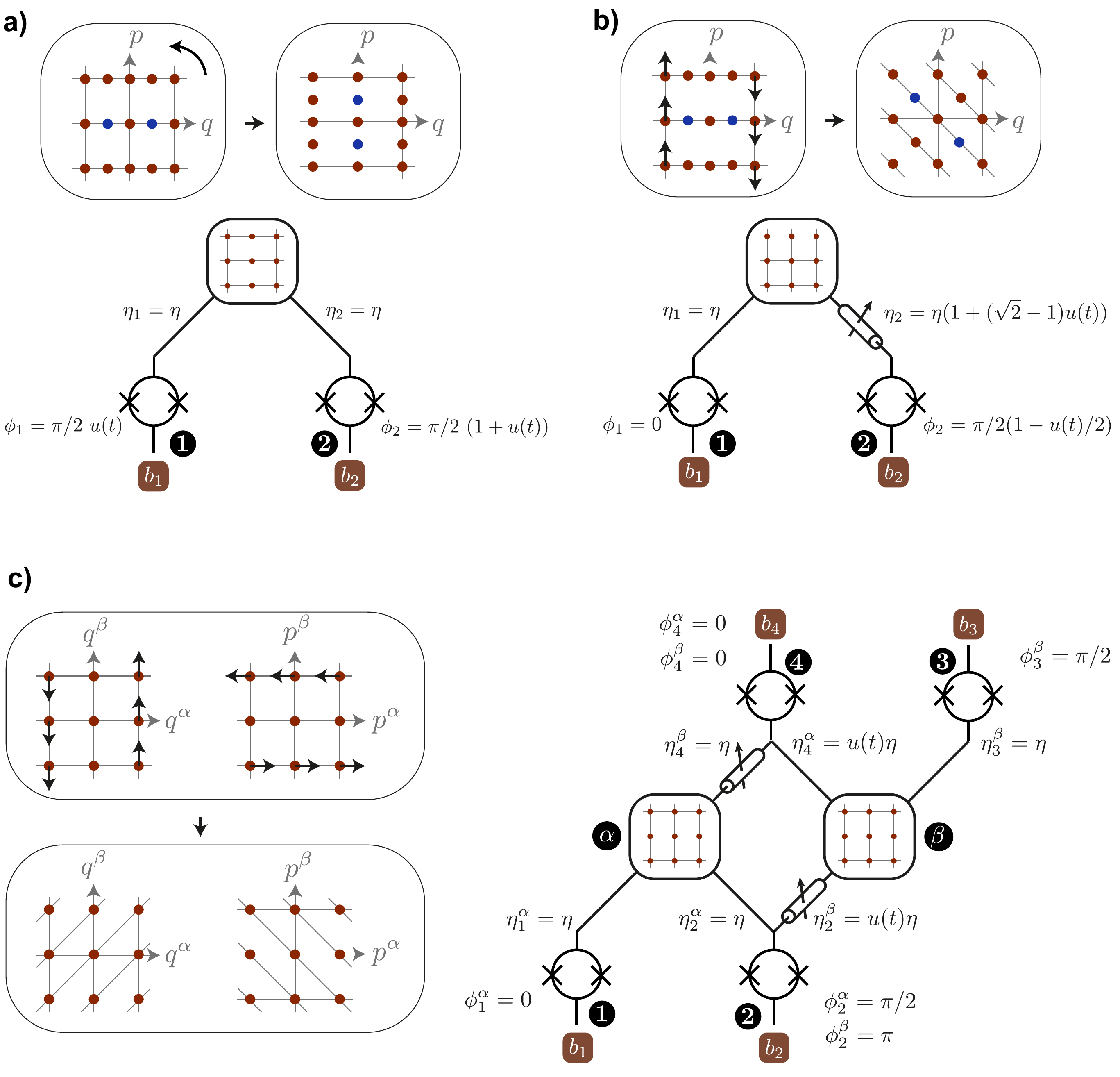}
		\caption{{\bf  Clifford gates by slow variation of the modular dissipation parameters}. The Hadamard {\bf (a)}, Phase {\bf (b)} and CNOT {\bf (c)} gates are each applied by continuously distorting the stabilized GKP lattice structure in phase-space. In boxes, the oscillator field states are represented by their standard deviation contours (red circles), along with the GKP lattice axes (blue and yellow) in the single mode  (a-b) or bipartite (c) phase space, before and after a gate. At the end of each gate, the distorted lattice aligns with the initial one. In our proposed architecture, each oscillator is connected to at least two Josephson rings, each ring being connected to a dissipative ancillary mode (brown box, see Fig.~\ref{fig:gatecircuit} for a detailed circuit) and responsible for the activation of a pair of Lindblad operators  $(\LL_{r,s},\LL_{r,d})$ (see  Eq.~\ref{eq:lindblad2}), where $r=q$ or $p$ before  and after the gate. Lattice distortion  is induced by rotating the target quadrature $r$ by an angle $\phi$ (controlled  by the timing  of the   pulses biasing the ring) and simultaneously scaling its length by a factor $\gateLengthFactor$ (controlled by the amplitude of the oscillator phase fluctuations  across the ring $\eta =  2\sqrt{\pi} \gateLengthFactor$, whose tunability is symbolized by a cylinder pierced by a diagonal arrow). The parameter $u:0\rightarrow 1$ is slowly varied during the gate so that the oscillator states remain at a fixed point of the dissipation at all time.
		}
		\label{fig:gates}
	\end{figure*}

\subsection{Clifford gates in the finite-energy GKP code}

Remarkably, the target evolutions proposed by GKP to implement Clifford gates in the infinite-energy code correspond to continuous  symplectic mappings of the target oscillator phase-space coordinates. In detail, for a  control parameter $u$ varying continuously from 0 to 1 during the gate,   these transformations read:\\

\textit{Hadamard gate}
\begin{equation}
    \begin{split}
        \mathcal{S}_u^H:\quad& \qop \rightarrow \mathrm{cos}(u\frac{\pi}{2}) \qop +\mathrm{sin}(u\frac{\pi}{2}) \Pop\\
        & \Pop \rightarrow -\mathrm{sin}(u\frac{\pi}{2}) \qop +\mathrm{cos}(u\frac{\pi}{2}) \Pop
    \end{split}
\end{equation}
The corresponding evolution is a quarter turn rotation of the target state $\UU^H_u=e^{iu \frac{\pi}{2}\adag \aop}$ (see Fig.~\ref{fig:gates}a).\\

\textit{Phase gate}
\begin{equation}
    \begin{split}
        \mathcal{S}^P_u:\quad& \qop \rightarrow \qop\\
        & \Pop \rightarrow \Pop -u \qop
    \end{split}
\end{equation}
 The corresponding evolution consists in squeezing and rotating  the target state $\UU^P_u=e^{iu \qop^2/2}$ (see Fig.~\ref{fig:gates}b).\\

 \textit{CNOT gate}
\begin{equation}
    \begin{split}
        \mathcal{S}^C_u:\quad& \qop^{\alpha} \rightarrow \qop^{\alpha}\\
        & {\Pop}^{\alpha} \rightarrow \Pop^{\alpha} -u \Pop^{\beta} \\
         & \qop^{\beta} \rightarrow \qop^{\beta} + u \qop^{\alpha}\\
        & {\Pop}^{\beta} \rightarrow \Pop^{\beta}
    \end{split}
\end{equation}
 Here the joint evolution of the  control and target oscillators labeled $\alpha$ and $\beta$ reads $\UU^P_u=e^{iu \qop^{\alpha} \Pop^{\beta}}$ (see  Fig.~\ref{fig:gates}c) and is the combination of two-mode squeezing and photon exchange (beam-splitter Hamiltonian).\\

 We now note that  the infinite-energy square code is entirely defined by its two stabilizers $\SSS_q=e^{ i\eta \qop}$ and $\SSS_p=e^{-i\eta \Pop}$. The code properties---namely the stabilizers and generalized Pauli operators commutation rules, the code states definition---are all inferred from the canonical commutation relation of the quadrature operators $[\qop,\Pop]=i$. Since symplectic transformations preserve commutation relations, the same modular functions of symplectically transformed variables $e^{ i\eta \mathcal{S}_u(\qop)}$ and $e^{ -i\eta \mathcal{S}_u(\Pop)}$, where $\mathcal{S}_u$ is one of the three aforementioned transformations,  are the stabilizers of another GKP code. In other words, Clifford gates are applied by continuously distorting the GKP lattice in phase-space so that the final lattice structure overlaps with the initial one, and that an exact  gate has been applied to the encoded qubit (see Fig.~\ref{fig:gates}). The same scheme is directly applicable to the finite-energy code, after normalizing all operators with $\E_{\Delta}=e^{-\Delta \adag \aop}$. The target evolutions now read $\VV^{\Delta}_u=\E_{\Delta} \UU_u \E^{-1}_{\Delta}$, and are in general non-unitary. As for the stabilizers of the distorted code, they read $\E_{\Delta} e^{ i\eta \mathcal{S}_u(\qop)} \E^{-1}_{\Delta}$ and $\E_{\Delta} e^{ -i\eta \mathcal{S}_u(\Pop)} \E^{-1}_{\Delta}$. Note that with this definition, the lattice structure is distorted, but the code states normalizing envelope remains Gaussian-symmetric. \\

\subsection{Clifford gates by slow variation of the modular dissipation parameters}

We now detail how to adapt the dissipation engineering technique described in Sec.~\ref{sec:dissipationengineering} to stabilize a finite-energy code distorted by $\mathcal{S}_u$. We consider the architecture depicted in  Fig.~\ref{fig:gates}, in which a target mode is connected  to two Josephson rings, each one coupled to a dissipative ancillary mode (brown box). Detailed circuits implementing this abstract architecture may be found in Fig.~\ref{fig:gatecircuit}.  Each ring activates one pair of Lindblad operators $(\LL_{r,s},\LL_{r,d})$ as defined in Eq.~\ref{eq:lindblad2}. For an idling logical qubit, these operators are modular functions of one of the oscillator quadratures $r=q$~or~$p$. When a gate is applied, the quadrature needs to be substituted with the symplectically transformed quadrature $\mathcal{S}_u(r)$, and $u$ varied slowly enough as to respect  the adiabaticity condition $\Gamma_c T_{\mathrm{gate}}\gg 1$. \\

First focusing on single-qubit gates, the transformed  quadrature is parametrized by its angle $\phi$ and length $\gateLengthFactor$ in phase-space.  Adjusting the value of $\phi$ only requires to time-shift the control pulses biasing the corresponding ring. Indeed, for an idling qubit, the Lindblad dissipators $\LL_{r,s}$ and $\LL_{r,d}$ are activated through the interaction Hamiltonian~\eqref{eq:hint} parametrized by   the angle $\phi_r$ ($\phi_r=0$ for $r=q$ or $\phi_r=\pi/2$ for $r=p$). In turn, this angle is determined by the phase of the pulse trains biasing the Josephson ring (see Eq.~\eqref{eq:ideal_controls}). One may generalize this approach and activate a Lindblad dissipator which is a modular function of a quadrature $\qop_a^{\phi}=e^{ i \phi\aop^{\dag}\aop} ~ \qop_a ~ e^{- i\phi \aop^{\dag}\aop} $ rotated by an arbitrary angle $\phi$.  On the other hand, adjusting the length $\gateLengthFactor$ of the symplectically transformed quadrature necessitates to adjust the spatial frequency of the modular interactions to $\eta= 2\sqrt{\pi}\gateLengthFactor$ in the Hamiltonian~\eqref{eq:hint}  ($\gateLengthFactor=1$ for an idling qubit). Physically, this parameter is set by the phase fluctuations $\eta$  of the target mode across the Josephson ring.  In Fig.~\ref{fig:gates}, we symbolize this control by a tunable coupler (cylinder pierced by an arrow) connecting the target mode with the ring. We introduce an actual circuit implementing this abstract model in  Sec.~\ref{sec:realcircuit}.
Note that the Hadamard gate only requires to vary the value of $\phi$
for both rings
\footnote{In fact, although the Hadamard gate provides a first simple example of our implementation
of Clifford gates, it could be implemented even more easily in practice:
the required $\pi/2$ rotation in phase-space can be performed by simply
adjusting the clock of the rotating frame in software.}
whereas the phase gate requires to vary the value of $\phi$ and $\eta$ simultaneously for a single ring (see Fig.~\ref{fig:gates}a-b).  \\

Similar controls are employed to apply a two-mode CNOT gate. Here, two of the four transformed quadratures $\mathcal{S}^C_u(r)$---with $r=q^{\alpha},~p^{\alpha},~q^{\beta}~\mathrm{or}~p^{\beta}$---combine a fixed contribution from one mode  and a varying contribution from the other one.  Therefore, applying a CNOT gate  requires  to adjust the phase-fluctuations of each mode across one of the rings employed to engineer dissipation in the other one. Thus, in Fig.~\ref{fig:gates}c, the coupling of ring 2---responsible for the activation of $(\LL_{q^{\alpha},s},\LL_{q^{\alpha},d})$ when the logical qubits are idle---to the mode $\beta$ is slowly ramped up during the gate. As a consequence, the ring witnesses increasing phase fluctuations $\eta^{\beta}_2$ from the oscillator $\beta$, while the phase fluctuations $\eta^{\alpha}_2$ from the mode $\alpha$ remain constant. Simultaneously, the coupling of ring 4 to the mode $\alpha$ is slowly ramped up such that $\eta^{\alpha}_4$ increases while  $\eta^{\beta}_4$  remains constant. Moreover, the signals biasing the rings 2 and 4 are enriched to mediate interactions of the form \eqref{eq:hint} between the target mode $\alpha$ and the ancillary mode $\beta$ and \textit{vice versa}. In practice, the added signals consist in trains of pulses at the frequency of the target modes modulating  carriers at the frequency of the ancillary modes. \\

Two important comments are in order about these gates and the proposed architecture. First, we remind the reader that once the evolution implementing a gate is complete, the GKP lattice  of each oscillator retrieves  its initial square structure. As a consequence, the control parameters $\phi$ and $\eta$, which have been varied throughout the gate, can be returned  to their initial values.
While the variation of the parameters
needs to be slow during the gate,
this last adjustment can be made on a much shorter timescale. The flux pulse trains biasing the ring being controlled should be interrupted during this stage in order not to inadvertently generate  a modular dissipation misaligned with the oscillator   GKP lattice. Second, when \textit{not} applying a CNOT gate, the coupling between one mode and the rings employed to engineer dissipation in the other one does \textit{not} need to be perfectly nullified in order to avoid cross-talks between the GKP qubits. Indeed, a ring experiencing small residual phase fluctuations from a mode  (\textit{e.g.} $\eta_2^{\beta}, \eta_4^{\alpha} \ll 2\sqrt{\pi}$ for idling qubits) may only generate small shifts of this mode, corrected by the modular dissipation. Based on similar arguments, we describe in the next section a method for  measuring the GKP qubit Pauli operators  which does not introduce spurious dephasing out of measurement times. 
\\

\subsection{Protected measurement of Pauli operators}
 \label{sec:readout}

In the previous section, we introduced an architecture in which a target mode embedding a GKP qubit is coupled to multiple rings with adjustable strength. The requirements to measure a Pauli operator of the GKP qubit are the same as to perform a phase gate, and are met by the architecture  depicted in Fig.~\ref{fig:gates}b. Indeed, biasing the ring 2 with a signal  $\xi_2(t) \propto \mathrm{cos}(\omega_{b_2} t ~-~\Theta_{b}) ~ \Sha_{\frac{\pi}{\omega_a}}(t-\frac{\Theta_a}{\omega_a}) $ (corresponding to the case $\delta_l=0$ in Eq.~\eqref{eq:ideal_controls}), one activates a modular interaction of the form $ \mathrm{cos}(\eta_2 \qop^{\Theta_a}_a)\qop^{\Theta_{b}}_{b_2} $---we use similar notations as in  Eq.~\eqref{eq:hint}  to denote rotated quadrature operators of the target mode and of the ancillary mode $b_2$ attached to the ring 2. The value of $\eta_2$ is adjustable \textit{in situ} by tuning the coupling of the ring 2 to the target mode.  We propose to measure the generalized Pauli operators $\Z=\mathrm{Sgn}\big(\mathrm{cos}(\sqrt{\pi}\qop)\big)$  through such an interaction:
 \begin{equation}
    \HH_{\mathrm{meas}}=\hbar g_{\mathrm{meas}}\mathrm{cos}(\eta_2 \qop_a) \qop_{b_2}
    \label{eq:Hmeas}
\end{equation}
where the  spatial frequency of the modular term $\eta_2=\sqrt{\pi}$ is  twice smaller than when engineering the modular dissipation stabilizing the GKP qubit.  $\X$ can be  measured via a similar interaction but for a $\pi/2$-rotation of the target mode quadrature, and $\Y$ may be measured by applying a phase-gate before measuring $\X$.  We refer the reader to Sec.~\ref{sec:modularengineering} for the expression of $g_{\mathrm{meas}}$ as a function of the physical system parameters.\\

The modular dissipation, mediated by the ring 1,  is kept on during the measurement, confining the target mode state onto the code manifold. If the confinement rate $\Gamma_c $ is much larger than the interaction rate $g_{\mathrm{meas}}$, one obtains an effective Zeno dynamics within the code manifold. Moreover, denoting $\mathcal{P}$ the projector onto the code manifold, we have $\mathcal{P}\HH_{\mathrm{meas}}\mathcal{P}^{\dagger}\simeq\tilde{g}_{\mathrm{meas}}\mathcal{P}\Z \mathcal{P}^{\dagger} \qop_b$ such that the value of $\Z$ is mapped to a displacement of the ancillary mode along $\Pop_b$. The effective interaction rate $\tilde{g}_{\mathrm{meas}}\lesssim g_{\mathrm{meas}}$ is slightly renormalized given that $\mathrm{cos}(\eta_2 \qop_a)\lesssim 1$ for finite-energy GKP states.\\

Letting the ancillary mode dissipate its excitations at rate $\kappa_{b_2}$ into a  transmission line   rather than into a resistor as pictured in Fig.~\ref{fig:dissipCircuit}a, one may retrieve this information through simple homodyne detection of the leaking field. The logical qubit is then continuously measured at rate $\Gamma_{\mathrm{meas}}\simeq 4 \eta_{det}\frac{\tilde{g}_{\mathrm{meas}}^2}{\kappa_{b_2}}$ where $\eta_{det}$ is the  detection efficiency. Note that the simple Zeno model we used requires that 
$\Gamma_{\mathrm{meas}} \ll \Gamma_c $, placing an upper bound on the measurement rate. Nevertheless, this rate may be orders of magnitude larger than the logical flip rate $\Gamma_L$ (see Sec.~\ref{sec:errorcorrec}) and since the  measurement is Quantum Non-Demolition, the signal can be integrated to yield a high fidelity readout.\\

Note that outside of measurement times, the coupling of the ring 2 to the target oscillator can be decreased such that $\eta_2 \ll 1$. As argued in the previous section, this ensures that  even if the ring Josephson energy is imperfectly cancelled, it   may only induce short, correctable displacements of the target mode. Therefore, no logical  information  unintentionally leaks out of the system and the logical qubit is not dephased outside of measurement times.

\subsection{Example circuit for Clifford gates and Pauli measurements}
\label{sec:realcircuit}

\begin{figure}[htbp]
	\centering
	\includegraphics[width=1\columnwidth]{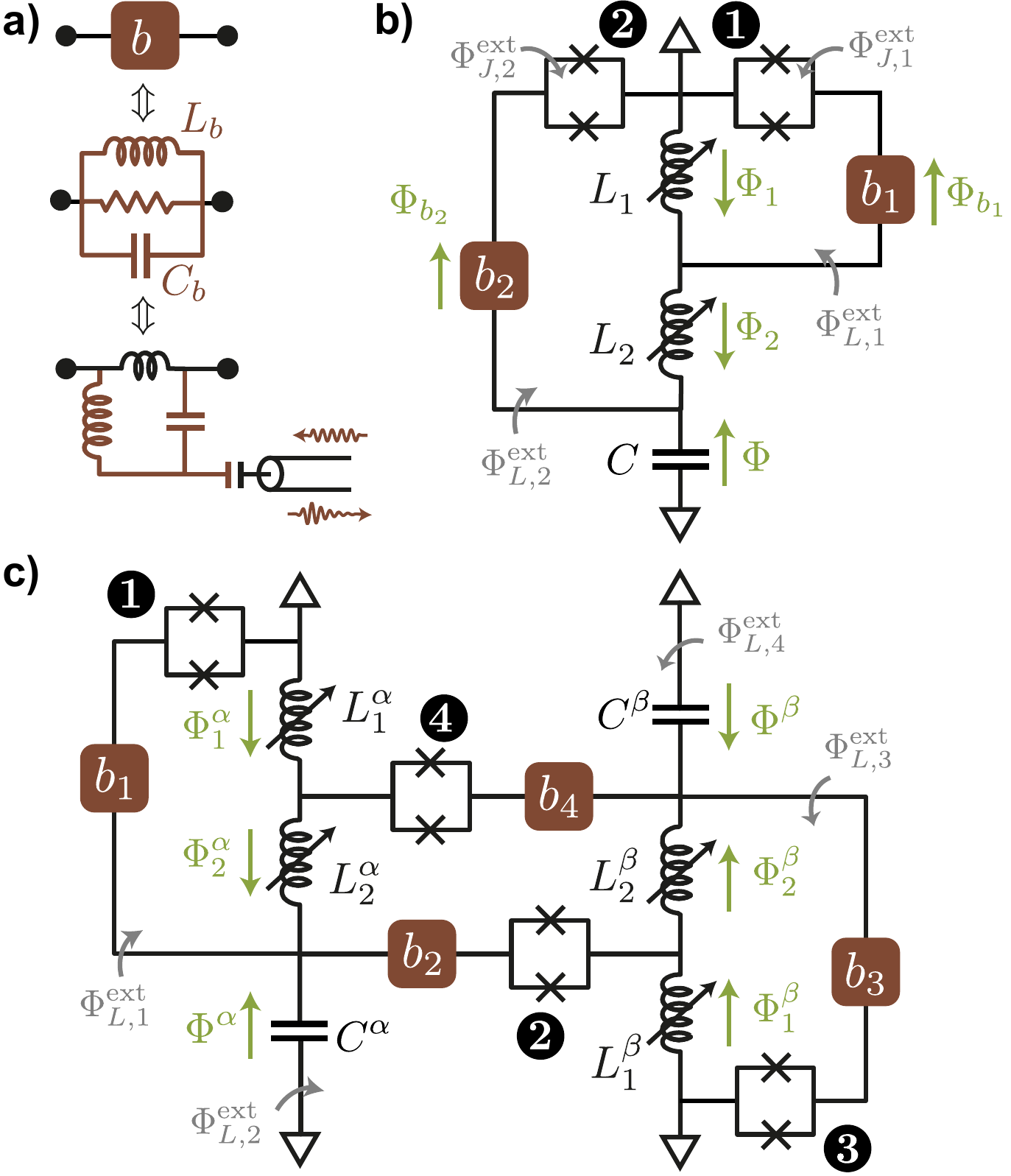}
	\caption{{\bf Example circuits}. {\bf a)} Symbol (top), Foster decomposition (middle) and realistic implementation (bottom) for the ancillary circuit  in series with each Josephson ring. It couples to a transmission line  through which drives are applied and excitations leak out (wriggled arrows), enabling dissipation engineering and Pauli measurements. {\bf b)} Circuit supporting a single target mode (phase variable $\Phi$) embedding a GKP qubit.  The ring and ancillary mode 2 support the full  phase drop $\Phi$ while the ring and ancillary mode 1 support a fraction of it set by the relative value of $L_1$ and $L_2$. Tuning independently the value of $L_1$ and $L_2$  allows maintaining constant  fluctuations of $\Phi$ across one ring  while adjusting the fluctuations across the other, as required for phase gates and Pauli measurements. {\bf c)} Circuit supporting two GKP qubits controllable with a CNOT gate. The target modes are defined across the capacitors $C^{\alpha}$ and $C^{\beta}$. The phase fluctuations of the mode $\alpha$ (respectively $\beta$) across the ring 1 and 2 (respectively ring 3 and 4) depend on the inductance $L_1^{\alpha}+ L_2^{\alpha}$ (respectively  $L_1^{\beta}+ L_2^{\beta}$) and are held constant at all time. The phase fluctuations of the mode $\beta$ (respectively $\alpha$) across the ring 2  (respectively ring 4) depend on the  relative value of $L_1^{\alpha}$ and $L_2^{\alpha}$ (respectively of $L_1^{\beta}$ and $L_2^{\beta}$) and are varied  during the CNOT gate. Fluxes threading the Josephson rings and phases of the ancillary modes are implicitly defined as in (b).}
  \label{fig:gatecircuit}
\end{figure}

In the previous sections, we considered abstract architectures allowing to tune \textit{in situ} the phase fluctuations of one or two target modes across several Josephson rings. Here, we introduce   example circuits implementing these abstract architectures.\\

The circuit depicted in Fig.~\ref{fig:gatecircuit}b implements the architecture of Fig.~\ref{fig:gates}b employed to perform a phase-gate and a Pauli operator measurement. A capacitor $C$ with phase variable $\Phi$ is shunted by two tunable inductors $L_1$ and $L_2$---for instance implemented by  chains of Josephson rings controlled with an external magnetic field (not shown). The Josephson ring 2 and the ancillary circuit $b_2$ are placed in parallel with both inductors, while the Josephson ring 1 and the ancillary circuit $b_1$ are placed in parallel with $L_1$ only  (ancillary circuits in brown, whose Foster decomposition and realistic implementation are detailed  in Fig.~\ref{fig:gatecircuit}a). We suppose that the effective Josephson energy of the ring 1 is much smaller than the  inductive energy $\varphi_0^2/L_1$. As a consequence, the same current flows through the inductors $L_1$ and $L_2$ and using  Kirchhoff's laws, one  finds the phase drop across each inductor  to be $\Phi_i=p_i \Phi$, where we introduced the participation ratios  $p_i=\frac{L_i}{L_1+L_2}$ for $i=1,2$. The circuit Lagrangian then reads
\begin{equation}
\begin{aligned}
\mathcal{L}&=\frac{C}{2}\dot{\Phi}^2 - \frac{1}{2} (\frac{p_1^2}{L_1}+\frac{p_2^2}{L_2}) \Phi^2 
+\mathcal{L}_{b_1}+\mathcal{L}_{b_2}\\
& + E_{J_1}\cos\big ( \frac{p_1 \Phi + \Phi_{b_1}- \Phi_{L,1}^{\mathrm{ext}}}{\varphi_0} \big) +  E_{J_2}\cos \big( \frac{\Phi + \Phi_{b_2} + \Phi_{L,2}^{\mathrm{ext}}}{\varphi_0} \big)
\label{eq:lagrangian}
\end{aligned}
\end{equation}
where  $\mathcal{L}_{b_i}= \frac{C_{b_i}}{2}\dot{\Phi}_{b_i}^2 - \frac{1}{2 L_{b_i}}  \Phi_{b_i}^2$ and $E_{J_i}$ is the effective Josephson energy of the ring $i$ set by the magnetic flux $\Phi_{J,i}^{\mathrm{ext}}$.  We now identify $\Phi$ as the phase coordinate of the target mode, which has a total inductance $L=(\frac{p_1^2}{L_1}+\frac{p_2^2}{L_2})^{-1}=L_1+L_2$ and is only coupled to the ancillary modes via the Josephson rings. Its phase fluctuations across the two rings read 
\begin{equation}
\begin{aligned}
\eta_1&=p_1 (\frac{L}{C})^{1/4}(\frac{2\pi}{R_Q})^{1/2}\\
\eta_2&= (\frac{L}{C})^{1/4}(\frac{2\pi}{R_Q})^{1/2}\\
\end{aligned}
\end{equation}
Therefore, tuning independently the value of the inductances $L_1$ and $L_2$, one controls $\eta_1$ and $\eta_2$ as required to perform a phase-gate and a Pauli operator measurement.\\

We now turn to the circuit depicted in Fig.~\ref{fig:gatecircuit}c, which implements the architecture of Fig.~\ref{fig:gates}c employed to perform a CNOT gate. Here, we assume that the effective Josephson energy of the rings 2 and 4  are much smaller than the  inductive energies $\varphi_0^2/L_2^{\alpha}$ and $\varphi_0^2/L_2^{\beta}$ so that the phase drop across the inductances read $\Phi_i^{\gamma}=p_i^{\gamma}\Phi^{\gamma}$ with $p_i^{\gamma}=\frac{L_i^{\gamma}}{L_1^{\gamma}+L_2^{\gamma}}$ for $i=1,2$ and $\gamma=\alpha, \beta$. Following the same line of reasoning as above, the phase coordinate of each target mode  $\Phi^{\gamma}$ is  defined across the capacitor $C^{\gamma}$ and the circuit Lagrangian reads
\begin{equation}
\begin{aligned}
	&\mathcal{L}=\sum_{\gamma=\alpha,\beta}\frac{C^{\gamma}}{2}(\dot{\Phi}^{\gamma})^2 - \frac{1}{2L^{\gamma}}  (\Phi^{\gamma})^2 +\sum_{j=1}^{4} \mathcal{L}_{b_i}\\
& +  E_{J_1}\cos\big ( \frac{ \Phi^{\alpha} + \Phi_{b_1}+ \Phi_{L,1}^{\mathrm{ext}}}{\varphi_0} \big) +   E_{J_3}\cos\big ( \frac{ \Phi^{\beta} + \Phi_{b_3}+ \Phi_{L,3}^{\mathrm{ext}}}{\varphi_0} \big)\\
& +  E_{J_2}\cos\big ( \frac{ \Phi^{\alpha}-  p^{\beta}_1\Phi^{\beta} +  \Phi_{b_2}+ \Phi_{L,2}^{\mathrm{ext}}}{\varphi_0} \big) \\   &+  E_{J_4}\cos\big ( \frac{ \Phi^{\beta}-  p^{\alpha}_1\Phi^{\alpha} +  \Phi_{b_4}+ \Phi_{L,4}^{\mathrm{ext}}}{\varphi_0} \big)
\end{aligned}
\end{equation}
where we use the same conventions as in Eq.~\eqref{eq:lagrangian} for the effective energy of the Josephson rings and the ancillary modes Lagrangians, and we have defined $L^{\gamma}=L_1^{\gamma}+L_2^{\gamma}$. The ring 1 (respectively the ring 3) only participates in the mode $\alpha$ (respectively $\beta$) and supports phase fluctuations 
\begin{equation}
\begin{aligned}
\eta^{\alpha}_1&= (\frac{L^{\alpha}}{C^{\alpha}})^{1/4}(\frac{2\pi}{R_Q})^{1/2}\\
\eta^{\beta}_3&= (\frac{L^{\beta}}{C^{\beta}})^{1/4}(\frac{2\pi}{R_Q})^{1/2}\\
\end{aligned}
\end{equation}
while the rings 2 and 4 participate in both modes and support phase fluctuations
\begin{equation}
\begin{aligned}
&\eta^{\alpha}_2= (\frac{L^{\alpha}}{C^{\alpha}})^{1/4}(\frac{2\pi}{R_Q})^{1/2} \qquad &\eta^{\beta}_2= p_1^{\beta} (\frac{L^{\alpha}}{C^{\alpha}})^{1/4}(\frac{2\pi}{R_Q})^{1/2}\\
&\eta^{\beta}_4= (\frac{L^{\beta}}{C^{\beta}})^{1/4}(\frac{2\pi}{R_Q})^{1/2} \qquad &\eta^{\alpha}_4= p_1^{\alpha} (\frac{L^{\alpha}}{C^{\alpha}})^{1/4}(\frac{2\pi}{R_Q})^{1/2}\\
\end{aligned}
\end{equation}
 In practice, the value of $L^{\alpha}$ and $L^{\beta}$ are held constant at all time such that $\eta_1^{\alpha}=\eta_2^{\alpha}=\eta_3^{\beta}=\eta_4^{\beta}=2\sqrt{\pi}$ and $p_1^{\alpha}$ and $p_1^{\beta}$ are varied during the CNOT gate, with their value matching that of the parameter $u$ in Fig.~\ref{fig:gates}c.\\
 
 To conclude this section, we point out that while the circuit depicted in Fig.~\ref{fig:gatecircuit}c does not allow for single-qubit phase gates, one can connect an additional Josephson ring to each mode---respectively placed in parallel with the inductors $L_1^{\alpha}$ and $L_1^{\beta}$ as in Fig.~\ref{fig:gatecircuit}b---to enable all Clifford operations and Pauli measurements in a single circuit. We also note  that the  assumption we made that the effective Josephson energy of some  rings is much smaller than the circuit inductive energies is not verified for the parameters proposed in Table~\ref{tableparam}. Indeed, when pulses with peak amplitude $\xi_{\mathrm{max}}$ are applied, the effective Josephson energy reaches $2E_J \xi_{\mathrm{max}} > \varphi_0^2 /L_a$. Unless imposing stringer constraints for the multi-qubit circuits  considered  in the current section,  the  participation ratios $p_1$, $p_1^{\alpha}$, $p_1^{\beta}$ defined above would be renormalized during pulses,  and   the magnitude of phase-fluctuations across some rings would vary. Alternative circuits circumventing this issue will be sought   in a future work.

\section{Conclusion and outlook}
In this paper, we have proposed a novel  scheme  to generate, error-correct and control GKP qubits.  The crucial difference with previous experimental demonstrations of GKP error-correction lies in  the \textit{modular} interactions  by which we couple the target mode---hosting the GKP qubit---and the ancillary  mode---leveraged to evacuate entropy from the system. These modular interactions prevent the back propagation of noise from the ancilla to the encoded qubit. In contrast,  the bi-linear coupling employed in state-of-the-art  experiments to map GKP error-syndromes to an ancillary qubit allows noise to back propagate, limiting the coherence of the encoded qubit~\cite{sivak2022real}. Furthermore, we propose a practical scheme to activate modular interactions in a high-impedance Josephson circuit,  and show how to combine  modular interactions with dissipation engineering techniques to autonomously error-correct the GKP qubits. Finally, we show how to measure and control GKP qubits with protected Clifford gates.\\

 We perform numerical simulations showing that with this approach, logical errors stemming from dominant error channels of both the target and the ancillary mode are exponentially suppressed as the engineered dissipation rate increases.   In a state-of-the-art circuit, the logical qubit lifetime could extend  orders of magnitude beyond the single photon dwell time in the embedding resonator, a feat never realized so far. Arguably, at this level of error suppression, quasi-particle poisoning, which opens an uncorrected error channel, could limit the device performances. Steady progress in understanding and controlling sources of quasi-particles in superconducting devices~\cite{cardani2021reducing,mannila2022superconductor,anthony2022stress} could conceivably overcome this roadblock in the near future.\\

The circuit we propose to embed   GKP qubits is remarkably simple (see Fig.~\ref{fig:gatecircuit}) and is fabricated in a parameter regime which, though demanding (see Table~\ref{tableparam}), should prove easier to achieve than alternative proposals to encode GKP qubits at the hardware level~\cite{brooks2013protected,groszkowski2018coherence,rymarz2021hardware}. Moreover,  circuit parameters do not necessitate fine-tuning so that our protocol is robust against fabrication disorder. Schematically, such robustness and ease of fabrication is made possible by transferring the  complexity of quantum error-correction  from   the hardware to the microwave control domain. Indeed, our system needs to be driven with a precise  microwave frequency comb spanning a 20~GHz range. Recent progress in digital synthesis of microwaves should prove instrumental in generating and delivering such a broadband signal with sufficient accuracy. Alternatively, direct on-chip synthesis of microwave frequency combs appears compatible with the circuits we consider~\cite{solinas2015radiation,solinas2015josephson}, and would drastically reduce control complexity.\\

On the long term, the relative simplicity of Clifford gates and the robustness of our multi-GKP qubit architecture to spurious microwave cross-talks paves the way for the concatenation of these bosonic qubits into a discrete variable code such as the surface code
 ~\cite{fukui2018high,vuillot2019quantum,terhal2020towards,noh2020fault,noh2022low}. Given that the coherence time of GKP qubits stabilized by modular dissipation should extend far beyond single and two-qubit gate time---which is set by the confinement rate onto the code manifold in our approach---the hope is that such a \textit{surface-GKP code} would operate well below threshold, implementing a fault-tolerant, universal quantum computer with minimum hardware overhead.

 \section*{Acknowledgments}
 We thank  W. C. Smith and  R. Lescanne for fruitful discussions on inductively shunted  circuits,
M. Burgelman for fruitful discussions about higher-order averaging methods,
 and M. H. Devoret, A. Eickbusch and S. Touzard for stimulating discussions on the GKP code.
We thank the maintainers of the CLEPS computing infrastructure from the Inria of Paris
for providing the computing means necessary
to speed up the parameter sweeps presented in the figures.  This project has received funding from the European Research Council (ERC) under the European Union’s Horizon 2020 research and innovation program (grant agreements No. 884762, No. 101042304 and No. 851740).
 A.S. and P.C.-I.  acknowledge support from the Agence Nationale de la Recherche (ANR) under grants HAMROQS and SYNCAMIL. The authors acknowledge funding from the Plan France 2030 through the project ANR-22-PETQ-0006.\\

During the final stages of preparation of this manuscript,
we became aware of the recent preprint~\cite{nathan2024selfcorrecting}. The authors propose to autonomously stabilize GKP grid states in a high-impedance circuit leveraging an abstract switch element activated with fast control pulses.

%% file: mt_ancilla_noise.tex
\subsection{Propagation of ancilla noise}
\label{sec:hot_ancilla}
The key advantage of the proposed protocol lies in its robustness
against imperfections of the ancilla.
First, in the limit of infinite-energy GKP states,
the target system is coupled to the ancilla through logical stabilizers only. Therefore, 
  spurious dynamics of the ancilla may only lead to an evolution of the target mode  generated by the stabilizers and thus cannot create logical errors.
Second, an implicit advantage of dissipative stabilization is that the predominant loss channel
of the ancilla,
namely photon loss,
is the main resource of the protocol: this is in stark contrast with phase estimation protocols
relying on low-order interactions, where the dominant dissipative processes acting on the ancilla
limits the performance of the protocol. \cite{campagne2020quantum,sivak2022real}

In order to study more closely the propagation of ancilla errors
for finite-energy GKP states,
we take a closer look at the key ingredient in \cref{sec:dissipationengineering}:
recall that our protocol relies on engineering dynamics of the form
\begin{equation}
	\label{eq:nominal_ae_equation}
\frac{d\rhoo}{dt} = -\tfrac{i}{\hbar} \left[ \HHop^{\textrm{int}}, \, \rhoo\right]
			+ \kappa_b \cD[\bop](\rhoo)
\end{equation}
with $\HHop^{\textrm{int}} = \hbar g \left( \LL \bop^\dag + \LL^\dag \bop\right)$
(corresponding to the Hamiltonian in \cref{eq:hint0} where we dropped the subscript index)
in order to engineer a dissipative evolution
\begin{equation}
	\frac{d\rhoo}{dt} = \Gamma \cD[\LL](\rhoo)
	\label{eq:ae_nominal_target}
\end{equation}
with $\Gamma = {4g^2}/{\kappa_b}$.
This strategy relies on a standard
adiabatic elimination procedure
 (see \cref{ssec:general_strategy}).
We can revisit this procedure when the ancilla mode $b$ is subject to additional error channels,
such as heating and dephasing.
To this end, we enrich \cref{eq:nominal_ae_equation} with two additional terms:
\begin{equation}
\begin{aligned}
\frac{d\rhoo}{dt} =&\; -\tfrac{i}{\hbar} \left[ \HHop^{\textrm{int}}, \, \rhoo\right]
			+ \kappa_b (1+\nth) \, \cD[\bop](\rhoo)\\
			&
						+ \kappa_b \, \nth \,  \cD[\bop^\dag](\rhoo)
			+ \kappa_\phi \, \cD[\bop^\dag \bop](\rhoo)
\end{aligned}
\end{equation}
with
$\kappa_\phi>0$ the dephasing rate
and
$\nth = \left( e^{\frac{\hbar \omega_b}{k_B T}} -1 \right)^{-1}$
the mean number of thermal photons following Bose-Einstein statistics 
(where $k_B$ denotes Boltzmann's constant and $T$ is the temperature, typically a few
tens of milliKelvins).
Under this new ancilla dynamics, \cref{eq:ae_nominal_target}
becomes
\begin{equation}
\begin{aligned}
\frac{d\rhoo}{dt}
	=&\; \tilde \Gamma \, \cD[\LL](\rhoo)
	+ \tilde \Gamma \, \tfrac{\nth}{1+\nth} \,\cD[\LL^\dag](\rhoo)\\
\end{aligned}
	\label{eq:ae_with_hot_ancilla}
\end{equation}
with
\begin{align}
	\tilde \Gamma = \frac{4g^2}{\kappa_b + \kappa_\phi} \, (1+\nth)
	= \Gamma \, \frac{\kappa_b}{\kappa_b+\kappa_\phi} \, (1+\nth).
	\label{eq:new_gamma_dephasing}
\end{align}
In particular, the only effect of ancilla dephasing at rate $\kappa_\phi$
is a renormalization of the engineered dissipation
rate $\Gamma$.
On the other end, the thermal population of the buffer leads to additional dissipators on the target
mode
\footnote{%
	This qualitative difference is easily understood by noting that dephasing does not
	change the steady state of the ancilla but kills the coherences created by the interaction
	Hamiltonian $\HHop^{\mathrm{int}}$,
	while the thermal population of the buffer changes its steady state
	to a thermal state
	$\rho_b^{\mathrm{th}} := \frac{1}{1+\nth} \left( \frac{\nth}{1+\nth}\right)^{\bop^\dag\bop}$.
}.

\begin{figure}[htbp]
		\centering
		\includegraphics[width=0.75\columnwidth]{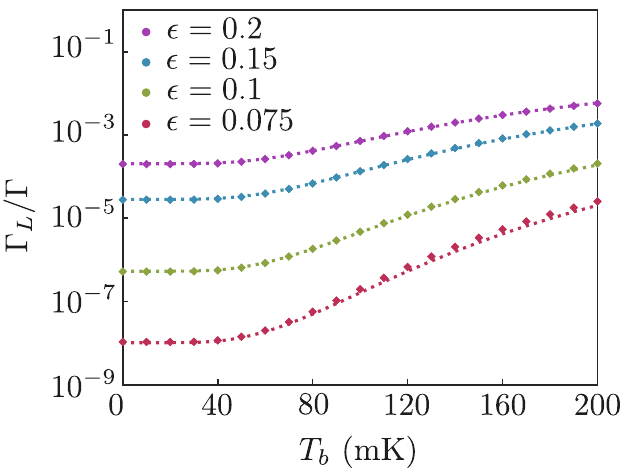}
		\caption{{\bf Propagation of ancilla noise.}
		Decay rate of the generalized Pauli operators $\X$ and $\Z$ under engineered modular dissipation taking the non-zero temperature of the ancilla into account. Colored dots correspond to decay rates extracted from numerical simulations of \cref{eq:simu_hot_ancilla}, while dashed lines follow the asymptotic ansatz in \cref{eq:logical_rate__hot_ancilla}. The value of $\nth$ is obtained following Bose-Einstein statistics
		at the ancilla resonance frequency $\omega_b/2\pi = 5~\unit{\giga\hertz}$ given in \cref{tableparam}.
		The choice of displayed temperature range is made considering that temperatures around a few tens of milliKelvins can be observed in state-of-the-art dilution cryostats, while temperatures around a few hundreds of milliKelvins can inform on robustness versus the exact temperature or thermalization issues. The ancilla dephasing rate $\kappa_\phi$ is not explicitly varied in the simulations since it can be taken into account through the renormalization of the engineered dissipation rate $\Gamma$ in \cref{eq:new_gamma_dephasing}.}
  \label{fig:hot_ancilla}
\end{figure}

As expected, the new dissipators are  function of the coupling operators $\LL$
and $\LL^\dag$ appearing in $\HHop^{\mathrm{int}}$.
For $\LL$ matching any of the operators $\LL_k$ in \cref{eq:lindissip},
we see that both $\LL$ and $\LL^\dag$ are stabilizers of the GKP code in the infinite-energy limit
(that is for $\epsilon=0$).
However, it is no longer the case in a finite-energy setting.
Remarkably, we can still compute explicit asymptotic expansions for both the rate
of convergence of the stabilizers $\Gamma_c$ and the logical decoherence rate $\Gamma_L^0$
(see \cref{sm__ssec__eigenvalues}).
More precisely, we consider the full
Lindblad master equation
\begin{equation}
	\frac{d\rhoo}{dt} =
		\tilde \Gamma \, \left(
		\sum_{k=0}^3 \cD[\LL_k](\rhoo) + \tfrac{\nth}{1+\nth} \cD[\LL_k^\dag](\rhoo)
		\right).
		\label{eq:simu_hot_ancilla}
\end{equation}
involving the four dissipators required for the GKP stabilization---with rates reduced by ancilla dephasing---alongside with the corresponding
four spurious dissipators stemming from  thermal excitations of the ancilla.
We show that, under this new dynamics,
the convergence rate of the stabilizers is reduced following
\begin{equation}
	\tilde \Gamma_c \sim \frac{\cA \epsilon \eta \tilde \Gamma}{1+\nth}
			= \Gamma_c \, \frac{\kappa_b}{\kappa_b+\kappa_\phi}.
\end{equation}
Similarly, the logical decoherence rate $\Gamma_L^0$ is replaced by
\begin{equation}
	\tilde \Gamma_L^0 = \frac{4}{\pi} \, \frac{\cA\epsilon\eta \tilde \Gamma}{1+\nth}
				\, e^{-\frac{4}{\cA\epsilon\eta(1+2\nth)}}.
	\label{eq:logical_rate__hot_ancilla}
\end{equation}
These asymptotic formulas are found to be in  good agreement
with direct numerical simulations
of \cref{eq:simu_hot_ancilla}, as shown in
\cref{fig:hot_ancilla}.

A striking feature of our analysis of ancilla noise propagation is that
the logical error rate $\tilde \Gamma_L^0$ in \cref{eq:logical_rate__hot_ancilla}
monotonically decreases as $\epsilon$ goes to zero -- or, equivalently, as the energy of the GKP states
goes to infinity
(in comparison, in \cref{sec:errorcorrec},
the study of low-weight noise channels directly affecting the target mode
revealed the existence of an optimal value of $\epsilon$ as a function of the target mode noise strength).
This feature is in full agreement with the intuitive notion that, in the infinite-energy GKP limit (corresponding to $\epsilon=0$),
ancilla noise does not propagate at all through modular interactions.
It is also numerically confirmed, in the parameter range of \cref{fig:hot_ancilla},
by the fact that the curves associated to different values of $\epsilon$ do not cross.
Crucially, we find that errors stemming from thermal excitations of the ancilla are negligible for typical operating temperatures of superconducting circuits ($T\sim 10-50$~mK).

%% file: appendix.tex
\newpage
\appendix
\addtocontents{toc}{\protect\setcounter{tocdepth}{0}}
\startcontents[supmat]
\printcontents[supmat]{l}{1}{\section*{Appendix}}
\addtocontents{toc}{\protect\setcounter{tocdepth}{3}}

\input{sm__hexagonal_code.tex}

\input{sm__analysis_of_modular_dissipation.tex}

\input{sm__dissipation_engineering.tex}

\input{sm__disorder_and_miscalibration.tex}
\input{sm__analysis__numeric_details.tex}

%% file: sm__hexagonal_code.tex
\section{Error correction by modular dissipation in the hexagonal GKP code}
\label{smsec:hexacode}

One can define GKP grid states associated to lattices in phase-space that are not necessarily
rectangular~\cite{gottesman2001encoding}.
Of particular interest is the hexagonal GKP codespace,
spanned by grid states supported along a hexagonal lattice in phase-space~\cite{grimsmo2021quantum}.
In particular, we will see that thanks to the symmetry of the hexagonal lattice,
eigenstates of the $\X, \Y$ and $\Z$ Pauli operators on the hexagonal GKP codespace
have the same lifetime,
whereas $\Y$ eigenstates decay twice faster than $\X$ or $\Z$ eigenstates
on a square GKP codespace.
The experimental stabilization of hexagonal GKP grid states has already been demonstrated
using stabilizer measurement via low-weight interactions
in superconducting circuits~\cite{campagne2020quantum}
and trapped ions~\cite{de2022error}
platforms.
Let us explain how our dissipative stabilization scheme can be adapted to the stabilization of the
hexagonal GKP codespace.

Similarly to the square case,
we can define
the hexagonal GKP codespace
as the common $+1$ eigenspace of the six commuting stabilizer operators
\begin{equation}
	\SSS_k = e^{i\eta \, \qop(\theta_k)}, \quad 0\leq k \leq 5
\end{equation}
for $\eta = 2\sqrt{\frac{2\pi}{\sqrt 3}}$,
$\theta_k= k\pi/3$
and $\qop(\theta) = e^{i\theta\N}\, \qop \, e^{-i\theta\N}=$ $\cos(\theta)\qop + \sin(\theta)\Pop.$
More precisely, the same codespace could be defined using only $\SSS_0$ and $\SSS_2$,
to which we add their images by successive $\pi/3$ rotations in phase space
to respect the symmetry of the hexagonal grid.
Note that $\eta$ is chosen such that
\(\eta^2 \sin(\pi/3) = 4\pi \)
as before.
The logical coordinates associated to any density operator
are defined as the expectation values of the three \emph{generalized Pauli} operators
\begin{equation}
	\begin{aligned}
		\Z &= \mathrm{Sgn}(\cos(\tfrac\eta2 \qop)),\\
		\X &= \mathrm{Sgn}(\cos(\tfrac\eta2 \qop(\tfrac{2\pi}3))),\\
		\Y &=i \X\Z.
		\end{aligned}
\end{equation}
Note that $\SSS_k^\dag = \SSS_{k+3 \textrm{ mod } 6}$
and $\X,\Y,\Z$
satisfy the Pauli algebra composition rules and commute with the stabilizers.
\\

We can introduce the corresponding finite-energy stabilizers
\begin{equation}
	\SSS_k^\Delta = \E_\Delta \SSS_k \E_\Delta^{-1}
\end{equation}
with $\E_\Delta = e^{-\Delta\oa^\dag\oa}$
and the associated Lindblad operators
\begin{equation}
	\MM_k = \SSS_k^\Delta - \II.
\end{equation}
These Lindblad operators being once again a combination of trigonometric and hyperbolic
functions of $\qop$ and $\Pop$,
we approximate them to first order in $\Delta$ by products of
trigonometric and linear functions of $\qop$ and $\Pop$ as
\begin{equation}
	\LL_k = \cA \Rop_{\frac{k\pi}3} \, e^{i\eta\qop}
		\left( \II - \epsilon\Pop\right) \Rop_{\frac{k\pi}3}^\dag - \II
	\label{eq:dissip_hexa_expression}
\end{equation}
with $\epsilon=\eta\sinh\Delta$ and $\cA = e^{-\epsilon\eta/2}$.

Finally, we propose to stabilize the hexagonal GKP codespace
using the following Lindblad-type dynamics with $6$ dissipators:
\begin{equation}
	\label{eq__lindblad_hexa_GKP}
	\frac{d}{dt}\rho =
	\Gamma
	\sum_{k=0}^5 \LL_k \rho \LL_k^\dag - \frac12\left(
		\LL_k^\dag\LL_k \rho + \rho\LL_k^\dag\LL_k\right).
\end{equation}
Formally, the only differences with the dynamics stabilizing the square GKP codespace
is that $\eta = 2\sqrt{\frac{2\pi}{\sqrt3}}$ (instead of $\eta = 2\sqrt\pi$),
we now use $6$ dissipators (instead of $4$),
related to each other by repeated rotations of $\pi/3$ (instead of $\pi/2$) in phase-space.

Crucially, the method proposed in \cref{sec:dissipationengineering}
for the engineering of the modular dissipators stabilizing the square GKP codespace
can be straightforwardly adapted to engineer these $6$ new dissipators.
Indeed, in both cases, we describe how to engineer one of the required dissipators;
the engineering of the remaining three (square case) or five (hexagonal case)
is easily deduced therefrom
(see~\cref{sec:dissipationengineering}, or~\cref{sm:sec__rwa} for more details).
\\

\begin{figure}[htbp]
	\centering
	\includegraphics[width=1\columnwidth]{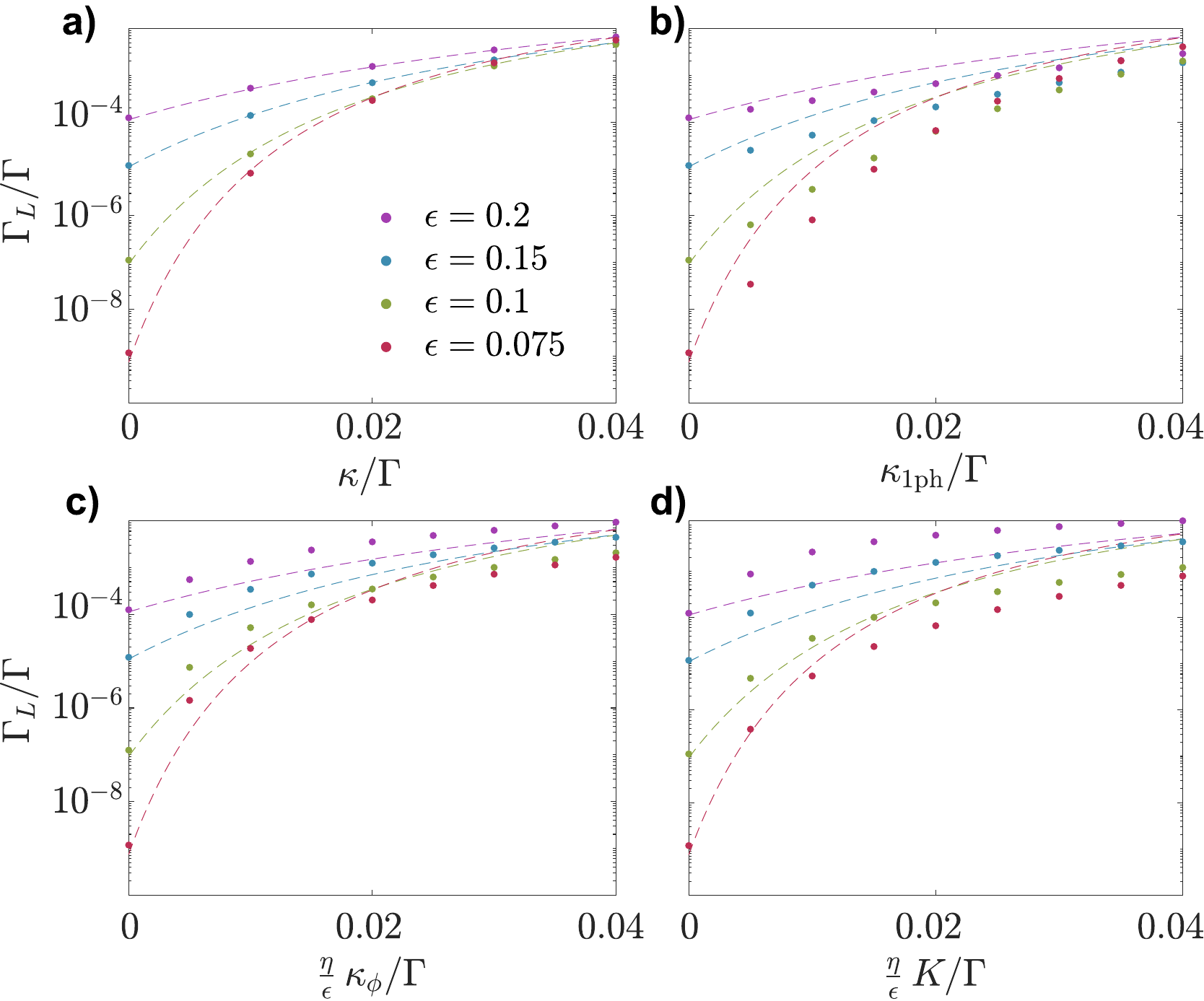}
	\caption{%
	{\bf GKP qubit protection by modular dissipation, hexagonal code}. %
	Logical error rates are extracted from numerical simulations (dots) %
	when varying the strength of some intrinsic noise channel %
	relative to  the modular dissipation rate $\Gamma$. %
	For all low-weight noise channels considered, %
	errors appear to be exponentially suppressed in the weak noise limit. %
	{\bf a)} Quadrature noise modeled by two Lindblad operators %
	$\sqrt{\kappa}\qop$ %
	and %
	$\sqrt{\kappa}\Pop$. %
	Dashed lines are predictions by %
	spectral analysis of the Lindblad superoperator (see~\cref{sm__ssec__eigenvalues}). %
	{\bf b)} Single-photon dissipation modeled by a Lindblad operator %
	$\sqrt{\kappa_{1\mathrm{ph}}}\aop$. %
	{\bf c)} Pure dephasing modeled by a  Lindblad operator %
	$\sqrt{\kappa_{\phi}}\aop^{\dag}\aop$. %
	{\bf d)} Kerr Hamiltonian perturbation of the form %
	$\frac{K}{2}(\aop^{\dag}\aop)^2$. %
	For (c-d), note the rescaling of the x-axis by %
	$2\bar{n}=\eta/\epsilon$. %
	For (b-d), dashed gray lines reproduce the dashed colored lines in (a), %
	un-rescaled, for comparison.%
	}
	\label{fig:gammavskappa_hexa}
\end{figure}

We can numerically compute the effective logical error rates
induced by additional low-weight noise channels entering the Lindblad dynamics
\cref{eq__lindblad_hexa_GKP}.
In \cref{fig:gammavskappa_hexa}, for typical noise channels,
we represent the logical error rate
extracted by spectral analysis of the Lindblad superoperator (dashed lines)
(see \cref{sm__ssec__eigenvalues}),
in quantitative agreement with a full
Lindblad master equation simulation (dots).
We observe results qualitatively similar to that of \cref{fig:gammavskappa}
(corresponding to the same comparison for the square case).
Note that the asymptotic logical error rates in the small noise regime
appear to be lower in these simulations
than the corresponding logical error rates in the square GKP simulations
presented in \cref{fig:gammavskappa}.
However, in realistic physical implementations, this effect would be partly compensated by
the fact that the dissipators appearing
in the Lindblad dynamics of~\cref{eq__lindblad_hexa_GKP}
would be activated sequentially to leverage a single ancillary mode (see \cref{sec:dissipationengineering}).
With this strategy, the effective modular dissipation rate $\Gamma$ is divided by the number
of dissipators to engineer, and is thus weaker by a factor $4/6$ in the hexagonal case.

%% file: sm__analysis_of_modular_dissipation.tex
\section{Analytical and numerical analysis of the modular dissipation}

\input{sm__analysis__eigenvalues.tex}

\input{sm__analysis__energybounds.tex}

%% file: sm__analysis__eigenvalues.tex
\subsection{Exponential convergence to the code manifold and explicit decoherence rates}
\label{sm__ssec__eigenvalues}

If one were able to directly engineer the Lindblad operators $\MM_k$ of~\cref{sec:modular dissip},
involving both trigonometric and hyperbolic functions of the quadrature operators $\qop$ and
$\Pop$,
it was shown in~\cite{sellem2022exponential} that the resulting Lindblad dynamics
would stabilize exactly the finite-energy GKP codespace,
with exponential convergence of any initial state towards the codespace.
This is no longer true with the approximate Lindblad operators $\LL_k$ in our proposal;
most notably, the approximate Lindblad operators fail to perfectly vanish on the GKP codespace,
which is consequently only metastable under our stabilization scheme.
In other words, even without any additional dissipation channel,
the encoded qubit suffers from \emph{intrinsic} residual logical decoherence.
We are able to explicitly compute the associated decoherence rates,
and the evolution of the encoded logical qubit,
without solving the Lindblad equation.
Additionally, we are able to extend this result to the case where additional dissipation
is added to the dynamics in the form of quadrature noise only.

Indeed, for both the square and the hexagonal GKP code,
the coordinates of the encoded logical qubit
are defined as expectation values of the generalized Pauli operators
$\X,\Y,\Z$.
Crucially, these operators are separable periodic observables of the form
\begin{equation}
	\label{eq:coperiodic_observables}
	\hop = h(\qop_1,\qop_2) = f(\tfrac\eta2\qop_1) \, g(\tfrac\eta2\qop_2)
\end{equation}
where $f$ and $g$ are real-valued $2\pi$-periodic functions
and
$\eta=2\sqrt\pi$, $\qop_1 = \qop$, $\qop_2 = \qop(\pi/2)=\Pop$ for the square GKP code,
while
$\eta=2\sqrt{\frac{2\pi}{\sqrt3}}$, $\qop_1 = \qop$,
$\qop_2 = \qop(2\pi/3)= \cos(2\pi/3)\qop + \sin(2\pi/3)\Pop = -\frac12\qop + \frac{\sqrt3}2\Pop$
for the hexagonal GKP code.
In the two cases,
\begin{equation}
	\left[ \eta\, \qop_1, \frac\eta2\, \qop_2 \right]
	= \left[ \frac\eta2\, \qop_1, \eta\, \qop_2 \right] = 2i\pi
\end{equation}
so that
by expanding the periodic functions $f$ and $g$
in Fourier series
and applying
the Baker-Campbell-Hausdorff formula
we get
\begin{equation}
	\label{eq:bch_eigen}
\left[f(\tfrac\eta2\qop_1), \, e^{i\eta\qop_2} \right]
= \left[g(\tfrac\eta2\qop_2), \, e^{i\eta\qop_1} \right]
=0.
\end{equation}

In the following subsections,
we show that,
for a density operator $\rho$
governed by the Lindblad equation proposed earlier to stabilize
the square (or hexagonal) GKP code,
we can leverage~\cref{eq:bch_eigen}
to compute explicitly the evolution of the average value
\( \trace(\hop\rho_t) \),
and thus in particular the evolution of the three generalized Pauli operators
$\X,\Y,\Z$,
without computing the solution $\rho_t$ of the Lindblad equation.
More precisely, we show that
\begin{equation}
	\label{eq:heisenberg_evolution}
	\frac d{dt} \trace(\hop\rho_t) = -\mathcal A \epsilon\eta\Gamma \,
	\trace(\mathcal L_\sigma(h)(\qop_1,\qop_2) \rho_t)
\end{equation}
where
$\mathcal A = e^{-\frac{\epsilon\eta}2}$,
$\Gamma>0$ is the engineered dissipation rate,
$\sigma>0$ is a parameter depending on $\eta,\epsilon$
and $\mathcal L_\sigma$ is an explicit differential operator on periodic functions
which depends on $\sigma$ and the geometry (square or hexagonal).
$\mathcal L_\sigma$ is diagonalizable
and has a positive and discrete spectrum $(\lambda_{k,\sigma})_{k\in\mathbb N}$.
As a consequence, when $h$ is an eigenfunction of $\mathcal L_\sigma$
associated to a given eigenvalue $\lambda$,
\cref{eq:heisenberg_evolution}
leads to
\begin{equation}
	\trace(\hop\rho_t) = e^{-\mathcal A\epsilon\eta\Gamma\,  \lambda \, t}
	\trace(\hop\rho_0).
\end{equation}
In the general case,
\begin{equation}
	\label{eq:evolution_observable_eigenvalues}
	\trace(\hop\rho_t) =
		\sum_{k=0}^{+\infty} c_k e^{-\mathcal A\epsilon\eta\Gamma \, \lambda_{k,\sigma} \, t}
		\trace(h_{k,\sigma}(\qop_1,\qop_2)\rho_0)
\end{equation}
where $h_{k,\sigma}$ is the eigenfunction of $\mathcal L_\sigma$
associated to the eigenvalue $\lambda_{k,\sigma}$
and $h = \sum_k c_k h_k$.
Computing the spectrum of $\mathcal L_\sigma$ is thus sufficient to
study the evolution of the coordinates of the encoded logical qubit,
and hence to derive decoherence rates as a function of $\sigma$.
The precise derivation of the differential operator $\mathcal L_\sigma$,
along with an asymptotic analysis of its spectral properties in the regime $\sigma \ll 1$,
can be found in the separate publication~\cite{sellem2023ifac}.
We recall the main results here for the sake of completeness.
We then explain how to go beyond that asymptotic analysis by numerically
computing the spectrum of $\mathcal L_\sigma$ for a given finite value of $\sigma>0$.

Remarkably, this spectral analysis can be adapted to take into account two
specific sources of noise:
quadrature noise affecting the system,
\emph{i.e.} additional dissipators in $\qop$ and $\Pop$ entering the Lindblad dynamics,
and additional dissipative terms obtained in \cref{sec:hot_ancilla}
when considering the effect of a noisy ancilla
in the dissipation engineering protocol.
In both cases, we show that
\cref{eq:heisenberg_evolution}
still holds true
with renormalized values of $\sigma$ and $\Gamma$
depending on the strength of the new dissipators.
Unfortunately, for other types of noise (namely photon-loss, dephasing and Kerr nonlinearities),
this shortcut,
whose proof depended on commutation relations between the Lindblad operators in the dynamics
and periodic observables,
no longer applies.
In these cases, we can only rely on a full simulation of the Lindblad dynamics.
We still find that these decoherence channels result in errors rates qualitatively similar
to those previously computed with our eigenvalue analysis in the case of quadrature noise
(see~\cref{fig:gammavskappa} and~\cref{fig:gammavskappa_hexa}).

These results are to be put into perspective
with the initial intuition behind the GKP bosonic code:
logical states, although they are designed for robustness against small phase-space shifts
thanks to the separation of their support in phase-space,
should also be robust to noise processes involving
only small polynomials in $\oa$ and $\oa^\dag$,
since the effects of such processes over a small time-interval can be approximated as
combinations of small phase-space shifts~\cite{gottesman2001encoding}.

\subsubsection{Square GKP}
\label{sec:eigen_square}
The generalized Pauli operators of the square GKP code
are defined from the code stabilizers
$\SSS_q = e^{i\eta\qop}$ and $\SSS_p=e^{-i\eta\Pop}$
(where $\eta=2\sqrt\pi$)
by
\begin{equation}
	\begin{split}
		\Z&
			= \mathrm{Sgn}(\cos(\tfrac{\eta}2 \qop)),\\
		\X&
			= \mathrm{Sgn}(\cos(\tfrac\eta2 \Pop)),\\
		\Y&= i \X \Z
	\end{split}
\end{equation}
which are products of periodic functions of $\qop$ and $\Pop$.

In the main text, we proposed to stabilize a finite-energy square GKP codespace
by engineering a dissipative Lindblad dynamics
\begin{equation}
	\label{eq:eigs_lindblad_square}
	\frac d{dt}\rho = \Gamma \, \sum_{k=0}^3 \cD[\LL_k](\rho)
\end{equation}
where
$\Gamma>0$ and
$\epsilon>0$ are parameters,
$\cA = e^{-\epsilon\eta/2}$,
\(
	\cD[\LL](\rho) = \LL\rho\LL^\dag - \frac12\left( \LL^\dag\LL \rho + \rho\LL^\dag\LL\right)
\)
and
\begin{equation}
	\LL_k = \cA
		e^{ik\tfrac{\pi}{2}\N}
		\, e^{i\eta\qop} \left( \II-\epsilon\Pop\right) \,
		e^{-ik\tfrac{\pi}{2}\N}
		- \II.
	\tag{\ref{eq:lindissip}}
\end{equation}

Let us now compute the evolution under~\cref{eq:eigs_lindblad_square}
of any
separable periodic observables of the form
\begin{equation}
	\hop = h(\qop,\Pop) = f(\tfrac\eta2\qop) \, g(\tfrac\eta2\Pop)
\end{equation}
where $f$ and $g$ are real-valued $2\pi$-periodic functions:
\begin{equation}
	\begin{aligned}
		\frac d{dt} \trace(\hop\rho_t)
		&= \trace(\hop \frac d{dt}\rho_t)\\
		&= \Gamma \sum_{k=0}^3 \trace\left( \hop \, \cD[\LL_k](\rho_t) \right)\\
		&= \Gamma \sum_{k=0}^3 \trace\left( \cD^*[\LL_k](\hop) \, \rho_t \right)
	\end{aligned}
\end{equation}
where
\begin{equation}
	\begin{aligned}
	\cD^*[\LL_k](\hop) :&=
		\LL_k^\dag \, \hop \, \LL_k
		-\frac12\left( \LL_k^\dag\LL_k \, \hop + \hop \, \LL_k^\dag \LL_k\right)\\
		&= \frac12 \left( \LL_k^\dag \, [\hop, \LL_k] + [\LL_k^\dag, \hop] \, \LL_k \right).
	\end{aligned}
\end{equation}
Using the relations of~\cref{eq:bch_eigen}
(along with the relation
\( [f({\bf A}), {\bf B}] = f'({\bf A}) \, [{\bf A}, {\bf B}] \)
whenever
\( [ {\bf A}, \, [{\bf A}, {\bf B}]] =0\)%
)
one is able to show that
\begin{equation}\label{eq:lindblad_heisenberg}\begin{aligned}
	\Gamma \sum_{k=0}^3 & \cD^*[\LL_k](\hop)\\
	=&
	-\mathcal A\epsilon\eta \Gamma \,
		\left(
		\sin(\eta\qop) f'(\tfrac\eta2\qop)
		- \frac{\mathcal A\epsilon\eta}4 f''(\tfrac\eta2\qop) \right) g(\tfrac\eta2\Pop)
	\\
	&-\mathcal A\epsilon\eta \Gamma \,
		f(\tfrac\eta2\qop)\left(
		\sin(\eta\Pop) g'(\tfrac\eta2\Pop)
		- \frac{\mathcal A\epsilon\eta}4 g''(\tfrac\eta2\Pop) \right)
	\\
	=& -\mathcal A\epsilon\eta \Gamma \,
	\mathcal L_\sigma (h)(\tfrac\eta2\qop, \tfrac\eta2\Pop)
\end{aligned}\end{equation}
where $\sigma = \frac{\mathcal A\epsilon\eta}4$
and $\mathcal L_\sigma$ is the differential operator defined by
\begin{equation}\begin{aligned}
	\label{eq:lsigmatensor}
	\mathcal L_\sigma &= \mathcal T_\sigma\otimes \II + \II\otimes \mathcal T_\sigma,\\
	(\mathcal T_\sigma f) (\theta) &= \sin(2\theta) f'(\theta) - \sigma f''(\theta)
\end{aligned}\end{equation}
(we refer to~\cite{sellem2023ifac} for the detailed computation).
\\
A straightforward computation shows, through integration by parts,
that for any $2\pi$-periodic functions $f$ and $g$:
\begin{equation}\label{eq:tsigma_sym}
	\begin{aligned}
	\langle f, \, \mathcal T_\sigma g \rangle_\sigma
		&= \langle \mathcal T_\sigma f, \, g\rangle_\sigma\\
	\langle f, \, \mathcal T_\sigma f \rangle_\sigma
		&= \sigma \, \langle f', \, f'\rangle_\sigma
	\end{aligned}
\end{equation}
for the scalar product $\langle \cdot, \cdot \rangle_\sigma$ defined as
\begin{equation}
	\langle f, g\rangle_\sigma =
		\int_{-\pi}^\pi e^{-\frac{1-\cos(2\theta)}{2\sigma}} f(\theta)^* g(\theta) d\theta.
\end{equation}
\cref{eq:tsigma_sym} can also be obtained by observing that the following relation holds:
\begin{equation}\label{eq:div_formula_lsigma}
	w_\sigma \mathcal T_\sigma(f) = - \sigma \left( w_\sigma f'\right)'
\end{equation}
where we introduced a weight function
\(
w_\sigma(\theta) = e^{-\frac{1-\cos(2\theta)}{2\sigma}}\)
(note that $\langle f,g\rangle_\sigma = \int w_\sigma f^* g$).
As a consequence, $\mathcal T_\sigma$ is self-adjoint for the scalar product
$\langle \cdot,\cdot\rangle_\sigma$
and non-negative;
classical results show that its spectrum is thus discrete, real and non-negative
(see \emph{e.g.} \cite{ZettlBook2005}).
Moreover,
using~\cref{eq:lsigmatensor},
we can easily deduce the eigenvalues of $\mathcal L_\sigma$
from that of $\mathcal T_\sigma$:
if
\begin{equation}
	\mu_{0,\sigma} \leq \mu_{1,\sigma} \leq \mu_{2,\sigma} \,  \ldots
\end{equation}
denote the eigenvalues of $\mathcal T_\sigma$ sorted in ascending order,
where each $\mu_{k,\sigma}$ is associated to an eigenfunction $f_{k,\sigma}$,
then the eigenvalues of $\mathcal L_\sigma$ are the sums
\begin{equation}
	\lambda_{k_1,k_2,\sigma} = \mu_{k_1,\sigma} + \mu_{k_2,\sigma},
\end{equation}
associated to the eigenfunctions
\begin{equation}
	\label{eq:eigfunc_sums}
	h_{k_1,k_2,\sigma} = f_{k_1,\sigma} \otimes f_{k_2,\sigma}.
\end{equation}
It is thus sufficient to compute the eigenvalues and eigenfunctions of $\mathcal T_\sigma$.
\\

Note that $\mu_{0,\sigma}=0$ is always an eigenvalue of $\mathcal T_\sigma$
associated to the eigenfunction $f \equiv 1$,
so that $\lambda_{0,\sigma}=0$ is an eigenvalue of $\mathcal L_\sigma$
associated to the eigenfunction $h\equiv 1\otimes 1 = 1$.
This simply correspond to the trivial identity
\begin{equation}
	\frac d{dt} \trace(\rho_t) = 0.
\end{equation}

In the limit $\sigma\rightarrow 0$,
the other eigenvalues of $\mathcal T_\sigma$ have been rigorously studied
in~\cite{sellem2023ifac} (using mathematical results from~\cite{michelSmallEigenvalues2019a}),
where the following asymptotics estimates were obtained:
\begin{equation}
	\label{eq:asymptotics_michel}
	\begin{aligned}
		\mu_{1,\sigma} &\sim \frac 4\pi e^{-1/\sigma},\\
		\mu_{2,\sigma} &\geq \upepsilon_0 > 0
	\end{aligned}
\end{equation}
where $\upepsilon_0$ is a fixed constant independent of $\sigma$.

In particular, for small enough $\sigma$, $2\mu_{1,\sigma} \ll \mu_{2,\sigma}$
so that the first eigenvalues of $\mathcal L_\sigma$ read:
\begin{equation}
	\label{eq:yworsefromeigenvalues}
	\begin{aligned}
		\lambda_{0,\sigma} &= \lambda_{0,0,\sigma} = 0,\\
		\lambda_{1,\sigma} &= \lambda_{0,1,\sigma} = 0+\mu_{1,\sigma}
			= \mu_{1,\sigma},\\
		\lambda_{2,\sigma} &= \lambda_{1,0,\sigma} = \mu_{1,\sigma}+0
			= \mu_{1,\sigma},\\
		\lambda_{3,\sigma} &= \lambda_{1,1,\sigma} = 2 \, \mu_{1,\sigma},
	\end{aligned}
\end{equation}
where the notation with single indices $\lambda_{k,\sigma}$
denotes eigenvalues of $\mathcal L_\sigma$ sorted in ascending order,
whereas the notation with doubles indices $\lambda_{k_1,k_2,\sigma}$
indicates how they are obtained as sums of eigenvalues of $\mathcal T_\sigma$.
\\

\begin{figure}[htbp]
	\centering
	\includegraphics[width=0.9\columnwidth]{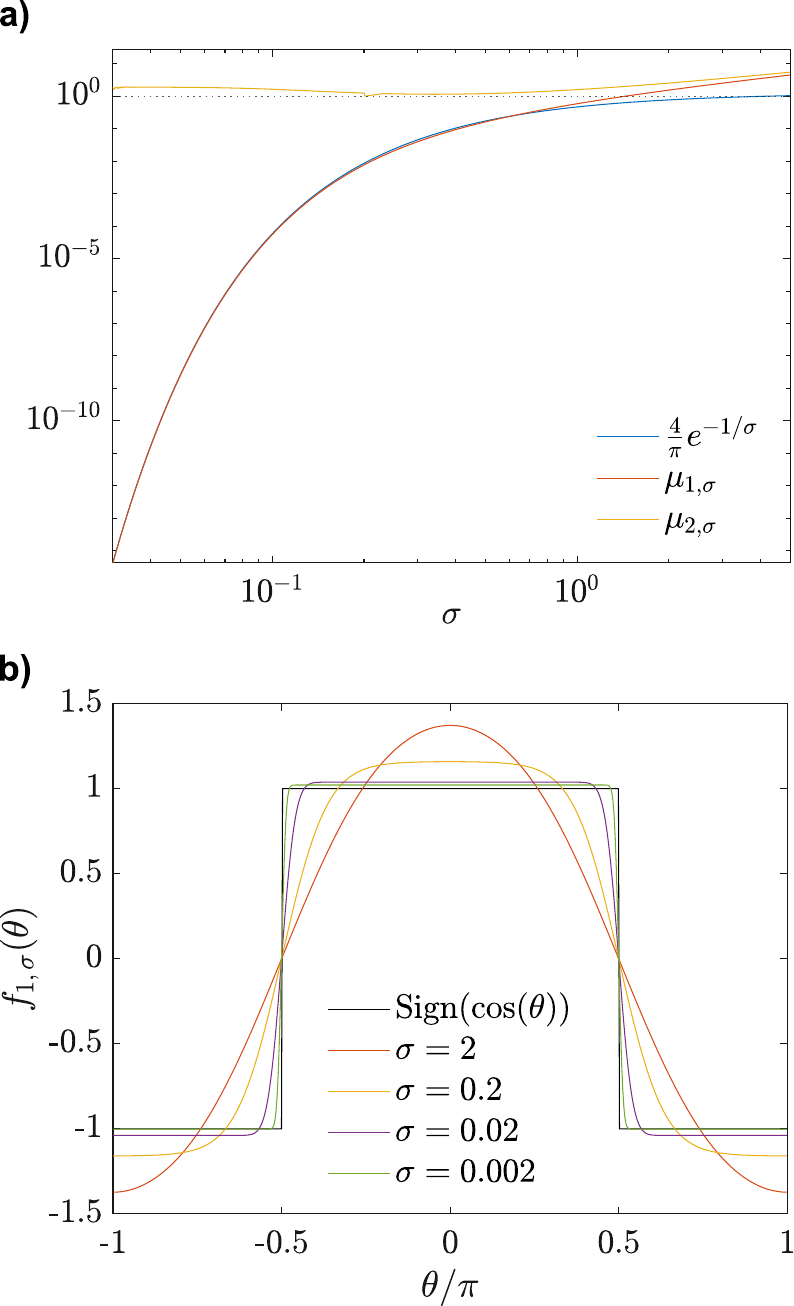}
	\caption{{\bf Spectral analysis of $\mathcal T_\sigma$ for the square GKP code}. %
	{\bf{a)}} First two non-zero eigenvalues of $\mathcal T_\sigma$
	where $\sigma = \frac{\mathcal A\epsilon\eta}4 = \frac{\epsilon\eta}4 e^{-\frac{\epsilon\eta}2}$.
	In the limit of small $\sigma$,
	we observe that the first non-zero eigenvalue
	$\mu_{1,\sigma}$ is well-approximated by the theoretical asymptotic value
	$\mu_{1,\sigma} = \frac4\pi \, e^{-1/\sigma}$
	obtained in~\cite{sellem2023ifac},
	while the second non-zero eigenvalue
	$\mu_{2,\sigma}$ is lower-bounded by $1$,
	opening an exponential spectral gap.
	{\bf{b)}} Numerically computed eigenfunction $f_{1,\sigma}$ associated to the first
	non-zero eigenvalue $\mu_{1,\sigma}$ for various values of $\sigma$.
	In the limit of small $\sigma$,
	we observe that $f_{1,\sigma}(\theta)\rightarrow \sign(\cos(\theta))$.
	In both figures, we used $K_{\mathrm{max}}=200$
	(corresponding to $2K_{\mathrm{max}}+1 = 401$ Fourier modes)
	and numerically checked that higher truncations do not change the results.
	}
	\label{fig:numerics_eigenvalues}
\end{figure}

Finally, apart from these asymptotic results,
we can turn to numerical diagonalization of $\mathcal T_\sigma$
to compute the exact value of $\lambda_{k,\sigma}$
for a fixed value of $\sigma$.
To that end, note that $\mathcal T_\sigma$ has a simple expression in the Fourier frame:
indeed, writing any (smooth) $2\pi$-periodic function $f$ as
\begin{equation}
	f(\theta) = \sum_{k\in\mathbb Z} f_k e^{ik\theta},
\end{equation}
we get
\begin{equation}
	\begin{aligned}
		(\mathcal T_\sigma f) (\theta)
			&= \sin(2\theta) f'(\theta) - \sigma f''(\theta)\\
			&= \sum_{k\in\mathbb Z} \left(
				ik \, \frac{e^{2i\theta} - e^{-2i\theta}}{2i}
				+ \sigma k^2\right) \, f_k \, e^{ik\theta}\\
			&= \sum_{k\in\mathbb Z} \left(
				\frac{k-2}2 f_{k-2} - \frac{k+2}2 f_{k+2} + \sigma k^2 f_k
				\right) e^{ik\theta}.
	\end{aligned}
\end{equation}
In the Fourier frame, $\mathcal T_\sigma$ can thus be approximated by a tridiagonal
matrix by truncating the sum above
to $-K_{\textrm{max}} \leq k \leq K_{\textrm{max}}$
(\emph{i.e.} bounding the maximum frequency
considered).
On~\cref{fig:numerics_eigenvalues} we compare the numerical eigenvalues obtained
by this procedure
to the explicit asymptotic expressions of~\cref{eq:asymptotics_michel},
together with the numerical eigenfunction $f_{1,\sigma}$
corresponding to the first non-zero eigenvalue $\mu_{1,\sigma}$ for different values of $\sigma$.
We observe that
\begin{equation}
	\label{eq:lim_f_1}
	f_{1,\sigma}(\theta) \xrightarrow[\sigma\rightarrow 0]{} \sign(\cos(\theta)).
\end{equation}
Hence, combining~\cref{eq:eigfunc_sums,eq:yworsefromeigenvalues,eq:lim_f_1}
(and estimating $\upepsilon_0$ from the numerically computed eigenvalues)
we obtain that the first eigenvalues and eigenfunctions of $\mathcal L_\sigma$
in the limit of small $\sigma$
are given by

\begin{equation*}
	\begin{aligned}
		\lambda_{0,\sigma} &= 0
		&&
			h_{0,\sigma}(\theta_1,\theta_2) = 1\\
		\lambda_{1,\sigma} &= \mu_{1,\sigma} \sim \frac4\pi e^{-1/\sigma}
		&&
			h_{1,\sigma}(\theta_1,\theta_2) = f_{1,\sigma}(\theta_1)\\
			& && \quad \quad \quad \quad \;
			\simeq \sign(\cos(\theta_1))\\
		\lambda_{2,\sigma} &= \mu_{1,\sigma} \sim \frac4\pi e^{-1/\sigma}
		&&
			h_{2,\sigma}(\theta_1,\theta_2) = f_{1,\sigma}(\theta_2)\\
			& && \quad \quad \quad \quad \;
				\simeq \sign(\cos(\theta_2))\\
		\lambda_{3,\sigma} &= 2 \, \mu_{1,\sigma} \sim \frac8\pi e^{-1/\sigma}
		&&
			h_{3,\sigma}(\theta_1,\theta_2)
			= f_{1,\sigma}(\theta_1) f_{1,\sigma}(\theta_2) \\
			& &&
			\simeq \sign(\cos(\theta_1)) \sign(\cos(\theta_2))
	\end{aligned}
\end{equation*}
\begin{equation}
	\label{eq:eigenvalues_and_functions_l1sigma_square}
	\begin{aligned}
		\lambda_{n,\sigma} &\geq \lambda_{4,\sigma} = \mu_{2,\sigma}
		\geq \upepsilon_0 \gtrsim 1
		&& \forall n\geq 4.
	\end{aligned}
\end{equation}
Crucially, coming back to periodic observables on the stabilized GKP qubit,
\cref{eq:eigenvalues_and_functions_l1sigma_square}
leads to the three observables
\begin{equation}
\begin{aligned}
	\hop_{1,\sigma}
		= h_{1,\sigma}(\tfrac\eta2 \qop,\tfrac\eta2\Pop)
		&\simeq \sign(\cos(\tfrac\eta2\qop)) = \Z\\
	\hop_{2,\sigma}
		= h_{2,\sigma}(\tfrac\eta2 \qop,\tfrac\eta2\Pop)
		&\simeq \sign(\cos(\tfrac\eta2\Pop)) = \X\\
	\hop_{3,\sigma}
		= h_{3,\sigma}(\tfrac\eta2 \qop,\tfrac\eta2\Pop)
		&\simeq \sign(\cos(\tfrac\eta2\qop))\, \sign(\cos(\tfrac\eta2\Pop))\\
		& = \Z\X = i\Y.
\end{aligned}
\end{equation}
\\
We can thus interpret
$\lambda_{1,\sigma}, \lambda_{2,\sigma}$ and $\lambda_{3,\sigma}$
as the decay rates of the logical coordinates associated to
$\Z,\X$ and $\Y$,
which are exponentially small in $1/\sigma$.
On the other hand,
due to the constant gap between $\lambda_{3,\sigma}$ and $\lambda_{4,\sigma}$,
all the contributions due to higher eigenvalues of $\mathcal L_\sigma$
vanish exponentially faster than the decay rates of $\X,\Y,\Z$.
In other words,
using~\cref{eq:evolution_observable_eigenvalues}
we can decompose the evolution of the expectation value $\trace(\hop\rho_t)$ of a separable
periodic observable $\hop = h(\tfrac\eta2\qop,\tfrac\eta2\Pop)$ into two stages:
\begin{enumerate}
	\item a fast transient on a typical timescale
$\tau_{conv} = \left( \mathcal A \epsilon\eta\Gamma \, \lambda_{4,\sigma} \right)^{-1}
\lesssim \left( \mathcal A \epsilon\eta\Gamma \right)^{-1}$
corresponding to the decay of the projection of $h$ onto
eigenfunctions of $\mathcal L_\sigma$ associated to eigenvalues $\lambda_{n,\sigma}$ with $n\geq4$.
We can consider this transient regime as a form of \emph{convergence to the codespace}
(even though the density operator $\rho_t$ need not converge to a two-dimensional subspace)
after which only the expectation value of $\X, \Y, \Z$
retain  memory of the prior state of the oscillator;
\item a slow decay of $\X, \Y, \Z$ on a typical timescale
	$\tau_{log} = \left( \mathcal A\epsilon\eta\Gamma \, \lambda_{1,\sigma} \right)^{-1}
		\simeq \frac\pi4\left( \mathcal A\epsilon\eta\Gamma \right)^{-1} e^{\frac{1}{\sigma}}
		\gg \tau_{conv}.$
\end{enumerate}
Note that through~\cref{eq:yworsefromeigenvalues}, we recover the fact that,
with our dissipative scheme for the stabilization of a square GKP code,
the expectation value of $\Y$ decays exactly twice faster than those of $\X$ and $\Z$.

\subsubsection{Hexagonal GKP}
\label{sec:eigen_hexa}
Let us detail how the eigenvalue analysis led in~\cite{sellem2023ifac}
is generalized to the hexagonal case.
Recall from~\cref{smsec:hexacode} the definition of
$\eta = 2\sqrt{\frac{2\pi}{\sqrt3}}$;
and the two adapted quadrature operators
$\qop_1=\qop, \qop_2 = \cos(2\pi/3)\qop + \sin(2\pi/3)\Pop = -\tfrac12 \qop + \tfrac{\sqrt3}2 \Pop$,
along with the corresponding generalized Pauli operators:
\begin{equation}
	\begin{split}
		\Z&
			= \mathrm{Sgn}(\cos(\tfrac{\eta}2 \qop_1)),\\
		\X&
			= \mathrm{Sgn}(\cos(\tfrac\eta2 \qop_2)),\\
		\Y&= i \X \Z
	\end{split}
\end{equation}
which are indeed products of periodic functions of $\qop_1$ and $\qop_2$.

In~\cref{smsec:hexacode}, we proposed to stabilize a finite-energy hexagonal GKP codespace
by engineering a dissipative Lindblad dynamics
\begin{equation}
	\label{eq:eigs_lindblad_hexa}
	\frac d{dt}\rho = \Gamma \, \sum_{k=0}^5 \cD[\LL_k](\rho)
	\tag{\ref{eq__lindblad_hexa_GKP}}
\end{equation}
where
$\Gamma>0$ and
$\epsilon>0$ are parameters,
$\cA = e^{-\epsilon\eta/2}$,
\(
	\cD[\LL](\rho) = \LL\rho\LL^\dag - \frac12\left( \LL^\dag\LL \rho + \rho\LL^\dag\LL\right)
\)
and
\begin{equation}
	\LL_k = \cA
		e^{ik\tfrac{\pi}{3}\N}
		\, e^{i\eta\qop} \left( \II-\epsilon\Pop\right) \,
		e^{-ik\tfrac{\pi}{3}\N}
		- \II.
		\tag{\ref{eq:dissip_hexa_expression}}
\end{equation}

Let us now compute the evolution under~\cref{eq:eigs_lindblad_hexa}
of any
separable periodic observables of the form
\begin{equation}
	\hop = h(\qop_1,\qop_2) = f(\tfrac\eta2\qop_1) \, g(\tfrac\eta2\qop_2)
\end{equation}
where $f$ and $g$ are real-valued $2\pi$-periodic functions:
\begin{equation}
	\begin{aligned}
		\frac d{dt} \trace(\hop\rho_t)
		&= \trace(\hop \frac d{dt}\rho_t)\\
		&= \Gamma \sum_{k=0}^5 \trace\left( \hop \, \cD[\LL_k](\rho_t) \right)\\
		&= \Gamma \sum_{k=0}^5 \trace\left( \cD^*[\LL_k](\hop) \, \rho_t \right)
	\end{aligned}
\end{equation}
where
\begin{equation}
	\begin{aligned}
	\cD^*[\LL_k](\hop) &=
		\LL_k^\dag \, \hop \, \LL_k
		-\frac12\left( \LL_k^\dag\LL_k \, \hop + \hop \, \LL_k^\dag \LL_k\right)\\
		&= \frac12 \left( \LL_k^\dag \, [\hop, \LL_k] + [\LL_k^\dag, \hop] \, \LL_k \right).
	\end{aligned}
\end{equation}
Using the relations of~\cref{eq:bch_eigen} together with
$[\qop_1,\Pop] = i$,
$[\qop_2,\Pop] = -\frac i2$,
$[\qop_1,\qop] = 0$
and
$[\qop_2,\qop] = -\frac{\sqrt3}2i$,
we obtain for instance

\begin{widetext}
	\vspace{2em}
\begin{equation}\begin{aligned}
	\LL_0^\dag \, [\hop, \LL_0]
	&= 
		\LL_0^\dag \, \left[ f(\tfrac\eta2\qop) \, g(\tfrac\eta2\qop_2),
					\cA e^{i\eta\qop}(\II-\epsilon\Pop) \right]\\
	&= \cA \, \LL_0^\dag f(\tfrac\eta2\qop) \,
		\left[ g(\tfrac\eta2\qop_2), e^{i\eta\qop}(\II-\epsilon\Pop) \right]
	  + \cA \, \LL_0^\dag \, \left[ f(\tfrac\eta2\qop), e^{i\eta\qop}(\II-\epsilon\Pop) \right]
	  	\, g(\tfrac\eta2\qop_2)\\
	&= -\epsilon \cA \, \LL_0^\dag \, e^{i\eta\qop} f(\tfrac\eta2\qop) \,
		\left[ g(\tfrac\eta2\qop_2), \Pop \right]
	  -\epsilon \cA \, \LL_0^\dag \, e^{i\eta\qop} \left[ f(\tfrac\eta2\qop), \Pop \right]
	  	\, g(\tfrac\eta2\qop_2)\\
	&= \frac{i\eta\epsilon\cA}{4} \, \LL_0^\dag \, e^{i\eta\qop} f(\tfrac\eta2\qop) \,
		g'(\tfrac\eta2\qop_2)
	  -\frac{i\eta\epsilon \cA}{2} \, \LL_0^\dag \, e^{i\eta\qop} f'(\tfrac\eta2\qop)
	  	\, g(\tfrac\eta2\qop_2)\\
	&= \frac{i\eta\epsilon\cA}{4} \, \LL_0^\dag \, e^{i\eta\qop}
		\left( f(\tfrac\eta2\qop) g'(\tfrac\eta2\qop_2) \right.
		\left. -2 f'(\tfrac\eta2\qop) g(\tfrac\eta2\qop_2) \right) \\
	&= \frac{i\eta\epsilon\cA}{4} \,
		\left( \cA \left( \II - \epsilon\Pop\right) - e^{i\eta\qop}\right)
		\left( f(\tfrac\eta2\qop) g'(\tfrac\eta2\qop_2) \right.
		\left. -2 f'(\tfrac\eta2\qop) g(\tfrac\eta2\qop_2) \right), \\
\end{aligned}\end{equation}
\begin{equation}\begin{aligned}
[\LL_0^\dag, \hop] \, \LL_0
	&= -\frac{i\eta\epsilon\cA}{4} \,
		\left( f(\tfrac\eta2\qop) g'(\tfrac\eta2\qop_2) \right.
		\left. -2 f'(\tfrac\eta2\qop) g(\tfrac\eta2\qop_2) \right)
		\left( \cA \left( \II - \epsilon\Pop\right) - e^{-i\eta\qop}\right),
\end{aligned}\end{equation}
\clearpage

\begin{equation}\begin{aligned}
\cD^*[\LL_0](\hop)
	&= \frac12 \LL_0^\dag \, [\hop, \LL_0] + \frac12 [\LL_0^\dag, \hop] \, \LL_0\\
	&= -\frac{i\eta\epsilon^2\cA^2}8 \,
		\left[ \Pop, f(\tfrac\eta2\qop) g'(\tfrac\eta2\qop_2) \right.
			\left. -2 f'(\tfrac\eta2\qop) g(\tfrac\eta2\qop_2) \right]\\
	  &\quad\quad\quad
	  -\frac{i\eta\epsilon\cA}8 e^{i\eta\qop}
		\left( f(\tfrac\eta2\qop) g'(\tfrac\eta2\qop_2) \right.
			\left. -2 f'(\tfrac\eta2\qop) g(\tfrac\eta2\qop_2) \right)\\
	  &\quad\quad\quad
	  +\frac{i\eta\epsilon\cA}8
		\left( f(\tfrac\eta2\qop) g'(\tfrac\eta2\qop_2) \right.
			\left. -2 f'(\tfrac\eta2\qop) g(\tfrac\eta2\qop_2) \right) e^{-i\eta\qop}\\
	&= -\frac{\eta^2\epsilon^2\cA^2}{8} \, f'(\tfrac\eta2\qop) g'(\tfrac\eta2\qop_2)
	   +\frac{\eta^2\epsilon^2\cA^2}{32} \, f(\tfrac\eta2\qop) g''(\tfrac\eta2\qop_2)
	   +\frac{\eta^2\epsilon^2\cA^2}{8} \, f''(\tfrac\eta2\qop) g(\tfrac\eta2\qop_2)\\
	  &\quad\quad\quad
	  -\frac{i\eta\epsilon\cA}8 e^{i\eta\qop}
		\left( f(\tfrac\eta2\qop) g'(\tfrac\eta2\qop_2) \right.
			\left. -2 f'(\tfrac\eta2\qop) g(\tfrac\eta2\qop_2) \right)\\
	  &\quad\quad\quad
	  +\frac{i\eta\epsilon\cA}8
		\left( f(\tfrac\eta2\qop) g'(\tfrac\eta2\qop_2) \right.
			\left. -2 f'(\tfrac\eta2\qop) g(\tfrac\eta2\qop_2) \right) e^{-i\eta\qop}\\
	&= \frac{\eta^2\epsilon^2\cA^2}{8} \, \left(
	    	f''(\tfrac\eta2\qop) g(\tfrac\eta2\qop_2)
		- f'(\tfrac\eta2\qop) g'(\tfrac\eta2\qop_2)
	   	+\frac14 f(\tfrac\eta2\qop) g''(\tfrac\eta2\qop_2)
	   \right)\\
	  &\quad\quad\quad
	  -\frac{i\eta\epsilon\cA}8 e^{i\eta\qop}
		\left( f(\tfrac\eta2\qop) g'(\tfrac\eta2\qop_2) \right.
			\left. -2 f'(\tfrac\eta2\qop) g(\tfrac\eta2\qop_2) \right)\\
	  &\quad\quad\quad
	  +\frac{i\eta\epsilon\cA}8 e^{-i\eta\qop}
		\left( f(\tfrac\eta2\qop) g'(\tfrac\eta2\qop_2) \right.
			\left. -2 f'(\tfrac\eta2\qop) g(\tfrac\eta2\qop_2) \right)\\
	&= \frac{\eta^2\epsilon^2\cA^2}{8} \, \left(
	    	f''(\tfrac\eta2\qop) g(\tfrac\eta2\qop_2)
		- f'(\tfrac\eta2\qop) g'(\tfrac\eta2\qop_2)
	   	+\frac14 f(\tfrac\eta2\qop) g''(\tfrac\eta2\qop_2)
	   \right)\\
	  &\quad\quad\quad
	  + \frac{\eta\epsilon\cA}4 \sin(\eta\qop)
		\left( f(\tfrac\eta2\qop) g'(\tfrac\eta2\qop_2) \right.
			\left. -2 f'(\tfrac\eta2\qop) g(\tfrac\eta2\qop_2) \right).
\end{aligned}\end{equation}
Applying a $\pi$-rotation in phase-space to the previous expression, we also obtain
$\cD^*[\LL_3](\hop) = \cD^*[\LL_0](\hop)$, so that
\begin{equation}\begin{aligned}
	\cD^*[\LL_0](\hop) + \cD^*[\LL_3](\hop)
	&= \frac{\eta^2\epsilon^2\cA^2}{4} \, \left(
	    	f''(\tfrac\eta2\qop) g(\tfrac\eta2\qop_2)
		- f'(\tfrac\eta2\qop) g'(\tfrac\eta2\qop_2)
	   	+\frac14 f(\tfrac\eta2\qop) g''(\tfrac\eta2\qop_2)
	   \right)\\
	  &\quad\quad\quad
	  + \frac{\eta\epsilon\cA}2 \sin(\eta\qop)
		\left( f(\tfrac\eta2\qop) g'(\tfrac\eta2\qop_2) \right.
			\left. -2 f'(\tfrac\eta2\qop) g(\tfrac\eta2\qop_2) \right).
\end{aligned}\end{equation}

Similar but slightly tedious computations lead to
$\cD^*[\LL_1](\hop) = \cD^*[\LL_4](\hop)$ with
\begin{equation}\begin{aligned}
	\cD^*[\LL_1](\hop) + \cD^*[\LL_4](\hop)
	&= \frac{\eta^2\epsilon^2\cA^2}{16} \, \left(
	    	f''(\tfrac\eta2\qop) g(\tfrac\eta2\qop_2)
		+2 f'(\tfrac\eta2\qop) g'(\tfrac\eta2\qop_2)
	   	+ f(\tfrac\eta2\qop) g''(\tfrac\eta2\qop_2)
	   \phantom{\frac14}\right)\\
	  &\quad\quad\quad
	  - \frac{\eta\epsilon\cA}2 \sin(\eta\qop+\eta\qop_2)
		\left( f(\tfrac\eta2\qop) g'(\tfrac\eta2\qop_2) \right.
			\left. + f'(\tfrac\eta2\qop) g(\tfrac\eta2\qop_2) \right),
\end{aligned}\end{equation}
as well as
$\cD^*[\LL_2](\hop) = \cD^*[\LL_5](\hop)$ with
\begin{equation}\begin{aligned}
	\cD^*[\LL_2](\hop) + \cD^*[\LL_5](\hop)
	&= \frac{\eta^2\epsilon^2\cA^2}{4} \, \left(
	    	\frac14 f''(\tfrac\eta2\qop) g(\tfrac\eta2\qop_2)
		- f'(\tfrac\eta2\qop) g'(\tfrac\eta2\qop_2)
	   	+ f(\tfrac\eta2\qop) g''(\tfrac\eta2\qop_2)
	   	\right)\\
	  &\quad\quad\quad
	  - \eta\epsilon\cA\, \sin(\eta\qop_2)
		\left( f(\tfrac\eta2\qop) g'(\tfrac\eta2\qop_2) \right.
			\left. -\frac12 f'(\tfrac\eta2\qop) g(\tfrac\eta2\qop_2) \right).
\end{aligned}\end{equation}
Finally, we obtain
\begin{equation}\begin{aligned}
	\sum_{k=0}^5 \cD^*[\LL_k](\hop)
	&= \frac{3\eta^2 \epsilon^2 \cA^2}8 \, \left(
	    	f''(\tfrac\eta2\qop) g(\tfrac\eta2\qop_2)
		- f'(\tfrac\eta2\qop) g'(\tfrac\eta2\qop_2)
	   	+ f(\tfrac\eta2\qop) g''(\tfrac\eta2\qop_2)
	   	\right)\\
	&\quad\quad\quad
	- \eta\epsilon\cA \,
		\left( \sin(\eta\qop) + \frac12 \sin(\eta\qop+\eta\qop_2)
			+ \frac12 \sin(\eta\qop_2) \right)
		f'(\tfrac\eta2\qop) g(\tfrac\eta2\qop_2) \\
	&\quad\quad\quad
	- \eta\epsilon\cA \,
		\left( \sin(\eta\qop_2) + \frac12 \sin(\eta\qop+\eta\qop_2)
			+ \frac12 \sin(\eta\qop) \right)
		f(\tfrac\eta2\qop) g'(\tfrac\eta2\qop_2).
\end{aligned}\end{equation}
We thus obtain the following evolution of $\hop$ in the Heisenberg picture:
\begin{equation}\label{eq:lindblad_heisenberg_hexa}\begin{aligned}
	\Gamma \sum_{k=0}^5 \cD^*[\LL_k](\hop)
	&= -\mathcal A\epsilon\eta \Gamma \,
	\mathcal L_\sigma (h)(\tfrac\eta2\qop_1, \tfrac\eta2\qop_2)
\end{aligned}\end{equation}
where $\sigma = \frac{3\mathcal A\epsilon\eta}8$
and $\mathcal L_\sigma$ is the differential operator defined by
\begin{equation}\begin{aligned}
	\label{eq:lsigma_hexa}
	(\mathcal L_\sigma h)(\theta_1,\theta_2) =&\,
	\left( \sin(2\theta_1) + \frac12 \sin(2\theta_1+2\theta_2) - \frac12\sin(2\theta_2)\right) %
	\frac{\partial h}{\partial {\theta_1}}\\
	& + \left( \sin(2\theta_2) + \frac12 \sin(2\theta_1+2\theta_2) - \frac12\sin(2\theta_1)\right)
	\frac{\partial h}{\partial {\theta_2}}\\
	&-\sigma \left(
	\frac{\partial^2 h}{\partial \theta_1^2}
	- \frac{\partial^2 h}{\partial \theta_1 \partial\theta_2}
	+ \frac{\partial^2 h}{\partial \theta_2^2}
	\right).
\end{aligned}\end{equation}
\end{widetext}
Here, the crucial difference with the square case
is that $\mathcal L_\sigma$ does not enjoy the nice decomposition of~\cref{eq:lsigmatensor}
which allowed us to reduce the study to that of a differential operator in one variable only.

Introducing an adapted scalar product
on periodic functions of two variables through
\begin{align}
	\label{eq:sp_2var_hexa}
	&\langle f,g\rangle_\sigma
		= \int_{-\pi}^\pi \int_{-\pi}^\pi
		w_\sigma(\theta_1,\theta_2)
				\, f(\theta_1,\theta_2)^* g(\theta_1,\theta_2)
				\, d\theta_1 d\theta_2 \nonumber\\
	&w_\sigma(\theta_1,\theta_2)
	= e^{-\frac{3-\cos(2\theta_1)-\cos(2\theta_1+2\theta_2)-\cos(2\theta_2)}{2\sigma}} 
\end{align}
one can check that $\mathcal L_\sigma$ is self-adjoint for the scalar product
$\langle \cdot,\cdot\rangle_\sigma$
and non-negative.
Indeed, it is a consequence of the following identity, analogous to~\cref{eq:div_formula_lsigma}:
\begin{equation}
	w_\sigma \mathcal L_\sigma(f) = - \sigma \, \mathrm{div}(w_\sigma A\nabla f)
\end{equation}
where we introduced the matrix
\begin{equation}
	A =
	\begin{bmatrix}
		1 & -1/2 \\
		-1/2 & 1
	\end{bmatrix}.
\end{equation}
The differential operator $\mathcal L_\sigma$ thus has a discrete, real and non-negative spectrum
that we can denote in ascending order
\begin{equation}
\lambda_{0,\sigma}
\leq
\lambda_{1,\sigma}
\leq
\lambda_{2,\sigma}
\leq\ldots
\end{equation}
Of course, for any $\sigma$, the first eigenvalue of $\mathcal L_\sigma$ still is
$\lambda_{0,\sigma} = 0$, corresponding to the constant eigenfunction $h\equiv 1$;
this simply corresponds to the conservation identity
\[ \frac d{dt} \trace(\rho_t) = 0.\]

In the limit $\sigma\rightarrow 0$,
the other eigenvalues of $\mathcal L_\sigma$ have also been rigorously studied
in~\cite{sellem2023ifac} (using mathematical results from~\cite{michelSmallEigenvalues2019a}),
where the following asymptotics estimates were obtained:
\begin{equation}
	\label{eq:asymptotics_michel_hexa}
	\begin{aligned}
		\lambda_{k,\sigma} &\sim \frac{12\sqrt3}\pi e^{-2/\sigma},\quad 1\leq k\leq 3\\
		\lambda_{4,\sigma} &\geq \upepsilon_0 > 0
	\end{aligned}
\end{equation}
where $\upepsilon_0$ is a fixed constant independent of $\sigma$.
Two remarks are in order when comparing~\cref{eq:asymptotics_michel_hexa}
and the corresponding result in the square case~\cref{eq:eigenvalues_and_functions_l1sigma_square}:
\begin{itemize}
	\item Looking at the exponential dependence on $1/\sigma$ in the
		first non-zero eigenvalue,
		one might think at first that we gain a quadratic factor in the hexagonal case.
		However, one must keep in mind that $\eta$ as well as
		the definition of $\sigma$ depend
		on the geometry.
		For a fixed value of $\epsilon$,
		expanding the definitions of $\eta$ and $\sigma$,
		we find
		$\lambda_{1,\textrm{square}} =
			\frac4\pi e^{-\frac2{\epsilon \sqrt\pi} e^{\epsilon\sqrt\pi}}
			\simeq \frac 4\pi e^{-\frac2{\epsilon \sqrt\pi}}$
		and
		$\lambda_{1,\textrm{hexa}} = \frac{12\sqrt3}\pi \,
				e^{  -(\frac{64}{27})^{1/4} \,
					\frac2{\epsilon\sqrt\pi}
					e^{\epsilon\sqrt{\frac{2\pi}{\sqrt3}}}}
				\simeq \frac{12\sqrt3}{\pi}
				e^{  -(\frac{64}{27})^{1/4} \, \frac2{\epsilon\sqrt\pi}}$.
		In others words, the relative scaling of the exponential dependence on $\epsilon$
		is given by
		\( \frac{2/\sigma_{\mathrm{hexa}}}{1/\sigma_{\mathrm{square}}}
			=(\frac{64}{27})^{1/4}  \simeq \frac54 \);
	\item In the square case, we observed that
		\(\lambda_{1,\sigma} = \lambda_{2,\sigma} = \frac12 \lambda_{3,\sigma}\);
		in terms of logical decay rates (see the section about the square case
		for the link between eigenvalues and logical rates),
		this meant that the logical $\Y$ coordinate decayed twice faster
		than the $\X$ and $\Z$ coordinates. This can be related to the fact
		that the Pauli $\X$ and $\Z$ operators correspond to phase-space shifts along
		the sides of the square grid cell, while $\Y = -i\Z\X$ correspond to a shift
		along the diagonal, which is longer by a factor $\sqrt2$.
		On the other hand, in the hexagonal case,
		we get \( \lambda_{1,\sigma} = \lambda_{2,\sigma} = \lambda_{3,\sigma}\),
		reflecting the symmetry of the hexagonal grid.
		Note that while the above analysis is specific to our dissipation scheme,
		a similar phenomenon was experimentally observed in~\cite{campagne2020quantum}
		where stabilization was achieved using phase estimation of the stabilizers.
\end{itemize}
\clearpage

\begin{figure}[htbp]
	\centering
	\includegraphics[width=0.95\columnwidth]{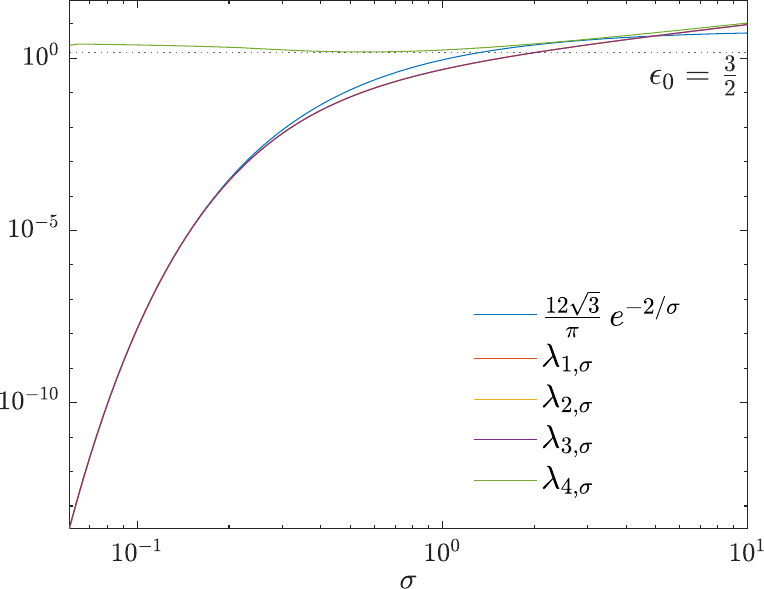}
	\caption{{\bf Spectral analysis of $\mathcal L_\sigma$ for the hexagonal GKP code}. %
	First four non-zero eigenvalues of $\mathcal L_\sigma$.
	In the limit of small $\sigma$,
	we observe that the first three (identical) non-zero eigenvalues
	$\lambda_{k,\sigma}, 0\leq k\leq 3$ are well-approximated by the theoretical asymptotic value
	$\lambda_{k,\sigma} = \frac{12\sqrt3}\pi \, e^{-2/\sigma}$
	obtained in~\cite{sellem2023ifac},
	while the fourth non-zero eigenvalue
	$\lambda_{4,\sigma}$ is lower-bounded by $\upepsilon_0 \simeq 3/2$ (estimated numerically),
	opening an exponential spectral gap.
	In this figure, we used $K_{\mathrm{max}}=200$
	(corresponding to $(2K_{\mathrm{max}}+1)^2 = 160801$ Fourier modes)
	and numerically checked that higher truncations do not change the results.
	}
	\label{fig:numerics_eigenvalues_hexa}
\end{figure}

Finally, apart from these asymptotic results, we turn to numerical diagonalization of $\mathcal L_\sigma$
to compute the exact value of $\lambda_{k,\sigma}$
for a fixed value of $\sigma$.
To that end, note that $\mathcal L_\sigma$ also has a simple expression in the Fourier frame:
indeed, take $h$ a (smooth) $2\pi$-periodic function of two variables
\begin{equation}
	h(\theta_1,\theta_2) = \sum_{k_1,k_2\in\mathbb Z}
			h_{k_1,k_2} \, e^{ik_1\theta_1} e^{ik_2\theta_2},
\end{equation}
we get
\begin{widetext}
\begin{equation}
	\begin{aligned}
		(\mathcal L_\sigma h) (\theta_1,\theta_2)
		&=\left( \sin(2\theta_1) + \frac12 \sin(2\theta_1+2\theta_2) - \frac12\sin(2\theta_2)\right) %
	\frac{\partial h}{\partial {\theta_1}}\\
	&\quad + \left( \sin(2\theta_2) + \frac12 \sin(2\theta_1+2\theta_2) - \frac12\sin(2\theta_1)\right)
	\frac{\partial h}{\partial {\theta_2}}\\
	&\quad -\sigma \left(
	\frac{\partial^2 h}{\partial \theta_1^2}
	- \frac{\partial^2 h}{\partial \theta_1 \partial\theta_2}
	+ \frac{\partial^2 h}{\partial \theta_2^2}
	\right)\\
			&= \sum_{k_1,k_2\in\mathbb Z} \left[ %
			\left(ik_1 \, \frac{e^{2i\theta_1} - e^{-2i\theta_1}}{2i} %
				+ \sigma k_1^2\right) %
			+ \left( %
				ik_2 \, \frac{e^{2i\theta_2} - e^{-2i\theta_2}}{2i} %
				+ \sigma k_2^2\right) \right.\\
			&\quad\quad\quad + \left(
				i\frac{k_1+k_2}2  \, \frac{e^{2i\theta_1}e^{2i\theta_2}
								-e^{-2i\theta_1}e^{-2i\theta_2}}{2i}
				- \sigma k_1 k_2\right)\\
			& \left. \quad\quad\quad - \left( %
				i\frac{k_1}2 \, \frac{e^{2i\theta_2} - e^{-2i\theta_2}}{2i} %
				+i\frac{k_2}2 \, \frac{e^{2i\theta_1} - e^{-2i\theta_1}}{2i} %
				\right) %
				\, \right] h_{k_1,k_2} \,
				e^{ik_1\theta_1} e^{ik_2\theta_2}\\
			&= \sum_{k_1,k_2\in\mathbb Z} \left[
				\left(
				\frac{k_1-2}2 f_{k_1-2,k_2} - \frac{k_1+2}2 f_{k_1+2,k_2}
					+ \sigma k_1^2 \, f_{k_1,k_2}
				\right) \right. \\
			&\quad \quad \quad + \left(
				\frac{k_2-2}2 f_{k_1,k_2-2} - \frac{k_2+2}2 f_{k_1,k_2+2}
					+ \sigma k_2^2 \, f_{k_1,k_2}
				\right) \\
			&\quad \quad \quad + \left(
				\frac{k_1+k_2-4}2 f_{k_1-2,k_2-2} - \frac{k_1+k_2+4}2 f_{k_1+2,k_2+2}
					+ \sigma \, k_1 k_2 f_{k_1,k_2}
				\right) \\
			&\quad \quad \quad + \left. \left( %
				k_1 \frac{f_{k_1,k_2-2} - f_{k_1,k_2+2}}2 %
				+k_2 \frac{f_{k_1-2,k_2} - f_{k_1+2,k_2}}2 %
				\right) %
				\right] e^{ik_1\theta_1} e^{ik_2\theta_2}.
	\end{aligned}
\end{equation}
\end{widetext}
In the Fourier frame, $\mathcal L_\sigma$ can thus be approximated by
truncating the sum above
to $-K_{\textrm{max}} \leq k_1,k_2 \leq K_{\textrm{max}}$
(\emph{i.e.} bounding the maximum frequency
considered).
On~\cref{fig:numerics_eigenvalues_hexa} we compare the numerical eigenvalues obtained
by this procedure %
\footnote{%
	We emphasize that, crucially, the Fourier representation of $\mathcal L_\sigma$ is sparse, %
	allowing us to numerically diagonalize the corresponding matrix of size $(2K_{\mathrm{max}}+1)^2$.}
to the explicit asymptotic expressions of~\cref{eq:asymptotics_michel_hexa}.
The analysis led at the end of~\cref{sec:eigen_square} to explain, in the square case,
the interpretation of the eigenvalues of $\mathcal L_\sigma$ in terms of logical decoherence rates
can be adapted straightforwardly to the hexagonal case.

\subsubsection{Protection against quadrature noise}
In presence of quadrature noise,
the previous exact eigenvalue analysis
can be adapted
by adding the contribution of dissipators along the
two quadrature operators $\qop$ and $\Pop$.
More precisely, for a given separable periodic observable
\(
	\hop = h(\tfrac\eta2\qop_1,\tfrac\eta2 \qop_2)
	= f(\tfrac\eta2\qop_1) \, g(\tfrac\eta2 \qop_2)
\)
we have to compute
\begin{equation}
	\begin{aligned}
		\frac d{dt} \trace &\Big( \hop
			\big( \kappa \cD[\qop](\rho_t) + \kappa \cD[\Pop](\rho_t)\big)
			\Big)\\
		&= \kappa \trace\left( \cD^*[\qop](\hop)\rho_t\right)
		+ \kappa \trace\left( \cD^*[\Pop](\hop)\rho_t\right).
	\end{aligned}
\end{equation}
\paragraph{Square GKP.}

For a square GKP grid, we have $\eta=2\sqrt\pi$, $\qop_1=\qop$, $\qop_2=\Pop$.
Using similar computations (see~\cite{sellem2023ifac})
we get
	\begin{equation}\label{eq:quadnoise_qop_heisenberg}\begin{aligned}
		\cD^*[\qop](\hop)
		&= \frac{\eta^2}8 f(\tfrac\eta2\qop) \, g''(\tfrac\eta2\Pop)
	\end{aligned}\end{equation}
	\begin{equation}\label{eq:quadnoise_pop_heisenberg}\begin{aligned}
		\cD^*[\Pop](\hop)
		&= \frac{\eta^2}8 f''(\tfrac\eta2\qop) \, g(\tfrac\eta2\Pop).
	\end{aligned}\end{equation}
Note that these terms already appeared in~\cref{eq:lindblad_heisenberg}
so that putting everything together we obtain

\begin{equation}
	\begin{aligned}
	\Gamma \sum_{k=0}^3& \cD^*[\LL_k](\hop)
		+ \kappa \left( \cD^*[\qop](\hop) + \cD^*[\Pop](\hop) \right)\\
		=& -\mathcal A\epsilon\eta \Gamma \,
			\left(
			\sin(\eta\qop) f'(\tfrac\eta2\qop)
			-\sigma
			f''(\tfrac\eta2\qop) \right) g(\tfrac\eta2\Pop) \\
		& -\mathcal A\epsilon\eta \Gamma \,
			f(\tfrac\eta2\qop)\left(
				\sin(\eta\Pop) g'(\tfrac\eta2\Pop)
			-\sigma
			g''(\tfrac\eta2\Pop) \right)
	\\
		=& -\mathcal A\epsilon\eta \Gamma \,
	\mathcal L_\sigma (h)(\tfrac\eta2\qop, \tfrac\eta2\Pop)
	\end{aligned}
\end{equation}
where the differential operator $\mathcal L_\sigma$ is still defined by~\cref{eq:lsigmatensor}
but its parameter $\sigma$ now also depends on $\kappa/\Gamma$
through
\begin{equation}\label{eq:sigma_square_noise}
	\sigma = \frac{\mathcal A\epsilon\eta}4 + \frac{\kappa\eta}{8\mathcal A\epsilon\Gamma}.
\end{equation}

As a consequence,
we emphasize that in presence of quadrature noise,
the optimal choice of energy truncation no longer consists in minimizing $\epsilon$.
Indeed, in the regime $\sigma\ll 1$,
the logical decoherence rate is given by
\begin{equation}\label{eq:asymptotic_Gamma_square}
	\begin{aligned}
	\Gamma_L = \cA \epsilon\eta\Gamma \, \lambda_{1,\sigma}
		&\simeq \frac4\pi \cA \epsilon\eta\Gamma \, e^{-1/{\sigma}}\\
		&= \frac4\pi \cA \epsilon\eta\Gamma \, e^{-\left( \frac{\mathcal A\epsilon\eta}4 + \frac{\kappa\eta}{8\mathcal A\epsilon\Gamma}\right)^{-1} }.
	\end{aligned}
\end{equation}
At leading order in $\kappa/\Gamma$,
this value is minimized for an optimal choice of $\epsilon$ given by
\begin{equation} \epsilon^\star = \sqrt{\frac\kappa{2\cA^2\Gamma}}\end{equation}
(if desired, one could refine this value by computing the next
coefficients of the asymptotic expansion of $\epsilon^\star$ in powers of $\sqrt{\frac\kappa\Gamma}$).
Using~\cref{eq:asymptotic_Gamma_square}, the corresponding logical decoherence rate is
\begin{equation}
	\Gamma^\star_L 
		\simeq \frac{4 \eta}\pi
			\, \sqrt{\frac{\kappa\Gamma}{2}}
			\, e^{-\frac2\eta \sqrt{\frac{2\Gamma}\kappa}}.
\end{equation}
We emphasize that the asymptotic expression of $\Gamma_L$ given by~\cref{eq:asymptotic_Gamma_square}
is only valid in the regime $\sigma\rightarrow 0$.
In particular, it cannot be used to estimate the decoherence rate
when $\epsilon\rightarrow0$ for a fixed value of $\kappa$,
as in that case $\sigma\rightarrow +\infty$.
In that opposite regime, the first non-zero eigenvalue $\lambda_{1,\sigma}$
of $\mathcal L_\sigma$ satisfies $\lambda_{1,\sigma} \simeq \sigma$
so that the corresponding logical decoherence rate is given, at first order, by
\begin{equation}
	\Gamma_L^{\epsilon\rightarrow 0} \simeq \cA\epsilon\eta\Gamma \sigma
		\simeq \frac{\kappa\eta^2}8.
\end{equation}
This decoherence rate is linear in $\kappa$ so that
the effect of the stabilization is entirely lost:
the logical decoherence rate is trivially piloted by the strength of the quadrature noise.

\paragraph{Hexagonal GKP.}
For a hexagonal GKP grid, we have $\eta=2\sqrt{\frac{2\pi}{\sqrt3}}$, $\qop_1=\qop$,
$\qop_2=\cos(\frac{2\pi}3)\qop + \sin(\frac{2\pi}3)\Pop = -\frac12 \qop + \frac{\sqrt3}2\Pop$, and in particular
$[\qop, \qop_1]=0$, $[\qop, \qop_2] = \frac{\sqrt3}2 i$,
$[\Pop,\qop_1] = -i$,
$[\Pop,\qop_2] = \frac i2$.
We thus get
\begin{widetext}
\begin{equation}
	\begin{aligned}
		\cD^*[\qop](\hop)
		&= \frac12 \qop [\hop,\qop] + \frac12[\qop,\hop]\qop\\
		&= \frac12 \qop \, f(\tfrac\eta2\qop) [g(\tfrac\eta2\qop_2),\qop]
			+ \frac12 f(\tfrac\eta2\qop) [\qop, g(\tfrac\eta2\qop_2)] \, \qop\\
		&= - \frac{i\eta\sqrt3}{8} \qop \, f(\tfrac\eta2\qop) g'(\tfrac\eta2\qop_2)
			+ \frac{i\eta\sqrt3}{8}  f(\tfrac\eta2\qop) g'(\tfrac\eta2\qop_2)\, \qop\\
		&=  \frac{i\eta\sqrt3}{8}  f(\tfrac\eta2\qop) [ g'(\tfrac\eta2\qop_2), \qop]\\
		&= \frac{3\eta^2}{32}  f(\tfrac\eta2\qop) g''(\tfrac\eta2\qop_2)
	\end{aligned}
\end{equation}
and
\begin{equation}
	\begin{aligned}
		\cD^*[\Pop](\hop)
		&= \frac12 \Pop [\hop,\Pop] + \frac12[\Pop,\hop]\Pop\\
		&=
			\frac12 \Pop \, [f(\tfrac\eta2\qop),\Pop] \, g(\tfrac\eta2\qop_2)
			+ \frac12 [\Pop, f(\tfrac\eta2\qop)] \, g(\tfrac\eta2\qop_2) \Pop
			+ \frac12 \Pop f(\tfrac\eta2\qop) \, [g(\tfrac\eta2\qop_2), \Pop]
			+ \frac12 f(\tfrac\eta2\qop) \, [\Pop, g(\tfrac\eta2\qop_2)] \Pop\\
		&=
			\frac{i\eta}{4} \Pop \, f'(\tfrac\eta2\qop) g(\tfrac\eta2\qop_2)
			- \frac{i\eta}{4} f'(\tfrac\eta2\qop) g(\tfrac\eta2\qop_2) \Pop
			- \frac{i\eta}{8} \Pop f(\tfrac\eta2\qop) g'(\tfrac\eta2\qop_2)
			+ \frac{i\eta}{8} f(\tfrac\eta2\qop) g'(\tfrac\eta2\qop_2) \Pop\\
		&=
			\frac{i\eta}4 [\Pop, f'(\tfrac\eta2\qop) g(\tfrac\eta2\qop_2)]
			-\frac{i\eta}8 [\Pop, f(\tfrac\eta2\qop) g'(\tfrac\eta2\qop_2)] \\
		&=
			\frac{i\eta}4  f'(\tfrac\eta2\qop)\, [\Pop, g(\tfrac\eta2\qop_2)]
			-\frac{i\eta}8 f(\tfrac\eta2\qop) \, [\Pop, g'(\tfrac\eta2\qop_2)] 
			+\frac{i\eta}4 [\Pop, f'(\tfrac\eta2\qop)]\, g(\tfrac\eta2\qop_2)]
			-\frac{i\eta}8 [\Pop, f(\tfrac\eta2\qop)]\, g'(\tfrac\eta2\qop_2)] \\
		&=
			-\frac{\eta^2}{8}  f'(\tfrac\eta2\qop)  g'(\tfrac\eta2\qop_2)
			+\frac{\eta^2}{32} f(\tfrac\eta2\qop) g''(\tfrac\eta2\qop_2)
			+\frac{\eta^2}{8} f''(\tfrac\eta2\qop) g(\tfrac\eta2\qop_2)
	\end{aligned}
\end{equation}
so that
\begin{equation}
	\begin{aligned}
		\cD^*[\qop](\hop) + \cD^*[\Pop](\hop)
		&= \frac{\eta^2}8 \Big(
			f''(\tfrac\eta2\qop) g(\tfrac\eta2\qop_2)
			- f'(\tfrac\eta2\qop)  g'(\tfrac\eta2\qop_2)
			+ f(\tfrac\eta2\qop) g''(\tfrac\eta2\qop_2)
			\Big)\\
		&= \frac{\eta^2}8 \Big(
			\frac{\partial^2 h}{\partial \theta_1^2}
			- \frac{\partial^2 h}{\partial \theta_1 \partial\theta_2}
			+ \frac{\partial^2 h}{\partial \theta_2^2}
			\Big) (\qop_1,\qop_2).
	\end{aligned}
\end{equation}
\end{widetext}

Note that these terms already appeared in~\cref{eq:lindblad_heisenberg_hexa}
so that putting everything together we obtain
\begin{equation}
	\begin{aligned}
	\Gamma \sum_{k=0}^5\, &\cD^*[\LL_k](\hop)
		+ \kappa \left( \cD^*[\qop](\hop) + \cD^*[\Pop](\hop) \right)\\
		&= -\mathcal A\epsilon\eta \Gamma \,
	\mathcal L_\sigma (h)(\tfrac\eta2\qop_1, \tfrac\eta2\qop_2)
	\end{aligned}
\end{equation}
where the differential operator $\mathcal L_\sigma$ is still defined by~\cref{eq:lsigma_hexa}
but its parameter $\sigma$ now also depends on $\kappa/\Gamma$
through
\begin{equation}\label{eq:sigma_hexa_noise}
	\sigma = \frac{3\mathcal A\epsilon\eta}8 + \frac{\kappa\eta}{8\mathcal A\epsilon\Gamma}.
\end{equation}
Once again, this entails that there is an optimal choice of energy truncation
in presence of quadrature noise,
approximately corresponding to
\begin{equation}
	\epsilon^\star = \sqrt{\frac\kappa{3\cA^2\Gamma}}.
\end{equation}
The corresponding logical decoherence rate is
\begin{equation}
	\Gamma^\star = \frac{12\eta}\pi \sqrt {\kappa \Gamma} \,
			e^{-\frac{8}{3\eta} \sqrt{\frac{3\Gamma}\kappa}}.
\end{equation}


\subsubsection{Propagation of ancilla noise}
\label{ssec:analysis_ancilla_noise}

When the ancillary mode used for dissipation engineering is noisy,
our eigenvalue analysis must take into account the additional dynamics induced on the target mode.
We explain this extension in the square GKP case;
it can be easily adapted to the hexagonal GKP case.
\newline

Following \cref{sec:hot_ancilla}, performing adiabatic elimination of
a finite-temperature ancilla subject
to heating and dephasing
leads to replace the stabilizing dynamics in \cref{eq:eigs_lindblad_square} by
\begin{equation}
	\frac{d\rhoo}{dt} =
		\tilde \Gamma \, \left(
		\sum_{k=0}^3 \cD[\LL_k](\rhoo) + \tfrac{\nth}{1+\nth} \cD[\LL_k^\dag](\rhoo)
		\right).
		\tag{\ref{eq:simu_hot_ancilla}}
\end{equation}
with $\tilde \Gamma = \Gamma \, \frac{\kappa_b}{\kappa_b + \kappa_\phi} \, (1+\nth)$,
where $\kappa_b$, $\kappa_\phi$ and $\nth$ respectively denote
the photon loss rate, dephasing rate and mean thermal photon number of the ancilla mode
(we refer to \cref{sec:hot_ancilla} for a more detailed presentation).
Thus, studying the evolution of a periodic observable
$\hop = h(\qop,\Pop) = f(\tfrac\eta2 \qop) \, g(\tfrac\eta2 \Pop)$ in the Heisenberg picture
now also requires computing $\cD[\LL_k^\dag]^*(\hop)$ with $0\leq k\leq 3$,
with
\begin{equation}
\cD[\LL_k^\dag]^*(\hop)
	= \frac12 \; \LL_k [ \hop, \LL_k^\dag] +  \frac12 \; [\LL_k, \hop] \LL_k^\dag.
\end{equation}

Note that the commutators $[\hop, \LL_k]$ and $[\hop, \LL_k^\dag]$ already
appeared in the computations of $\cD[\LL_k]^*(\hop)$
(see \cref{sec:eigen_square} and \cite{sellem2023ifac})
and can be reused here, yielding

\begin{align}
	\cD[\LL_0^\dag]&^*(\hop)
	= \cD[\LL_2^\dag]^*(\hop)\notag\\
	&= - \frac{\cA \epsilon \eta}2 \;
		\left(
		-\sin(\eta \qop) \, f'(\tfrac\eta2 \qop) - \frac{\cA\epsilon\eta}4 f''(\tfrac\eta2 \qop)
		\right) \, g(\tfrac\eta2 \Pop),
\end{align}

\begin{align}
	\cD[\LL_1^\dag]&^*(\hop)
	= \cD[\LL_3^\dag]^*(\hop)\notag\\
	&= - \frac{\cA \epsilon \eta}2 \, f(\tfrac\eta2 \qop) \;
		\left(
		-\sin(\eta \Pop) \, g'(\tfrac\eta2 \Pop) - \frac{\cA\epsilon\eta}4 g''(\tfrac\eta2 \Pop)
		\right).
\end{align}

Combining these expressions with \cref{eq:simu_hot_ancilla,eq:lindblad_heisenberg},
we find

\begin{widetext}
\begin{align}
		&\tilde\Gamma \, \left(
			\sum_{k=0}^3 \cD[\LL_k]^*(\hop)
			+ \frac{\nth}{1+\nth} \cD[\LL_k^\dag]^* (\hop)
				\right) \notag \\
	&\quad\quad = -\mathcal A\epsilon\eta \tilde\Gamma \,
	\left[
		\left(
		\sin(\eta\qop) f'(\tfrac\eta2\qop)
		- \frac{\mathcal A\epsilon\eta}4 f''(\tfrac\eta2\qop) \right) g(\tfrac\eta2\Pop)
		+
		f(\tfrac\eta2\qop)\left(
		\sin(\eta\Pop) g'(\tfrac\eta2\Pop)
		- \frac{\mathcal A\epsilon\eta}4 g''(\tfrac\eta2\Pop) \right)
	\right] \notag \\
&\quad\quad\quad -\mathcal A\epsilon\eta \tilde\Gamma  \, \tfrac{\nth}{1+\nth}\,
	\left[
		\left(
		-\sin(\eta\qop) f'(\tfrac\eta2\qop)
		- \frac{\mathcal A\epsilon\eta}4 f''(\tfrac\eta2\qop) \right) g(\tfrac\eta2\Pop)
		+
		f(\tfrac\eta2\qop)\left(
		-\sin(\eta\Pop) g'(\tfrac\eta2\Pop)
		- \frac{\mathcal A\epsilon\eta}4 g''(\tfrac\eta2\Pop) \right)
	\right]\\
&\quad\quad = -\mathcal A\epsilon\eta \frac{\tilde\Gamma}{1+\nth} \,
	\Big[
		\left(
		\sin(\eta\qop) f'(\tfrac\eta2\qop)
		- \sigma f''(\tfrac\eta2\qop) \right) g(\tfrac\eta2\Pop)
		+
		f(\tfrac\eta2\qop)\left(
		\sin(\eta\Pop) g'(\tfrac\eta2\Pop)
		- \sigma g''(\tfrac\eta2\Pop) \right)
	\Big]
	\label{eq:nth_heisenberg}\\
	&\quad\quad = -\cA \epsilon\eta \, \frac{\tilde\Gamma}{1+\nth} \;
		\mathcal L_{\sigma}(h)(\tfrac\eta2 \qop, \tfrac\eta2 \Pop)
\end{align}
\end{widetext}
with $\sigma = \frac{\cA \epsilon\eta}{4} (1+2\nth)$ depending on the value of $\nth$.
The evolution of $\hop$ in the Heisenberg picture can thus be directly deduced from the
spectral analysis of the differential operator $\mathcal L_{\sigma}$.
\newline

Finally, note that we separated the treatment of ancilla noise from that of quadrature noise
for clarity sake,
but we could take both into account simultaneously.
Adding quadrature noise to the Lindblad dynamics of \cref{eq:simu_hot_ancilla} above, we obtain
\begin{equation}
	\begin{aligned}
		\frac{d\rhoo}{dt} =\,&
		\tilde \Gamma \, \left(
		\sum_{k=0}^3 \cD[\LL_k](\rhoo) + \tfrac{\nth}{1+\nth} \cD[\LL_k^\dag](\rhoo)
		\right)\\
		&+ \kappa \Big( \cD[\qop](\rho) + \cD[\Pop](\rho) \Big)
	\end{aligned}
\end{equation}
with $\kappa>0$ the strength of quadrature noise.
In the Heisenberg picture, adding the contributions of $\cD[\qop]^*(\hop)$ and $\cD[\Pop]^*(\hop)$
(computed in \cref{eq:quadnoise_qop_heisenberg,eq:quadnoise_pop_heisenberg})
into \cref{eq:nth_heisenberg},
we obtain
\begin{equation}
	\begin{aligned}
		\tilde\Gamma \,& \left(
			\sum_{k=0}^3 \cD[\LL_k]^*(\hop)
			+ \frac{\nth}{1+\nth} \cD[\LL_k^\dag]^* (\hop)
				\right)
		\\&\quad  + \kappa \Big( \cD[\qop]^*(\hop) + \cD[\Pop]^*(\hop) \Big)\\
		&= -\cA \epsilon\eta \, \frac{\tilde\Gamma}{1+\nth}  \;
		\mathcal L_{\sigma}(h)(\tfrac\eta2 \qop, \tfrac\eta2 \Pop)
	\end{aligned}
\end{equation}
with the value of $\sigma$ now depending both on $\nth$ and $\kappa/\tilde\Gamma$ through
\begin{equation}
	\sigma = \frac{\cA \epsilon\eta}{4} \, (1+2 \nth)
			+ \frac{\kappa\eta (1+\nth)}{8\cA \epsilon \tilde\Gamma}.
\end{equation}

Note that, when only taking the effect of ancilla noise into account (that is for $\kappa/\tilde\Gamma=0$),
the optimal choice of energy truncation consists in minimizing $\epsilon$
(or equivalently going to the infinite-energy GKP limit).
Indeed, in the regime $\sigma\ll 1$,
the logical decoherence rate is given by
\begin{equation}\label{eq:asymptotic_Gamma_square_nth}
	\begin{aligned}
	\Gamma_L = \cA \epsilon\eta \, \frac{\tilde\Gamma}{1+\nth} \, \lambda_{1,\sigma}
	&\simeq \frac4\pi \cA \epsilon\eta \, \frac{\tilde\Gamma}{1+\nth} \, e^{-1/{\sigma}}\\
	&= \frac4\pi \cA \epsilon\eta \, \frac{\tilde\Gamma}{1+\nth}
	\, e^{-\frac4{{\mathcal A\epsilon\eta}  (1+2\nth)} }.
	\end{aligned}
\end{equation}
We thus see that the impact of ancilla noise and target noise
on achievable logical decoherence rates are
fundamentally different:
the former vanishes for large-energy GKP states while the latter imposes an optimal energy truncation.

%% file: sm__analysis__energybounds.tex
\subsection{Explicit energy estimates}
\label{ssec:energy_estimates}
In~\cref{sm__ssec__eigenvalues},
we analyzed the evolution of periodic operators in the Heisenberg picture.
To support the claim that the dissipation stabilizes finite-energy grid states,
we also need to verify that
energy is bounded under this dissipative dynamics.
To this end, we showed in a separate publication that
one can compute explicit bounds on the
energy (average photon number)
\( \langle \N \rangle = \trace(\N\rho) \)
along trajectories of a quantum system stabilized by the
modular dissipators proposed above;
in this subsection, we recall and gather the results obtained in~\cite{sellem2023ifac}.
When performing numerical dynamical simulations in the Fock basis,
such energy estimates give a rationale for choosing an adapted truncation.
\\

For any integer $M\geq 1$, let us introduce the generalized family of $2M$ Lindblad operators
defined by
\begin{equation}
	\LL_k = e^{i\theta_k\N} \, \left( \mathcal A
	e^{i\eta\qop}\left( \II-\epsilon\Pop \right)
	- \II \right)
	e^{-i\theta_k\N},
 \quad 0\leq k\leq 2M-1
\end{equation}
where $\theta_k = \frac{ik\pi}M$ and $\mathcal A = e^{-\eta\epsilon/2}$.
For $M=2$, we find the four Lindblad operators used to stabilize a square GKP code,
while
for $M=3$, we find the six Lindblad operators used to stabilize a hexagonal GKP code.
\\
We have the following energy estimate:
\begin{estimate}
	\label{thm__energy}
	Assume that $\trace(\N \rho_0) <+\infty$,
	$\epsilon\eta/2 < 0.4$
	and the evolution of $\rho_t$ is governed by the Lindblad equation
	\begin{equation}
		\begin{split}
			\frac{\mathrm{d}\rho_t}{\mathrm{d}t}
			&=\sum^{2M-1}_{k=0}\Gamma \mathcal{D}[\LL_k]\rho_t\\
			&=\sum^{2M-1}_{k=0} \Gamma \big(
			\LL_k \rho_t \LL_k^{\dag}
			-\frac{1}{2}(\LL_k^{\dag}\LL_k \rho_t +  \rho_t \LL_k^{\dag}\LL_k) \big).
			\label{eq:lindblad_M}
		\end{split}
	\end{equation}

	Then, for all $t\geq 0$ :
	\begin{equation}
		\begin{split}
			\trace(\N\rho_t) &\leq \,
			e^{-\lambda(\epsilon,\eta) t} \trace(\N\rho_0) 
			+ \left(1-e^{-\lambda(\epsilon,\eta) t}\right) C(\epsilon,\eta).
		\end{split}
		\label{eq__energyboundvalue}
	\end{equation}
	where
	$\lambda(\epsilon,\eta)$ and $C(\epsilon,\eta)$
	are positive constants defined in~\cite{sellem2023ifac},
	which satisfy the following asymptotics when $\epsilon\rightarrow 0^+$:
	\begin{equation}
		\begin{split}
			\lambda(\epsilon, \eta) &\sim 2 \,  M \Gamma \, \epsilon \eta,\\
			C(\epsilon, \eta) &\sim \frac\eta{2\epsilon}.
		\end{split}
	\end{equation}

\end{estimate}

We emphasize that the puzzling hypothesis $\epsilon\eta/2 < 0.4$ is only a sufficient condition
to ensure $((2-\epsilon\eta/2)\cA-1)>0$ and can be safely mentally replaced by
\emph{for small enough $\epsilon$} while reading the proof.
In particular, it is satisfied in every numerical simulation presented here.
As a consequence of this estimate, any solution
starting from an initial state satisfying
$\trace(\N\rho_0) \leq C(\epsilon,\eta)$
satisfies
$\trace(\N \rho_t)\leq C(\epsilon,\eta)$ along the whole trajectory.

Finally, note that, while the previous computations are
valid for any value of $M$,
the only choices allowing
the resulting dynamics to stabilize a logical qubit
are $M=2$ (corresponding to a square grid)
and $M=3$ (corresponding to a hexagonal grid).
\\

The full proof of~\cref{thm__energy} can be found in~\cite{sellem2023ifac}.
Let us only recall one core idea of the proof, which will also prove instrumental
in~\cref{sm_ssec_instability_lindblad2dissip}
to understand the
\emph{instability} of other possible candidate dynamics for the stabilization of GKP qubits.

	Using the definition of the operators $\LL_k$,
	we compute the evolution of $\N$ as
	\begin{equation}
		\label{eq__evol_n_heisenberg}
		\begin{split}
			\frac d{dt}\trace(\N\rho_t) &= %
			\sum_{k=0}^{2M-1} \Gamma \trace(\cD^*[\LL_k](\N)\rho_t) \\
			&= \sum_{k=0}^{2M-1} \Gamma \,
			\trace(e^{\frac{ik\pi}{M}\N} \,
			\cD^*[\LL_0](\N) \, %
			e^{-\frac{ik\pi}{M}\N} \rho_t)\quad
		\end{split}
	\end{equation}
	where
	\begin{equation}
		\begin{split}
			\cD^*[\LL_0](\N) &:= \LL_0^\dag \N \LL_0 - \frac12\left(
			\LL_0^\dag \LL_0 \N + \N \LL_0^\dag \LL_0\right) \\
			&= \frac12  \left( \LL_0^\dag \big[ \N, \LL_0\big]
			+ \big[\N, \LL_0^\dag\big]\LL_0 \right) \\
			&= \frac12 \, \LL_0^\dag \left[ \N, \LL_0\right] +\hc
		\end{split}
	\end{equation}

We can then show~\cite{sellem2023ifac} that
\begin{align}
	\cD^*[\LL_0](\N)
	&=  \epsilon^2 \eta\cA^2 \Pop^3 \notag \\
	&\quad + \frac{\epsilon^2+\eta^2}{2}\cA^2 \II + \eta (1-\epsilon\eta) \cA^2\Pop \nonumber\\
	&\quad -\epsilon\eta (2-\frac{\epsilon\eta}2)\cA^2\Pop^2
	+\frac{\epsilon\eta^3}{4}\cA \cos(\eta\qop) \nonumber\\
	&\quad -\epsilon\cA \qop\sin(\eta\qop)
	 -\eta (1+\frac{\epsilon\eta}2)\cA \cos(\eta\qop)\Pop \nonumber\\
	&\quad -i \frac{\eta^2}4 (2+\epsilon\eta)\cA \sin(\eta\qop) \nonumber\\
	&\quad + \epsilon\eta\cA \, \Pop \cos(\eta\qop) \Pop.
\end{align}

Crucially, the ominous cubic leading term, that could lead to instability,
cancels out when we sum the contributions of all the dissipators in~\cref{eq__evol_n_heisenberg},
as each Lindblad operator is paired to its image by a $\pi$ rotation in phase-space.
The leading coefficient of the next quadratic term is negative, which allows us to recover
the desired stability property.

\subsection{Instability of a dynamics enforced by two dissipators}
\label{sm_ssec_instability_lindblad2dissip}
As explained in~\cref{sec:dissipationengineering},
realizing an effective Lindblad dynamics with several engineered dissipators
can be done either by coupling the target mode to
as many ancillary modes as there are dissipators to engineer,
or by resorting to a Trotterization procedure where one activates sequentially each dissipator.
In the first case, the complexity of the experiment is proportional to the number of dissipators
to engineer.
In the second case, the achievable engineered dissipation rate $\Gamma$ is inversely proportional
to that number.
It is thus natural to wonder whether the dynamics we propose,
featuring respectively four dissipators for the stabilization of square GKP states
and six dissipators for the stabilization of hexagonal GKP states,
are optimal in the number of dissipators to engineer.
For the remaining of this section, we will focus on the square GKP case;
our arguments are straightforwardly adapted to the hexagonal case.
We examine several ideas for the stabilization of the square GKP codespace
using only two dissipators,
and find that each of them leads to instable dynamical behavior.

\subsubsection{Candidate dynamics}
Recall from~\cref{sec:gkp} that the 
infinite-energy square GKP codespace can be defined as the common +1-eigenspace
of the two stabilizer operators
$\SSS_q = e^{i\eta\qop}$
and
$\SSS_p = e^{-i\eta\Pop}$
which correspond to shift operators in phase space.
Similarly, the
finite-energy square GKP codespace can be defined as the common +1-eigenspace of the two
stabilizer operators
$\SSS_q^\Delta = e^{-\Delta \oa^\dag \oa} \, \SSS_q \, e^{\Delta \oa^\dag \oa}$
and
$\SSS_p^\Delta = e^{-\Delta \oa^\dag \oa} \, \SSS_p \, e^{\Delta \oa^\dag \oa}$.
The stabilizing dynamics we introduced in~\cref{sec:modular dissip}
reads
\begin{equation}
	\frac{d\rho_t}{dt} := \mathcal L_4(\rho_t)
		= \Gamma \sum_{k=0}^3 \cD[\LL_k](\rho_t),
\end{equation}
where the four Lindblad operators
\begin{align*}
	\LL_k = \cA \Rop_{\frac{k\pi}2} \,
		e^{i\eta\qop} (1-\epsilon\Pop) \,
		\Rop_{\frac{k\pi}2}^\dag - \II,
		\\
		0\leq k\leq 3,
		\quad \epsilon = \eta \sinh(\Delta)
\end{align*}
correspond to first order approximations of $\SSS_q^\Delta-\II$, $\SSS_p^\Delta - \II$
as well as their image by a $\pi$ rotation in phase-space.
This rotation in phase-space can be understood as initially adding the two adjoint stabilizer operators
$\SSS_q^\dag = e^{-i\eta\qop}$ and $\SSS_p^\dag = e^{i\eta\Pop}$
in the definition of the GKP codespace, which seems redundant.
It is thus tempting to think that only the two first dissipators are required,
and that one could engineer the simpler candidate Lindblad dynamics
\begin{equation}
	\frac{d\rho_t}{dt} := \mathcal L_2(\rho_t)
			= \Gamma \big( \cD[\LL_0](\rho_t) + \cD[\LL_1](\rho_t) \big).
\end{equation}
A related idea would be to consider instead the symmetric sums of Lindblad operators
\begin{align*}
	\LL_{q,s} &= (\LL_0+\LL_2)/\sqrt2\\
		&= \sqrt2 \big( \cA \left(\cos(\eta\qop) -i\epsilon\sin(\eta\qop)\Pop\right)
					- \II \big)\\
	\LL_{p,s} &= (\LL_1+\LL_3)/\sqrt2\\
		&= \sqrt2 \big( \cA \left(\cos(\eta\Pop) +i\epsilon\sin(\eta\Pop)\qop\right)
					- \II \big),\\
\end{align*}
which amounts to defining the GKP codespace through the stabilizers
$\cos(\eta\qop)$ and $\cos(\eta\Pop)$
instead of $e^{i\eta\qop}$ and $e^{-i\eta\Pop}$.
Another candidate Lindblad dynamics with only two dissipators is thus
\begin{equation}
	\frac{d\rho_t}{dt} := \mathcal L_{2,s}(\rho_t)
			= \Gamma \big( \cD[\LL_{q,s}](\rho_t) + \cD[\LL_{p,s}](\rho_t) \big).
\end{equation}

Finally, we note that another Lindblad dynamics with two dissipators
was proposed in~\cite{royer2020stabilization}:
\begin{equation}
	\frac{d\rho_t}{dt} := \mathcal L_{2,\log{}}(\rho_t)
			= \Gamma \big( \cD[\LL_{0,\log{}}](\rho_t) + \cD[\LL_{1,\log{}}](\rho_t) \big)
\end{equation}
where
\begin{align*}
	\LL_{0,\log{}} &= -\frac{i}{2\sqrt{\pi\cosh(\Delta)\sinh(\Delta)}}\log(\SSS_q^\Delta) \\
	&= \frac{\qop_{[m]}}{\sqrt{2\tanh(\Delta)}} + i \Pop \sqrt{\frac{\tanh(\Delta)}2} \\
	\LL_{1,\log{}} &= -\frac{i}{2\sqrt{\pi\cosh(\Delta)\sinh(\Delta)}}\log(\SSS_p^\Delta) \\
	&= \frac{\Pop_{[m]}}{\sqrt{2\tanh(\Delta)}} - i \qop \sqrt{\frac{\tanh(\Delta)}2}
\end{align*}
where $m = \frac{2\pi}{\eta\cosh(\Delta)}$
and $\qop_{[m]} = \qop \mod m$, $\Pop_{[m]}= \Pop \mod m$ are modular quadrature operators.
This choice of dissipators can be intuitively understood as such:
to build a Lindblad operator that cancels on the codespace,
given a stabilizer operator $\SSS$,
the authors use $\log(\SSS)$ instead of $\SSS-\II$.
Note that, while the strategy
presented in the main text to engineer the dynamics $\mathcal L_4$
would also be immediately suitable for the engineering of the two candidate dynamics
$\mathcal L_2$ and $\mathcal L_{2,s}$,
there is currently no clear way to engineer the dynamics $\mathcal L_{2,\log{}}$.
For this reason, we focus mainly on the analysis of $\mathcal L_2$ and $\mathcal L_{2,s}$,
but still included $\mathcal L_{2,\log}$ in our numerical comparisons to assess its
theoretical merits.
Note however that, following \cite{royer2020stabilization},
the dynamics $\mathcal L_{2,\log}$ can be further approximated by discrete-time dynamics
using Trotter decompositions. Depending on the exact decomposition chosen, this leads to three
dynamics known as the Sharpen-Trim, small-Big-small and Big-small-Big protocols,
which are at the core of previous experimental realizations of GKP states
\cite{fluhmann2018encoding,campagne2020quantum,de2022error,sivak2022real}.

\subsubsection{Numerical study of the stability}
To detect potential dynamical instabilities,
we numerically simulate the evolution of a density operator $\rho_t$ initialized in
vacuum ($\rho_0 = |0\rangle\langle0|$).
The exact choice of initial state is arbitrary: in order to ease the detection of instabilities,
one should preferably choose an initial state far from the codespace
(as states in the codespace would be metastable states of any reasonable candidate dynamics).

\begin{figure}
	\includegraphics[width=0.95\columnwidth]{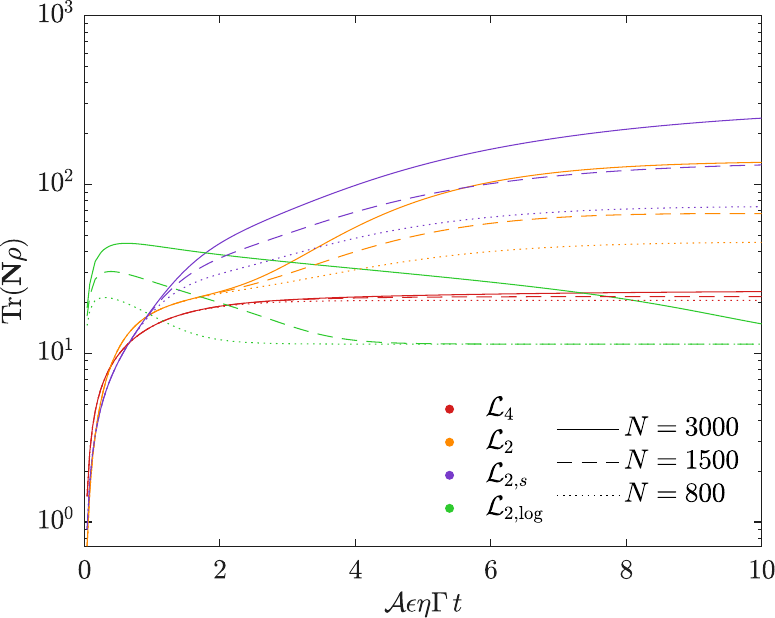}
	\caption{{\bf Instability of the candidates two-dissipators dynamics.} %
	For each proposed dynamics, we plot the evolution of the average photon number %
	$\langle \N \rangle (t) = \trace(\N \rho_t)$ %
	when the system is initialized in $\rho_0 = |0\rangle\langle 0|$. %
	The dynamics $\mathcal L_4$ corresponds to the four-dissipators Lindblad equation %
	we propose for the stabilization of the square GKP codespace. %
	The three other dynamics $\mathcal L_2, \mathcal L_{2,s}$ and $\mathcal L_{2,\log}$ %
	correspond to the candidate two-dissipators dynamics introduced in the current section. %
	Here, %
	we chose a parameter $\epsilon=0.15$ %
	and the numerical simulations are performed in a Hilbert space truncated %
	to the first $N$ Fock states, with respectively %
	$N=800$ (dotted lines), %
	$N=1500$ (dashed lines) and %
	$N=3000$ (full lines). %
	We observe that the curves associated to $\mathcal L_2, \mathcal L_{2,s}$ %
	and $\mathcal L_{2,\log}$ do not seem to converge even at these rather high truncation numbers, %
	and that their maximal value actually increases with the truncation. %
	This phenomenon hints at the intrinsic instability of these three dynamics, %
	that only seem bounded for a given truncation due to the effects of the truncation itself.%
	}
	\label{fig:instable_simu_twodissip}
\end{figure}

On~\cref{fig:instable_simu_twodissip},
we plot the evolution of
the mean photon number $\langle \N \rangle(t) = \trace(\N \rho_t)$
for the four studied dynamics: $\mathcal L_4, \mathcal L_2, \mathcal L_{2,s}$ and
$\mathcal L_{2,\log}$.
Using the stability results obtained in~\cref{ssec:energy_estimates},
we already know that the mean photon number
remains bounded when $\rho_t$ is governed by the four dissipators dynamics $\mathcal L_4$,
where the exact bound depends on the maximum between the initial energy and a fixed constant
depending only on $\epsilon$ and $\Gamma$;
this result is verified numerically.
On the other hand, when $\rho_t$ is evolved with $\mathcal L_2$ and $\mathcal L_{2,s}$,
the mean photon number grows well beyond its value in the codespace.
We emphasize that one should not conclude that this energy still stays bounded, only with a higher bound:
in fact, we observe that the maximum value grows when increasing
the necessary truncation of the Hilbert space
used in simulation, from which one can only conclude that the true trajectory is unbounded.
Finally, when $\rho_t$ is evolved with $\mathcal L_{2,\log}$,
the mean photon number features an initial bump, before eventually decreasing to a value close to the
one observed with $\mathcal L_4$.
However, once again, we observe that the height of the initial bump increases
with the truncation of the Hilbert space,
suggesting that this dynamics is also instable and merely artificially
constrained by the numerical truncation.

\subsubsection{Qualitative understanding}
\label{sec:energy_qualitative_analysis}
In addition to the formal stability result of~\cref{ssec:energy_estimates},
we can propose more qualitative insight into the stability of $\mathcal L_4$
and the instability of the other dynamics, especially $\mathcal L_2$ and $\mathcal L_{2,s}$.

In general, when considering any time-independent Lindblad equation of the form
\begin{equation}
	\frac{d\rho_t}{dt}
		= -i [\HH,\rho_t] + \sum_{k\in K} \cD[\LL_k](\rho_t)
\end{equation}
one can, at least formally,
write the solution at time $t$ using the integral representation
\begin{widetext}
\begin{equation}
	\rho_t = \sum_{n=0}^{+\infty} \sum_{k_1,\ldots,k_n\in K}
		\int_{\substack{\vspace{0.5em}\\0\leq t_1 \leq \ldots \leq t_n \leq t}}
		\hspace{-0.8em}
		e^{(t-t_n)\GG} \, \LL_{k_n} \, 
		e^{(t_n-t_{n-1})\GG} \, \LL_{k_{n-1}} \ldots 
		e^{(t_2-t_1)\GG} \, \LL_{k_1} \,
		e^{t_1 \GG} \,
		\rho_0 \,
		e^{t_1 \GG^\dag} \,
		\LL_{k_1}^\dag \ldots
		\LL_{k_n}^\dag \, e^{(t-t_n)\GG^\dag} \,
		\small{dt_1 \ldots dt_n}
\end{equation}
\end{widetext}
where $\GG := -i\HH - \frac12\sum_{k\in K} \LL_k^\dag\LL_k$.
Note that in this representation formula, the term corresponding to a given value of $n$
in the first sum can be intuitively understood as an average
over all possible jump times $t_1,\ldots,t_n$
of a no-jump trajectory (generated by $\GG$) of length $t$
interrupted by jump events at times $t_1,\ldots,t_n$.
The full formula leads to an additional average over all possible number $n$ of jump events.

When designing a Lindblad dynamics for the stabilization of a given subspace
of the ambient Hilbert space,
one should thus not only consider the effect of the Lindblad operators (generating the jumps)
and Hamiltonian of the dynamics,
but also that of the no-jump generator $\GG$.
In particular, by construction, all eigenvalues of $\GG$ have a negative real part;
the no-jump contributions of the form $e^{s\GG}$ in the previous integrals can thus be understood
as exponential convergences towards the eigenspace associated to the eigenvalue of $\GG$
with maximum real part (that is, the closest to $0$ on the real axis).

\paragraph*{Stability of $\mathcal L_4$.}
In the case of the four-dissipators dynamics $\mathcal L_4$,
the jump events can, at first order in $\epsilon$,
be seen as introducing energy-truncated shifts in phase space
along the four cardinal directions
(given the definition of the Lindblad operators $\LL_k$ in~\cref{sec:modular dissip}
as energy-truncated version of the GKP stabilizers).
Additionally, tedious but straightforward computations give that the no-jump generator is
\begin{widetext}
\begin{equation}\label{eq:nojump_4dissip}
	\GG = -\frac\Gamma2 \, \sum_{k=0}^3 \LL_k^\dag \LL_k
	    = -2\Gamma \, \Big(
	    	\frac{\cA^2 \epsilon^2}2 \left( \qop^2 + \Pop^2\right)
		-\cA (1+\tfrac{\epsilon\eta}2) \left(\cos(\eta\qop) + \cos(\eta\Pop)\right)
		+ (1+\cA^2)\II \Big).
\end{equation}
\end{widetext}
Neglecting the fine details of each constant, which are irrelevant to our qualitative
discussion, we recognize the opposite of the so-called finite-energy GKP Hamiltonian,
already introduced \emph{e.g.} in~\cite{rymarzHardwareEncodingGrid2021}.
For $\epsilon=0$, this Hamiltonian boils down to the infinite-energy GKP Hamiltonian
$\HH_\infty := 2 - \cos(\eta\qop) - \cos(\eta\Pop)$
of the original GKP paper~\cite{gottesman2001encoding},
whose ground states are exactly the infinite-energy square GKP states.
For finite $\epsilon>0$ it can be understood as a regularization of $\HH_\infty$
by a confining quadratic potential $\frac{\epsilon^2}2 (\qop^2+\Pop^2)$;
the resulting ground states
approximately coincide with the finite-energy GKP states.

\paragraph*{Instability of $\mathcal L_2$.}
The instability of the two-dissipators dynamics $\mathcal L_2$
is easily understood when looking at the possible jump events.
Indeed, instead of introducing shifts along each of the four cardinal directions,
the only two remaining Lindblad operators only introduce shifts to the right and the top
of phase-space.
The resulting instability can be formally proven
by adapting the analysis of~\cref{ssec:energy_estimates}.
In this case, we get
\begin{align}\label{eq:N_heisenberg_2exp}
	\frac{d}{dt} \trace(\N \rho_t)
		&= \Gamma \trace\left( \cD^*[\LL_0](\N) \rho_t + \cD^*[\LL_1](\N)\rho_t \right) \notag \\
		&= \Gamma \trace\Big( \cD^*[\LL_0](\N)\rho_t
		\nonumber \\ &\quad \quad \quad \;
		+ e^{i\tfrac\pi2 \N} \, \cD^*[\LL_0](\N) \, e^{-i\tfrac\pi2 \N} \rho_t
			\Big).
\end{align}
We previously obtained that
\begin{align}
	\cD^*[\LL_0](\N)
		&=  \epsilon^2 \eta\cA^2 \Pop^3 \notag \\
		&\quad + \frac{\epsilon^2+\eta^2}{2}\cA^2 \II + \eta (1-\epsilon\eta) \cA^2\Pop\notag\\
		&\quad
				-\epsilon\eta (2-\frac{\epsilon\eta}2)\cA^2\Pop^2
		  +\frac{\epsilon\eta^3}{4}\cA \cos(\eta\qop) \notag\\
		&\quad -\epsilon\cA \qop\sin(\eta\qop)
		  -\eta (1+\frac{\epsilon\eta}2)\cA \cos(\eta\qop)\Pop \notag\\
		&\quad  -i \frac{\eta^2}4 (2+\epsilon\eta)\cA \sin(\eta\qop)\notag\\
		&\quad + \epsilon\eta\cA \, \Pop \cos(\eta\qop) \Pop.
\end{align}
In presence of the four dissipators, the leading term in $\Pop^3$ would cancel out
when added to its image by a $\pi$ rotation in phase space.
This is no longer the case when we consider only the dissipators of $\mathcal L_2$:
when evaluating~\cref{eq:N_heisenberg_2exp},
the derivative of $\N$ features a leading cubic term proportional to $\qop^3 + \Pop^3$
that can explain the growth of $\langle \N\rangle$ along trajectories.

\paragraph*{Instability of $\mathcal L_{2,s}$.}
The case of the symmetric dynamics $\mathcal L_{2,s}$ is more subtle
to analyze,
as the two dissipators introduce symmetric jump events,
seemingly countering the previous flaw of $\mathcal L_2$.
However, focusing now on the no-jump generator $\GG$, we find that
\begin{widetext}
\begin{align}
	\GG &= -\frac\Gamma 2 \, \big( \LL_{q,s}^\dag \LL_{q,s} + \LL_{p,s}^\dag \LL_{p,s} \big) \\
	&= -2\Gamma \, \Big(
		\frac{\cA^2\epsilon^2}2 \left( \Pop\sin^2(\eta\qop)\Pop+\qop\sin^2(\eta\Pop)\qop \right)
		- \cA(1+\tfrac{\epsilon\eta}2)\left(\cos(\eta\qop) + \cos(\eta\Pop)\right) \\ \notag
		&\quad \quad \quad \quad \quad + \II + \frac{\cA^2}2 \left( \cos^2(\qop) + \cos^2(\Pop) \right)
		+ \frac{\cA^2\epsilon\eta}2 \left( \cos(2\eta\qop) + \cos(2\eta\Pop) \right)
		\Big)
\end{align}
\end{widetext}
which should be compared to~\cref{eq:nojump_4dissip}.
Crucially, in that case, the quadratic term
\(\frac{\epsilon^2}2 \left( \Pop\sin^2(\eta\qop)\Pop+\qop\sin^2(\eta\Pop)\qop \right)\)
is no longer confining, as it vanishes periodically in phase space.
Hence, $\GG$ does not correspond to a proper finite-energy regularization of the infinite-energy
GKP Hamiltonian $\HH_\infty$.

%% file: sm__dissipation_engineering.tex
\section{Dissipation engineering}
\label{sm:sec__rwa}
We go back to the multimode circuit proposed in~\cref{fig:dissipCircuit} of the main text.
For arbitrary values of the flux biases
$\Phi^{\mathrm{ext}}_J$
and
$\Phi^{\mathrm{ext}}_L$,
the corresponding circuit Hamiltonian reads
\begin{equation}
	\begin{aligned}
		\HH_0(t) &=
		\omega_a \aop^\dag \aop
		+  \omega_b \bop^\dag \bop
		- E_J \cos\left( \frac{\phiop - \Phi^{\mathrm{ext}}_L(t)}{\varphi_0}\right)\\
		&\quad- E_J \cos\left( \frac{\phiop - \Phi^{\mathrm{ext}}_L(t) - \Phi^{\mathrm{ext}}_J(t)}
									{\varphi_0}\right)
	\end{aligned}
\end{equation}
where we took $\hbar=1$ for simplicity
and defined, as previously,
$\phiop = \varphi_0(\eta_a \qop_a + \eta_b\qop_b)$
with
$\eta_a=\sqrt{2\pi Z_a/R_Q}$
and  $\eta_b=\sqrt{2\pi Z_b/R_Q}\ll 1$.
We then define
\begin{align}
	\xi(t) &= \sin\left(\frac{\Phi^{\mathrm{ext}}_J(t) - \varphi_0 \pi}2\right),\label{eq:def_xi}\\
	\zeta(t) &= \frac{\Phi^{\mathrm{ext}}_L(t) + \Phi^{\mathrm{ext}}_J(t)/2}{\varphi_0}
	\label{eq:def_zeta}
\end{align}
so that the previous Hamiltonian can be recast as
\begin{align}
	\HH_0(t) &=
		\omega_a \aop^\dag \aop
		+  \omega_b \bop^\dag \bop
		+ 2 E_J \xi(t) \cos\left( \frac{\phiop}{\varphi_0}-\zeta(t)\right)\notag\\
		&=
		\omega_a \aop^\dag \aop
		+  \omega_b \bop^\dag \bop \notag\\
		&\quad + 2 E_J \xi(t) \cos\left( \eta_a \qop_a + \eta_b \qop_b -\zeta(t)\right).
\label{eq:circuit_h_xi_zeta}
\end{align}
Note that we recover the expression of~\cref{eq:ho} in the main text for the choice
$\zeta(t) = \pi/4$,
corresponding to a constant phase relation between the two biases:
$\Phi^{\mathrm{ext}}_L(t) = -\frac{\Phi^{\mathrm{ext}}_J(t)}2 + \varphi_0 \frac\pi4$.
This choice will be explained further in this section.

In the rotating frame of both modes, the Hamiltonian of~\cref{eq:circuit_h_xi_zeta}
gives rise to
the interaction Hamiltonian
\begin{equation}
	\begin{aligned}
	\HH(t) &= e^{it  \left( \omega_a \aop^\dag\aop + \omega_b \bop^\dag\bop \right)}
		\, \HH_0(t) \,
		e^{-it  \left( \omega_a \aop^\dag\aop + \omega_b \bop^\dag\bop \right)} \\
		&= 2 E_J \xi(t) \cos(\eta_a \qop_a(t) + \eta_b \qop_b(t) - \zeta(t)).
	\label{eq:hamiltonian_int_rwa}
	\end{aligned}
\end{equation}
where we defined rotating quadratures
\begin{equation}\begin{aligned}
	\qop_a(t) &= e^{i\omega_a t \, \aop^\dag\aop} \, \qop_a \, e^{-i\omega_a t \, \aop^\dag\aop}\\
		&= \cos(\omega_a t) \, \qop_a + \sin(\omega_a t) \, \Pop_a, \\
	\qop_b(t) &= e^{i\omega_b t \, \bop^\dag\bop} \, \qop_b \, e^{-i\omega_b t \, \bop^\dag\bop}\\
		&= \cos(\omega_b t) \, \qop_b + \sin(\omega_b t) \, \Pop_b.\\
\end{aligned}\end{equation}
The target mode $a$, used to encode the logical information, should ideally be free of any
intrinsic dissipation channel;
on the other hand, the ancillary mode $b$ is voluntarily lossy,
as we will want to adiabatically eliminate it later on.
In a first step, without taking into account yet any additional imperfections,
we thus model the evolution of our system with the Lindblad master equation
\begin{align}
	\frac{d\rho_t}{dt} &= -i [ \HH(t), \rho_t] + \kappa_b \cD[\bop](\rho_t) \notag\\
			&= -i \Big[ 2E_J \xi(t) \cos(\eta_a \qop_a(t) + \eta_b \qop_b(t) - \zeta(t)),\, \rho_t\Big]\notag \\
			&\quad	+ \kappa_b \cD[\bop](\rho_t)
	\label{eq:lindblad_circuit}
\end{align}
where $\kappa_b>0$ is the dissipation rate of mode b.

Recall that we want to engineer the following Lindbladian evolution on mode $a$ only:
\begin{equation}
	\frac{d\rho_t}{dt}
		= \sum_{k=0}^3 \cD[\LL_k](\rho_t)
		= \sum_{r\in\{q,p\}, l\in\{s,d\}} \cD[\LL_{r,l}](\rho_t)
	\label{eq:lindblad_target4dissip}
\end{equation}
where
we introduced two families of Lindblad operators that give rise to the same Lindblad equation:
\begin{equation}\begin{aligned}
	\LL_0 &= \cA e^{i\eta_a\qop_a} \left( \II - \epsilon\, \Pop_a\right) - \II,\\
	\LL_1 &= \cA e^{i\eta_a\Pop_a} \left( \II + \epsilon\, \qop_a\right) - \II,\\
	\LL_2 &= \cA e^{-i\eta_a\qop_a} \left( \II + \epsilon\, \Pop_a\right) - \II,\\
	\LL_3 &= \cA e^{-i\eta_a\Pop_a} \left( \II - \epsilon\, \qop_a\right) - \II,\\
\end{aligned}
\label{eq:exp_dissip}
\end{equation}
and their symmetric and antisymmetric sums
\begin{equation}\begin{aligned}
	\LL_{q,s} &= (\LL_0+\LL_2)/\sqrt2 \\
	&= \sqrt2 \left( \cA \left(\cos(\eta_a\qop_a) -i\epsilon\sin(\eta_a\qop_a)\Pop_a\right) - \II\right),\\
	\LL_{p,s} &= (\LL_1+\LL_3)/\sqrt2\\
	&= \sqrt2 \left( \cA \left(\cos(\eta_a\Pop_a) +i\epsilon\sin(\eta_a\Pop_a)\qop_a\right) - \II\right),\\
	\LL_{q,d} &= (\LL_0-\LL_2)/{i\sqrt2}\\
	&= \sqrt2 \cA \left(\sin(\eta_a\qop_a) +i\epsilon\cos(\eta_a\qop_a)\Pop_a\right),\\
	\LL_{p,d} &= (\LL_1-\LL_3)/{i\sqrt2}\\
	&= \sqrt2 \cA \left(\sin(\eta_a\Pop_a) -i\epsilon\cos(\eta_a\Pop_a)\qop_a\right).\\
\end{aligned}
\label{eq:sym_dissip}
\end{equation}

To go from~\cref{eq:lindblad_circuit} to~\cref{eq:lindblad_target4dissip}
we make use of two combined types of approximation:
the Rotating Wave Approximation,
allowing us to replace the time-dependent Hamiltonian $\HH(t)$
by an effective constant Hamiltonian $\HH_{\mathrm{RWA}}$,
and Adiabatic Elimination,
allowing us to derive an effective dynamics of mode $a$ in the limit where
mode $b$ is strongly dissipative.
Before diving into the details of our specific problem,
we recall the working principle of these two techniques
and the useful references and formulas that we use.

\subsection{Approximation formulas}
\label{ssec:approx_formulae}
\bigpar{Rotating Wave Approximation (RWA)}
As we will see in the next sections,
we need a formalism able to accommodate control functions $u(t)$
that are almost-periodic rather than periodic,
\emph{i.e.} that can be written as a sum of periodic functions with different frequencies.
In this setting,
we use the following first
order approximation result
that can be found in~\cite[Chapter 2]{lecturenotesmirrahimirouchon}.

Assume that $u(t)$ is a quasi-periodic signal
\begin{equation}
	u(t) = \sum_j \left( u_j e^{i\omega_j t} + u_j^* e^{-i\omega_j t} \right)
\end{equation}
and consider the following controlled Hamiltonian evolution,
for some constant Hamiltonians $\HH_c$ and $\HH_1$:
\begin{equation}
	\frac{d\rho_t}{dt}
		= -i[\HH_c + u(t) \HH_1,\rho_t].
\end{equation}
In the rotating frame given by $\HH_c$,
the interaction Hamiltonian is
\begin{equation}\label{eq:hinter_rwa}
	\HH(t) = u(t) \, e^{it\HH_c} \, \HH_1 \, e^{-it\HH_c}
\end{equation}
which is also a quasi-periodic operator
(involving frequencies that are linear combinations of the frequencies in $u$
and eigenvalues of $\HH_c$).
We want to approximate the solution to the equation
\begin{equation}
	\frac{d\rho_t}{dt}
		= -i[\HH(t), \rho_t]
\end{equation}
by the solution to another equation with a constant Hamiltonian $\HH_{\mathrm{RWA}}$:
\begin{equation}
	\frac{d\rho_t}{dt}
		= -i[\HH_{\mathrm{RWA}}, \rho_t].
\end{equation}

Then, at first order, we can use
\begin{equation}
	\HH_{\mathrm{RWA}}^{(1)} = \overline{\HH}
		= \lim_{T\rightarrow +\infty} \frac1T \int_0^T \HH(t) dt
	\label{eq:rwa1_recipe}
\end{equation}
where the overline means taking the time-average as defined in the right-hand side.
More precisely, we can introduce a small parameter $\epsilon_{\mathrm{RWA}}$ such that
the above approximation is valid at order
$\epsilon_{\mathrm{RWA}}$
on a timescale $T_{\mathrm{RWA}} = 1/\epsilon_{\mathrm{RWA}}$.
In our case, defining
\( u_{\max{}} = \max_j |u_j| \)
and $\omega_{\mathrm{min}}$ the minimum non-zero frequency
appearing in the quasi-periodic Hamiltonian $\HH$,
(which, given~\cref{eq:hinter_rwa}, is a linear combination of a frequency $\omega_j$
appearing in the control input $u$ and eigenvalues of the constant Hamiltonian $\HH_c$),
the relevant figure of merit is given by
\begin{equation}
	\epsilon_{\mathrm{RWA}} = \frac{u_{\max{}}}{\omega_{\min}}.
\end{equation}

We emphasize that, in all generality, this result does not apply \emph{as is}
to the full Lindblad evolution given by~\cref{eq:lindblad_circuit}
because of the dissipation on mode $b$,
which should be taken into account when performing the averaging analysis.
We explicitly neglect any such potential coupling
between the RWA and the dissipation
under the assumption of a strict separation of timescales,
that is assuming that $\kappa_b\ll \omega_{\mathrm{min}}$
(intuitively, this entails that the effect of dissipation can be neglected
on the typical timescale of the slowest periodic terms in the Hamiltonian part of the dynamics).
With the parameters of the main text, $\omega_{\mathrm{min}}$ is of the same order
as $\omega_a$,
so that the previous constraint reads $\kappa_b \ll \omega_a$.
\bigparskip

\bigpar{Adiabatic Elimination}
Generally speaking,
adiabatic elimination covers a set of techniques used
to simplify the study of dissipative system featuring separated dissipation timescales,
by eliminating the rapidly dissipating degrees of freedom
and deriving the effective dynamics of the remaining degrees of freedom.
In the context of open quantum systems governed by Lindblad equations,
it is thus adapted to the case where the dynamics of the system can be written
in the form
\begin{equation}
	\frac{d\rho_t}{dt}
		= \mathcal L_{\mathrm{fast}}(\rho_t)
		+ \epsilon_{\mathrm{AE}} \, \mathcal L_{\mathrm{slow}} (\rho_t)
\end{equation}
where, for $\epsilon_{\mathrm{AE}}=0$, that is considering the effect of $\mathcal L_{\mathrm{fast}}$ only,
the system converges to a stationary regime.
In view of~\cref{eq:lindblad_circuit},
the fast part of the dynamics would be the intrinsic dissipation of mode $b$,
$\mathcal L_{\mathrm{fast}}(\rho_t) = \kappa_b \cD[\bop](\rho_t)$
which, if considered alone, makes the system converge to a state where mode $b$ is in vacuum,
that is $\rho_\infty = \rho_{\infty,a}\otimes |0_b\rangle\langle 0_b|$;
while the slow part of the dynamics would be the Hamiltonian coupling between the two modes:
$\epsilon_{\mathrm{AE}} \, \mathcal L_{\mathrm{slow}} (\rho_t)
 = -i[\HH(t), \rho_t]$.
This choice assumes that the coupling is much weaker than the natural dissipation of mode $b$;
the relevant small parameter for adiabatic elimination is given by
\begin{equation}
	\epsilon_{\mathrm{AE}} = \frac{u_{\max{}}}{\kappa_b}.
\end{equation}
In practice, instead of considering the true time-dependent Hamiltonian $\HH(t)$,
we will rather perform adiabatic elimination on the system obtained
after the rotating waves approximation,
thus considering
$\epsilon_{\mathrm{AE}} \, \mathcal L_{\mathrm{slow}} (\rho_t)
= -i[\HH_{\mathrm{RWA}}, \rho_t]$.
We emphasize that, in all generality, the interplay of these two approximations is unclear.
We choose to perform, independently, RWA before adiabatic elimination,
by relying once again on the assumption of strict timescale separation $\kappa_b \ll \omega_{\mathrm{min}}$
that we announced in the previous paragraph.
Note that this assumption introduces a hierarchy
$u_{\max{}} \ll \kappa_b \ll \omega_{\mathrm{min}}$,
and in particular $\epsilon_{\mathrm{RWA}} \ll \epsilon_{\mathrm{AE}}$.

A crucial property of the dynamics we consider is that it describes a bipartite quantum system,
made of two coupled harmonic oscillators,
but the \emph{fast} part of the dynamics, to be eliminated, is acting only on one of the systems
(the ancillary mode $b$).
Specific adiabatic elimination formulas for this setting can be found
in~\cite{azouitGenericAdiabatic2017},
which extensively studied the specific case of bipartite quantum systems
(see also the related PhD thesis~\cite{azouitAdiabaticElimination2017}
for a more extensive and pedagogical presentation).
We recall here the main results that will be useful in our analysis.

Consider the following Lindblad equation on two coupled systems $a$ and $b$:
\begin{equation}
	\frac{d\rho}{dt}
		= -i g [\HH, \rho] + \kappa_b \cD[\bop](\rho)
	\label{eq:ae_setting}
\end{equation}
and assume the following form of the coupling Hamiltonian:
\begin{equation}
	\HH = \sum_{1\leq k\leq n_H} \Aop_k \otimes \Bop_k,
	\label{eq:ae_setting_H}
\end{equation}
where $\Aop_k$ and $\Bop_k$ are
operators acting respectively on system $a$ and $b$,
not necessarily Hermitian but such that the whole sum is.
Define the Gram matrix $G$ whose coefficients are given by
\begin{equation}
	G_{k,k'} = 
			\langle 0_b | \, \Bop_k^\dag (\bop^\dag \bop)^{-1} \Bop_{k'} \, |0_b\rangle
\end{equation}
with $(\bop^\dag \bop)^{-1}$ the Moore-Penrose inverse of $\bop^\dag \bop$,
defined in the Fock basis by
\begin{align*}
	(\bop^\dag \bop)^{-1} |n\rangle &= \frac1n |n\rangle, \quad n\geq 1,\\
	(\bop^\dag \bop)^{-1} |0\rangle &= 0.
\end{align*}
Define also $\Lambda$ a Cholesky square-root of the Gram matrix $G$, that is
$G = \Lambda^\dag \Lambda$.
Then, up to third order terms (in $\epsilon_{\mathrm{AE}} = \frac g{\kappa_b}$),
we can perform adiabatic elimination of mode $b$ in~\cref{eq:ae_setting},
yielding the Lindblad equation
\begin{equation}
	\begin{aligned}
	\frac{d\rho_{a}}{dt}
		&= -i \, g \left[ \sum_{1\leq k\leq n_H} \langle 0_b | \Bop_k |0_b\rangle \, \Aop_k, %
				\rho_{a} \right]\\
		&\quad	+ \frac{4g^2}{\kappa_b} \, \sum_{1\leq k\leq n_H} \cD[\LL_k](\rho_a)
	\label{eq:ae_2ndorder}
	\end{aligned}
\end{equation}
with $\LL_k = \sum_{1\leq k'\leq n_H} \Lambda_{k,k'} \Aop_{k'}$.
More precisely, it was shown in~\cite{azouitGenericAdiabatic2017}
that if $\rho_a$ is a solution to~\cref{eq:ae_2ndorder},
then one can build
a Kraus map $\mathcal K_{\mathrm{AE}}$ close to identity
such that
$\rho = \mathcal K_{\mathrm{AE}} \left( \rho_a \otimes |0_b\rangle\langle 0_b| \right)$
is a solution to the original equation~\cref{eq:ae_setting} up to third-order terms
(note in particular that in general, this is \emph{not} equivalent to
taking the partial trace of $\rho$ with respect to system $b$;
in fact, they were able to prove that beyond first-order,
the partial trace generally does not follow a proper Lindblad evolution).
\newline
\input{sm__beyond_rwa_discussion.tex}

\subsection{General strategy}
\label{ssec:general_strategy}
Let us sketch the strategy to engineer the desired four-dissipators dynamics
of~\cref{eq:lindblad_target4dissip}
from the physically accessible controlled dynamics of~\cref{eq:lindblad_circuit},
leveraging both adiabatic elimination and the rotating wave approximation.
\newline

\bigpar{Adiabatic elimination}
Assume for now that we want to engineer only one of the Lindblad operators
appearing in~\cref{eq:lindblad_target4dissip},
that we will write $\LL$.

Using adiabatic elimination, we can engineer instead the dynamics
\begin{equation}
	\frac{d\rho}{dt} = -i g \left[ \LL \bop^\dag + \LL^\dag \bop, \rho\right]
				+ \kappa_b \cD[\bop](\rho)
	\label{eq:eng_adiab_onedissip}
\end{equation}
with $g>0$ a coupling parameter.
Indeed, this fits perfectly in the setting of~\cref{eq:ae_setting,eq:ae_setting_H}
with $n_H=2$ and
\begin{align*}
	\Aop_1 &= \LL, \quad \Bop_1 = \bop^\dag,\\
	\Aop_2 &= \LL^\dag, \quad \Bop_2 = \bop.
\end{align*}
The Gram matrix $G$ is particularly easy to compute in this case, and we find
\begin{equation}
	G = G^\dag G =
	\begin{bmatrix}
		1 & 0\\ 0 & 0
	\end{bmatrix}
\end{equation}
so that $\Lambda = G$;
moreover $\langle 0_b | \Bop_1 | 0_b\rangle = \langle 0_b | \Bop_2 | 0_b\rangle = 0$.
Using~\cref{eq:ae_2ndorder}, we obtain the following equation after adiabatic elimination:
\begin{equation}
	\frac{d\rho}{dt} = \Gamma \, \cD[\LL](\rho)
	\label{eq:one_dissip_ae_result}
\end{equation}
with $\Gamma = \frac{4g^2}{\kappa_b}$.

Let us quickly mention a slight variation on this idea that will turn out useful later on.
In the Hamiltonian coupling term of~\cref{eq:eng_adiab_onedissip},
we can consider a photon-number dependent correction on the ancillary mode $b$:
\begin{equation}
	\begin{aligned}
		\frac{d\rho}{dt} &= -i g \left[ \LL \left( \bop^\dag \, \mu(\bop^\dag\bop)^\dag \right)
					+ \LL^\dag \left(  \mu(\bop^\dag\bop) \, \bop\right),
						\rho\right]\\
				&\quad + \kappa_b \cD[\bop](\rho)
	\label{eq:eng_adiab_onedissip_modif}
	\end{aligned}
\end{equation}
where $\mu$ is some complex-valued function.
This still fits in the previous setting,
with now
$\Bop_1 = \bop^\dag \, \mu(\bop^\dag\bop)^\dag$,
$\Bop_2 = \mu(\bop^\dag\bop) \, \bop$.
After adiabatic elimination using~\cref{eq:ae_2ndorder},
we obtain the same Lindblad equation as in~\cref{eq:one_dissip_ae_result}
but with a modified rate
$\Gamma =  |\mu(0)|^2 \, \frac{4g^2}{\kappa_b}$.
Such photon-number dependent corrections
in the coupling with $b$ are thus straightforward to accommodate in this formalism.
In particular, if one is only interested in the effective dynamics after adiabatic elimination of
mode $b$,
it is equivalent to engineer the dynamics given by~\cref{eq:eng_adiab_onedissip}
with a coupling strength $g$
or the dynamics given by~\cref{eq:eng_adiab_onedissip_modif}
with a renormalized coupling strength
$g_\mu := g/|\mu(0)|$.
\bigparskip

\bigpar{From one to multiple dissipators}
At this stage,
the technique of the previous paragraph
only allows for the engineering of a single dissipator.
To adapt it to the
engineering of the full dynamics with four dissipators,
one solution is to use four ancillary modes $b_k, 0\leq k \leq 3$, and engineer
\begin{equation}
	\frac{d\rho}{dt} =
		-ig \left[ \sum_{0\leq k\leq 3} (\LL_k \bop_k^\dag + \LL_k^\dag \bop_k), \rho\right]
		+ \kappa_b \sum_{0\leq k\leq 3}\cD[\bop_k](\rho).
\end{equation}
It is straightforward to see that adiabatic elimination can be extended to that case and yields
\begin{equation}
	\frac{d\rho}{dt} = \Gamma \, \sum_{0\leq k\leq 3} \cD[\LL_k](\rho).
\end{equation}
A more hardware-efficient solution, requiring only one ancillary mode $b$,
consists in activating stroboscopically each dissipator.
We introduce a periodic switching function $k(t)$ that is piecewise constant
and cycles through $k\in\{0,1,2,3\}$
with a switching time $T_{\mathrm{switch}}$.
Consider the evolution
\begin{equation}
	\frac{d\rho}{dt} =
		-ig \left[ \LL_{k(t)} \bop^\dag + \LL_{k(t)}^\dag \bop, \rho\right]
		+ \kappa_b \cD[\bop](\rho).
\end{equation}
Assuming $T_{\mathrm{switch}}$ to be much larger than $1/\kappa_b$,
we can perform piecewise adiabatic elimination to get
\begin{equation}
	\frac{d\rho}{dt} = \Gamma \, \cD[\LL_{k(t)}](\rho).
\end{equation}
Assuming now the switching period to be much shorter than $1/\Gamma$,
a first-order (in $\Gamma T_{\mathrm{switch}}$) Trotter approximation yields
\begin{equation}
	\frac{d\rho}{dt} = \frac{\Gamma}4 \, \sum_{0\leq k\leq 3} \cD[\LL_k](\rho)
\end{equation}
which is exactly the desired evolution, but with a reduced $\Gamma/4$ engineered dissipation rate.
Note that for the generalization to hexagonal GKP states
proposed in~\cref{smsec:hexacode},
where the engineering of six dissipators is required,
one has to choose between using six ancillary modes
or using a similar Trotter decomposition with an effective dissipation rate of $\Gamma/6$.

From now on, we assume that one of these two solutions is adopted,
and focus on the engineering of only one dissipator.
\bigparskip

\bigpar{Rotating wave approximation}
Let us now focus on how to engineer the evolution of~\cref{eq:eng_adiab_onedissip}
from that of~\cref{eq:lindblad_circuit},
that is how to engineer the Hamiltonian
\begin{equation}
	\HH_{\mathrm{AE}} = g \left( \LL \bop^\dag + \LL^\dag \bop \right),
\end{equation}
where $\LL$ is one of the Lindblad operators in the target dynamics,
given
\begin{equation}
	\HH(t) = 2E_J \xi(t) \cos(\eta_a \qop_a(t) + \eta_b\qop_b(t) - \zeta(t)).
\end{equation}

We first remark that, for any complex number $\upepsilon$,
a contribution of the form $\upepsilon\bop^\dag + \upepsilon^*\bop$
in the interaction Hamiltonian
can be simply implemented as a resonant drive on the ancillary mode $b$.
Consequently, we can always ignore scalar terms in the Lindblad operator $\LL$ to engineer
by exploiting the decomposition
\begin{equation}
	\HH_{\mathrm{AE}} = g \left( (\LL-\tfrac\upepsilon g\II) \bop^\dag
					+ (\LL-\tfrac\upepsilon g\II)^\dag \bop \right)
			    + (\upepsilon\bop^\dag + \upepsilon^*\bop)
\end{equation}
and assuming that the rightmost term is engineered with a resonant drive on mode $b$.
For the remaining of this section,
unless explicitly stated otherwise,
we will thus allow ourselves to implicitly identify the problem of engineering the Lindblad
operators
$\LL_k = e^{i\frac{k\pi}{2}\aop^\dag\aop} \, \cA e^{i\eta\qop_a}(\II-\epsilon\Pop_a)
			\, e^{- i\frac{k\pi}{2}\aop^\dag\aop} - \II$,
introduced in~\cref{eq:exp_dissip},
with that
of engineering
$\LL_k+\II = e^{i\frac{k\pi}{2}\aop^\dag\aop} \, \cA e^{i\eta\qop_a}(\II-\epsilon\Pop_a)
			\, e^{- i\frac{k\pi}{2}\aop^\dag\aop}$;
similarly, for the equivalent Lindblad operators of~\cref{eq:sym_dissip},
we will replace the symmetric operators $\LL_{r,s}$ by $\LL_{r,s} + \sqrt2 \II$.

Finally, using~\cref{eq:rwa1_recipe}, one must find a quasi-periodic
control signals $\xi(t)$
such that
\begin{equation}
\overline{\HH} = \lim_{T\rightarrow +\infty} \frac1T \int_0^T \HH(t) dt
		= g(\LL \, \bop^\dag + \LL^\dag \,\bop)
\end{equation}
(where, anticipating slightly the results of~\cref{ssec:constant_zeta},
we announce that we will be able to choose a constant value of $\zeta(t)$).
In the following sections, we explain how to choose such control signals,
taking into account experimental limitations.
We refer to~\cref{smsec:numerics_rwa1} for numerical estimations of the logical decoherence rates
associated to imperfections of these control signals.


\subsection{Driving with frequency combs}
For the sake of pedagogy,
we first consider arbitrary time-dependent
control signals $\xi$ and $\zeta$.
We can rewrite the interaction Hamiltonian in~\cref{eq:hamiltonian_int_rwa}
as
\begin{align}
	\HH(t) &= u(t) e^{i\eta_a \qop_a(t)} e^{i\eta_b\qop_b(t)} +\hc\\
	u(t) &= E_j \, \xi(t) e^{-i\zeta(t)}. \label{eq:u_xi_exp_izeta}
\end{align}
It is thus enough to design the complex-valued control signal $u$,
from which we can easily deduce $\xi$ and $\zeta$
(respectively from the amplitude and phase of $u$).
\bigparskip

\bigpar{Equivalent expression of the target Lindblad operators}
We introduce another point of view on the target Lindblad operators
which will help clarify our choice of control signals.
Let us first consider the Lindblad operator from~\cref{eq:exp_dissip}
\begin{equation}
	\LL_0 = \cA e^{i\eta_a \qop_a} \left( \II-\epsilon \Pop_a\right)-\II.
\end{equation}
As previously explained, we ignore any scalar term in Lindblad operators,
as we engineer them separately through direct drives on the ancilla mode.
We thus focus on
\begin{equation}
	\widetilde \LL_0 := \LL_0 + \II = \cA e^{i\eta_a \qop_a} \left( \II-\epsilon \Pop_a\right)
			= \cA e^{i\eta_a \qop_a} - \cA \epsilon e^{i\eta_a \qop_a} \Pop_a.
\end{equation}
We can write
\begin{equation}
	e^{i\eta_a\qop_a} = \left( e^{i\eta_a\qop_a(t)}\right) \rvert_{t=0}.
\end{equation}
Additionally, computing the derivative of the time-dependent operator
$e^{i\eta_a\qop_a(t)}$, we get:
\begin{align}
	\frac{d}{dt} \Big( &e^{i\eta_a \qop_a(t)} \Big)
	= \frac{d}{dt} \left( e^{i\eta_a \left( \cos(\omega_a t)\qop_a + \sin(\omega_a t)\Pop_a\right)} \right)\notag\\
	&= \frac{d}{d t} \left( e^{i\omega_a t\,\aop^\dag \aop} \, e^{i\eta_a \qop_a} \, e^{-i\omega_a t\,\aop^\dag \aop} \right)\notag\\
	&= i\omega_a \, e^{i\omega_a t\,\aop^\dag \aop} [ \aop^\dag \aop, e^{i\eta_a \qop_a} ] e^{-i\omega_a t\,\aop^\dag \aop}\notag\\
	&= \tfrac{i\omega_a\eta_a}{2} \, e^{i\omega_a t\,\aop^\dag \aop}
	\left( \Pop_a \, e^{i\eta_a \qop_a} +  e^{i\eta_a \qop_a} \Pop_a\right)
	e^{-i\omega_a t\,\aop^\dag \aop}\notag\\
	&= \tfrac{i\omega_a\eta_a}{2} \, e^{i\omega_a t\,\aop^\dag \aop}
	\left( [\Pop_a, e^{i\eta_a \qop_a}] +  2 e^{i\eta_a \qop_a} \Pop_a\right)
	e^{-i\omega_a t\,\aop^\dag \aop}\notag\\
	&= i\omega_a \eta_a \, e^{i\omega_a t\,\aop^\dag \aop}
	\, e^{i\eta_a \qop_a} \left( \Pop_a + \tfrac{\eta_a}2 \II\right)
	e^{-i\omega_a t\,\aop^\dag \aop}\notag\\
	&= i\omega_a \eta_a \, e^{i\eta_a \qop_a(t)} \left( \Pop_a(t) + \tfrac{\eta_a}2 \II\right)
\label{eq:deriv_eietaq}
\end{align}
with $\Pop_a(t) = \cos(\omega_a t)\Pop_a - \sin(\omega_a t)\qop_a$,
so that
\begin{equation}
	\begin{aligned}
	e^{i\eta_a \qop_a}\, \Pop_a
		&= -\tfrac{i}{\omega_a\eta_a} \, \left( \tfrac{d}{dt} e^{i\eta_a \qop_a(t)}\right) \rvert_{t=0}
		\\ &\quad
		- \tfrac{\eta_a}2 \left( e^{i\eta_a\qop_a(t)}\right) \rvert_{t=0}
	\end{aligned}
\end{equation}
and finally
\begin{equation}
	\widetilde\LL_0 = \cA (1+\tfrac{\epsilon\eta_a}2)
				\left( e^{i\eta_a\qop_a(t)}\right) \rvert_{t=0}
			+ \tfrac{i\epsilon\cA}{\omega_a \eta_a}
				\left( \tfrac{d}{dt} e^{i\eta_a \qop_a(t)}\right) \rvert_{t=0}.
	\label{eq:L0_expr_deriv}
\end{equation}
Similarly, for the rotated Lindblad operators $\LL_k, 1\leq k\leq 3$, we get
\begin{equation}
	\begin{aligned}
		\widetilde\LL_k &= \cA (1+\tfrac{\epsilon\eta_a}2)
				\left( e^{i\eta_a\qop_a(t)}\right) \rvert_{(t=\tfrac{k\pi}{2\omega_a})}
			\\&\quad
			+ \tfrac{i\epsilon\cA}{\omega_a \eta_a}
				\left( \tfrac{d}{dt} e^{i\eta_a \qop_a(t)}\right) \rvert_{(t=\tfrac{k\pi}{2 \omega_a})}.
	\label{eq:Lk_expr_deriv}
	\end{aligned}
\end{equation}
We can get similar expressions for the symmetric and antisymmetric
Lindblad operators in~\cref{eq:sym_dissip}
as linear combinations of the previous ones:
\begin{equation}\begin{aligned}
	\widetilde \LL_{q,s}
			&:= \LL_{q,s} + \sqrt2 \II
				= \frac{ \widetilde \LL_0 + \widetilde \LL_2}{\sqrt2}\\
			&= \tfrac\cA{\sqrt2} (1+\tfrac{\epsilon\eta_a}2)
				\left( e^{i\eta_a\qop_a(t)}\right) \rvert_{t=0}
			\\&\quad\quad
			+ \tfrac\cA{\sqrt2} (1+\tfrac{\epsilon\eta_a}2)
				\left( e^{i\eta_a\qop_a(t)}\right) \rvert_{t=\frac\pi{\omega_a}}\\
			&\quad\quad
				+ \tfrac{i\epsilon\cA}{\sqrt2\omega_a \eta_a}
				\left( \tfrac{d}{dt} e^{i\eta_a \qop_a(t)}\right) \rvert_{t=0}
			\\&\quad\quad
				+ \tfrac{i\epsilon\cA}{\sqrt2\omega_a \eta_a}
				\left( \tfrac{d}{dt} e^{i\eta_a \qop_a(t)}\right) \rvert_{t=\frac\pi{\omega_a}}\\[0.5em]
			&= \sqrt2\cA (1+\tfrac{\epsilon\eta_a}2)
					\, \cos\left(\eta_a\qop_a(t)\right) \rvert_{t=0}
			\\&\quad\quad
			+ \tfrac{i\sqrt2 \epsilon\cA}{\omega_a \eta_a}
				\,\left(\tfrac{d}{dt} \cos\left(\eta_a\qop_a(t)\right) \right)
										\rvert_{t=0},
\end{aligned}
\label{eq:Lsym_expr_deriv1}
\end{equation}
\begin{equation}\begin{aligned}
	\widetilde \LL_{p,s}
			&= \sqrt2\cA (1+\tfrac{\epsilon\eta_a}2)
					\, \cos\left(\eta_a\qop_a(t)\right) \rvert_{t=\frac\pi{2 \omega_a}}
			\\&\quad\quad
			+ \tfrac{i\sqrt2 \epsilon\cA}{\omega_a \eta_a}
				\,\left(\tfrac{d}{dt} \cos\left(\eta_a\qop_a(t)\right) \right)
										\rvert_{t=\frac\pi{2 \omega_a}},
\end{aligned}
\label{eq:Lsym_expr_deriv2}
\end{equation}
\begin{equation}\begin{aligned}
	\LL_{q,d}
			&= \sqrt2\cA (1+\tfrac{\epsilon\eta_a}2)
					\, \sin\left(\eta_a\qop_a(t)\right) \rvert_{t=0}
			\\&\quad\quad
			+ \tfrac{i\sqrt2 \epsilon\cA}{\omega_a \eta_a}
				\,\left(\tfrac{d}{dt} \sin\left(\eta_a\qop_a(t)\right) \right)
										\rvert_{t=0},
\end{aligned}
\label{eq:Lsym_expr_deriv3}
\end{equation}
\begin{equation}\begin{aligned}
	\LL_{p,d}
			&= \sqrt2\cA (1+\tfrac{\epsilon\eta_a}2)
					\, \sin\left(\eta_a\qop_a(t)\right) \rvert_{t=\frac\pi{2 \omega_a}}
			\\&\quad\quad
			+ \tfrac{i\sqrt2 \epsilon\cA}{\omega_a \eta_a}
				\,\left(\tfrac{d}{dt} \sin\left(\eta_a\qop_a(t)\right) \right)
										\rvert_{t=\frac\pi{2 \omega_a}}.
\end{aligned}
\label{eq:Lsym_expr_deriv4}
\end{equation}
\vspace{3em}

\bigpar{Two-mode coupling with modulated frequency combs}
Let us now denote by $\LL$ any of the previously considered Lindblad operators
(possibly stripped of any scalar term),
that is either $\widetilde \LL_k$ for $0\leq k\leq 3$
or $\widetilde \LL_{r,l}$ for $r\in\{q,p\}$ and $l\in\{s,d\}$.
Following the general strategy exposed in~\cref{ssec:general_strategy},
we need to find a complex-valued control signal $u$ such that
\begin{equation}
	\overline{\HH(t)}
	= \overline{ u(t) e^{i\eta_a\qop_a(t)} e^{i\eta_b\qop_b(t)}+\hc}
	= g \left( \LL \bop^\dag + \LL^\dag\bop\right).
\end{equation}
Recall from the previous exposition of adiabatic elimination that we
can slightly relax this requirement to
\begin{equation}
	\overline{\HH(t)}
	= g_\mu \, \left( \LL \Bop^\dag + \LL^\dag \Bop \right)
	\label{eq:rwa_cond_Bop_mu}
\end{equation}
where $\Bop = \mu(\bop^\dag \bop) \bop$ for some function $\mu$
and $g_\mu = g/|\mu(0)|$.

Assuming $\omega_a$ and $\omega_b$ to be incommensurate,
\cref{eq:rwa_cond_Bop_mu} can be solved by
finding a separable control signal
\begin{equation}
	u(t) = g_\mu \, u_a(t) \, u_b(t)
	\label{eq:u_separable}
\end{equation}
where
$u_a$ is $\tfrac{2\pi}{\omega_a}$ periodic,
$u_b$ is $\tfrac{2\pi}{\omega_b}$ periodic,
and such that
\begin{equation}
	\overline{u_a(t) e^{i\eta_a\qop_a(t)}} = \LL,
	\quad \overline{u_b(t) e^{i\eta_b\qop_b(t)}} = \Bop^\dag.
\end{equation}
Let us solve for $u_b$ first.
We use the following operator decomposition of $e^{i\eta\qop_b}$,
obtained in~\cref{eq:op_decomposition_eietaq,eq:coefs_decomposition_eietaq}
of~\cref{smsec:numerical_details}:
\begin{equation}
	\begin{aligned}
		e^{i\eta_b\qop_b} &= \phi_0(\bop^\dag \bop; \eta_b)
	\\&\quad
	+ \sum_{k=1}^{+\infty}
	i^k \left(
	\phi_k(\bop^\dag \bop; \eta_b) \, \bop^k
	+ \bop^{\dag k} \, \phi_k(\bop^\dag \bop; \eta_b)
	\right)
	\label{eq:opdecomp_eietabqb}
	\end{aligned}
\end{equation}
where the $\phi_k$ are real-valued functions defined by
\begin{equation}
	\phi_k(n; \eta) = (-i)^k \sqrt{\frac{n!}{(n+k)!}} \, \bra n e^{i\eta\qop} \ket{n+k}.
\end{equation}
We remind the reader that, so far, this decomposition
can simply be understood as regrouping the coefficients of $e^{i\eta_b\qop_b}$
in the Fock basis along each diagonal,
and refer to~\cref{smsec:numerical_details} for details.
Combining~\cref{eq:opdecomp_eietabqb}
with the relations
\begin{align*}
	&e^{i\eta_b\qop_b(t)}
	= e^{i\omega_b t \, \bop^\dag \bop}
		\, e^{i\eta_b\qop_b}
		\, e^{-i\omega_b t \, \bop^\dag \bop}\\
	&e^{i\omega_b t \, \bop^\dag \bop} \, \bop \, e^{-i\omega_b t \, \bop^\dag \bop}
	= e^{-i\omega_b t} \, \bop
\end{align*}
we can extend the previous operator decomposition into
\begin{equation}
	\begin{aligned}
		e^{i\eta_b\qop_b(t)} &=
	\phi_0(\bop^\dag \bop; \eta_b)
				\\&\quad
	+ \sum_{k=1}^{+\infty}
	i^k \Big(
	\phi_k(\bop^\dag \bop; \eta_b) \, \bop^k \, e^{-ik\omega_b t}
\\&\quad\quad\quad\quad\quad
	+ e^{ik\omega_b t} \, \bop^{\dag k} \, \phi_k(\bop^\dag \bop; \eta_b)
	\Big).
	\label{eq:opdecomp_eietabqb_time}
	\end{aligned}
\end{equation}
In particular, we get
\begin{equation}
	\overline{ u_b(t) e^{i\eta_b\qop_b(t)} }
	= i \, \bop^\dag \phi_1(\bop^\dag \bop; \eta_b)
	= \Bop^\dag
	\label{eq:bdag_rwab}
\end{equation}
for $u_b(t) = e^{-i\omega_b t}$ and $\Bop := -i \, \phi_1(\bop^\dag \bop; \eta_b) \bop$.
With this operator $\Bop$, we find
\begin{equation}
	g_\mu = g/|\phi_1(0; \eta_b)| = g/ |\langle 0_b | e^{i\eta_b \qop_b} | 1_b\rangle |
			= g \tfrac{\sqrt2}{\eta_b} e^{\eta_b^2/4}.
\end{equation}
\vspace{0.5em}

Let us now solve for $u_a$. We can directly read the desired control signal
from the expression of the target Lindblad operators
obtained in~\cref{eq:L0_expr_deriv,eq:Lk_expr_deriv,%
eq:Lsym_expr_deriv1,eq:Lsym_expr_deriv2,eq:Lsym_expr_deriv3,eq:Lsym_expr_deriv4%
}.
Indeed, defining the Dirac comb of period $T = \frac{2\pi}{\omega}$
as
\begin{equation}
		\Sha_{T} (t) = \sum_{k\in\mathbb Z} \delta(t-k T)
				= \frac1T \sum_{k\in\mathbb Z} e^{\frac{2ik\pi}T t},
\end{equation}
\cref{eq:L0_expr_deriv}
can be recast as
\begin{equation}\begin{aligned}
	\widetilde\LL_0 
			&= \overline{u_{a,0}(t) e^{i\eta_a\qop_a(t)}},\\
	u_{a,0}(t) &:= \tfrac{2\pi}{\omega_a} \cA \left(
			(1+\tfrac{\epsilon\eta_a}2)
			\Sha_{\frac{2\pi}{\omega_a}}(t)
			- \tfrac{i\epsilon}{\omega_a \eta_a}
			\Sha'_{\frac{2\pi}{\omega_a}}(t)
			\right)
\end{aligned}
\label{eq:u_a_0_dirac}
\end{equation}
where $\Sha'$ denotes the time-derivative of $\Sha$.
Using~\cref{eq:u_separable},
the full control signal is thus given by
\begin{widetext}
\begin{equation}
	u_0(t) = g_\mu u_a(t) u_b(t)
		= \tfrac{2\sqrt2\pi}{\eta_b\omega_a} \, e^{\eta_b^2/4} \,g \cA
		\left(
		(1+\tfrac{\epsilon\eta_a}2)
		\Sha_{\frac{2\pi}{\omega_a}}(t)
		- \tfrac{i\epsilon}{\omega_a \eta_a}
		\Sha'_{\frac{2\pi}{\omega_a}}(t)
		\right)
		\, e^{-i\omega_b t}.
		\label{eq:u0_complex}
\end{equation}
Similarly, we get
\begin{equation}\begin{aligned}
	\widetilde\LL_k 
			&= \overline{u_{a,k}(t) e^{i\eta_a\qop_a(t)}},\\
	u_{a,k}(t) &:= \tfrac{2\pi}{\omega_a}\left(
			\cA (1+\tfrac{\epsilon\eta_a}2)
			\Sha_{\frac{2\pi}{\omega_a}}(t-\tfrac{k\pi}{2 \omega_a})
			- \tfrac{i\epsilon\cA}{\omega_a \eta_a}
			\Sha'_{\frac{2\pi}{\omega_a}}(t-\tfrac{k\pi}{2 \omega_a})
			\right).
\end{aligned}\end{equation}
Finally,
we obtain control signals for the engineering of the operators $\LL_{r,l}$
as linear combinations of the previous ones;
we can compactly express the result as
\begin{equation}\begin{aligned}
	\widetilde\LL_{r,l}
			&= \overline{u_{a,r,l}(t) e^{i\eta_a\qop_a(t)}},\\
	u_{a,r,l}(t) &:= \tfrac{\sqrt2\pi\cA}{\omega_a}
			(1+\tfrac{\epsilon\eta_a}2)
			\left(
			\Sha_{\frac{2\pi}{\omega_a}}(t-t_r)
			+ (-1)^{\delta_l} \Sha_{\frac{2\pi}{\omega_a}}(t-t_r - \tfrac\pi{\omega_a})\right)\\
			&\quad\quad
			- \tfrac{i\sqrt2\pi\epsilon\cA}{\omega_a^2 \eta_a}
			\left(
			\Sha'_{\frac{2\pi}{\omega_a}}(t-t_r)
			+ (-1)^{\delta_l} \Sha'_{\frac{2\pi}{\omega_a}}(t-t_r-\tfrac\pi{\omega_a})
			\right)
\end{aligned}
\label{eq:u_rl_dirac_ideal}
\end{equation}
\end{widetext}
where we introduced $\delta_l$ defined as $\delta_s=0, \delta_d=1$
and $t_r$ defined as $t_q=0, t_p = \frac\pi{2 \omega_a}$.

\subsection{Taking experimental constraints into account}
\label{ssec:exp_limit_u}
The control signals obtained so far,
albeit quasi-periodic,
feature harmonics of unbounded frequency and amplitude,
as seen from the Fourier series
\( \Sha_T(t) = \frac1T \sum_{k\in\mathbb Z} e^{ik\omega t} \)
and
\( \Sha_T'(t) = \frac{i\omega}T \sum_{k\in\mathbb Z} k \, e^{ik\omega t} \).
We thus need to study their approximation by a signal of limited
bandwidth and amplitude;
in particular,
through the definition given in~\cref{eq:def_xi},
we see that $|\xi(t)|=|u(t)|/E_J$ cannot excess $1$.

\vspace{1em}
\bigpar{Finite-bandwidth of the control signals}
To get rid of the derivative of a Dirac comb of period $T = 2\pi/\omega$,
we approximate it by a (symmetric) finite difference as
\begin{equation}
	\Sha_T'(t) \simeq \frac{\Sha_T(t+\delta) - \Sha_T(t-\delta)}{2\delta}
\end{equation}
for an arbitrary parameter $\delta>0$.
Note that in the Fourier domain,
this amounts to approximating
the quantity $k\omega$
(appearing in the Fourier coefficients of $\Sha_T'$)
by ${\sin(k\omega\delta)}/{\delta}$,
which is a bounded function of $k$.
For instance, this leads to replacing the control signal proposed in~\cref{eq:u0_complex}
by
\begin{widetext}
\begin{equation}
	u_{0}^{(\delta)}(t) = \frac{2\sqrt2\pi}{\eta_b\omega_a} \, e^{\eta_b^2/4} \, g \cA\, \left(
			\left( 1 + \frac{\epsilon\eta_a}2 \right) \Sha_{\frac{2\pi}{\omega_a}}(t)
			- \frac{i\epsilon}{2\eta_a\omega_a\delta} \Sha_{\frac{2\pi}{\omega_a}}(t+\delta)
			+ \frac{i\epsilon}{2\eta_a\omega_a\delta} \Sha_{\frac{2\pi}{\omega_a}}(t-\delta)
		\right) \, e^{-i\omega_b t}.
	\label{eq:first_try_ua0_delta}
\end{equation}
\end{widetext}

We can slightly adjust this finite difference approximation
by revisiting the analysis led in~\cref{eq:deriv_eietaq}
when replacing exact time-derivatives by finite differences.
Up to second order terms in $\delta$ and using the Baker-Campbell-Hausdorff formula,
we get
\begin{equation}
	\begin{aligned}
		&e^{i\eta_a\qop_a(t+\delta)}
		= e^{i\eta_a \left( \cos(\omega_a t + \omega_a \delta) \qop_a
					+ \sin(\omega_a t + \omega_a \delta) \Pop_a \right) }\\
		&\quad\quad
		\simeq e^{i\eta_a \qop_a(t) + i\eta_a \omega_a\delta \left( -\sin(\omega_a t) \qop_a
					+ \cos(\omega_a t) \Pop_a \right) }\\
		&\quad\quad
		= e^{i\eta_a \qop_a(t) + i\eta_a \omega_a\delta \,\Pop_a(t) }\\
		&\quad\quad
		= e^{\frac{i\eta_a^2\omega_a\delta}{2}}
			e^{i\eta_a \qop_a(t)} \, e^{i\eta_a \omega_a\delta \,\Pop_a(t)}\\
		&\quad\quad
		\simeq e^{\frac{i\eta_a^2\omega_a\delta}{2}}
			e^{i\eta_a \qop_a(t)} \,
			\left( \II + i\eta_a \omega_a\delta \,\Pop_a(t) \right)
	\end{aligned}
\end{equation}
\\

so that
\begin{align}
	&\frac{%
		e^{-\frac{i\eta_a^2\omega_a\delta}{2}} \, e^{i\eta_a\qop_a(t+\delta)}
		- e^{\frac{i\eta_a^2\omega_a\delta}{2}} \, e^{i\eta_a\qop_a(t-\delta)}
	}{2\delta}
		\notag\\[0.2em]&\quad \quad \simeq i\eta_a\omega_a \, e^{i\eta_a \qop_a(t)} \, \Pop_a(t).
\end{align}
As a consequence, instead of scaling the “centered” Dirac comb by $1+\epsilon\eta_a/2$
in~\cref{eq:first_try_ua0_delta},
we can adjust the phases of the off-centered Dirac combs coming from the finite-difference
approximation, leading to
\begin{widetext}
\begin{equation}
	u_{0}^{(\delta)}(t) = \frac{2\sqrt2\pi}{\eta_b\omega_a}  \, e^{\eta_b^2/4} \, g \cA \left(
			\Sha_{\frac{2\pi}{\omega_a}}(t)
			- \frac{i\epsilon\gamma}{2\eta_a\omega_a\delta} \Sha_{\frac{2\pi}{\omega_a}}(t+\delta)
			+ \frac{i\epsilon\gamma^*}{2\eta_a\omega_a\delta} \Sha_{\frac{2\pi}{\omega_a}}(t-\delta)
		\right) \, e^{-i\omega_b t}
\end{equation}
\end{widetext}
where we introduced a unitary complex number $\gamma = e^{\frac{i\eta_a^2\omega_a\delta}{2}}$.
Then, all Dirac combs
(including those stemming from the previous finite difference approximation)
are truncated to a finite number of harmonics in the Fourier domain as
\footnote{Note that $\Sha_T^{(N)}$ has $2N+1$ non-zero exponential Fourier coefficients,
or $N+1$ trigonometric Fourier coefficients as
$\Sha_T^{(N)}(t) := \frac 1T  + \frac2T \sum_{k=1}^N \cos(k\omega t)$.
To study the effect of the control bandwidth, our convention is to simply use N, giving
the maximum frequency, as the figure of merit.}

\vspace{-2em}
\begin{equation}
	\Sha_T(t) \simeq \Sha_T^{(N)}(t) := \frac 1T \sum_{k=-N}^N e^{ik\omega t}.
	\label{eq:trunc_dcomb}
\end{equation}
For instance, the previous control signal $u_{0}^{(\delta)}$
becomes
\begin{widetext}
\begin{equation}
	u_{0}^{(N,\delta)}(t) := \frac{2\sqrt2\pi}{\eta_b\omega_a}  \, e^{\eta_b^2/4} \, g \cA \left(
			\Sha_{\frac{2\pi}{\omega_a}}^{(N)}(t)
			- \frac{i\epsilon\gamma}{2\eta_a\omega_a\delta}
				\Sha_{\frac{2\pi}{\omega_a}}^{(N)}(t+\delta)
			+ \frac{i\epsilon\gamma^*}{2\eta_a\omega_a\delta}
				\Sha_{\frac{2\pi}{\omega_a}}^{(N)}(t-\delta)
		\right) e^{-i\omega_b t}.
	\label{eq:control_finiteN_L0}
\end{equation}
\end{widetext}

\bigpar{Bounded amplitude of the control signals}
Going back to the definition of a truncated Dirac comb in~\cref{eq:trunc_dcomb},
we see that its peak value is given by
\begin{equation}
	|\Sha_T^{(N)}(t)|\leq \tfrac{2N+1}T.
\end{equation}
For $N$ large enough, so that we can consider that at most one of the Dirac combs
in~\cref{eq:control_finiteN_L0} takes a non-negligible value at any given time,
we obtain the bound
\begin{equation}
	|u(t)| \leq (2N+1) \, \frac{\sqrt2 g\cA}{\eta_b} \, e^{\eta_b^2/4}
			\, \max(1,\frac{\epsilon}{2\eta_a\omega_a\delta}).
\end{equation}
In particular, while taking $N$ large and $\delta$ small is desirable to accurately
approximate the ideal control signal of~\cref{eq:u0_complex},
it also limits the achievable coupling rate $g$.
In practice, in the main text (and in every simulation presented),
we chose the finite difference parameter
\begin{equation}
	\delta = \frac{\epsilon}{2\eta_a\omega_a}
\end{equation}
so that all truncated Dirac combs have the same amplitude,
and the achievable coupling rate scales as $1/(2N+1)$;
numerical simulations are then used to find a balance between increasing the truncation number $N$
for accuracy and keeping a strong enough effective coupling $g$
(see~\cref{sec:harmonics} of the main text and~\cref{smsec:numerics_rwa1} for details of the simulations).
With this choice, the previous expression of the control signal is simplified to
\begin{widetext}
\begin{equation}
	u_{0}^{(N)}(t) := \frac{2\sqrt2\pi}{\eta_b\omega_a}  \, e^{\eta_b^2/4} \, g \cA \left(
			\Sha_{\frac{2\pi}{\omega_a}}^{(N)}(t)
			- i\gamma
				\Sha_{\frac{2\pi}{\omega_a}}^{(N)}(t+\tfrac{\epsilon}{2\eta_a\omega_a})
			+ i\gamma^*
				\Sha_{\frac{2\pi}{\omega_a}}^{(N)}(t-\tfrac{\epsilon}{2\eta_a\omega_a})
		\right) e^{-i\omega_b t}.
	\label{eq:control_finiteN_L0_fixeddiffstep}
\end{equation}
\end{widetext}
The previous analysis can be straightforwardly adapted
to deduce control signals $u_{k}^{(N)}$ corresponding to the engineering
of $\widetilde \LL_k$,
or
$u_{r,l}^{(N)}$ corresponding to the engineering of $\widetilde \LL_{r,l}$.

\subsection{Equivalent simpler scheme with a constant flux relation}
\label{ssec:constant_zeta}
Going back to~\cref{eq:def_xi,eq:def_zeta,eq:u_xi_exp_izeta},
linking the control signal $u$ to the actual circuit flux biases
$\Phi^{\mathrm{ext}}_J$
and
$\Phi^{\mathrm{ext}}_L$,
we see that the amplitude of $u$, piloted by $\xi$, depends only on $\Phi^{\mathrm{ext}}_J$;
while its phase, piloted by $\zeta$, depends only on
the linear combination~$\Phi^{\mathrm{ext}}_L+\Phi^{\mathrm{ext}}_J/2$.
With the control signals designed so far,
both control amplitude and phase vary rapidly as a function of time.
However, we will show that we can translate any complex-valued control signal 
to another control signal 
with constant phase
and still yielding the same interaction Hamiltonian after rotating wave approximation.
More precisely, using the previous notations,
the choice made in the main text corresponds to imposing $\zeta(t) = \pi/4$ at all times.
\newline

Let us denote the desired constant-phase control
as $u_F(t) e^{-i\pi/4}$ where $u_F$ is a real-valued function to be determined.
Let us then write the controlled interaction Hamiltonians corresponding to $u$ and $u_F$:
\begin{align}
	\HH(t) &= u(t) \, e^{i\eta_a\qop_a(t)} e^{i\eta_b\qop_b(t)} + \hc\\
	\HH_F(t) &= u_F(t) \, e^{i\eta_a\qop_a(t)} e^{i\eta_b\qop_b(t)} e^{-i\pi/4} + \hc \notag\\
		&= 2 \, u_F(t) \, \cos(\eta_a \qop_a(t) + \eta_b\qop_b(t) - \pi/4).
\end{align}
Rewriting the previous Hamiltonians as
\begin{equation}\begin{aligned}
	\HH(t) &=
		2 \, \Re(u(t)) \cos(\eta_a \qop_a(t)+\eta_b\qop_b(t))\\
		&\quad
		-2 \, \Im(u(t)) \sin(\eta_a \qop_a(t)+\eta_b\qop_b(t)),\\
	\HH_F(t) &=
		\sqrt2 \, u_F(t) \cos(\eta_a \qop_a(t)+\eta_b\qop_b(t))\\
		&\quad
		+\sqrt2 \, u_F(t) \sin(\eta_a \qop_a(t)+\eta_b\qop_b(t)),
\end{aligned}\end{equation}
we first see that one cannot hope for a pointwise equality
$\HH_F(t) = \HH(t)$ as this would entail that
\[
	\Re(u(t)) = -\Im(u(t)) = u_F(t)/\sqrt2,
\]
which would be satisfied only if $u$ already had a constant phase of $-\pi/4$;
this is not the case for the control signals designed in the previous sections.

One can thus only hope for an equality in average as required in~\cref{eq:rwa1_equality_ur}.
However, since we are only interested in the effective Hamiltonian
obtained after the rotating wave approximation,
we only need an equality in average:
\begin{equation}
	\overline{\HH} = \overline{\HH_F}.
	\label{eq:rwa1_equality_ur}
\end{equation}
To go further, we need to exploit the specific structure of the
control signals that we want to engineer.
Note from the analysis of the previous sections that
all complex-valued control signals that we considered
have the generic form
\begin{equation}
	u(t) = \sum_{r\in\mathbb Z} u_r e^{i \, r\omega_a t} e^{ -i\, \omega_b t}
\end{equation}
so that their real and imaginary part can be decomposed as
\begin{equation}\begin{aligned}
	\Re(u(t)) &=
		\Re\left( \sum_{r\in\mathbb Z} u_r e^{ir\omega_a t} \right) \cos(\omega_b t)\\
	&\quad	+ \Im\left( \sum_{r\in\mathbb Z} u_r e^{ir\omega_a t} \right) \sin(\omega_b t),\\
	\Im(u(t)) &=
		\Im\left( \sum_{r\in\mathbb Z} u_r e^{ir\omega_a t} \right) \cos(\omega_b t)\\
	&\quad
	- \Re\left( \sum_{r\in\mathbb Z} u_r e^{ir\omega_a t} \right) \sin(\omega_b t).
\end{aligned}
\label{eq:decomp_real_imag_uc}
\end{equation}
In particular, $\Re(u)$ and $\Im(u)$ involve only frequencies of the form
$\omega = r\omega_a \pm \omega_b, \, r\in\mathbb Z$.
However, using once again the operator decomposition
of~\cref{eq:op_decomposition_eietaq,eq:opdecomp_eietabqb_time},
we see that
\begin{equation}\begin{aligned}
	\overline{e^{i(2r\omega_a\pm\omega_b)t} \sin(\eta_a \qop_a(t) + \eta_b\qop_b(t)}) &= 0,\\
	\overline{e^{i((2r+1)\omega_a\pm\omega_b)t} \cos(\eta_a \qop_a(t) + \eta_b\qop_b(t)}) &= 0\\
\end{aligned}\end{equation}
so that the components of $u$ containing respectively even or odd multiples of $\omega_a$
are decoupled,
in the sense that only the even multiples of $\omega_a$ can introduce
resonant terms when multiplied by $\cos(\eta_a \qop_a(t) + \eta_b\qop_b(t))$,
while only the odd multiples can introduce
resonant terms when multiplied by $\sin(\eta_a \qop_a(t) + \eta_b\qop_b(t))$.

A suitable choice of real-valued control signal is thus given by
\begin{equation}
	u_F(t) = \sqrt2 \, \Big( \Re(u(t))\rvert_{even} + \Im(u(t))\rvert_{odd}\Big)
	\label{eq:choice_ur}
\end{equation}
where
only the frequencies with even multiples of $\omega_a$ are kept within $\Re(u(t))$
while
only the frequencies with odd multiples of $\omega_a$ are kept within $\Im(u(t))$:
\begin{equation}\begin{aligned}
	\Re(u(t))\rvert_{even} &:=
		\Re\left( \sum_{r\in\mathbb Z} u_{2r} \, e^{i(2r)\omega_a t} \right) \cos(\omega_b t)\\
		&\quad
		+ \Im\left( \sum_{r\in\mathbb Z} u_{2r} \, e^{i(2r)\omega_a t} \right) \sin(\omega_b t),\\
	\Im(u(t))\rvert_{odd} &:=
		\Im\left( \sum_{r\in\mathbb Z} u_{2r+1} \, e^{i(2r+1)\omega_a t} \right) \cos(\omega_b t)\\
		&\quad
		- \Re\left( \sum_{r\in\mathbb Z} u_{2r+1} \, e^{i(2r+1)\omega_a t} \right) \sin(\omega_b t).
\end{aligned}\end{equation}

A few comments are in order at this stage:
\begin{itemize}
	\item While the above choice of real-valued control $u_F$
		guarantees that $\overline\HH = \overline\HH_F$,
		which is sufficient to see that $\HH$ and $\HH_F$
		give rise to the same average Hamiltonian after a first-order RWA approximation,
		they would \emph{not} lead to equivalent corrections
in a second-order RWA analysis, as $\HH - \overline\HH \neq \HH_F - \overline\HH_F$;
we plan to study in more detail the impact of second-order corrections in forthcoming work.
	\item We could choose a phase other than $\pi/4$ in $\HH_F$.
		Indeed, for $\zeta\in\mathbb R$ such that $\cos(\zeta)\neq 0$ and $\sin(\zeta)\neq0$
		(that is $\zeta \neq 0 \, [\mathrm{mod} \, \tfrac\pi2]$),
		we could consider
		\( \HH_F(t) = u_F(t) \cos(\eta_q \qop_a(t) + \eta_b \qop_b(t) - \zeta)\).
		In that case, we would have to amend~\cref{eq:choice_ur} as follows:
		\( u_F(t) = \frac{1}{\cos(\zeta)} \, \Re(u(t))\rvert_{even}
				+\frac{1}{\sin(\zeta)} \Im(u(t))\rvert_{odd}
				\).
		We picked a balanced choice $\zeta = \pi/4$ leading to
		$\cos(\zeta)=\sin(\zeta)=1/\sqrt2$,
		but emphasize that a detectable miscalibration of $\zeta$ could thus be compensated for
		in software.
	\item We mention a slight generalization of~\cref{eq:decomp_real_imag_uc}
		that will simplify some forthcoming computations:
		for any phase $\nu\in\mathbb R$,
\begin{equation}\begin{aligned}
	\Re(u(t)) &=
		\Re\left( e^{i\nu} \sum_{r\in\mathbb Z} u_r e^{ir\omega_a t} \right)
			\cos(\omega_b t - \nu)\\
		&\quad
		+ \Im\left( e^{i\nu} \sum_{r\in\mathbb Z} u_r e^{ir\omega_a t} \right)
			\sin(\omega_b t - \nu),\\
	\Im(u(t)) &=
		\Im\left( e^{i\nu}\sum_{r\in\mathbb Z} u_r e^{ir\omega_a t} \right)
			\cos(\omega_b t - \nu)\\
		&\quad
		- \Re\left( e^{i\nu} \sum_{r\in\mathbb Z} u_r e^{ir\omega_a t} \right)
			\sin(\omega_b t - \nu).
	\label{eq:real_imag_decomp_phase}
\end{aligned}\end{equation}
\end{itemize}
\vspace{16pt}

\bigpar{Reconstructing the control signals of the main text}
We wrap up this section by showing
how we combine the techniques we exposed to find the control signals
proposed in~\cref{eq:ideal_controls} of the main text,
which correspond to the engineering of the symmetrized and anti-symmetrized
Lindblad operators $\LL_{r,l}$.
As usual, we will focus only on the engineering of the corresponding operators
$\widetilde \LL_{r,l}$,
where we remind the reader that $\widetilde \LL_{r,s} = \LL_{r,s} + \sqrt2 \II$
and $\widetilde \LL_{r,d} = \LL_{r,d}$;
we assume that the remaining scalar terms are engineered through direct drives on the ancillary mode.
\newline

Using~\cref{eq:u_separable,eq:u_rl_dirac_ideal},
we first build an ideal (unbounded bandwidth and amplitude) control signal
\begin{widetext}
\begin{equation}\begin{aligned}
	u_{r,l}(t) &:= \tfrac{2\pi\cA g  \, e^{\eta_b^2/4} }{\eta_b\omega_a}\left(
			(1+\tfrac{\epsilon\eta_a}2)
			\left(
			\Sha_{\frac{2\pi}{\omega_a}}(t-t_r)
			+ (-1)^{\delta_l} \Sha_{\frac{2\pi}{\omega_a}}(t-t_r - \tfrac\pi{\omega_a})\right)
				\right.\\
			&\quad\quad\quad\quad\quad
			\left.
			- \tfrac{i\epsilon}{\omega_a \eta_a}
			\left(
			\Sha'_{\frac{2\pi}{\omega_a}}(t-t_r)
			+ (-1)^{\delta_l} \Sha'_{\frac{2\pi}{\omega_a}}(t-t_r-\tfrac\pi{\omega_a})
			\right)
			\right)
			e^{-i\omega_b t}
\end{aligned}
\end{equation}
\end{widetext}
where we remind the reader that we previously defined the coefficients $t_r$ and $\delta_l$
as $t_q=0, t_p=\frac\pi{2\omega_a}$ and $\delta_s = 0, \delta_d = 1$.
We then transform $u_{r,l}$ to a control signal of bounded bandwidth and amplitude
as explained in~\cref{ssec:exp_limit_u}, obtaining
\begin{equation}
	u_{r,l}(t) = \frac{2\pi\cA g \, e^{\eta_b^2/4}}{\eta_b\omega_a} \sum_{j=0,+1,-1}
		u^{(a)}_{r,l,j}(t) e^{-i\omega_b t}
\end{equation}
where 
\begin{widetext}
\begin{equation}\begin{aligned}
	u^{(a)}_{r,l,0}(t) &= \Sha_{\frac{2\pi}{\omega_a}} (t-t_r)
			+ (-1)^{\delta_l}\Sha_{\frac{2\pi}{\omega_a}} (t-t_r-\tfrac{\pi}{\omega_a}),\\
	u^{(a)}_{r,l,-1}(t) &=
	-i e^{i\epsilon\eta/4}\left(
		\Sha_{\frac{2\pi}{\omega_a}} (t-t_r+\tfrac{\epsilon}{2\eta_a\omega_a})
		+(-1)^{\delta_l} \Sha_{\frac{2\pi}{\omega_a}} (t-t_r+\tfrac{\epsilon}{2\eta_a\omega_a} - \tfrac\pi{\omega_a})\right),\\
	u^{(a)}_{r,l,+1}(t) &=
	i e^{-i\epsilon\eta/4} \left(
		\Sha_{\frac{2\pi}{\omega_a}} (t-t_r-\tfrac{\epsilon}{2\eta_a\omega_a})
	+ (-1)^{\delta_l} \Sha_{\frac{2\pi}{\omega_a}} (t-t_r-\tfrac{\epsilon}{2\eta_a\omega_a} - \tfrac\pi{\omega_a})\right),
\end{aligned}\end{equation}
\end{widetext}
where each Dirac comb should now be understood as a truncated Dirac comb
$\Sha^{(N)}$ as defined in~\cref{ssec:exp_limit_u};
we drop the superscript $(N)$ from now on to alleviate the notations.
We can unify the previous expressions by introducing new parameters
$\theta_a = \frac{\epsilon}{2\eta_a}$,
$\theta_b = \frac\pi2 - \frac{\epsilon\eta_a}4$
and $\phi_r = \omega_a t_r$;
we obtain
\begin{widetext}
\begin{equation}
	u^{(a)}_{r,l,j}(t) = e^{i(j\theta_b)}
	\left(
		\Sha_{\frac{2\pi}{\omega_a}} \left(t-\frac{j\theta_a+\phi_r}{\omega_a}\right)
		+(-1)^{\delta_l} \Sha_{\frac{2\pi}{\omega_a}} \left(t-\frac{j\theta_a+\phi_r+\pi}{\omega_a}\right)\right).
\end{equation}
Each of these control signals is then replaced by a control signal with constant phase
using~\cref{eq:choice_ur,eq:real_imag_decomp_phase}.
Since $e^{i(-j\theta_b)}\, u^{(a)}_{r,l,j}$ is already real-valued and even,
we replace
$u^{(a)}_{r,l,j}(t) e^{-i\omega_b t} $
by
\begin{equation}\begin{aligned}
	u_{r,l,j}(t)
		&= \sqrt2 \left(
		\Sha_{\frac{2\pi}{\omega_a}} \left(t-\frac{j\theta_a+\phi_r}{\omega_a}\right)
		+(-1)^{\delta_l} \Sha_{\frac{2\pi}{\omega_a}} \left(t-\frac{j\theta_a+\phi_r+\pi}{\omega_a}\right)\right)
			\, \cos(\omega_b t -j\theta_b).
\end{aligned}\end{equation}

We see that we obtained the control signal of the main text
\begin{equation}\begin{aligned}
	u_{r,l}(t) &= E_J \xi_{r,l}(t),\\
	\xi_{r,l}(t) &= \sum_{j=0,+1,-1} \xi_{r,l}^j(t)\\
			 &= \sum_{j=0,+1,-1} \xi_1 \left(
		\Sha_{\frac{2\pi}{\omega_a}} \left(t-\frac{j\theta_a+\phi_r}{\omega_a}\right)
		+(-1)^{\delta_l} \Sha_{\frac{2\pi}{\omega_a}} \left(t-\frac{j\theta_a+\phi_r+\pi}{\omega_a}\right)\right)
		\, \cos(\omega_b t - j\theta_b)
\end{aligned}\end{equation}
\clearpage
\end{widetext}

where we introduced a single notation $\xi_1$ to encompass the amplitude of the control signal;
its relation to the corresponding engineered coupling rate is given by
\begin{equation}
	g= \frac{E_J \eta_b \, \omega_a \xi_1}{2\sqrt2 \pi \cA} \, e^{-\eta_b^2/4},
\end{equation}
or, when we want to specify $\hbar$:
\begin{equation}
	g= \frac{E_J \eta_b \, \omega_a \xi_1}{2\sqrt2\pi \hbar\cA}  \, e^{-\eta_b^2/4}.
\end{equation}
In the main text, given the value $\eta_b=0.3$ proposed in~\cref{table},
we neglected the near unit correction $e^{-\eta_b^2/4} \simeq 0.98$
when estimating the effective coupling strength $g$ in~\cref{sec:modularengineering}.

%% file: sm__beyond_rwa_discussion.tex
\bigpar{Beyond RWA and adiabatic elimination}
Two different routes can be envisioned to take into account corrections beyond a first-order RWA: higher-order averaging methods and non-perturbative methods.

The simplest non-perturbative method is the bruteforce simulation of the full time-dependent Lindblad equation. However, it is usually unfeasible due to the shear computing power and memory required.
In our case, the direct simulation of a two-mode and rapidly oscillating Lindblad equation, with a memory mode living in a Hilbert space truncated to a few thousands Fock states, is definitely out of reach.
Under additional restrictive assumptions, ad-hoc non-perturbative methods
can sometimes lead to tractable numerically exact methods.
A prime example is Markov-Floquet theory \cite{grifoniDrivenQuantumTunneling1998},
which can be applied for the analysis of periodic Hamiltonians and received considerable attention in recent years for its ability to analyze
strongly driven quantum systems in regimes where low-order perturbation methods fail
(see \emph{e.g.}
\cite{%
verneyStructuralInstabilityDriven2019,%
cohenReminiscenceClassicalChaos2023,%
burgelmanStructurallyStableSubharmonic2022,%
petrescuAccurateMethodsAnalysis2023%
}).
To the best of our knowledge, generic equivalent methods do not exist yet
for the analysis of both dissipative and non-periodic systems
(as opposed to periodic Hamiltonian systems).
In our paper, these two difficulties are treated perturbatively:
we use a RWA formalism that does not require periodic signals (more precisely, it is valid for quasi-periodic signals), while the dissipation is treated through adiabatic elimination.
The development of ad-hoc non-perturbative techniques in this setting
is certainly to be considered an intriguing open question
for future research and would have positive implications far beyond our subject.

On the other hand, staying within the realm of perturbation theory as proposed in our paper comes with
the benefits of yielding analytical, closed-form expressions.
On top of the usual advantages of analytical models,
and the interpretability of the results they provide,
we should emphasize a very pragmatic consequence in the specific case of the analysis performed in our paper:
for the design of the stabilization scheme, rotating waves approximations are performed on the full two-mode Hamiltonian including the ancilla mode used for stabilization, which is later to be eliminated (using adiabatic elimination).
If one were to either turn to a fully numerical treatment of the RWA or replace it with another numerical method altogether,
it would in turn require the development of numerical
tools for the automated treatment of adiabatic elimination too.
To the best of our knowledge, such numerical tools are not readily available yet, although their development would represent an interesting research direction of independent interest.

In the near future and given the currently existing theory,
the soundest extension to the research presented here would be to exploit higher-order RWA formulae. In particular, second-order averaging formulae can be found \emph{e.g.} in
\cite{lecturenotesmirrahimirouchon} and
would still provide closed-form expressions.
They could \emph{a priori} be extended to arbitrary orders.
Such development could greatly benefit from the development
of dedicated symbolic computer algebra systems, as explored for instance in
\cite{%
venkatramanStaticEffectiveHamiltonian2022,%
xiaoDiagrammaticMethodCompute2023%
} --
systems which are of course of independent interest.

Finally, regarding the extension of adiabatic elimination beyond second order,
the picture is less clear.
For a given open quantum system coupled to a dissipative ancillary bath,
the existence of a reduced quantum model accurate beyond second order is generically an open question,
with both positive
\cite{sarletteQuantumAdiabaticElimination2020}
and negative
\cite{tokiedaCompletePositivityViolation2023}
known instances.

%% file: sm__disorder_and_miscalibration.tex
\section{Realistic circuit fabrication and control}
\label{sec:disorder_miscab}

\subsection{Array modes of the superinductor}
\label{sec:array}

\begin{figure}[htbp]
		\centering
		\includegraphics[width=1\columnwidth]{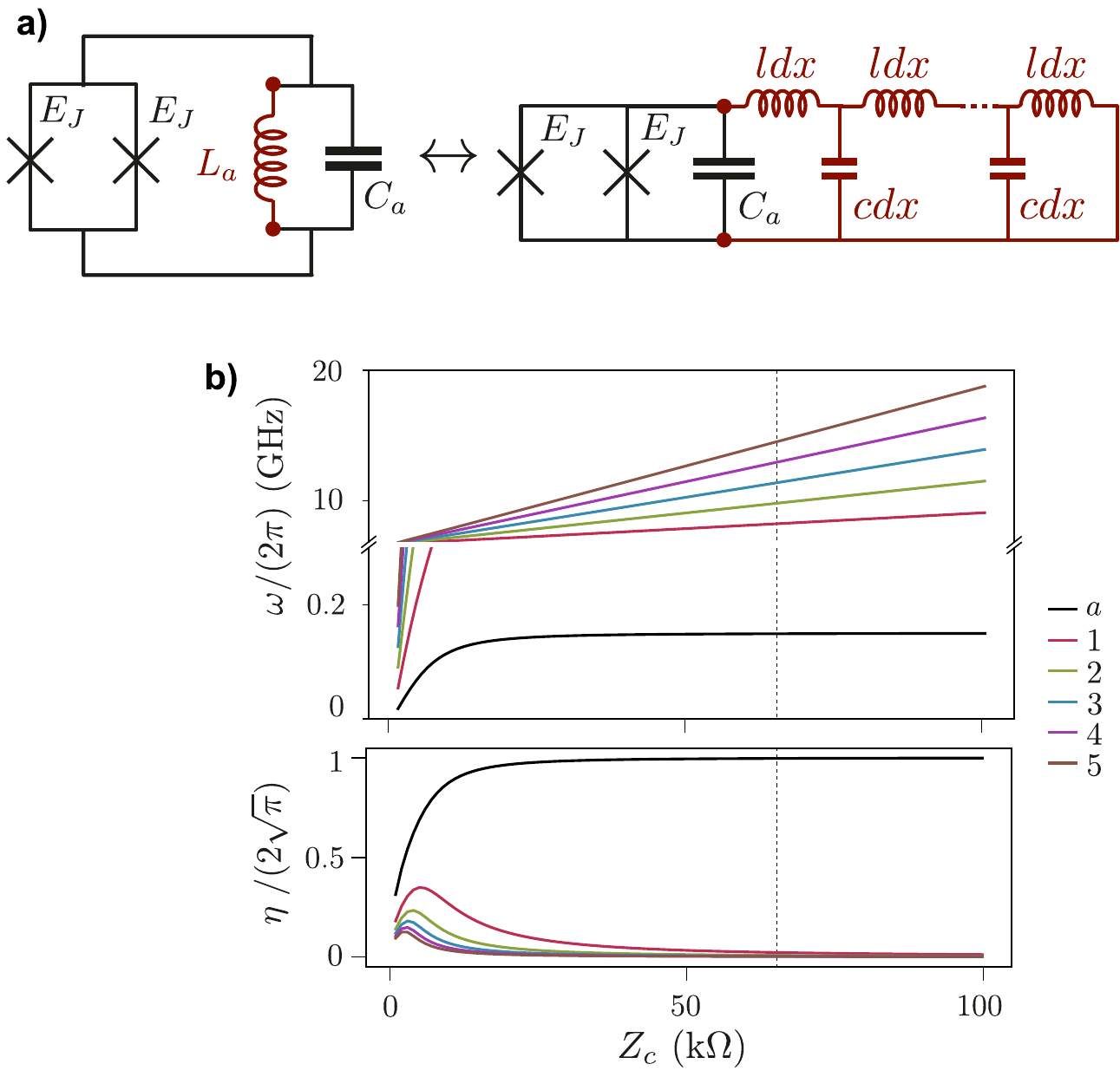}
		\caption{{\bf Array modes of the superinductor. } {\bf a)} A realistic superinductor is modeled by a transmission line with large inductance per unit length $l$ and  stray capacitance to ground per unit length $c$ (in red). Such a realistic model for the target mode inductance $L_a$ (ancillary mode used for dissipation engineering not represented) leads to spurious resonances known as array modes.  {\bf b)}  Frequency (top) and vacuum phase  fluctuations across the Josephson ring (bottom) of the target mode (black) and the  first five array modes (in color), all loaded by the target mode shunt capacitance $C_a$, as a function of the characteristic impedance $Z_c=\sqrt{l/c}$ of the inductor. Other circuit parameters are those of Tab.~\ref{table}, and in particular, the total inductance $L_a=14~\mu\mathrm{H}$ is fixed. The dashed line represents the highest reported value of $Z_c$ in the literature~\cite{pechenezhskiy2020superconducting}.  }	\label{fig:zcharac}
	\end{figure}

The superconducting gap of aluminium    places a hard limit  around 90~GHz on the maximum frequency at which Josephson junctions built from this material can be driven. In practice, most laboratory equipment and circuitQED architectures have a narrower working bandwidth of 20~GHz. Moreover, in our proposal, we have neglected the intrinsic capacitance of the Josephson junctions forming the circuit ring, which is equivalent to assuming that the circuit is only biased  below each junction   plasma frequency $\omega_J=\sqrt{8 E_J E_{c_J} }/\hbar$, where $E_J$ is the Josephson energy of each junction and  $E_{c_J}$ its charging energy. This plasma frequency typically lies in the 10---50~GHz range, with its exact value depending on the thickness of the oxide barrier of the junction. In Tab.~\ref{table}, we choose to limit the frequency comb bandwidth to 0---20~GHz and set the target resonator to 150~MHz in order to fit a hundred  harmonics of the target mode frequency in this limited bandwidth (assuming the comb to be centered at 5~GHz, which is  the ancillary mode frequency). Even though less conservative hypotheses may be considered for the junctions plasma frequency and bandwidth of the control electronics, the hard limit mentioned above prompts the need for such low frequency target mode. Since the mode impedance is set to $Z_a=2R_Q$, an inductance $L_a$ in the tens of $\mu$H range needs to be employed (see Tab.~\ref{table}).   \\

Currently, the most promising technologies for such superinductors are chains of Josephson junctions~\cite{masluk2012microwave}, disordered superconducting films~\cite{grunhaupt2019granular} and planar superconducting coils~\cite{peruzzo2020surpassing}. These metallic structures typically have a $\sim$mm size and suffer from stray capacitance to ground. In Fig.~\ref{fig:dissipCircuit}a, we model such a realistic superinductor as a continuous transmission line of length $\lambda$, inductance per unit length $l$ (with $L_a=l \lambda$) and capacitance per unit length $c$  (for the sake of simplicity, the ancillary mode involved in dissipation engineering is not represented).  This  circuit  hosts spurious resonances known as array modes of the superinductor~\cite{viola2015collective}, which have two  advert effects. First, by diluting the target mode inductive energy
over multiple inductors, which are not directly connected to the ring,  they tend to decrease its vacuum phase fluctuations $\eta_a$ across the Josephson ring. Second,
array modes (labeled by an integer $k\geq 1$) will appear in the circuit Hamiltonian \eqref{eq:ho} as spurious ancillary modes, the generalized phase operator across the ring becoming
\begin{equation}
{\bf \Phi}=\varphi_0 \big( \eta_a \qop_a^0 + \eta_b \qop_b^0+\sum_k \eta_k \qop_k^0 \big),
\end{equation}
where $\qop_{\bullet}^0$ designates a quadrature operator in the laboratory frame as in \eqref{eq:ho} and $\eta_{\bullet}$ represents the phase fluctuations of a mode across the ring. Even though we have not quantitatively investigated the impact of  such modes on the GKP qubit lifetime, their proliferation  with non-negligible phase fluctuations across the ring (\emph{i.e.} $\eta_k \gtrsim \eta_b$) will lead to  frequency collisions and inadvertent activation of high-order multimode processes, invalidating  the two-mode picture presented in Sec.~\ref{sec:modularengineering} and Sec.~\ref{sm:sec__rwa}.\\

Since the total inductance $L_a=\lambda l$  is fixed (see Tab.~\ref{tableparam}), the superinductor is fully characterized by its characteristic impedance $Z_c=\sqrt{l/c}$. This figure of merit sets both the frequency $\omega_k$ of  array modes and their vacuum phase fluctuations across the ring $\eta_k$. In Fig.~\ref{fig:zcharac}b, we represent these values for the first five array modes, extracted with the method  of energy participation ratios~\cite{minev2021energy} in the spirit of \cite{smith2016quantization},  as a function of $Z_c$. All other circuit parameters are those proposed in Tab.~\ref{table}. We find that the array modes frequency increases and their fluctuations across the ring decreases with $Z_c$, with phase fluctuations becoming negligible for  $Z_c \gg Z_a = 13~\mathrm{k}\Omega$. Quantitatively, for the  characteristic impedance of $65~\mathrm{k}\Omega$ recently reported for a Josephson junction chain released from its substrate~\cite{pechenezhskiy2020superconducting} (dashed line in Fig.~\ref{fig:zcharac}b), we find that fewer than 10 array modes  lie in the frequency comb bandwidth, each with phase fluctuations $\eta_k \ll \eta_b$, justifying their omission in our model.\\

An important remark is in order here. In the inductor model of Fig.~\ref{fig:dissipCircuit}a, we have assumed the device inductance to be purely linear. For an implementation based on a chain of Josephson junctions, this model is only accurate if two conditions are met (we refer the reader to Refs.~\cite{manucharyan2012superinductance,maleeva2018circuit}   for a detailed analysis). First, the number $N_J $ of  junctions in the chain should be sufficiently large and the  individual inductance of the junctions sufficiently small---remember that $N_J\frac{\varphi_0^2}{E_J}=L_a$---that phase-slips through the array occur at a negligible rate (requiring $E_J \gg E_{c_J}$) and that the Kerr non-linearity induced on the target mode is negligible (scaling in $1/N_J^2$). Second, we have neglected the intrinsic shunt capacitance of each junction. When included in the circuit model, these capacitances curve the dispersion relation of array modes, whose frequency saturates at the plasma frequency $\omega_p$. Our model is thus only correct if $\omega_p$ is much larger than the frequency comb bandwidth. Superinductors based on  nanometric scale tracks  of granular aluminium, which effectively behave as long chains of large junctions with high plasma frequency ($\omega_J\sim70~$GHz in Ref.~\cite{maleeva2018circuit}) appear to meet these requirements. We consider that the record characteristic impedance mentioned above (dashed line in Fig.~\ref{fig:zcharac}b) will probably be surpassed in this type of architecture in the near future~\cite{kamenov2020granular}.

\vspace{-0.8em}
\subsection{Excessive target mode impedance}
\label{ssec:excessive_impedance}
\begin{figure}[htbp]
		\centering
		\includegraphics[width=1\columnwidth]{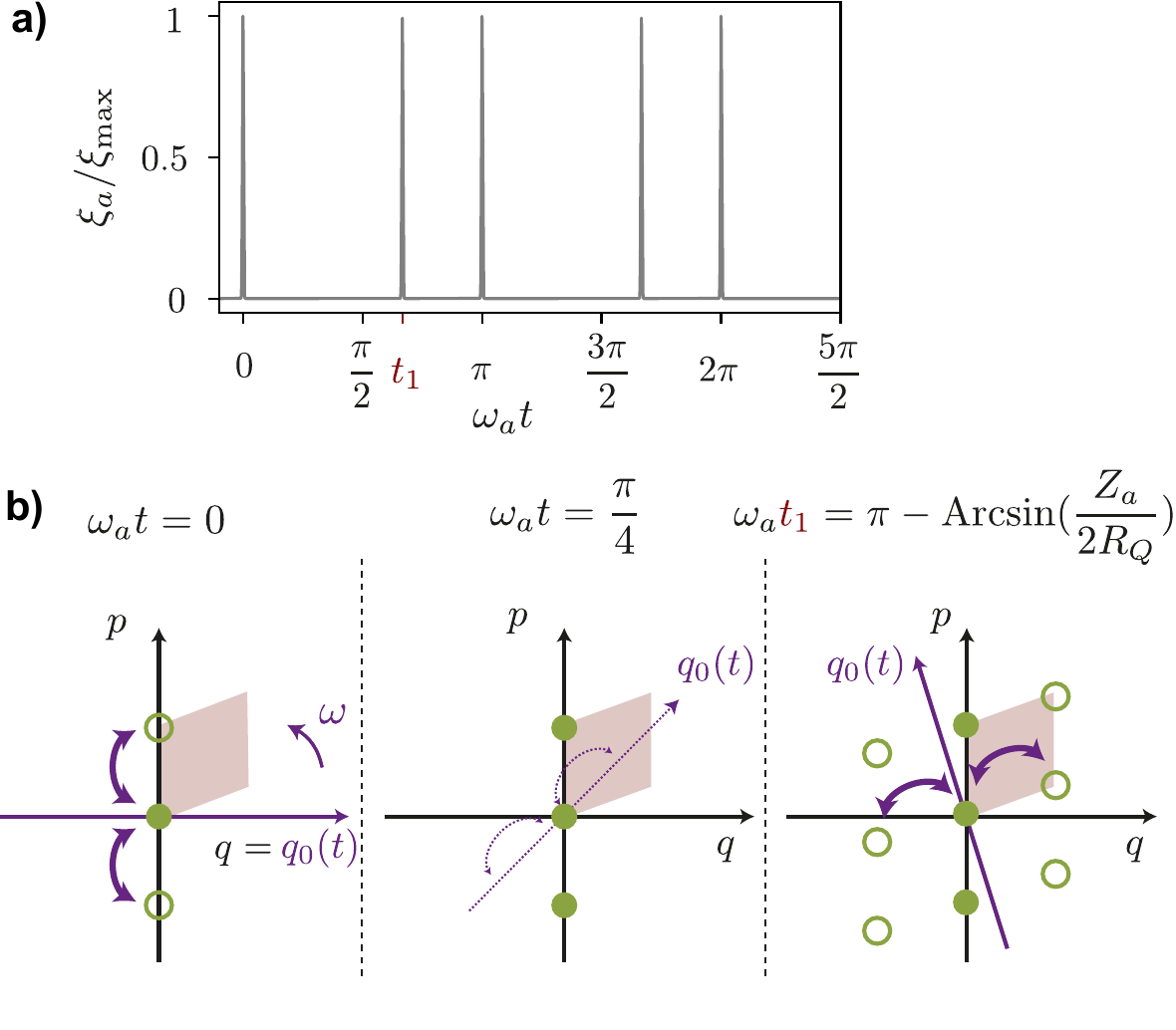}
		\caption{{\bf Adapting the GKP lattice to the target mode impedance } {\bf a)} When the target mode impedance $Z_a$ exceeds $2R_Q$, the timing of the bias pulses needs to be adapted so that  {\bf b)} in the target mode phase-space with axes $(q,p)$ rotating at $\omega_a$, Josephson tunneling  triggers coherent displacements by $\pm \sqrt{2\pi Z_a/R_Q}$ along a diamond-shaped lattice with unit cell area $4\pi$ (in pink).  The method is presented here for the case of Hamiltonian engineering. }	\label{fig:diamonds}
	\end{figure}

We remind the reader that the target mode impedance $Z_a=2R_Q$ was chosen so that, in reduced phase-space coordinates $(q,p)$, the target mode vacuum phase fluctuations across the Josephson ring  $\eta_a=\sqrt{2\pi Z_a /R_Q}$  match the square GKP code lattice unit cell length. The GKP lattice can however be continuously distorted as long as the unit cell area in phase-space remains $4\pi$. Explicitly, the modular operators $e^{i\tilde{\eta}\qop}$ and $e^{i \tilde{\eta} \sop }$, where $\tilde{\eta}>\eta$, $\sop=\Rop_{\theta} \qop \Rop^{\dag}_{\theta} $ and $\mathrm{Sin}(\theta)=(\eta/\tilde{\eta})^2$, are the stabilizers of a diamond shaped lattice GKP. As pictured in Fig.~\ref{fig:diamonds}, when the impedance of the target mode exceeds $2R_Q$, one simply adapts the timing of the bias pulses to stabilize such a non-square GKP code.  Fig.~\ref{fig:diamonds} represents such a situation in the case of Hamiltonian engineering as described in Sec.~\ref{sec:hamiltonian}, but is directly adaptable when engineering modular dissipation. Note that in the latter case, the normalizing envelope of the stabilized grid states envelope remains a rotational-symmetric Gaussian.\\

\begin{figure}[htbp]
	\centering
	\includegraphics[width=0.95\columnwidth]{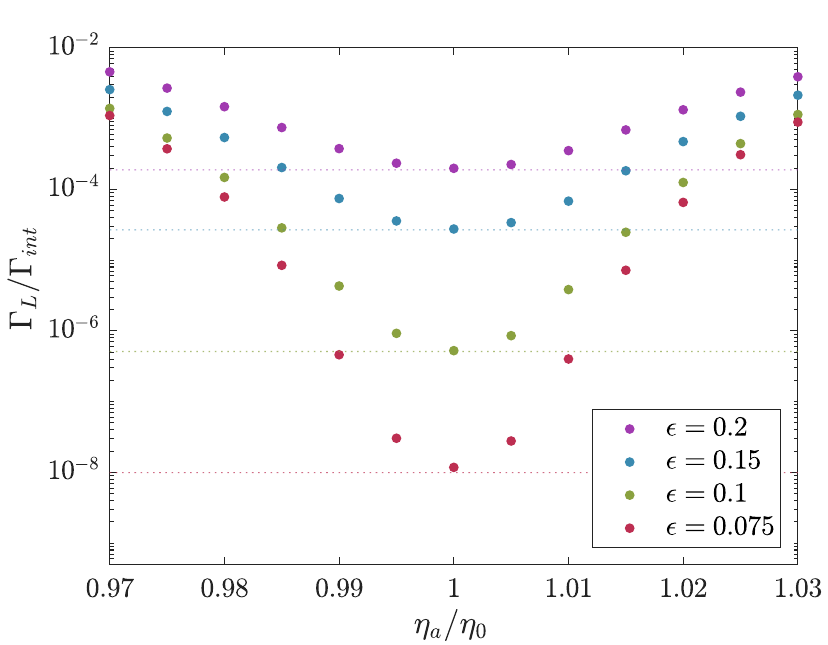}
	\caption{Logical error rate as a function of the relative error $\eta_a/\eta$ %
	on the value of $\eta_a$ with respect to its ideal value $\eta=2\sqrt\pi$. %
	Colored dots correspond to numerically extracted decay rates, %
	while dotted lines indicate the theoretical asymptotic value of the decay rates only stemming %
	from the finite energy of the stabilized GKP code.}
	\label{fig:square_wrongeta}
\end{figure}

Crucially, the angle $\theta$ only  needs to be  adjusted within some realistic margin of tolerance to avoid spurious logical errors. This tolerance depends on the  dissipation parameter $\epsilon$. Intuitively, $\epsilon$ sets the extension of code states in phase-space.
Errors appear when this extension is sufficient to sense the oscillating pseudo-potential from the modular dissipation going out of phase with the GKP lattice.
In order to quantitatively estimate the required precision,
we perturb the Lindblad operators entering the Lindblad dynamics of a stabilized GKP qubit
by choosing a value $\eta_a$ slightly deviating from its ideal value $\eta=2\sqrt\pi$;
this deviation models the residual uncertainty on the value of $\eta_a$
after adjustment of the chosen unit cell.
We then numerically compute the logical decay rate $\Gamma_L$  of the generalized Pauli operators $\X$ and $\Z$. On~\cref{fig:square_wrongeta}, we present the dependence of this decay rate on the ratio
$\eta_a/\eta$.
We find that the value of $\eta_a$
(or equivalently $\theta$ when actively compensating for a known bias)
needs to be adjusted at the $10^{-2}$ to $10^{-3}$ level
to preserve the  performance of the stabilization scheme;
as expected, the smaller values of $\epsilon$ lead to the most stringent requirements on $\theta$.

\subsection{Josephson junctions asymmetry}

\begin{figure}[htbp]
		\centering
		\includegraphics[width=1\columnwidth]{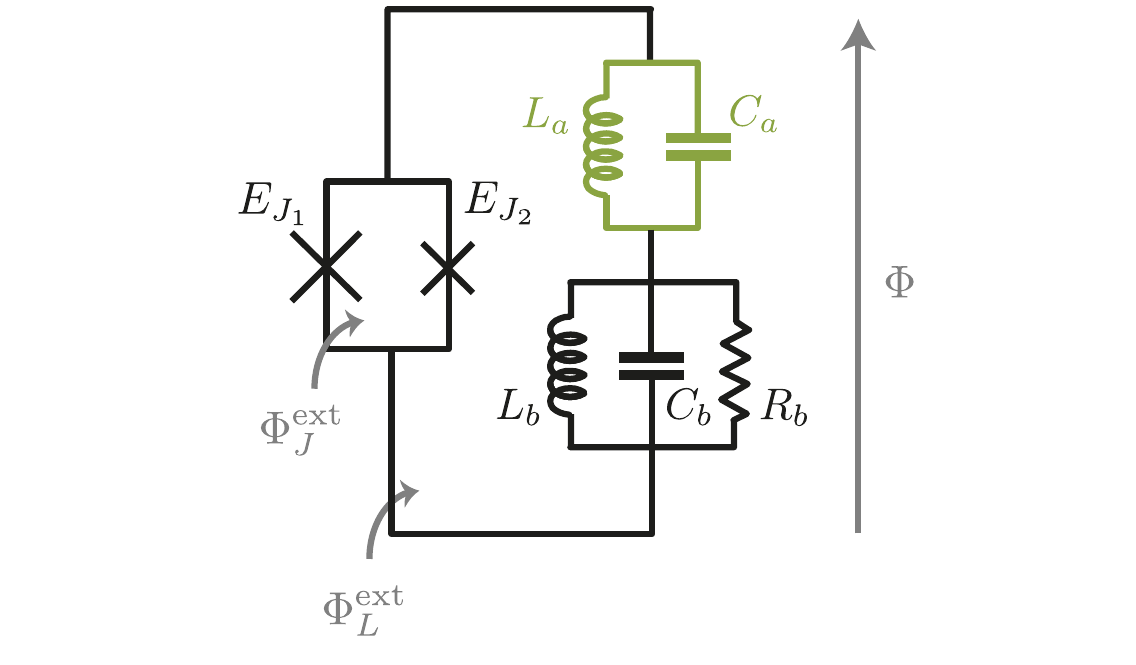}
		\caption{{\bf Compensating for Josephson junctions asymmetry.} We recall the circuit proposed to engineer modular dissipation in Fig.~\ref{fig:dissipCircuit}a A Josephson ring is placed in parallel of a high-impedance target resonator (green) and a low-impedance, dissipative, ancillary resonator (black). The  Josephson tunneling amplitude  $2E_J \xi(t)$ and  phase are adjusted with the control fluxes $\Phi_{J,L}^{\mathrm{ext}}$ biasing the circuit.  Here, we consider the possibility that the ring junctions have slightly different energies $E_{J_1}$ and $E_{J_2}$.  }	\label{fig:JJasym}
	\end{figure}
The energy of Josephson junctions  is never perfectly reproducible, with a typical mismatch of the order of a percent between two nominally identical junctions in the same device.  When the two  junctions forming the ring of the circuit depicted in Fig.~\ref{fig:JJasym} have slightly different energies $E_{J_1}$ and $E_{J_2}$, the amplitude of Josephson tunneling cannot be perfectly cancelled by threading the ring with half a flux quantum ($\Phi^{\mathrm{ext}}_J=\varphi_0 \pi$), as proposed in Sec.~\ref{sec:modularengineering}. We remind the reader that in our protocol, tunneling is only triggered by fast flux pulses  when the Josephson phase operator  $\Phi$ aligns with  the  GKP  lattice axes in the target oscillator rotating frame: imperfect cancellation of Josephson tunneling in between pulses may lead to long shifts of the oscillator state along a random axis and cause logical errors.\\

We propose to mitigate this advert effect by adjusting the circuit DC flux bias so that the spurious Josephson tunneling term becomes non-resonant and drops out in the RWA. Letting $E_{\Sigma}=E_{J_1}+E_{J_2}$ and  $E_{\Delta}=E_{J_1}-E_{J_2}$, we thus set
\begin{equation}
\begin{split}
\Phi^{\mathrm{ext}}_J=&\varphi_0(\pi-2A-2B )\\
\Phi^{\mathrm{ext}}_L=&-\frac{\Phi^{\mathrm{ext}}_J}{2}+\varphi_0 \frac{\pi}{4}
\end{split}
\end{equation}
where
$B=\mathrm{Arcsin}(\xi)$ is the same AC bias signal as described in Sec.~\ref{sec:modularengineering} and $A=\mathrm{Arctan}(d)$ is a small DC offset that depends on the junctions asymmetry $d=\frac{E_{\Delta}}{E_{\Sigma}}$. Denoting $ \boldsymbol{\varphi}={\bf \Phi}/\varphi_0=\eta_a \qop_a+\eta_b \qop_b$ the reduced phase  across the  ring, the  ring contribution to the circuit Hamiltonian  reads
\begin{equation}
\begin{split}
\HH_J
	=& -E_{J_1}\mathrm{cos}\left(\frac{\phiop-\Phi^{\mathrm{ext}}_J-\Phi^{\mathrm{ext}}_L}{\varphi_0}\right)
-E_{J_2}\mathrm{cos}\left(\frac{\phiop-\Phi^{\mathrm{ext}}_L}{\varphi_0}\right)\\
	=& -E_{J_1}\mathrm{cos}(\boldsymbol{\varphi}-\frac{3\pi}{4}+A+B) \\
	& -E_{J_2}\mathrm{cos}(\boldsymbol{\varphi}+\frac{\pi}{4}-A-B)\\
	=& - E_{\Sigma}\mathrm{cos}(\boldsymbol{\varphi}-\frac{\pi}{4})\mathrm{sin}(A+B) \\
	& - E_{\Delta}\mathrm{sin}(\boldsymbol{\varphi}-\frac{\pi}{4})\mathrm{cos}(A+B)
\end{split}
\end{equation}
Expanding the cosine and sine terms and using that $E_{\Sigma}\mathrm{sin}(A)=E_{\Delta}\mathrm{cos}(A)$, we find that
\begin{equation}
	\begin{aligned}
      \HH_J
		=& -E_{\Sigma} \big(\frac{1+d^4}{1+d^2} \big)^{\frac{1}{2}} \xi \mathrm{cos}(\boldsymbol{\varphi}-\frac{\pi}{4}+e) \\
		&-E_{\Sigma}\big(\frac{2d^2}{1+d^2}(1-\xi^2)\big)^{\frac{1}{2}}\mathrm{sin}(\boldsymbol{\varphi})
	\end{aligned}
\end{equation}
where $e=\mathrm{Arctan}(d^2)$. We now remark that, since the $\mathrm{sin}(\boldsymbol{\varphi})$ operator in the rightmost term only contains  operators of the form $(\aop+\adag)^k(\bop+\bdag)^l$ with $k+l$ an odd number, while the time-varying prefactor $(1-\xi^2)^{1/2}$ has non-zero Fourier coefficients only for $\omega=k'\omega_a+l'\omega_b$ with $k'+l'$ an even number, this spurious term does not contribute in the RWA. As for the leftmost term, it is similar to the Hamiltonian~\eqref{eq:H2}, but for a prefactor close to~1 and a phase offset close to 0 when the junction asymmetry is small. These corrections only slightly modify the dissipation rates of the four engineered dissipators, which can be compensated for by adjusting the relative amplitudes of the four bias signals. Note that the RWA is valid for $\sqrt{2}E_{\Delta}\ll\omega_a$. In Tab~\ref{tableparam}, we choose the value of  $E_J$ so that $\sqrt{2}E_{\Delta}\simeq \omega_a/10$ for junctions with asymmetry $d=1\%$ (corresponding to an energy mismatch of $2\%$ as quoted in the main text).

\subsection{Miscalibration of the control signals}
\label{sm:control_miscab}
As explained in~\cref{sec:harmonics},
two main limitations prevent us from using the theoretical control signals
defined in~\cref{eq:ideal_controls},
which are linear combination of periodic Dirac combs.
The restriction to periodic control signals with finite bandwidth
was already studied in~\cref{sm:sec__rwa}.
However, even for a finite bandwidth signal,
one also has to take into account the uncertainty introduced
by the unknown dispersion of the feedlines that carry the signals,
generated at room temperature, to the superconducting circuit.
In practice, this dispersion relation has to be determined experimentally in a preliminary
calibration procedure.
To quantify the relative precision required in this calibration step,
we study the impact of imperfect calibration, modeled as random noise affecting
the control signals.
More precisely, we replace each desired periodic controls
$\xi(t)$
by an imperfectly calibrated signal
$\tilde\xi(t)$
in which the Fourier coefficients of the target signal
are multiplied by independent random  coefficients close to 1:

\[\xi_k \mapsto \tilde \xi_k = s_k\, \xi_k\]
with
\( s_k \sim 1 + \frac{\sigma_{\textrm{control}}}{\sqrt2} \left(\mathcal N(0,1) + i\mathcal N(0,1) \right). \)
The noise coefficients $s_k$ are complex-valued, independent Gaussian coefficients
with mean $1$ and variance $\sigma_{\textrm{control}}^2$.

The miscalibrated control signal $\tilde\xi$ is still
a periodic signal, which we can feed into the RWA analysis presented
in~\cref{sm:sec__rwa}
to determine the effective dynamics at first order.
For a given realization of the random noise coefficients,
we can thus compute the logical decoherence rate
associated to the dynamics
\begin{equation}
	\frac {d\rho_t}{dt} = \Gamma \sum_{k=0}^3 \mathcal D[\tilde \LL_k]\rho_t
\end{equation}
where the desired Lindblad operators $\LL_k$ are replaced by the
effective operators obtained through RWA with miscalibrated control signals.

\begin{figure}[htbp!]
	\centering
	\includegraphics[width=1\columnwidth]{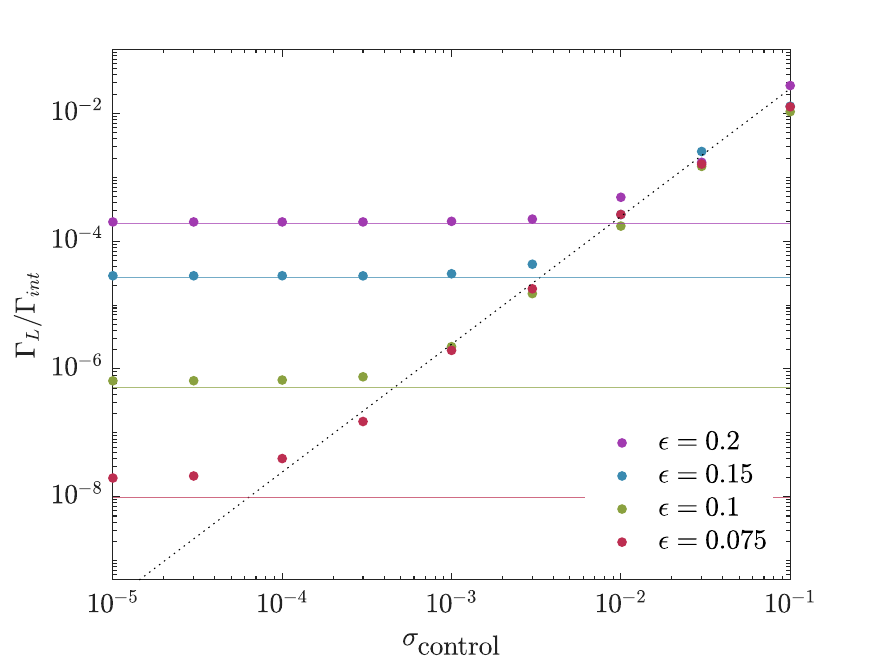}
	\caption{%
		{\bf Dependence of the logical decoherence rate %
		on the uncertainty on the complex amplitude of the harmonics of the control.} %
	The dots correspond to decoherence rates extracted from numerical simulations, %
	where the full lines correspond to the asymptotic ideal rates %
	obtained in~\cref{sm__ssec__eigenvalues}. %
	The black dotted line is proportional to $\sigma_{\textrm{control}}^2$ %
	to serve as a guide for the eye. %
	Beware that this figure uses a logarithmic scale, as opposed to a semi-logarithmic scale. %
	The observed dependence is thus polynomial %
	(approximately quadratic in $\sigma_{\textrm{control}}$, %
	\emph{i.e.} linear in the variance of the uncertainty) %
	and not exponential. %
	This figure presents the results obtained for \emph{one} %
	random realization of the noise coefficients, %
	rather than an average over multiple realizations.%
	}
	\label{fig:control_noise}
\end{figure}

On~\cref{fig:control_noise},
we represent the dependence of the logical decoherence rate
on the standard deviation $\sigma_{\textrm{control}}$
modeling the relative precision of the calibration.
Contrary to many figures in the paper, we do not observe an exponential decay
of the decoherence rate with $\sigma_{\textrm{control}}$.
This indicates that a rather precise calibration of the dispersion of the feedlines
is required;
numerically, we estimate that a relative accuracy around $0.1\%$,
which we still consider experimentally realistic,
is sufficient to maintain the logical decoherence rate
several orders of magnitude below the engineered dissipation rate.

%% file: sm__analysis__numeric_details.tex
\section{Details of numerical simulations}
\label{smsec:numerical_details}

\subsection{Scope of the simulations}
In this paper, we do not try to simultaneously study
all sources of errors, resulting from experimental limitations
(finite bandwidth of the control signals,
stroboscopic implementations of each dissipator, etc.)
as well as imperfections of the device (loss channels, Hamiltonian perturbations,
uncertainty on the parameters, etc.).
For each type of error, we rather
study its impact on logical performance under the assumption that everything else
is perfect: for instance, controls with limited bandwidth, which
yield imperfect stabilizing dissipators, are studied
without taking into account loss channels such as photon loss, etc.
whereas these channels, in turn, are studied with perfect dissipators.
Our goal is to understand precisely the contribution of each type of possible
errors in isolation.
This allows us to roughly identify a parameter regime,
that is a set of (possibly over-optimistic) constraints on all the experimental parameters entering the
dynamics, under which each source of error, taken separately,
leads to a reasonable degradation of the logical performance of a GKP qubit.
In particular, it should be understood that our goal is not to precisely
simulate a given experiment (which does not exist yet!), but to help in its future design,
by identifying possible experimental challenges in implementing
our proposal and quantifying the experimental developments that will be required to enter
this favorable parameter regime
(in terms of \emph{e.g.} device imperfections,
precision of calibrations, microwave control, etc.).

\subsection{Numerical scheme}
Several solutions coexist to compute the solution of Lindblad master equations,
that we can schematically sort into two categories:
\begin{enumerate}
	\item Use already available general-purpose routines
		for the simulation of dynamical systems
		(such as Runge-Kutta methods, implicit Adams methods, etc.).
		This is the solution used in popular quantum libraries such as
		QuTiP~\cite{johansson2012qutip,johansson2013qutip},
		QuantumOptics.jl~\cite{quantumopticsjl2018}
		or the newly announced Qiskit Dynamics~\cite{qiskitdynamics2021}.

		On the one hand, this solution leverages already existing, tested and optimized
		high-order schemes.
		On the other hand, it ignores the structure of the problem,
		in particular that the solution $\rho$ must be a density operator.
		Thus, numerical errors can generate negativities in the computed solution%
		~\cite{RIESCH2019290},
		notably when the true solution of the problem features zero or small eigenvalues.
		We argue that this is in fact a generic property of any simulation
		of a qubit embedded in a bosonic mode, as the
		density operator of the system lives in an infinite-dimensional Hilbert space,
		truncated numerically to a given (possibly high) dimension,
		but encodes a qubit living in $\mathbb C^2$.
	\item Use structure-preserving numerical schemes designed for the simulation of density operators%
		~\cite{steinbachgarrawayknight1995,%
		rouchonralph2015,%
		lucao2021}%
		~\cite[Appendix B]{jordanAnatomyFluorescence2016}.
		To the best of our knowledge,
		these schemes are not readily available in common quantum libraries,
		but have already been used in the literature.
\end{enumerate}

Preliminary versions of the simulations presented in this paper
relied on the QuTiP library
to compute the solution of Lindblad master equations in the Fock basis.
However, this solution turned out to be impractical
due to the high dimension needed to accurately simulate
the evolution of GKP states in the Fock basis.
In particular, the most demanding simulations required up to $N = 5000$ Fock states
to observe numerical convergence
(notably, to capture logical error rates as low as $10^{-8}$
from the numerical simulations).
Moreover, the Lindblad operators entering the dynamics,
such as $\LL_0 = e^{-\epsilon\eta/2} e^{i\eta\qop}(\II-\epsilon\Pop) - \II$,
do not have a sparse representation in the Fock basis
(this is in stark contrast to other bosonic encodings, for instance the cat qubit encoding,
where several stabilization schemes were proposed using only low-order polynomials in $\oa, \oa^\dag$,
the sparse representation of which allows for efficient simulations even in high dimensions).

Additionally, we found that QuTiP suffers from a design flaw:
to compute the solution of a Lindblad equation of the form
\( \frac d{dt}\rho = \mathcal L\rho\) in dimension $N$,
where $\rho$ is an $N\times N$ matrix,
it first rewrites it to the equivalent equation
\( \frac{d}{dt} \hat \rho = \hat{\mathcal L}\hat\rho\)
in dimension $N^2$, where $\hat\rho$ is the vectorized representation of $\rho$
(column stacked in an $N^2$ vector)
and $\hat{\mathcal L}$ is the Liouvillian of the problem.
While formally equivalent, this method requires $O(N^4)$ storage capacity
and the evaluation of $\hat{\mathcal L}\hat\rho$ has a time complexity of $O(N^4)$
(matrix-vector product in dimension $N^2$),
while the evaluation of $\mathcal L\rho$ has a time-complexity of only $O(N^3)$
(matrix-matrix product in dimension $N$) and requires an $O(N^2)$ storage capacity.
For $N=5000$ in a dense problem, this design choice prohibits the use of QuTiP routines.
\\
We thus implemented the first-order structure-preserving scheme
proposed in
\cite[Appendix B]{jordanAnatomyFluorescence2016}.
It can be seen as a fully-linear refinement over the $\mathcal A^{1}$ scheme proposed
in~\cite{lucao2021}.
Whilst this is only a first-order scheme,
in practice,
we found that it was fast enough to achieve numerical convergence in reasonable
runtimes on a laptop for all simulations presented in the paper%
~\footnote{parameter sweeps where performed on a cluster to parallelize the simulations,
but each simulation could run on a laptop.}.
Note that in principle, explicit structure-preserving schemes of arbitrary order
can be developed following~\cite{lucao2021}.
\\

For the sake of pedagogy, let us explain how to derive this scheme in the case of
a generic, time-independent Lindblad equation of the form
\begin{equation}
	\label{eq:generic_lindblad}
	\frac d{dt} \rho_t = -i \, [\HHop,\rho_t] + \sum_{j=1}^J \cD[\LL_j](\rho_t).
\end{equation}
A naive approach for the simulation of \cref{eq:generic_lindblad}
is to use Euler's explicit scheme of order $1$:
\begin{equation}
	\rho_{n+1} = \rho_n
		+ dt \left(
			-i \, [\HHop,\rho_t] + \sum_{j=1}^J \cD[\LL_j](\rho_t)
			\right)
\end{equation}
where $\rho_n \simeq \rho(ndt)$.
This scheme is not structure-preserving.
However, we can approximate it with a quantum channel up to second-order terms:
	\begin{align}
		\rho_{n+1} &= \sum_{j=0}^J \MM_j \, \rho_n \, \MM_j^\dag \nonumber\\
		&= \rho_n\
		+ dt \left(
			-i \, [\HHop,\rho_t] + \sum_{j=1}^J \cD[\LL_j](\rho_t)
			\right) + O(dt^2)\\
		\MM_0 &= \II + dt\left( -i\HHop - \frac12 \sum_{j=1}^J \LL_j^\dag \LL_j\right)\\
		\MM_j &= \sqrt{dt} \LL_j, \quad 1\leq j\leq J.
	\end{align}
This does not describe an exact quantum channel as $\sum_{j=0}^J \MM_j^\dag \MM_j = 1 + O(dt^2)$.
The (non-linear) structure-preserving scheme $\mathcal A^1$ in \cite{lucao2021} compensates this by explicitly
enforcing the conservation of the trace:
\begin{equation}
	\begin{aligned}
		\rho_{n+1} &= \mathcal A^1(\rho_n) \\&=
	\frac1{\trace\left( \left( \sum_{j=0}^J \MM_j^\dag \MM_j \right) \rho_n \right) } \,
	\sum_{j=0}^J \MM_j \, \rho_n \, \MM_j^\dag.
	\end{aligned}
\end{equation}
We use instead the fully-linear version proposed in \cite[Appendix B]{jordanAnatomyFluorescence2016}:
\begin{equation}\label{eq:sqtrick}
	\begin{aligned}
		\rho_{n+1} &= \sum_{j=0}^J \NN_j \, \rho_n \, \NN_j^\dag\\
		\NN_j &= \MM_j \left( \sum_{j=0}^J \MM_j^\dag \MM_j \right)^{-1/2}.
	\end{aligned}
\end{equation}
Note that $\sum_{j=0}^J \NN_j^\dag \NN_j = \II$ and $\NN_j = \MM_j + O(dt^2)$,
which ensures the consistency of the scheme.
Additionally, this normalization step needs to be performed only once
(when the operators $\HH$ and $\LL_j$ are time-independent)
so that its cost is negligible.
We highlight that this is a general procedure, that could be applied to any numerical scheme
expressed as a pseudo quantum channel, that is of the form
\( \rho_{n+1} = \sum_j M_j \rho M_j^\dagger\)
where $\sum_j M_j^\dag M_j \neq \II$.
The first-order scheme considered here could thus be replaced by any
of the higher-order, positivity-preserving scheme proposed in~\cite{lucao2021},
modified according to~\cref{eq:sqtrick} to get a fully-linear scheme of the same order.
In our experiments, we find that this linear, first-order scheme converges faster with $dt\rightarrow 0$
than the non-linear $\mathcal A^1$ scheme (both are first-order schemes but with different prefactors).
\\

In practice, for the numerical simulation of Lindblad equation in the Fock basis,
we have to choose a dimension truncation $N$
and a time-step $dt$.
The time-step $dt$ is chosen to ensure numerical convergence of the scheme on a given simulation
(\emph{i.e.} $dt$ is decreased until the results reach a stationary value).
On the other hand, for the simulation of a GKP dynamics,
the precise estimates of \cref{ssec:energy_estimates}
indicate that the truncation should satisfy $N\gg \eta/\epsilon$.
We additionally use the fact that we can explicitly compute logical errors rates
induced by quadrature noise only (see \cref{sm__ssec__eigenvalues}) to determine the exact truncation:
$N$ is increased until the simulated errors rates match those predicted by this exact analysis.
We then use the same truncation in our other simulations where exact errors rates are not available
due to the presence of other decoherence channels.
We find that relatively high truncation are needed to
accurately compute the exponentially small logical error
rates presented in our figures;
our simulations used truncations up to $N=5000$.

\subsection{Exact computations of the Lindblad operators in the Fock basis}
\label{sm:sss_compute_eietaq}
We can compute explicitly (without any numerical approximation)
the matrix elements in the Fock basis of any operator in the form
\[ \LL_\theta = e^{i\theta\NN}
	\left( e^{i\eta\qop} \left( \II-\epsilon\Pop\right) - \II \right)
	e^{-i\theta\NN} \]
where $\theta\in\mathbb R$, $\eta >0$, $\epsilon > 0$
(rather than, for instance, computing $e^{i\eta\qop}$
by expressing $\qop$ in
a given truncation of the Fock basis,
then computing its numerical matrix exponential,
accumulating truncation errors along the way).
This canvas encompasses all Lindblad operators proposed in our schemes for the stabilization
of both square and hexagonal GKP codes;
additionally, it is also sufficient for the computation of their approximate version obtained
through modular dissipation engineering with control signals of limited bandwidth
(see \cref{sm:sec__rwa}).
The operators $\Pop = \frac{\oa-\oa^\dag}{\sqrt 2i}$
and $\Rop_\theta = e^{i\theta\NN}$
have simple expression in the Fock basis:
\begin{equation}
	\begin{aligned}
		\bra m \Pop \ket n &= -\frac{i}{\sqrt2}\left(%
			\sqrt{n} \, \delta_{m+1,n}%
			- \sqrt{m}\, \delta_{m,n+1} \right),\\
		\bra m \Rop_\theta \ket n &= e^{i\theta n} \, \delta_{m,n}.
	\end{aligned}
\end{equation}
On the other hand, the matrix elements of $e^{i\eta\qop}$
are obtained through recurrence relations.
The initialization is obtained by identifying $e^{i\eta\qop}$ to a displacement operator as
\( e^{i\eta\qop} = \D(\frac{i\eta}{\sqrt 2})\), so that:
	\begin{align}
		\bra 0 e^{i\eta\qop} \ket n
		&= \bra n e^{i\eta\qop} \ket 0
		= \bra n \D\left(\tfrac{i\eta}{\sqrt 2}\right) \ket 0 
		= \braket{n | (\tfrac{i\eta}{\sqrt{2}})_c} \nonumber \\
		&= e^{-\tfrac{\eta^2}4} \left( \frac{i \eta}{\sqrt2} \right) ^n \frac1{\sqrt{n!}}%
	\end{align}
where $\ket{(\tfrac{i\eta}{\sqrt2})_c}$ denotes the coherent states of amplitude $\frac{i\eta}{\sqrt2}$.
The remaining matrix elements can then be computed through the following relations,
where $\alpha = \frac{i\eta}{\sqrt{2}}$:

\begin{widetext}
\begin{equation}
	\label{eq:recurrence_eietaq}
	\begin{aligned}
		\bra{m+1} e^{i\eta\qop} \ket{n+1}%
			&= \bra{m+1} \D(\alpha) \ket{n+1}\\
			&= \frac1{\sqrt{(m+1)(n+1)}} \,%
				\bra{m} \aop \, \D(\alpha) \, \adag \ket{n}\\
			&= \frac1{\sqrt{(m+1)(n+1)}} \,%
				\bra{m} \aop(\adag-\alpha^*) \, \D(\alpha) \ket{n}\\
			&= \frac1{\sqrt{(m+1)(n+1)}} \, \Big(%
				(m+1) \bra{m} \D(\alpha) \ket{n}%
				-\alpha^* \bra{m} \aop \, \D(\alpha) \ket{n}
				\Big)\\
			&= \frac1{\sqrt{(m+1)(n+1)}} \, \Big(%
				(m+1) \bra{m} \D(\alpha) \ket{n}%
				-\alpha^* \bra{m} \D(\alpha)\, (\aop + \alpha) \ket{n}
				\Big)\\
			&= \frac1{\sqrt{(m+1)(n+1)}} \, \Big(%
				(m+1-|\alpha|^2) \bra{m} \D(\alpha) \ket{n}%
				-\alpha^* \bra{m} \D(\alpha)\, \aop \ket{n}
				\Big)\\
			&= \frac1{\sqrt{(m+1)(n+1)}} \, \Big(%
				(m+n+1-|\alpha|^2) \bra{m} \D(\alpha) \ket{n}%
				- \bra{m} \D(\alpha)\,(\adag+\alpha^*) \aop \ket{n}
				\Big)\\
			&= \frac1{\sqrt{(m+1)(n+1)}} \, \Big(%
				(m+n+1-|\alpha|^2) \bra{m} \D(\alpha) \ket{n}%
				- \bra{m} \adag \, \D(\alpha)\, \aop \ket{n}
				\Big)\\
			&= \frac1{\sqrt{(m+1)(n+1)}} \, \Big(%
				(m+n+1-|\alpha|^2) \bra{m} \D(\alpha) \ket{n}%
				- \sqrt{mn} \bra{m-1} \D(\alpha) \ket{n-1}
				\Big)\\
			&= \frac{m+n+1-\eta^2/2}{\sqrt{(m+1)(n+1)}} \, %
				\bra{m} e^{i\eta\qop} \ket{n}%
				- \sqrt{\frac{mn}{(m+1)(n+1)}} \, \bra{m-1} e^{i\eta\qop} \ket{n-1}
	\end{aligned}
\end{equation}
\end{widetext}
(where the last term appears only when $m,n>0$).
\\

Since all coefficients of this recurrence relation are real,
we can show that for any $n,m$,
\begin{equation}
	\label{eq:sym_l_theta_coeffs}
	\begin{aligned}
		\bra m e^{i\eta\qop} \ket n &= \bra n e^{i\eta\qop} \ket m,\\
		\bra m e^{i\eta\qop} \ket n &\in \mathbb R%
			\quad\textrm{ if } m-n \textrm{ even},\\
		\bra m e^{i\eta\qop} \ket n &\in i\mathbb R%
			\quad\textrm{ if } m-n \textrm{ odd}.
	\end{aligned}
\end{equation}
We emphasize that using~\cref{eq:recurrence_eietaq},
each diagonal of the matrix of $\LL_\theta$ in the Fock basis can be computed independently;
in practice, when dealing with control signals of limited bandwidth,
we need only compute as many diagonals as the number of harmonics in the control signal
(see \cref{sm:sec__rwa}).
Additionally, computing the Hermitian (respectively anti-Hermitian) part of $\LL_\theta$
as in~\cref{sec:dissipationengineering}
amounts to computing separately the even (respectively odd) diagonals of $\LL_\theta$.
\\

Finally, using~\cref{eq:sym_l_theta_coeffs}
we can reformulate the previous matrix decomposition along diagonals
as the following operator decomposition:
\begin{equation}
	e^{i\eta\qop} = \phi_0(\NN; \eta) + \sum_{k=1}^{+\infty}
		i^k \left(
			\phi_k(\NN; \eta) \, \oa^k
			+ \oa^{\dag k} \, \phi_k(\NN; \eta)
			\right)
	\label{eq:op_decomposition_eietaq}
\end{equation}
where the $\phi_k$ are real-valued functions defined by
\begin{equation}
	\phi_k(n; \eta) = (-i)^k \sqrt{\frac{n!}{(n+k)!}} \, \bra n e^{i\eta\qop} \ket{n+k}.
	\label{eq:coefs_decomposition_eietaq}
\end{equation}

\subsection{Extraction of logical error rates}
The logical coordinates associated to a given density operator $\rho$
are defined as the expectation values of the three
generalized Pauli $\X,\Y,\Z$ operators.
In other words, the \emph{encoded} logical qubit is defined as
\begin{equation}
	\rho_L = \frac{\II + \trace(\X\rho)\sigma_X + \trace(\Y\rho)\sigma_Y + \trace(\Z\rho)\sigma_Z}2
\end{equation}
where $\sigma_X,\sigma_Y,\sigma_Z$ are the usual Pauli operators on $\mathbb C^2$.

In~\cref{sm__ssec__eigenvalues} we show that, for both stabilization schemes proposed here
(corresponding to square and hexagonal GKP codes),
in presence of quadrature noise only,
the expectation value of periodic observables (such as $\X,\Y,\Z$)
evolve according to two timescales:
an initial fast transient regime with a typical timescale $\tau_{conv} \sim \frac1{\epsilon\eta \Gamma}$
(where $\Gamma>0$ is the engineered dissipation rate),
that we can interpret as a fast convergence to a coding state,
followed by a slow decay with an exponentially larger typical timescale
$\tau_{log} \gg \tau_{conv}$,
that we can interpret as decoherence of the encoded logical qubit.
In presence of generic decoherence channels
(such as \emph{e.g.} photon loss, Kerr effect or dephasing)
we thus extract the logical error rates by the following procedure:
\begin{enumerate}
	\item prepare an initial density operator $\rho_0$ with non-zero logical coordinates
		$\trace(\X\rho_0), \trace(\Y\rho_0), \trace(\Z\rho_0)$;
	\item let $\rho_t$ evolve from $\rho_0$ following the Lindblad dynamics under study,
		during a simulation time $T_{simu} \gg \tau_{conv}$;
	\item compute the evolution of the logical coordinates
		$\trace(\X\rho_t), \trace(\Y\rho_t), \trace(\Z\rho_t)$
		along the trajectory
		\footnote{In practice, the expectation values defining logical coordinates are
		computed at each time-step rather than after computing the whole trajectory,
		to avoid storage of the full history of $\rho_t$.}%
		;
	\item fit the post-transient dynamics with an exponential function of time.
\end{enumerate}

For the initialization of the procedure
with a density operator yielding a non-zero expectation value of $\X,\Y,\Z$,
a numerically cheap strategy is to exploit the fact that
finite-energy square GKP states
are approximate ground states of the so-called square GKP Hamiltonian
\[
	\HH_{GKP,\textrm{square}} = \frac{\epsilon^2}2 \left( \qop^2 + \Pop^2\right)%
	- \left(\cos(\eta\qop) + \cos(\eta\Pop)\right)
\]
while finite-energy hexagonal GKP states
are approximate ground states of the so-called hexagonal GKP Hamiltonian
\begin{align*}
	\HH_{GKP,\textrm{hexagonal}} &= \frac{\epsilon^2}2 \left( \qop^2 + \Pop^2\right) \\
	&\quad - \frac23 \left(\cos(\eta\qop) + \cos(\eta\qop_{\frac \pi 3})
		+ \cos(\eta\qop_{\frac{2\pi}3})\right)
\end{align*}
where $\qop_\theta = \cos(\theta)\qop + \sin(\theta)\Pop$
(note that the value of $\eta$ depends on the geometry).
These two Hamiltonians are easy to compute numerically in the Fock basis using the tools of
the previous section,
and we use the eigenvectors corresponding to the lowest-lying eigenvalues
as initial states.
This choice is all the more motivated by the stability analysis
led in~\cref{sec:energy_qualitative_analysis},
where we established that, with our dissipative stabilization scheme,
this GKP Hamiltonian governs the “no-jump” part of the Lindblad evolution.
\\

On~\cref{fig:rate_extraction}, we illustrate this procedure for the computation of the
logical error rate of a square GKP qubit subjected to photon loss,
as studied in the top-right panel of~\cref{fig:gammavskappa},
corresponding to the Lindblad equation
\begin{equation}
	\label{eq:rate_extraction_photonloss}
	\begin{aligned}
		\frac d{dt} \rho &=
			\Gamma \sum_{k=0}^3 \cD[\LL_k](\rho)
			+ \kappa_{1\textrm{ph}} \, \cD[\oa](\rho),\\
		\LL_k &= \mathcal A \,
			e^{ik\tfrac\pi2 \N} \,
			e^{i\eta\qop} \, \left( \II - \epsilon\Pop \right)
			e^{-ik\tfrac\pi2 \N} \,
			- \II,\\
		\rho_0 &= \ket{\psi_0}\bra{\psi_0}
	\end{aligned}
\end{equation}
where $\eta = 2\sqrt{\pi}$,
$\epsilon$ is the energy-regularization parameter,
$\mathcal A = e^{-\epsilon\eta/2}$,
$\ket{\psi_0}$ is the eigenvector of $\HH_{GKP,\textrm{square}}$
corresponding to its lowest eigenvalue,
$\Gamma>0$ is the engineered dissipation rate
and $\kappa_{1\textrm{ph}}$ is the photon loss rate.

We plot the evolution of $\trace(\X\rho_t)$ along the solution of \cref{eq:rate_extraction_photonloss}
A semilogarithmic display of the same quantity clearly highlights that the decay is well-approximated
by an exponential after an initial transient regime on the order of $\tau_{\mathrm{conv}}$.
However, fitting the post-transient trajectory to an exponential function
requires running the simulation long enough to determine the asymptotic value of $\trace(\X\rho_t)$.
This solution is impractical when trying to extract exponentially low decoherence rates
as it would require running the simulations for exponentially long durations.
A computationally efficient alternative exploits the fact that the asymptotic
value of any observable does not depend on the initial condition $\rho_0$ of the simulation, but only
on the steady-state $\rho_\infty$ of the Lindbladian.
Accordingly, it can be easily eliminated by computing a second trajectory $\tilde \rho_t$
initialized at a different point $\tilde\rho_0$,
and fitting $\trace(\X\rho_t) - \trace(\X\tilde\rho_t)$ to an exponential function.
In our simulations, we typically choose $\tilde\rho_0 = \ket{\psi_1}\bra{\psi_1}$
with $\ket{\psi_1}$ the eigenvector of $\HH_{GKP, \textrm{square}}$ corresponding to its second
lowest eigenvalue.
The logarithmic derivative of
$| \trace(\X\rho_t) - \trace(\X\tilde\rho_t) |$
is found to be nearly stationary after the transient regime,
confirming the above analysis: its stationary value
$\Gamma_L$ can be extracted from simulations satisfying the constraint $T_{\textrm{simu}} \gg \tau_{\textrm{conv}}$
(instead of the prohibitive naive constraint $T_{\textrm{simu}} \gg 1/\Gamma_L$).
\\

Finally, note that the square GKP code has
asymmetrical logical error rates associated to
$\X,\Y,\Z$
satisfying
\begin{equation}
	\Gamma_{L}^{X} = \Gamma_{L}^{Z} = \frac12 \Gamma_{L}^{Y}
\end{equation}
since a $\Z$ (respectively $\X$) error
corresponds to a shift of length $\sqrt\pi$ along the $q$ (respectively $p$) axis
in phase-space,
while a $\Y$ error correspond to their composition,
that is a longer shift of length $\sqrt{2\pi}$ along the diagonal.
In the figures, the plotted logical error rates
correspond to the value of $\Gamma_{L}^{X}$,
from which we can immediately deduce the other two.
For the hexagonal GKP code, on the other hand,
these three values are identical
thanks to the symmetry of the code.
\clearpage

\begin{widetext}

\begin{figure}[p]
	\centering
	\includesvg[width=0.9\columnwidth]{new_rate_extraction.svg}
	\caption{
	{\bf Extraction of logical error rates}. %
	{\bf{a)}} Evolution
	of the logical $x$-coordinate for a square GKP qubit subject to photon loss %
	following~\cref{eq:rate_extraction_photonloss}. %
	Here we fix $\epsilon=0.1$ and $\kappa_{1\textrm{ph}}/\Gamma = 4\times 10^{-2}$ %
	and we plot the solution associated to two initial conditions:
	$\rho_0 = \ket{\psi_0} \bra{\psi_0}$
	and
	$\tilde \rho_0 = \ket{\psi_1} \bra{\psi_1}$,
	where $\ket{\psi_0}, \ket{\psi_1}$ are the eigenvectors of $\HH_{GKP, square}$
	associated to its two lowest eigenvalues.
	{\bf{b-c)}}
	After a transient regime of typical timescale $\tau_{conv} = 1/\mathcal A\epsilon\eta\Gamma$
	(shown in subfigure b by zooming in on early times),
	we can fit the evolution of the logical coordinate
	to an exponential decay (subfigure c).
	In general, the logical coordinate need not necessarily converge to $0$,
	but rather to a fixed value depending on the noise channel under consideration (asymptotic value of the dashed black line in subfigure c).
	An efficient procedure to extract the logical decoherence rate $\Gamma_L$
	--that is the exponential decay parameter-- without determining this asymptotic
	value consists in studying the difference of two trajectories
	$\trace(\X\rho_t) - \trace(\X\tilde\rho_t)$
	(red line in subfigure c);
	it is here found to be $\Gamma_L \simeq 0.01 \Gamma$.
	{\bf{d-e)}} In practice,
	after the transient regime,
	the evolution of logical coordinates
	is governed by the logical decoherence rate,
	allowing to extract $\Gamma_L$ from simulations on a duration
	$T_{\mathrm{simu}} \gg \tau_{\mathrm{conv}}$.
	This is confirmed by observing that the logarithmic derivative of
	$\trace(\X\rho_t) - \trace(\X\tilde\rho_t)$
	converges to $\Gamma_L/\Gamma$ on a timescale commensurate to $\tau_{\mathrm{conv}}$
	(subfigure d) and stays roughly constant from there on (subfigure e).
	}
	\label{fig:rate_extraction}
\end{figure}
\end{widetext}

\subsection{Simulations with imperfect control signals}
\label{smsec:numerics_rwa1}
To account for realistic experimental conditions,
two kinds of constraints were imposed on the control signals.
First, the accessible bandwidth is limited, so that perfect Dirac combs are not accessible in experiments;
accordingly,
\cref{sec:harmonics,sm:sec__rwa} introduced finite-bandwidth control signals by truncating
the Fourier representation of ideal control signals.
Additionally, in \cref{sm:control_miscab}, we introduced random perturbation
of these finite-bandwidth control signals
to account for miscalibration.
Let us now explain how these effects are implemented in the numerical simulations to study
the robustness of our dissipative stabilization scheme.

We see from the results of~\cref{sm:sec__rwa}
and \cref{sm:control_miscab}
that all considered control signals
have the form
\begin{equation}
	u(t) = g \sum_{r=-N}^N u_r e^{i(r\omega_a - \omega_b)t},
\end{equation}
and we want to compute the effective Hamiltonian
\begin{equation}
	\HH_{\mathrm{RWA}}^{(1)} = \overline{ u(t) e^{i\eta_a\qop_a(t)} e^{i\eta_b\qop_b(t)} + \hc}
\end{equation}
from the Fourier coefficients $(u_r)_{-N\leq r\leq N}$.
Using the operator decomposition of~\cref{eq:op_decomposition_eietaq},
we get

\begin{widetext}
\begin{equation}\begin{aligned}
	e^{i\eta_a\qop_a} e^{i\eta_b\qop_b}
	&=
	\left( \phi_0(\N_a; \eta_a) + \sum_{{k_a}=1}^{+\infty}
	i^{k_a} \left( \phi_{k_a}(\N_a; \eta_a) \, \aop^{k_a}  + \aop^{\dag {k_a}} \, \phi_{k_a}(\N_a; \eta_a)  \right)
	\right) \\
	&\quad\quad\quad\quad\quad
	\left( \phi_0(\N_b; \eta_b) + \sum_{{k_b}=1}^{+\infty}
	i^{k_b} \left( \phi_{k_b}(\N_b; \eta_b) \, \bop^{k_b}  + \bop^{\dag {k_b}} \, \phi_{k_b}(\N_b; \eta_b)  \right)
	\right)
\end{aligned}\end{equation}

where $\N_a = \aop^\dag \aop$ and $\N_b = \bop^\dag \bop$;
thus

\begin{equation}\label{eq:rwa_fourier_control}\begin{aligned}
	u(t) e^{i\eta_a\qop_a(t)} e^{i\eta_b\qop_b(t)}
	&= g
	\left( \sum_{r=-N}^N u_r e^{i(r\omega_a - \omega_b)t} \right) \\
	&\quad\quad\quad\quad\quad
	\left( \phi_0(\N_a; \eta_a) + \sum_{{k_a}=1}^{+\infty}
	i^{k_a} \left( \phi_{k_a}(\N_a; \eta_a) \, \aop^{k_a} e^{-i{k_a}\omega_a t}
			+ e^{i{k_a}\omega_a t} \aop^{\dag {k_a}} \, \phi_{k_a}(\N_a; \eta_a)  \right)
	\right) \\
	&\quad\quad\quad\quad\quad
	\left( \phi_0(\N_b; \eta_b) + \sum_{{k_b}=1}^{+\infty}
	i^{k_b} \left( \phi_{k_b}(\N_b; \eta_b) \, \bop^{k_b} e^{-i{k_b}\omega_b t}
			+ e^{i{k_b}\omega_b t} \bop^{\dag {k_b}} \, \phi_{k_b}(\N_b; \eta_b)  \right)
	\right).
\end{aligned}\end{equation}
Extracting the resonant terms from the previous expression,
and assuming as always that $\omega_a$ and $\omega_b$ are incommensurate,
we get

\begin{equation}\begin{aligned}
	&\overline{u(t) e^{i\eta_a\qop_a(t)} e^{i\eta_b\qop_b(t)}} \\
	&\quad = ig \, \overline{
	\left( \sum_{r=-N}^N u_r e^{ir\omega_a t} \right)
	\left( \phi_0(\N_a; \eta_a) + \sum_{{k_a}=1}^{+\infty}
	i^{k_a} \left( \phi_{k_a}(\N_a; \eta_a) \, \aop^{k_a} e^{-i{k_a}\omega_a t}
			+ e^{i{k_a}\omega_a t} \aop^{\dag {k_a}} \, \phi_{k_a}(\N_a; \eta_a)  \right)
	\right) }
	\bop^{\dag} \, \phi_{1}(\N_b; \eta_b)\\
	&\quad = ig \,
	\left( u_0 \phi_0(\N_a; \eta_a) + \sum_{{k_a}=1}^{N}
	i^{k_a} \left( u_{k_a} \phi_{k_a}(\N_a; \eta_a) \, \aop^{k_a}
			+ u_{-k_a} \aop^{\dag {k_a}} \, \phi_{k_a}(\N_a; \eta_a)  \right)
	\right)
	\bop^{\dag} \, \phi_{1}(\N_b; \eta_b)\\
	&\quad = g \Aop_u \, \Bop^\dag
\end{aligned}\end{equation}

where
\begin{equation}\begin{aligned}
	\Bop &= -i \phi_{1}(\N_b; \eta_b) \, \bop\\
	\Aop_u &= u_0 \phi_0(\N_a; \eta_a)
		+\sum_{{k_a}=1}^{N}
		i^{k_a} \Big( u_{k_a} \phi_{k_a}(\N_a; \eta_a) \, \aop^{k_a}
		+ u_{-k_a} \aop^{\dag {k_a}} \, \phi_{k_a}(\N_a; \eta_a)\Big).
\end{aligned}\end{equation}
\clearpage
\end{widetext}

Note that, in the Fock basis, $\Aop_u$ can be computed
by multiplying the diagonal, sub-diagonal and sur-diagonal of index at most $N$
of $e^{i\eta_a \qop_a}$
by the corresponding Fourier coefficient of $u$;
this diagonal decomposition can be computed efficiently as shown in~\cref{smsec:numerical_details}.

All in all, we obtain
\begin{equation}
	\HH_{\mathrm{RWA}} = g \left( \Aop_u \Bop^\dag + \Aop_u^\dag \Bop\right)
\end{equation}
that we can compute for the four control signals $u_0,\ldots, u_3$ corresponding
to the four dissipators to engineer.
Adding a resonant drive wherever required
(to all Lindblad operators if we take the family $(\LL_k)_{0\leq k\leq 3}$,
or only to the symmetric Lindblad operators
if we take the family of symmetric and antisymmetric Lindblad operators
$(\LL_{r,l})_{r\in\{q,p\}, \, l\in\{s,d\}}$%
),
we get an effective Hamiltonian
\begin{equation}\begin{aligned}
	\HH &= g \left ( (\Aop_u-\sigma \II) \Bop^\dag + (\Aop_u-\sigma\II)^\dag \Bop\right),
		\quad \sigma\in\{0,1\}\\
	&= g \left( \LL_u \Bop^\dag + \LL_u^\dag \Bop\right),
\end{aligned}\end{equation}
where $\sigma\in\mathbb R$ is the strength of the drive.
We can directly feed this Hamiltonian into~\cref{eq:eng_adiab_onedissip_modif}
in order to obtain an effective Lindblad equation after adiabatic elimination of mode $b$:
\begin{equation}
	\frac{d\rho_t}{dt} = \Gamma \, \cD[\LL_u](\rho_t)
\end{equation}
with $\Gamma = \frac{4g^2}{\kappa_b} |\phi_{1}(0; \eta_b)|^2
	= \frac{4g^2}{\kappa_b} \frac{\eta_b^2}2 \, e^{-\eta_b^2/2}$.
We can then numerically simulate the dynamics
\begin{equation}
	\frac{d\rho_t}{dt} = \Gamma \, \sum_{k=0}^3 \cD[\LL_{u,k}](\rho_t)
\end{equation}
to compute logical decoherence rates associated to the proposed finite-bandwidth control signals;
the results are shown in~\cref{fig:comb} of the main text.
\newline

Note that one could also easily compute the off-resonant terms in \cref{eq:rwa_fourier_control}.
We leave for future research the exploitation of these formulae for the numerical computation
of higher-order corrections in the RWA as proposed in \cref{sm:sec__rwa}.

%% file: main.bbl
\begin{thebibliography}{130}%
\makeatletter
\providecommand \@ifxundefined [1]{%
 \@ifx{#1\undefined}
}%
\providecommand \@ifnum [1]{%
 \ifnum #1\expandafter \@firstoftwo
 \else \expandafter \@secondoftwo
 \fi
}%
\providecommand \@ifx [1]{%
 \ifx #1\expandafter \@firstoftwo
 \else \expandafter \@secondoftwo
 \fi
}%
\providecommand \natexlab [1]{#1}%
\providecommand \enquote  [1]{``#1''}%
\providecommand \bibnamefont  [1]{#1}%
\providecommand \bibfnamefont [1]{#1}%
\providecommand \citenamefont [1]{#1}%
\providecommand \href@noop [0]{\@secondoftwo}%
\providecommand \href [0]{\begingroup \@sanitize@url \@href}%
\providecommand \@href[1]{\@@startlink{#1}\@@href}%
\providecommand \@@href[1]{\endgroup#1\@@endlink}%
\providecommand \@sanitize@url [0]{\catcode `\\12\catcode `\$12\catcode
  `\&12\catcode `\#12\catcode `\^12\catcode `\_12\catcode `\%12\relax}%
\providecommand \@@startlink[1]{}%
\providecommand \@@endlink[0]{}%
\providecommand \url  [0]{\begingroup\@sanitize@url \@url }%
\providecommand \@url [1]{\endgroup\@href {#1}{\urlprefix }}%
\providecommand \urlprefix  [0]{URL }%
\providecommand \Eprint [0]{\href }%
\providecommand \doibase [0]{https://doi.org/}%
\providecommand \selectlanguage [0]{\@gobble}%
\providecommand \bibinfo  [0]{\@secondoftwo}%
\providecommand \bibfield  [0]{\@secondoftwo}%
\providecommand \translation [1]{[#1]}%
\providecommand \BibitemOpen [0]{}%
\providecommand \bibitemStop [0]{}%
\providecommand \bibitemNoStop [0]{.\EOS\space}%
\providecommand \EOS [0]{\spacefactor3000\relax}%
\providecommand \BibitemShut  [1]{\csname bibitem#1\endcsname}%
\let\auto@bib@innerbib\@empty
\bibitem [{\citenamefont {Gottesman}(1997)}]{gottesman1997stabilizer}%
  \BibitemOpen
  \bibfield  {author} {\bibinfo {author} {\bibfnamefont {D.}~\bibnamefont
  {Gottesman}},\ }\href@noop {} {\emph {\bibinfo {title} {Stabilizer codes and
  quantum error correction}}}\ (\bibinfo  {publisher} {California Institute of
  Technology},\ \bibinfo {year} {1997})\BibitemShut {NoStop}%
\bibitem [{\citenamefont {Kitaev}(2003)}]{kitaev2003fault}%
  \BibitemOpen
  \bibfield  {author} {\bibinfo {author} {\bibfnamefont {A.~Y.}\ \bibnamefont
  {Kitaev}},\ }\bibfield  {title} {\bibinfo {title} {Fault-tolerant quantum
  computation by anyons},\ }\href@noop {} {\bibfield  {journal} {\bibinfo
  {journal} {Annals of Physics}\ }\textbf {\bibinfo {volume} {303}},\ \bibinfo
  {pages} {2} (\bibinfo {year} {2003})}\BibitemShut {NoStop}%
\bibitem [{\citenamefont {Freedman}\ and\ \citenamefont
  {Meyer}(2001)}]{freedman2001projective}%
  \BibitemOpen
  \bibfield  {author} {\bibinfo {author} {\bibfnamefont {M.~H.}\ \bibnamefont
  {Freedman}}\ and\ \bibinfo {author} {\bibfnamefont {D.~A.}\ \bibnamefont
  {Meyer}},\ }\bibfield  {title} {\bibinfo {title} {Projective plane and planar
  quantum codes},\ }\href@noop {} {\bibfield  {journal} {\bibinfo  {journal}
  {Foundations of Computational Mathematics}\ }\textbf {\bibinfo {volume}
  {1}},\ \bibinfo {pages} {325} (\bibinfo {year} {2001})}\BibitemShut {NoStop}%
\bibitem [{\citenamefont {Bravyi}\ and\ \citenamefont
  {Kitaev}(1998)}]{bravyi1998quantum}%
  \BibitemOpen
  \bibfield  {author} {\bibinfo {author} {\bibfnamefont {S.~B.}\ \bibnamefont
  {Bravyi}}\ and\ \bibinfo {author} {\bibfnamefont {A.~Y.}\ \bibnamefont
  {Kitaev}},\ }\bibfield  {title} {\bibinfo {title} {Quantum codes on a lattice
  with boundary},\ }\href@noop {} {\bibfield  {journal} {\bibinfo  {journal}
  {arXiv preprint quant-ph/9811052}\ } (\bibinfo {year} {1998})}\BibitemShut
  {NoStop}%
\bibitem [{\citenamefont {Bombin}\ and\ \citenamefont
  {Martin-Delgado}(2006)}]{bombin2006topological}%
  \BibitemOpen
  \bibfield  {author} {\bibinfo {author} {\bibfnamefont {H.}~\bibnamefont
  {Bombin}}\ and\ \bibinfo {author} {\bibfnamefont {M.~A.}\ \bibnamefont
  {Martin-Delgado}},\ }\bibfield  {title} {\bibinfo {title} {Topological
  quantum distillation},\ }\href@noop {} {\bibfield  {journal} {\bibinfo
  {journal} {Physical review letters}\ }\textbf {\bibinfo {volume} {97}},\
  \bibinfo {pages} {180501} (\bibinfo {year} {2006})}\BibitemShut {NoStop}%
\bibitem [{\citenamefont {Gottesman}\ \emph {et~al.}(2001)\citenamefont
  {Gottesman}, \citenamefont {Kitaev},\ and\ \citenamefont
  {Preskill}}]{gottesman2001encoding}%
  \BibitemOpen
  \bibfield  {author} {\bibinfo {author} {\bibfnamefont {D.}~\bibnamefont
  {Gottesman}}, \bibinfo {author} {\bibfnamefont {A.}~\bibnamefont {Kitaev}},\
  and\ \bibinfo {author} {\bibfnamefont {J.}~\bibnamefont {Preskill}},\
  }\bibfield  {title} {\bibinfo {title} {Encoding a qubit in an oscillator},\
  }\href@noop {} {\bibfield  {journal} {\bibinfo  {journal} {Phys. Rev. A}\
  }\textbf {\bibinfo {volume} {64}},\ \bibinfo {pages} {012310} (\bibinfo
  {year} {2001})}\BibitemShut {NoStop}%
\bibitem [{\citenamefont {Grimsmo}\ and\ \citenamefont
  {Puri}(2021)}]{grimsmo2021quantum}%
  \BibitemOpen
  \bibfield  {author} {\bibinfo {author} {\bibfnamefont {A.~L.}\ \bibnamefont
  {Grimsmo}}\ and\ \bibinfo {author} {\bibfnamefont {S.}~\bibnamefont {Puri}},\
  }\bibfield  {title} {\bibinfo {title} {Quantum error correction with the
  {Gottesman-Kitaev-Preskill} code},\ }\href@noop {} {\bibfield  {journal}
  {\bibinfo  {journal} {PRX Quantum}\ }\textbf {\bibinfo {volume} {2}},\
  \bibinfo {pages} {020101} (\bibinfo {year} {2021})}\BibitemShut {NoStop}%
\bibitem [{\citenamefont {Cochrane}\ \emph {et~al.}(1999)\citenamefont
  {Cochrane}, \citenamefont {Milburn},\ and\ \citenamefont
  {Munro}}]{Cochrane1999cats}%
  \BibitemOpen
  \bibfield  {author} {\bibinfo {author} {\bibfnamefont {P.~T.}\ \bibnamefont
  {Cochrane}}, \bibinfo {author} {\bibfnamefont {G.~J.}\ \bibnamefont
  {Milburn}},\ and\ \bibinfo {author} {\bibfnamefont {W.~J.}\ \bibnamefont
  {Munro}},\ }\bibfield  {title} {\bibinfo {title} {{Macroscopically distinct
  quantum-superposition states as a bosonic code for amplitude damping}},\
  }\href {https://doi.org/10.1103/PhysRevA.59.2631} {\bibfield  {journal}
  {\bibinfo  {journal} {Phys. Rev. A}\ }\textbf {\bibinfo {volume} {59}},\
  \bibinfo {pages} {2631} (\bibinfo {year} {1999})}\BibitemShut {NoStop}%
\bibitem [{\citenamefont {Mirrahimi}\ \emph {et~al.}(2014)\citenamefont
  {Mirrahimi}, \citenamefont {Leghtas}, \citenamefont {Albert}, \citenamefont
  {Touzard}, \citenamefont {Schoelkopf}, \citenamefont {Jiang},\ and\
  \citenamefont {Devoret}}]{mirrahimi2014dynamically}%
  \BibitemOpen
  \bibfield  {author} {\bibinfo {author} {\bibfnamefont {M.}~\bibnamefont
  {Mirrahimi}}, \bibinfo {author} {\bibfnamefont {Z.}~\bibnamefont {Leghtas}},
  \bibinfo {author} {\bibfnamefont {V.~V.}\ \bibnamefont {Albert}}, \bibinfo
  {author} {\bibfnamefont {S.}~\bibnamefont {Touzard}}, \bibinfo {author}
  {\bibfnamefont {R.~J.}\ \bibnamefont {Schoelkopf}}, \bibinfo {author}
  {\bibfnamefont {L.}~\bibnamefont {Jiang}},\ and\ \bibinfo {author}
  {\bibfnamefont {M.~H.}\ \bibnamefont {Devoret}},\ }\bibfield  {title}
  {\bibinfo {title} {Dynamically protected cat-qubits: a new paradigm for
  universal quantum computation},\ }\href@noop {} {\bibfield  {journal}
  {\bibinfo  {journal} {New J. Phys.}\ }\textbf {\bibinfo {volume} {16}},\
  \bibinfo {pages} {045014} (\bibinfo {year} {2014})}\BibitemShut {NoStop}%
\bibitem [{\citenamefont {Michael}\ \emph {et~al.}(2016)\citenamefont
  {Michael}, \citenamefont {Silveri}, \citenamefont {Brierley}, \citenamefont
  {Albert}, \citenamefont {Salmilehto}, \citenamefont {Jiang},\ and\
  \citenamefont {Girvin}}]{michael2016new}%
  \BibitemOpen
  \bibfield  {author} {\bibinfo {author} {\bibfnamefont {M.~H.}\ \bibnamefont
  {Michael}}, \bibinfo {author} {\bibfnamefont {M.}~\bibnamefont {Silveri}},
  \bibinfo {author} {\bibfnamefont {R.}~\bibnamefont {Brierley}}, \bibinfo
  {author} {\bibfnamefont {V.~V.}\ \bibnamefont {Albert}}, \bibinfo {author}
  {\bibfnamefont {J.}~\bibnamefont {Salmilehto}}, \bibinfo {author}
  {\bibfnamefont {L.}~\bibnamefont {Jiang}},\ and\ \bibinfo {author}
  {\bibfnamefont {S.~M.}\ \bibnamefont {Girvin}},\ }\bibfield  {title}
  {\bibinfo {title} {New class of quantum error-correcting codes for a bosonic
  mode},\ }\href@noop {} {\bibfield  {journal} {\bibinfo  {journal} {Phys. Rev.
  X}\ }\textbf {\bibinfo {volume} {6}},\ \bibinfo {pages} {031006} (\bibinfo
  {year} {2016})}\BibitemShut {NoStop}%
\bibitem [{\citenamefont {Hu}\ \emph {et~al.}(2019)\citenamefont {Hu},
  \citenamefont {Ma}, \citenamefont {Cai}, \citenamefont {Mu}, \citenamefont
  {Xu}, \citenamefont {Wang}, \citenamefont {Wu}, \citenamefont {Wang},
  \citenamefont {Song}, \citenamefont {Zou} \emph {et~al.}}]{hu2019quantum}%
  \BibitemOpen
  \bibfield  {author} {\bibinfo {author} {\bibfnamefont {L.}~\bibnamefont
  {Hu}}, \bibinfo {author} {\bibfnamefont {Y.}~\bibnamefont {Ma}}, \bibinfo
  {author} {\bibfnamefont {W.}~\bibnamefont {Cai}}, \bibinfo {author}
  {\bibfnamefont {X.}~\bibnamefont {Mu}}, \bibinfo {author} {\bibfnamefont
  {Y.}~\bibnamefont {Xu}}, \bibinfo {author} {\bibfnamefont {W.}~\bibnamefont
  {Wang}}, \bibinfo {author} {\bibfnamefont {Y.}~\bibnamefont {Wu}}, \bibinfo
  {author} {\bibfnamefont {H.}~\bibnamefont {Wang}}, \bibinfo {author}
  {\bibfnamefont {Y.}~\bibnamefont {Song}}, \bibinfo {author} {\bibfnamefont
  {C.-L.}\ \bibnamefont {Zou}}, \emph {et~al.},\ }\bibfield  {title} {\bibinfo
  {title} {Quantum error correction and universal gate set operation on a
  binomial bosonic logical qubit},\ }\href@noop {} {\bibfield  {journal}
  {\bibinfo  {journal} {Nat. Phys.}\ }\textbf {\bibinfo {volume} {15}},\
  \bibinfo {pages} {503} (\bibinfo {year} {2019})}\BibitemShut {NoStop}%
\bibitem [{\citenamefont {Leghtas}\ \emph {et~al.}(2015)\citenamefont
  {Leghtas}, \citenamefont {Touzard}, \citenamefont {Pop}, \citenamefont {Kou},
  \citenamefont {Vlastakis}, \citenamefont {Petrenko}, \citenamefont {Sliwa},
  \citenamefont {Narla}, \citenamefont {Shankar}, \citenamefont {Hatridge}
  \emph {et~al.}}]{leghtas2015confining}%
  \BibitemOpen
  \bibfield  {author} {\bibinfo {author} {\bibfnamefont {Z.}~\bibnamefont
  {Leghtas}}, \bibinfo {author} {\bibfnamefont {S.}~\bibnamefont {Touzard}},
  \bibinfo {author} {\bibfnamefont {I.~M.}\ \bibnamefont {Pop}}, \bibinfo
  {author} {\bibfnamefont {A.}~\bibnamefont {Kou}}, \bibinfo {author}
  {\bibfnamefont {B.}~\bibnamefont {Vlastakis}}, \bibinfo {author}
  {\bibfnamefont {A.}~\bibnamefont {Petrenko}}, \bibinfo {author}
  {\bibfnamefont {K.~M.}\ \bibnamefont {Sliwa}}, \bibinfo {author}
  {\bibfnamefont {A.}~\bibnamefont {Narla}}, \bibinfo {author} {\bibfnamefont
  {S.}~\bibnamefont {Shankar}}, \bibinfo {author} {\bibfnamefont {M.~J.}\
  \bibnamefont {Hatridge}}, \emph {et~al.},\ }\bibfield  {title} {\bibinfo
  {title} {Confining the state of light to a quantum manifold by engineered
  two-photon loss},\ }\href@noop {} {\bibfield  {journal} {\bibinfo  {journal}
  {Science}\ }\textbf {\bibinfo {volume} {347}},\ \bibinfo {pages} {853}
  (\bibinfo {year} {2015})}\BibitemShut {NoStop}%
\bibitem [{\citenamefont {Fl{\"u}hmann}\ \emph {et~al.}(2019)\citenamefont
  {Fl{\"u}hmann}, \citenamefont {Nguyen}, \citenamefont {Marinelli},
  \citenamefont {Negnevitsky}, \citenamefont {Mehta},\ and\ \citenamefont
  {Home}}]{fluhmann2018encoding}%
  \BibitemOpen
  \bibfield  {author} {\bibinfo {author} {\bibfnamefont {C.}~\bibnamefont
  {Fl{\"u}hmann}}, \bibinfo {author} {\bibfnamefont {T.~L.}\ \bibnamefont
  {Nguyen}}, \bibinfo {author} {\bibfnamefont {M.}~\bibnamefont {Marinelli}},
  \bibinfo {author} {\bibfnamefont {V.}~\bibnamefont {Negnevitsky}}, \bibinfo
  {author} {\bibfnamefont {K.}~\bibnamefont {Mehta}},\ and\ \bibinfo {author}
  {\bibfnamefont {J.}~\bibnamefont {Home}},\ }\bibfield  {title} {\bibinfo
  {title} {Encoding a qubit in a trapped-ion mechanical oscillator},\
  }\href@noop {} {\bibfield  {journal} {\bibinfo  {journal} {Nature}\ }\textbf
  {\bibinfo {volume} {566}},\ \bibinfo {pages} {513} (\bibinfo {year}
  {2019})}\BibitemShut {NoStop}%
\bibitem [{\citenamefont {Puri}\ \emph {et~al.}(2019)\citenamefont {Puri},
  \citenamefont {Grimm}, \citenamefont {Campagne-Ibarcq}, \citenamefont
  {Eickbusch}, \citenamefont {Noh}, \citenamefont {Roberts}, \citenamefont
  {Jiang}, \citenamefont {Mirrahimi}, \citenamefont {Devoret},\ and\
  \citenamefont {Girvin}}]{puri2018stabilized}%
  \BibitemOpen
  \bibfield  {author} {\bibinfo {author} {\bibfnamefont {S.}~\bibnamefont
  {Puri}}, \bibinfo {author} {\bibfnamefont {A.}~\bibnamefont {Grimm}},
  \bibinfo {author} {\bibfnamefont {P.}~\bibnamefont {Campagne-Ibarcq}},
  \bibinfo {author} {\bibfnamefont {A.}~\bibnamefont {Eickbusch}}, \bibinfo
  {author} {\bibfnamefont {K.}~\bibnamefont {Noh}}, \bibinfo {author}
  {\bibfnamefont {G.}~\bibnamefont {Roberts}}, \bibinfo {author} {\bibfnamefont
  {L.}~\bibnamefont {Jiang}}, \bibinfo {author} {\bibfnamefont
  {M.}~\bibnamefont {Mirrahimi}}, \bibinfo {author} {\bibfnamefont {M.~H.}\
  \bibnamefont {Devoret}},\ and\ \bibinfo {author} {\bibfnamefont {S.~M.}\
  \bibnamefont {Girvin}},\ }\bibfield  {title} {\bibinfo {title} {Stabilized
  cat in a driven nonlinear cavity: A fault-tolerant error syndrome detector},\
  }\href {https://doi.org/10.1103/PhysRevX.9.041009} {\bibfield  {journal}
  {\bibinfo  {journal} {Phys. Rev. X}\ }\textbf {\bibinfo {volume} {9}},\
  \bibinfo {pages} {041009} (\bibinfo {year} {2019})}\BibitemShut {NoStop}%
\bibitem [{\citenamefont {Ma}\ \emph {et~al.}(2020)\citenamefont {Ma},
  \citenamefont {Zhang}, \citenamefont {Wong}, \citenamefont {Noh},
  \citenamefont {Rosenblum}, \citenamefont {Reinhold}, \citenamefont
  {Schoelkopf},\ and\ \citenamefont {Jiang}}]{ma2020path}%
  \BibitemOpen
  \bibfield  {author} {\bibinfo {author} {\bibfnamefont {W.-L.}\ \bibnamefont
  {Ma}}, \bibinfo {author} {\bibfnamefont {M.}~\bibnamefont {Zhang}}, \bibinfo
  {author} {\bibfnamefont {Y.}~\bibnamefont {Wong}}, \bibinfo {author}
  {\bibfnamefont {K.}~\bibnamefont {Noh}}, \bibinfo {author} {\bibfnamefont
  {S.}~\bibnamefont {Rosenblum}}, \bibinfo {author} {\bibfnamefont
  {P.}~\bibnamefont {Reinhold}}, \bibinfo {author} {\bibfnamefont {R.~J.}\
  \bibnamefont {Schoelkopf}},\ and\ \bibinfo {author} {\bibfnamefont
  {L.}~\bibnamefont {Jiang}},\ }\bibfield  {title} {\bibinfo {title}
  {Path-independent quantum gates with noisy ancilla},\ }\href@noop {}
  {\bibfield  {journal} {\bibinfo  {journal} {Physical Review Letters}\
  }\textbf {\bibinfo {volume} {125}},\ \bibinfo {pages} {110503} (\bibinfo
  {year} {2020})}\BibitemShut {NoStop}%
\bibitem [{\citenamefont {Shi}\ \emph {et~al.}(2019)\citenamefont {Shi},
  \citenamefont {Chamberland},\ and\ \citenamefont {Cross}}]{shi2019fault}%
  \BibitemOpen
  \bibfield  {author} {\bibinfo {author} {\bibfnamefont {Y.}~\bibnamefont
  {Shi}}, \bibinfo {author} {\bibfnamefont {C.}~\bibnamefont {Chamberland}},\
  and\ \bibinfo {author} {\bibfnamefont {A.}~\bibnamefont {Cross}},\ }\bibfield
   {title} {\bibinfo {title} {Fault-tolerant preparation of approximate {GKP}
  states},\ }\href@noop {} {\bibfield  {journal} {\bibinfo  {journal} {New
  Journal of Physics}\ }\textbf {\bibinfo {volume} {21}},\ \bibinfo {pages}
  {093007} (\bibinfo {year} {2019})}\BibitemShut {NoStop}%
\bibitem [{\citenamefont {Siegele}\ and\ \citenamefont
  {Campagne-Ibarcq}(2023)}]{siegele2023robust}%
  \BibitemOpen
  \bibfield  {author} {\bibinfo {author} {\bibfnamefont {C.}~\bibnamefont
  {Siegele}}\ and\ \bibinfo {author} {\bibfnamefont {P.}~\bibnamefont
  {Campagne-Ibarcq}},\ }\bibfield  {title} {\bibinfo {title} {Robust
  suppression of noise propagation in {{Gottesman-Kitaev-Preskill}} error
  correction},\ }\href {https://doi.org/10.1103/PhysRevA.108.042427} {\bibfield
   {journal} {\bibinfo  {journal} {Phys. Rev. A}\ }\textbf {\bibinfo {volume}
  {108}},\ \bibinfo {pages} {042427} (\bibinfo {year} {2023})}\BibitemShut
  {NoStop}%
\bibitem [{\citenamefont {Rosenblum}\ \emph {et~al.}(2018)\citenamefont
  {Rosenblum}, \citenamefont {Reinhold}, \citenamefont {Mirrahimi},
  \citenamefont {Jiang}, \citenamefont {Frunzio},\ and\ \citenamefont
  {Schoelkopf}}]{rosenblum2018fault}%
  \BibitemOpen
  \bibfield  {author} {\bibinfo {author} {\bibfnamefont {S.}~\bibnamefont
  {Rosenblum}}, \bibinfo {author} {\bibfnamefont {P.}~\bibnamefont {Reinhold}},
  \bibinfo {author} {\bibfnamefont {M.}~\bibnamefont {Mirrahimi}}, \bibinfo
  {author} {\bibfnamefont {L.}~\bibnamefont {Jiang}}, \bibinfo {author}
  {\bibfnamefont {L.}~\bibnamefont {Frunzio}},\ and\ \bibinfo {author}
  {\bibfnamefont {R.~J.}\ \bibnamefont {Schoelkopf}},\ }\bibfield  {title}
  {\bibinfo {title} {Fault-tolerant detection of a quantum error},\ }\href@noop
  {} {\bibfield  {journal} {\bibinfo  {journal} {Science}\ }\textbf {\bibinfo
  {volume} {361}},\ \bibinfo {pages} {266} (\bibinfo {year}
  {2018})}\BibitemShut {NoStop}%
\bibitem [{\citenamefont {Von~Neumann}(1996)}]{Neumann1955}%
  \BibitemOpen
  \bibfield  {author} {\bibinfo {author} {\bibfnamefont {J.}~\bibnamefont
  {Von~Neumann}},\ }\href@noop {} {\emph {\bibinfo {title} {Mathematical
  Foundations of Quantum Mechanics}}}\ (\bibinfo  {publisher} {Princeton Univ.
  Press},\ \bibinfo {address} {Princeton},\ \bibinfo {year} {1996})\BibitemShut
  {NoStop}%
\bibitem [{\citenamefont {Aharonov}\ \emph {et~al.}(1969)\citenamefont
  {Aharonov}, \citenamefont {Pendleton},\ and\ \citenamefont
  {Petersen}}]{Aharonov1969}%
  \BibitemOpen
  \bibfield  {author} {\bibinfo {author} {\bibfnamefont {Y.}~\bibnamefont
  {Aharonov}}, \bibinfo {author} {\bibfnamefont {H.}~\bibnamefont
  {Pendleton}},\ and\ \bibinfo {author} {\bibfnamefont {A.}~\bibnamefont
  {Petersen}},\ }\bibfield  {title} {\bibinfo {title} {{Modular variables in
  quantum theory}},\ }\href {https://doi.org/10.1007/BF00670008} {\bibfield
  {journal} {\bibinfo  {journal} {Int. J. Theor. Phys.}\ }\textbf {\bibinfo
  {volume} {2}},\ \bibinfo {pages} {213} (\bibinfo {year} {1969})}\BibitemShut
  {NoStop}%
\bibitem [{\citenamefont {Popescu}(2010)}]{Popescu2010}%
  \BibitemOpen
  \bibfield  {author} {\bibinfo {author} {\bibfnamefont {S.}~\bibnamefont
  {Popescu}},\ }\bibfield  {title} {\bibinfo {title} {{Dynamical quantum
  non-locality}},\ }\href {https://doi.org/10.1038/nphys1619} {\bibfield
  {journal} {\bibinfo  {journal} {Nat. Phys.}\ }\textbf {\bibinfo {volume}
  {6}},\ \bibinfo {pages} {151} (\bibinfo {year} {2010})}\BibitemShut {NoStop}%
\bibitem [{\citenamefont {Fl{\"u}hmann}\ \emph {et~al.}(2018)\citenamefont
  {Fl{\"u}hmann}, \citenamefont {Negnevitsky}, \citenamefont {Marinelli},\ and\
  \citenamefont {Home}}]{fluhmann2018sequential}%
  \BibitemOpen
  \bibfield  {author} {\bibinfo {author} {\bibfnamefont {C.}~\bibnamefont
  {Fl{\"u}hmann}}, \bibinfo {author} {\bibfnamefont {V.}~\bibnamefont
  {Negnevitsky}}, \bibinfo {author} {\bibfnamefont {M.}~\bibnamefont
  {Marinelli}},\ and\ \bibinfo {author} {\bibfnamefont {J.~P.}\ \bibnamefont
  {Home}},\ }\bibfield  {title} {\bibinfo {title} {Sequential modular position
  and momentum measurements of a trapped ion mechanical oscillator},\
  }\href@noop {} {\bibfield  {journal} {\bibinfo  {journal} {Phys. Rev. X}\
  }\textbf {\bibinfo {volume} {8}},\ \bibinfo {pages} {021001} (\bibinfo {year}
  {2018})}\BibitemShut {NoStop}%
\bibitem [{\citenamefont {Travaglione}\ and\ \citenamefont
  {Milburn}(2002)}]{travaglione2002preparing}%
  \BibitemOpen
  \bibfield  {author} {\bibinfo {author} {\bibfnamefont {B.}~\bibnamefont
  {Travaglione}}\ and\ \bibinfo {author} {\bibfnamefont {G.~J.}\ \bibnamefont
  {Milburn}},\ }\bibfield  {title} {\bibinfo {title} {Preparing encoded states
  in an oscillator},\ }\href@noop {} {\bibfield  {journal} {\bibinfo  {journal}
  {Phys. Rev. A}\ }\textbf {\bibinfo {volume} {66}},\ \bibinfo {pages} {052322}
  (\bibinfo {year} {2002})}\BibitemShut {NoStop}%
\bibitem [{\citenamefont {Pirandola}\ \emph {et~al.}(2006)\citenamefont
  {Pirandola}, \citenamefont {Mancini}, \citenamefont {Vitali},\ and\
  \citenamefont {Tombesi}}]{Pirandola2006}%
  \BibitemOpen
  \bibfield  {author} {\bibinfo {author} {\bibfnamefont {S.}~\bibnamefont
  {Pirandola}}, \bibinfo {author} {\bibfnamefont {S.}~\bibnamefont {Mancini}},
  \bibinfo {author} {\bibfnamefont {D.}~\bibnamefont {Vitali}},\ and\ \bibinfo
  {author} {\bibfnamefont {P.}~\bibnamefont {Tombesi}},\ }\bibfield  {title}
  {\bibinfo {title} {{Continuous variable encoding by ponderomotive
  interaction}},\ }\href {https://doi.org/10.1140/epjd/e2005-00306-3}
  {\bibfield  {journal} {\bibinfo  {journal} {Eur. Phys. J. D}\ }\textbf
  {\bibinfo {volume} {37}},\ \bibinfo {pages} {283} (\bibinfo {year}
  {2006})}\BibitemShut {NoStop}%
\bibitem [{\citenamefont {Terhal}\ and\ \citenamefont
  {Weigand}(2016)}]{terhal2016encoding}%
  \BibitemOpen
  \bibfield  {author} {\bibinfo {author} {\bibfnamefont {B.}~\bibnamefont
  {Terhal}}\ and\ \bibinfo {author} {\bibfnamefont {D.}~\bibnamefont
  {Weigand}},\ }\bibfield  {title} {\bibinfo {title} {Encoding a qubit into a
  cavity mode in circuit qed using phase estimation},\ }\href@noop {}
  {\bibfield  {journal} {\bibinfo  {journal} {Phys. Rev. A}\ }\textbf {\bibinfo
  {volume} {93}},\ \bibinfo {pages} {012315} (\bibinfo {year}
  {2016})}\BibitemShut {NoStop}%
\bibitem [{\citenamefont {Motes}\ \emph {et~al.}(2017)\citenamefont {Motes},
  \citenamefont {Baragiola}, \citenamefont {Gilchrist},\ and\ \citenamefont
  {Menicucci}}]{motes2017encoding}%
  \BibitemOpen
  \bibfield  {author} {\bibinfo {author} {\bibfnamefont {K.~R.}\ \bibnamefont
  {Motes}}, \bibinfo {author} {\bibfnamefont {B.~Q.}\ \bibnamefont
  {Baragiola}}, \bibinfo {author} {\bibfnamefont {A.}~\bibnamefont
  {Gilchrist}},\ and\ \bibinfo {author} {\bibfnamefont {N.~C.}\ \bibnamefont
  {Menicucci}},\ }\bibfield  {title} {\bibinfo {title} {Encoding qubits into
  oscillators with atomic ensembles and squeezed light},\ }\href@noop {}
  {\bibfield  {journal} {\bibinfo  {journal} {Phys. Rev. A}\ }\textbf {\bibinfo
  {volume} {95}},\ \bibinfo {pages} {053819} (\bibinfo {year}
  {2017})}\BibitemShut {NoStop}%
\bibitem [{\citenamefont {Weigand}\ and\ \citenamefont
  {Terhal}(2020)}]{weigand2020realizing}%
  \BibitemOpen
  \bibfield  {author} {\bibinfo {author} {\bibfnamefont {D.~J.}\ \bibnamefont
  {Weigand}}\ and\ \bibinfo {author} {\bibfnamefont {B.~M.}\ \bibnamefont
  {Terhal}},\ }\bibfield  {title} {\bibinfo {title} {Realizing modular
  quadrature measurements via a tunable photon-pressure coupling in circuit
  {QED}},\ }\href@noop {} {\bibfield  {journal} {\bibinfo  {journal} {Physical
  Review A}\ }\textbf {\bibinfo {volume} {101}},\ \bibinfo {pages} {053840}
  (\bibinfo {year} {2020})}\BibitemShut {NoStop}%
\bibitem [{\citenamefont {Royer}\ \emph {et~al.}(2020)\citenamefont {Royer},
  \citenamefont {Singh},\ and\ \citenamefont
  {Girvin}}]{royer2020stabilization}%
  \BibitemOpen
  \bibfield  {author} {\bibinfo {author} {\bibfnamefont {B.}~\bibnamefont
  {Royer}}, \bibinfo {author} {\bibfnamefont {S.}~\bibnamefont {Singh}},\ and\
  \bibinfo {author} {\bibfnamefont {S.}~\bibnamefont {Girvin}},\ }\bibfield
  {title} {\bibinfo {title} {Stabilization of finite-energy
  {Gottesman-Kitaev-Preskill} states},\ }\href@noop {} {\bibfield  {journal}
  {\bibinfo  {journal} {Physical Review Letters}\ }\textbf {\bibinfo {volume}
  {125}},\ \bibinfo {pages} {260509} (\bibinfo {year} {2020})}\BibitemShut
  {NoStop}%
\bibitem [{\citenamefont {Campagne-Ibarcq}\ \emph {et~al.}(2020)\citenamefont
  {Campagne-Ibarcq}, \citenamefont {Eickbusch}, \citenamefont {Touzard},
  \citenamefont {Zalys-Geller}, \citenamefont {Frattini}, \citenamefont
  {Sivak}, \citenamefont {Reinhold}, \citenamefont {Puri}, \citenamefont
  {Shankar}, \citenamefont {Schoelkopf} \emph {et~al.}}]{campagne2020quantum}%
  \BibitemOpen
  \bibfield  {author} {\bibinfo {author} {\bibfnamefont {P.}~\bibnamefont
  {Campagne-Ibarcq}}, \bibinfo {author} {\bibfnamefont {A.}~\bibnamefont
  {Eickbusch}}, \bibinfo {author} {\bibfnamefont {S.}~\bibnamefont {Touzard}},
  \bibinfo {author} {\bibfnamefont {E.}~\bibnamefont {Zalys-Geller}}, \bibinfo
  {author} {\bibfnamefont {N.~E.}\ \bibnamefont {Frattini}}, \bibinfo {author}
  {\bibfnamefont {V.~V.}\ \bibnamefont {Sivak}}, \bibinfo {author}
  {\bibfnamefont {P.}~\bibnamefont {Reinhold}}, \bibinfo {author}
  {\bibfnamefont {S.}~\bibnamefont {Puri}}, \bibinfo {author} {\bibfnamefont
  {S.}~\bibnamefont {Shankar}}, \bibinfo {author} {\bibfnamefont {R.~J.}\
  \bibnamefont {Schoelkopf}}, \emph {et~al.},\ }\bibfield  {title} {\bibinfo
  {title} {Quantum error correction of a qubit encoded in grid states of an
  oscillator},\ }\href@noop {} {\bibfield  {journal} {\bibinfo  {journal}
  {Nature}\ }\textbf {\bibinfo {volume} {584}},\ \bibinfo {pages} {368}
  (\bibinfo {year} {2020})}\BibitemShut {NoStop}%
\bibitem [{\citenamefont {de~Neeve}\ \emph {et~al.}(2022)\citenamefont
  {de~Neeve}, \citenamefont {Nguyen}, \citenamefont {Behrle},\ and\
  \citenamefont {Home}}]{de2022error}%
  \BibitemOpen
  \bibfield  {author} {\bibinfo {author} {\bibfnamefont {B.}~\bibnamefont
  {de~Neeve}}, \bibinfo {author} {\bibfnamefont {T.-L.}\ \bibnamefont
  {Nguyen}}, \bibinfo {author} {\bibfnamefont {T.}~\bibnamefont {Behrle}},\
  and\ \bibinfo {author} {\bibfnamefont {J.~P.}\ \bibnamefont {Home}},\
  }\bibfield  {title} {\bibinfo {title} {Error correction of a logical grid
  state qubit by dissipative pumping},\ }\href@noop {} {\bibfield  {journal}
  {\bibinfo  {journal} {Nature Physics}\ }\textbf {\bibinfo {volume} {18}},\
  \bibinfo {pages} {296} (\bibinfo {year} {2022})}\BibitemShut {NoStop}%
\bibitem [{\citenamefont {Sivak}\ \emph {et~al.}(2023)\citenamefont {Sivak},
  \citenamefont {Eickbusch}, \citenamefont {Royer}, \citenamefont {Singh},
  \citenamefont {Tsioutsios}, \citenamefont {Ganjam}, \citenamefont {Miano},
  \citenamefont {Brock}, \citenamefont {Ding}, \citenamefont {Frunzio} \emph
  {et~al.}}]{sivak2022real}%
  \BibitemOpen
  \bibfield  {author} {\bibinfo {author} {\bibfnamefont {V.}~\bibnamefont
  {Sivak}}, \bibinfo {author} {\bibfnamefont {A.}~\bibnamefont {Eickbusch}},
  \bibinfo {author} {\bibfnamefont {B.}~\bibnamefont {Royer}}, \bibinfo
  {author} {\bibfnamefont {S.}~\bibnamefont {Singh}}, \bibinfo {author}
  {\bibfnamefont {I.}~\bibnamefont {Tsioutsios}}, \bibinfo {author}
  {\bibfnamefont {S.}~\bibnamefont {Ganjam}}, \bibinfo {author} {\bibfnamefont
  {A.}~\bibnamefont {Miano}}, \bibinfo {author} {\bibfnamefont
  {B.}~\bibnamefont {Brock}}, \bibinfo {author} {\bibfnamefont
  {A.}~\bibnamefont {Ding}}, \bibinfo {author} {\bibfnamefont {L.}~\bibnamefont
  {Frunzio}}, \emph {et~al.},\ }\bibfield  {title} {\bibinfo {title} {Real-time
  quantum error correction beyond break-even},\ }\href@noop {} {\bibfield
  {journal} {\bibinfo  {journal} {Nature}\ }\textbf {\bibinfo {volume} {616}},\
  \bibinfo {pages} {50} (\bibinfo {year} {2023})}\BibitemShut {NoStop}%
\bibitem [{\citenamefont {Lachance-Quirion}\ \emph {et~al.}(2024)\citenamefont
  {Lachance-Quirion}, \citenamefont {Lemonde}, \citenamefont {Simoneau},
  \citenamefont {St-Jean}, \citenamefont {Lemieux}, \citenamefont {Turcotte},
  \citenamefont {Wright}, \citenamefont {Lacroix}, \citenamefont
  {Fr\'echette-Viens}, \citenamefont {Shillito}, \citenamefont {Hopfmueller},
  \citenamefont {Tremblay}, \citenamefont {Frattini}, \citenamefont
  {Camirand~Lemyre},\ and\ \citenamefont {St-Jean}}]{nord_quantique_exp}%
  \BibitemOpen
  \bibfield  {author} {\bibinfo {author} {\bibfnamefont {D.}~\bibnamefont
  {Lachance-Quirion}}, \bibinfo {author} {\bibfnamefont {M.-A.}\ \bibnamefont
  {Lemonde}}, \bibinfo {author} {\bibfnamefont {J.~O.}\ \bibnamefont
  {Simoneau}}, \bibinfo {author} {\bibfnamefont {L.}~\bibnamefont {St-Jean}},
  \bibinfo {author} {\bibfnamefont {P.}~\bibnamefont {Lemieux}}, \bibinfo
  {author} {\bibfnamefont {S.}~\bibnamefont {Turcotte}}, \bibinfo {author}
  {\bibfnamefont {W.}~\bibnamefont {Wright}}, \bibinfo {author} {\bibfnamefont
  {A.}~\bibnamefont {Lacroix}}, \bibinfo {author} {\bibfnamefont
  {J.}~\bibnamefont {Fr\'echette-Viens}}, \bibinfo {author} {\bibfnamefont
  {R.}~\bibnamefont {Shillito}}, \bibinfo {author} {\bibfnamefont
  {F.}~\bibnamefont {Hopfmueller}}, \bibinfo {author} {\bibfnamefont
  {M.}~\bibnamefont {Tremblay}}, \bibinfo {author} {\bibfnamefont {N.~E.}\
  \bibnamefont {Frattini}}, \bibinfo {author} {\bibfnamefont {J.}~\bibnamefont
  {Camirand~Lemyre}},\ and\ \bibinfo {author} {\bibfnamefont {P.}~\bibnamefont
  {St-Jean}},\ }\bibfield  {title} {\bibinfo {title} {Autonomous quantum error
  correction of {{Gottesman-Kitaev-Preskill}} states},\ }\href
  {https://doi.org/10.1103/PhysRevLett.132.150607} {\bibfield  {journal}
  {\bibinfo  {journal} {Phys. Rev. Lett.}\ }\textbf {\bibinfo {volume} {132}},\
  \bibinfo {pages} {150607} (\bibinfo {year} {2024})}\BibitemShut {NoStop}%
\bibitem [{\citenamefont {Kitaev}(1995)}]{kitaev1995quantum}%
  \BibitemOpen
  \bibfield  {author} {\bibinfo {author} {\bibfnamefont {A.~Y.}\ \bibnamefont
  {Kitaev}},\ }\bibfield  {title} {\bibinfo {title} {Quantum measurements and
  the abelian stabilizer problem},\ }\href@noop {} {\bibfield  {journal}
  {\bibinfo  {journal} {arXiv preprint quant-ph/9511026}\ } (\bibinfo {year}
  {1995})}\BibitemShut {NoStop}%
\bibitem [{\citenamefont {Svore}\ \emph {et~al.}(2013)\citenamefont {Svore},
  \citenamefont {Hastings},\ and\ \citenamefont {Freedman}}]{svore2013faster}%
  \BibitemOpen
  \bibfield  {author} {\bibinfo {author} {\bibfnamefont {K.~M.}\ \bibnamefont
  {Svore}}, \bibinfo {author} {\bibfnamefont {M.~B.}\ \bibnamefont
  {Hastings}},\ and\ \bibinfo {author} {\bibfnamefont {M.}~\bibnamefont
  {Freedman}},\ }\bibfield  {title} {\bibinfo {title} {Faster phase
  estimation},\ }\href@noop {} {\bibfield  {journal} {\bibinfo  {journal}
  {Quantum Inf. Comput.}\ }\textbf {\bibinfo {volume} {14}},\ \bibinfo {pages}
  {306} (\bibinfo {year} {2013})}\BibitemShut {NoStop}%
\bibitem [{\citenamefont {Blais}\ \emph {et~al.}(2021)\citenamefont {Blais},
  \citenamefont {Grimsmo}, \citenamefont {Girvin},\ and\ \citenamefont
  {Wallraff}}]{blais2021circuit}%
  \BibitemOpen
  \bibfield  {author} {\bibinfo {author} {\bibfnamefont {A.}~\bibnamefont
  {Blais}}, \bibinfo {author} {\bibfnamefont {A.~L.}\ \bibnamefont {Grimsmo}},
  \bibinfo {author} {\bibfnamefont {S.}~\bibnamefont {Girvin}},\ and\ \bibinfo
  {author} {\bibfnamefont {A.}~\bibnamefont {Wallraff}},\ }\bibfield  {title}
  {\bibinfo {title} {Circuit quantum electrodynamics},\ }\href@noop {}
  {\bibfield  {journal} {\bibinfo  {journal} {Reviews of Modern Physics}\
  }\textbf {\bibinfo {volume} {93}},\ \bibinfo {pages} {025005} (\bibinfo
  {year} {2021})}\BibitemShut {NoStop}%
\bibitem [{\citenamefont {Cohen}\ \emph {et~al.}(2017)\citenamefont {Cohen},
  \citenamefont {Smith}, \citenamefont {Devoret},\ and\ \citenamefont
  {Mirrahimi}}]{cohen2017degeneracy}%
  \BibitemOpen
  \bibfield  {author} {\bibinfo {author} {\bibfnamefont {J.}~\bibnamefont
  {Cohen}}, \bibinfo {author} {\bibfnamefont {W.~C.}\ \bibnamefont {Smith}},
  \bibinfo {author} {\bibfnamefont {M.~H.}\ \bibnamefont {Devoret}},\ and\
  \bibinfo {author} {\bibfnamefont {M.}~\bibnamefont {Mirrahimi}},\ }\bibfield
  {title} {\bibinfo {title} {Degeneracy-preserving quantum nondemolition
  measurement of parity-type observables for cat qubits},\ }\href@noop {}
  {\bibfield  {journal} {\bibinfo  {journal} {Physical review letters}\
  }\textbf {\bibinfo {volume} {119}},\ \bibinfo {pages} {060503} (\bibinfo
  {year} {2017})}\BibitemShut {NoStop}%
\bibitem [{\citenamefont {Mooij}\ and\ \citenamefont
  {Nazarov}(2006)}]{mooij2006superconducting}%
  \BibitemOpen
  \bibfield  {author} {\bibinfo {author} {\bibfnamefont {J.}~\bibnamefont
  {Mooij}}\ and\ \bibinfo {author} {\bibfnamefont {Y.~V.}\ \bibnamefont
  {Nazarov}},\ }\bibfield  {title} {\bibinfo {title} {Superconducting nanowires
  as quantum phase-slip junctions},\ }\href@noop {} {\bibfield  {journal}
  {\bibinfo  {journal} {Nature Physics}\ }\textbf {\bibinfo {volume} {2}},\
  \bibinfo {pages} {169} (\bibinfo {year} {2006})}\BibitemShut {NoStop}%
\bibitem [{\citenamefont {Astafiev}\ \emph {et~al.}(2012)\citenamefont
  {Astafiev}, \citenamefont {Ioffe}, \citenamefont {Kafanov}, \citenamefont
  {Pashkin}, \citenamefont {Arutyunov}, \citenamefont {Shahar}, \citenamefont
  {Cohen},\ and\ \citenamefont {Tsai}}]{astafiev2012coherent}%
  \BibitemOpen
  \bibfield  {author} {\bibinfo {author} {\bibfnamefont {O.}~\bibnamefont
  {Astafiev}}, \bibinfo {author} {\bibfnamefont {L.}~\bibnamefont {Ioffe}},
  \bibinfo {author} {\bibfnamefont {S.}~\bibnamefont {Kafanov}}, \bibinfo
  {author} {\bibfnamefont {Y.~A.}\ \bibnamefont {Pashkin}}, \bibinfo {author}
  {\bibfnamefont {K.~Y.}\ \bibnamefont {Arutyunov}}, \bibinfo {author}
  {\bibfnamefont {D.}~\bibnamefont {Shahar}}, \bibinfo {author} {\bibfnamefont
  {O.}~\bibnamefont {Cohen}},\ and\ \bibinfo {author} {\bibfnamefont {J.~S.}\
  \bibnamefont {Tsai}},\ }\bibfield  {title} {\bibinfo {title} {Coherent
  quantum phase slip},\ }\href@noop {} {\bibfield  {journal} {\bibinfo
  {journal} {Nature}\ }\textbf {\bibinfo {volume} {484}},\ \bibinfo {pages}
  {355} (\bibinfo {year} {2012})}\BibitemShut {NoStop}%
\bibitem [{\citenamefont {Le}\ \emph {et~al.}(2019)\citenamefont {Le},
  \citenamefont {Grimsmo}, \citenamefont {M{\"u}ller},\ and\ \citenamefont
  {Stace}}]{le2019doubly}%
  \BibitemOpen
  \bibfield  {author} {\bibinfo {author} {\bibfnamefont {D.~T.}\ \bibnamefont
  {Le}}, \bibinfo {author} {\bibfnamefont {A.}~\bibnamefont {Grimsmo}},
  \bibinfo {author} {\bibfnamefont {C.}~\bibnamefont {M{\"u}ller}},\ and\
  \bibinfo {author} {\bibfnamefont {T.}~\bibnamefont {Stace}},\ }\bibfield
  {title} {\bibinfo {title} {Doubly nonlinear superconducting qubit},\
  }\href@noop {} {\bibfield  {journal} {\bibinfo  {journal} {Physical Review
  A}\ }\textbf {\bibinfo {volume} {100}},\ \bibinfo {pages} {062321} (\bibinfo
  {year} {2019})}\BibitemShut {NoStop}%
\bibitem [{\citenamefont {Brooks}\ \emph {et~al.}(2013)\citenamefont {Brooks},
  \citenamefont {Kitaev},\ and\ \citenamefont
  {Preskill}}]{brooks2013protected}%
  \BibitemOpen
  \bibfield  {author} {\bibinfo {author} {\bibfnamefont {P.}~\bibnamefont
  {Brooks}}, \bibinfo {author} {\bibfnamefont {A.}~\bibnamefont {Kitaev}},\
  and\ \bibinfo {author} {\bibfnamefont {J.}~\bibnamefont {Preskill}},\
  }\bibfield  {title} {\bibinfo {title} {Protected gates for superconducting
  qubits},\ }\href@noop {} {\bibfield  {journal} {\bibinfo  {journal} {Physical
  Review A}\ }\textbf {\bibinfo {volume} {87}},\ \bibinfo {pages} {052306}
  (\bibinfo {year} {2013})}\BibitemShut {NoStop}%
\bibitem [{\citenamefont {Groszkowski}\ \emph {et~al.}(2018)\citenamefont
  {Groszkowski}, \citenamefont {Di~Paolo}, \citenamefont {Grimsmo},
  \citenamefont {Blais}, \citenamefont {Schuster}, \citenamefont {Houck},\ and\
  \citenamefont {Koch}}]{groszkowski2018coherence}%
  \BibitemOpen
  \bibfield  {author} {\bibinfo {author} {\bibfnamefont {P.}~\bibnamefont
  {Groszkowski}}, \bibinfo {author} {\bibfnamefont {A.}~\bibnamefont
  {Di~Paolo}}, \bibinfo {author} {\bibfnamefont {A.}~\bibnamefont {Grimsmo}},
  \bibinfo {author} {\bibfnamefont {A.}~\bibnamefont {Blais}}, \bibinfo
  {author} {\bibfnamefont {D.}~\bibnamefont {Schuster}}, \bibinfo {author}
  {\bibfnamefont {A.}~\bibnamefont {Houck}},\ and\ \bibinfo {author}
  {\bibfnamefont {J.}~\bibnamefont {Koch}},\ }\bibfield  {title} {\bibinfo
  {title} {Coherence properties of the 0-$\pi$ qubit},\ }\href@noop {}
  {\bibfield  {journal} {\bibinfo  {journal} {New Journal of Physics}\ }\textbf
  {\bibinfo {volume} {20}},\ \bibinfo {pages} {043053} (\bibinfo {year}
  {2018})}\BibitemShut {NoStop}%
\bibitem [{\citenamefont {Manucharyan}\ \emph {et~al.}(2009)\citenamefont
  {Manucharyan}, \citenamefont {Koch}, \citenamefont {Glazman},\ and\
  \citenamefont {Devoret}}]{manucharyan2009fluxonium}%
  \BibitemOpen
  \bibfield  {author} {\bibinfo {author} {\bibfnamefont {V.~E.}\ \bibnamefont
  {Manucharyan}}, \bibinfo {author} {\bibfnamefont {J.}~\bibnamefont {Koch}},
  \bibinfo {author} {\bibfnamefont {L.~I.}\ \bibnamefont {Glazman}},\ and\
  \bibinfo {author} {\bibfnamefont {M.~H.}\ \bibnamefont {Devoret}},\
  }\bibfield  {title} {\bibinfo {title} {Fluxonium: Single {Cooper}-pair
  circuit free of charge offsets},\ }\href@noop {} {\bibfield  {journal}
  {\bibinfo  {journal} {Science}\ }\textbf {\bibinfo {volume} {326}},\ \bibinfo
  {pages} {113} (\bibinfo {year} {2009})}\BibitemShut {NoStop}%
\bibitem [{\citenamefont {Pechenezhskiy}\ \emph {et~al.}(2020)\citenamefont
  {Pechenezhskiy}, \citenamefont {Mencia}, \citenamefont {Nguyen},
  \citenamefont {Lin},\ and\ \citenamefont
  {Manucharyan}}]{pechenezhskiy2020superconducting}%
  \BibitemOpen
  \bibfield  {author} {\bibinfo {author} {\bibfnamefont {I.~V.}\ \bibnamefont
  {Pechenezhskiy}}, \bibinfo {author} {\bibfnamefont {R.~A.}\ \bibnamefont
  {Mencia}}, \bibinfo {author} {\bibfnamefont {L.~B.}\ \bibnamefont {Nguyen}},
  \bibinfo {author} {\bibfnamefont {Y.-H.}\ \bibnamefont {Lin}},\ and\ \bibinfo
  {author} {\bibfnamefont {V.~E.}\ \bibnamefont {Manucharyan}},\ }\bibfield
  {title} {\bibinfo {title} {The superconducting quasicharge qubit},\
  }\href@noop {} {\bibfield  {journal} {\bibinfo  {journal} {Nature}\ }\textbf
  {\bibinfo {volume} {585}},\ \bibinfo {pages} {368} (\bibinfo {year}
  {2020})}\BibitemShut {NoStop}%
\bibitem [{\citenamefont {Koch}\ \emph {et~al.}(2007)\citenamefont {Koch},
  \citenamefont {Terri}, \citenamefont {Gambetta}, \citenamefont {Houck},
  \citenamefont {Schuster}, \citenamefont {Majer}, \citenamefont {Blais},
  \citenamefont {Devoret}, \citenamefont {Girvin},\ and\ \citenamefont
  {Schoelkopf}}]{koch2007charge}%
  \BibitemOpen
  \bibfield  {author} {\bibinfo {author} {\bibfnamefont {J.}~\bibnamefont
  {Koch}}, \bibinfo {author} {\bibfnamefont {M.~Y.}\ \bibnamefont {Terri}},
  \bibinfo {author} {\bibfnamefont {J.}~\bibnamefont {Gambetta}}, \bibinfo
  {author} {\bibfnamefont {A.~A.}\ \bibnamefont {Houck}}, \bibinfo {author}
  {\bibfnamefont {D.~I.}\ \bibnamefont {Schuster}}, \bibinfo {author}
  {\bibfnamefont {J.}~\bibnamefont {Majer}}, \bibinfo {author} {\bibfnamefont
  {A.}~\bibnamefont {Blais}}, \bibinfo {author} {\bibfnamefont {M.~H.}\
  \bibnamefont {Devoret}}, \bibinfo {author} {\bibfnamefont {S.~M.}\
  \bibnamefont {Girvin}},\ and\ \bibinfo {author} {\bibfnamefont {R.~J.}\
  \bibnamefont {Schoelkopf}},\ }\bibfield  {title} {\bibinfo {title}
  {Charge-insensitive qubit design derived from the {Cooper} pair box},\
  }\href@noop {} {\bibfield  {journal} {\bibinfo  {journal} {Physical Review
  A}\ }\textbf {\bibinfo {volume} {76}},\ \bibinfo {pages} {042319} (\bibinfo
  {year} {2007})}\BibitemShut {NoStop}%
\bibitem [{\citenamefont {Conrad}\ \emph {et~al.}(2022)\citenamefont {Conrad},
  \citenamefont {Eisert},\ and\ \citenamefont {Arzani}}]{conrad2022gottesman}%
  \BibitemOpen
  \bibfield  {author} {\bibinfo {author} {\bibfnamefont {J.}~\bibnamefont
  {Conrad}}, \bibinfo {author} {\bibfnamefont {J.}~\bibnamefont {Eisert}},\
  and\ \bibinfo {author} {\bibfnamefont {F.}~\bibnamefont {Arzani}},\
  }\bibfield  {title} {\bibinfo {title} {{Gottesman-Kitaev-Preskill} codes: A
  lattice perspective},\ }\href@noop {} {\bibfield  {journal} {\bibinfo
  {journal} {Quantum}\ }\textbf {\bibinfo {volume} {6}},\ \bibinfo {pages}
  {648} (\bibinfo {year} {2022})}\BibitemShut {NoStop}%
\bibitem [{\citenamefont {Rymarz}\ \emph
  {et~al.}(2021{\natexlab{a}})\citenamefont {Rymarz}, \citenamefont {Bosco},
  \citenamefont {Ciani},\ and\ \citenamefont
  {DiVincenzo}}]{rymarz2021hardware}%
  \BibitemOpen
  \bibfield  {author} {\bibinfo {author} {\bibfnamefont {M.}~\bibnamefont
  {Rymarz}}, \bibinfo {author} {\bibfnamefont {S.}~\bibnamefont {Bosco}},
  \bibinfo {author} {\bibfnamefont {A.}~\bibnamefont {Ciani}},\ and\ \bibinfo
  {author} {\bibfnamefont {D.~P.}\ \bibnamefont {DiVincenzo}},\ }\bibfield
  {title} {\bibinfo {title} {{Hardware-encoding grid states in a nonreciprocal
  superconducting circuit}},\ }\href@noop {} {\bibfield  {journal} {\bibinfo
  {journal} {Physical Review X}\ }\textbf {\bibinfo {volume} {11}},\ \bibinfo
  {pages} {011032} (\bibinfo {year} {2021}{\natexlab{a}})}\BibitemShut
  {NoStop}%
\bibitem [{\citenamefont {Chapman}\ \emph {et~al.}(2017)\citenamefont
  {Chapman}, \citenamefont {Rosenthal}, \citenamefont {Kerckhoff},
  \citenamefont {Moores}, \citenamefont {Vale}, \citenamefont {Mates},
  \citenamefont {Hilton}, \citenamefont {Lalumiere}, \citenamefont {Blais},\
  and\ \citenamefont {Lehnert}}]{chapman2017widely}%
  \BibitemOpen
  \bibfield  {author} {\bibinfo {author} {\bibfnamefont {B.~J.}\ \bibnamefont
  {Chapman}}, \bibinfo {author} {\bibfnamefont {E.~I.}\ \bibnamefont
  {Rosenthal}}, \bibinfo {author} {\bibfnamefont {J.}~\bibnamefont
  {Kerckhoff}}, \bibinfo {author} {\bibfnamefont {B.~A.}\ \bibnamefont
  {Moores}}, \bibinfo {author} {\bibfnamefont {L.~R.}\ \bibnamefont {Vale}},
  \bibinfo {author} {\bibfnamefont {J.}~\bibnamefont {Mates}}, \bibinfo
  {author} {\bibfnamefont {G.~C.}\ \bibnamefont {Hilton}}, \bibinfo {author}
  {\bibfnamefont {K.}~\bibnamefont {Lalumiere}}, \bibinfo {author}
  {\bibfnamefont {A.}~\bibnamefont {Blais}},\ and\ \bibinfo {author}
  {\bibfnamefont {K.}~\bibnamefont {Lehnert}},\ }\bibfield  {title} {\bibinfo
  {title} {Widely tunable on-chip microwave circulator for superconducting
  quantum circuits},\ }\href@noop {} {\bibfield  {journal} {\bibinfo  {journal}
  {Physical Review X}\ }\textbf {\bibinfo {volume} {7}},\ \bibinfo {pages}
  {041043} (\bibinfo {year} {2017})}\BibitemShut {NoStop}%
\bibitem [{\citenamefont {Lecocq}\ \emph {et~al.}(2017)\citenamefont {Lecocq},
  \citenamefont {Ranzani}, \citenamefont {Peterson}, \citenamefont {Cicak},
  \citenamefont {Simmonds}, \citenamefont {Teufel},\ and\ \citenamefont
  {Aumentado}}]{lecocq2017nonreciprocal}%
  \BibitemOpen
  \bibfield  {author} {\bibinfo {author} {\bibfnamefont {F.}~\bibnamefont
  {Lecocq}}, \bibinfo {author} {\bibfnamefont {L.}~\bibnamefont {Ranzani}},
  \bibinfo {author} {\bibfnamefont {G.}~\bibnamefont {Peterson}}, \bibinfo
  {author} {\bibfnamefont {K.}~\bibnamefont {Cicak}}, \bibinfo {author}
  {\bibfnamefont {R.}~\bibnamefont {Simmonds}}, \bibinfo {author}
  {\bibfnamefont {J.}~\bibnamefont {Teufel}},\ and\ \bibinfo {author}
  {\bibfnamefont {J.}~\bibnamefont {Aumentado}},\ }\bibfield  {title} {\bibinfo
  {title} {Nonreciprocal microwave signal processing with a field-programmable
  {Josephson} amplifier},\ }\href@noop {} {\bibfield  {journal} {\bibinfo
  {journal} {Physical Review Applied}\ }\textbf {\bibinfo {volume} {7}},\
  \bibinfo {pages} {024028} (\bibinfo {year} {2017})}\BibitemShut {NoStop}%
\bibitem [{\citenamefont {Barzanjeh}\ \emph {et~al.}(2017)\citenamefont
  {Barzanjeh}, \citenamefont {Wulf}, \citenamefont {Peruzzo}, \citenamefont
  {Kalaee}, \citenamefont {Dieterle}, \citenamefont {Painter},\ and\
  \citenamefont {Fink}}]{barzanjeh2017mechanical}%
  \BibitemOpen
  \bibfield  {author} {\bibinfo {author} {\bibfnamefont {S.}~\bibnamefont
  {Barzanjeh}}, \bibinfo {author} {\bibfnamefont {M.}~\bibnamefont {Wulf}},
  \bibinfo {author} {\bibfnamefont {M.}~\bibnamefont {Peruzzo}}, \bibinfo
  {author} {\bibfnamefont {M.}~\bibnamefont {Kalaee}}, \bibinfo {author}
  {\bibfnamefont {P.}~\bibnamefont {Dieterle}}, \bibinfo {author}
  {\bibfnamefont {O.}~\bibnamefont {Painter}},\ and\ \bibinfo {author}
  {\bibfnamefont {J.~M.}\ \bibnamefont {Fink}},\ }\bibfield  {title} {\bibinfo
  {title} {Mechanical on-chip microwave circulator},\ }\href@noop {} {\bibfield
   {journal} {\bibinfo  {journal} {Nature communications}\ }\textbf {\bibinfo
  {volume} {8}},\ \bibinfo {pages} {1} (\bibinfo {year} {2017})}\BibitemShut
  {NoStop}%
\bibitem [{\citenamefont {Mahoney}\ \emph {et~al.}(2017)\citenamefont
  {Mahoney}, \citenamefont {Colless}, \citenamefont {Pauka}, \citenamefont
  {Hornibrook}, \citenamefont {Watson}, \citenamefont {Gardner}, \citenamefont
  {Manfra}, \citenamefont {Doherty},\ and\ \citenamefont
  {Reilly}}]{mahoney2017chip}%
  \BibitemOpen
  \bibfield  {author} {\bibinfo {author} {\bibfnamefont {A.}~\bibnamefont
  {Mahoney}}, \bibinfo {author} {\bibfnamefont {J.}~\bibnamefont {Colless}},
  \bibinfo {author} {\bibfnamefont {S.}~\bibnamefont {Pauka}}, \bibinfo
  {author} {\bibfnamefont {J.}~\bibnamefont {Hornibrook}}, \bibinfo {author}
  {\bibfnamefont {J.}~\bibnamefont {Watson}}, \bibinfo {author} {\bibfnamefont
  {G.}~\bibnamefont {Gardner}}, \bibinfo {author} {\bibfnamefont
  {M.}~\bibnamefont {Manfra}}, \bibinfo {author} {\bibfnamefont
  {A.}~\bibnamefont {Doherty}},\ and\ \bibinfo {author} {\bibfnamefont
  {D.}~\bibnamefont {Reilly}},\ }\bibfield  {title} {\bibinfo {title} {On-chip
  microwave quantum hall circulator},\ }\href@noop {} {\bibfield  {journal}
  {\bibinfo  {journal} {Physical Review X}\ }\textbf {\bibinfo {volume} {7}},\
  \bibinfo {pages} {011007} (\bibinfo {year} {2017})}\BibitemShut {NoStop}%
\bibitem [{\citenamefont {Conrad}(2021)}]{conrad2021twirling}%
  \BibitemOpen
  \bibfield  {author} {\bibinfo {author} {\bibfnamefont {J.}~\bibnamefont
  {Conrad}},\ }\bibfield  {title} {\bibinfo {title} {Twirling and hamiltonian
  engineering via dynamical decoupling for {Gottesman-Kitaev-Preskill} quantum
  computing},\ }\href@noop {} {\bibfield  {journal} {\bibinfo  {journal}
  {Physical Review A}\ }\textbf {\bibinfo {volume} {103}},\ \bibinfo {pages}
  {022404} (\bibinfo {year} {2021})}\BibitemShut {NoStop}%
\bibitem [{\citenamefont {Lescanne}\ \emph {et~al.}(2020)\citenamefont
  {Lescanne}, \citenamefont {Villiers}, \citenamefont {Peronnin}, \citenamefont
  {Sarlette}, \citenamefont {Delbecq}, \citenamefont {Huard}, \citenamefont
  {Kontos}, \citenamefont {Mirrahimi},\ and\ \citenamefont
  {Leghtas}}]{lescanne2020exponential}%
  \BibitemOpen
  \bibfield  {author} {\bibinfo {author} {\bibfnamefont {R.}~\bibnamefont
  {Lescanne}}, \bibinfo {author} {\bibfnamefont {M.}~\bibnamefont {Villiers}},
  \bibinfo {author} {\bibfnamefont {T.}~\bibnamefont {Peronnin}}, \bibinfo
  {author} {\bibfnamefont {A.}~\bibnamefont {Sarlette}}, \bibinfo {author}
  {\bibfnamefont {M.}~\bibnamefont {Delbecq}}, \bibinfo {author} {\bibfnamefont
  {B.}~\bibnamefont {Huard}}, \bibinfo {author} {\bibfnamefont
  {T.}~\bibnamefont {Kontos}}, \bibinfo {author} {\bibfnamefont
  {M.}~\bibnamefont {Mirrahimi}},\ and\ \bibinfo {author} {\bibfnamefont
  {Z.}~\bibnamefont {Leghtas}},\ }\bibfield  {title} {\bibinfo {title}
  {Exponential suppression of bit-flips in a qubit encoded in an oscillator},\
  }\href@noop {} {\bibfield  {journal} {\bibinfo  {journal} {Nature Physics}\
  }\textbf {\bibinfo {volume} {16}},\ \bibinfo {pages} {509} (\bibinfo {year}
  {2020})}\BibitemShut {NoStop}%
\bibitem [{\citenamefont {Cahill}\ and\ \citenamefont
  {Glauber}(1969)}]{Cahill1969}%
  \BibitemOpen
  \bibfield  {author} {\bibinfo {author} {\bibfnamefont {K.~E.}\ \bibnamefont
  {Cahill}}\ and\ \bibinfo {author} {\bibfnamefont {R.~J.}\ \bibnamefont
  {Glauber}},\ }\bibfield  {title} {\bibinfo {title} {{Ordered expansions in
  boson amplitude operators}},\ }\href
  {https://doi.org/10.1103/PhysRev.177.1857} {\bibfield  {journal} {\bibinfo
  {journal} {Phys. Rev.}\ }\textbf {\bibinfo {volume} {177}},\ \bibinfo {pages}
  {1857} (\bibinfo {year} {1969})}\BibitemShut {NoStop}%
\bibitem [{\citenamefont {Cohen}(2017)}]{cohen2017autonomous}%
  \BibitemOpen
  \bibfield  {author} {\bibinfo {author} {\bibfnamefont {J.}~\bibnamefont
  {Cohen}},\ }\emph {\bibinfo {title} {Autonomous quantum error correction with
  superconducting qubits}},\ \href@noop {} {Ph.D. thesis},\ \bibinfo  {school}
  {Universit{\'e} Paris sciences et lettres} (\bibinfo {year}
  {2017})\BibitemShut {NoStop}%
\bibitem [{\citenamefont {Menicucci}(2014)}]{menicucciRefFiniteGKP}%
  \BibitemOpen
  \bibfield  {author} {\bibinfo {author} {\bibfnamefont {N.~C.}\ \bibnamefont
  {Menicucci}},\ }\bibfield  {title} {\bibinfo {title} {Fault-tolerant
  measurement-based quantum computing with continuous-variable cluster
  states},\ }\href {https://doi.org/10.1103/PhysRevLett.112.120504} {\bibfield
  {journal} {\bibinfo  {journal} {Phys. Rev. Lett.}\ }\textbf {\bibinfo
  {volume} {112}},\ \bibinfo {pages} {120504} (\bibinfo {year}
  {2014})}\BibitemShut {NoStop}%
\bibitem [{\citenamefont {Matsuura}\ \emph {et~al.}(2020)\citenamefont
  {Matsuura}, \citenamefont {Yamasaki},\ and\ \citenamefont
  {Koashi}}]{matsuura2020equivalence}%
  \BibitemOpen
  \bibfield  {author} {\bibinfo {author} {\bibfnamefont {T.}~\bibnamefont
  {Matsuura}}, \bibinfo {author} {\bibfnamefont {H.}~\bibnamefont {Yamasaki}},\
  and\ \bibinfo {author} {\bibfnamefont {M.}~\bibnamefont {Koashi}},\
  }\bibfield  {title} {\bibinfo {title} {Equivalence of approximate
  {{Gottesman-Kitaev-Preskill}} codes},\ }\href@noop {} {\bibfield  {journal}
  {\bibinfo  {journal} {Physical Review A}\ }\textbf {\bibinfo {volume}
  {102}},\ \bibinfo {pages} {032408} (\bibinfo {year} {2020})}\BibitemShut
  {NoStop}%
\bibitem [{\citenamefont {Sellem}\ \emph {et~al.}(2022)\citenamefont {Sellem},
  \citenamefont {Campagne-Ibarcq}, \citenamefont {Mirrahimi}, \citenamefont
  {Sarlette},\ and\ \citenamefont {Rouchon}}]{sellem2022exponential}%
  \BibitemOpen
  \bibfield  {author} {\bibinfo {author} {\bibfnamefont {L.-A.}\ \bibnamefont
  {Sellem}}, \bibinfo {author} {\bibfnamefont {P.}~\bibnamefont
  {Campagne-Ibarcq}}, \bibinfo {author} {\bibfnamefont {M.}~\bibnamefont
  {Mirrahimi}}, \bibinfo {author} {\bibfnamefont {A.}~\bibnamefont
  {Sarlette}},\ and\ \bibinfo {author} {\bibfnamefont {P.}~\bibnamefont
  {Rouchon}},\ }\bibfield  {title} {\bibinfo {title} {Exponential convergence
  of a dissipative quantum system towards finite-energy grid states of an
  oscillator},\ }in\ \href {https://doi.org/10.1109/CDC51059.2022.9992722}
  {\emph {\bibinfo {booktitle} {2022 IEEE 61st Conference on Decision and
  Control (CDC)}}}\ (\bibinfo {year} {2022})\ pp.\ \bibinfo {pages}
  {5149--5154}\BibitemShut {NoStop}%
\bibitem [{\citenamefont {Duivenvoorden}\ \emph {et~al.}(2017)\citenamefont
  {Duivenvoorden}, \citenamefont {Terhal},\ and\ \citenamefont
  {Weigand}}]{Duivenvoorden2017}%
  \BibitemOpen
  \bibfield  {author} {\bibinfo {author} {\bibfnamefont {K.}~\bibnamefont
  {Duivenvoorden}}, \bibinfo {author} {\bibfnamefont {B.~M.}\ \bibnamefont
  {Terhal}},\ and\ \bibinfo {author} {\bibfnamefont {D.}~\bibnamefont
  {Weigand}},\ }\bibfield  {title} {\bibinfo {title} {{Single-mode displacement
  sensor}},\ }\href {https://doi.org/10.1103/PhysRevA.95.012305} {\bibfield
  {journal} {\bibinfo  {journal} {Phys. Rev. A}\ }\textbf {\bibinfo {volume}
  {95}},\ \bibinfo {pages} {012305} (\bibinfo {year} {2017})}\BibitemShut
  {NoStop}%
\bibitem [{Note1()}]{Note1}%
  \BibitemOpen
  \bibinfo {note} {The relatively low mode frequency $\omega _a = 2\pi \times
  150$~MHz proposed later in \protect \cref {tableparam} can lead to
  non-negligible thermal population at typical cryogenic temperatures. In order
  to take this effect into account, one could consider a one-photon gain
  process on top of one-photon loss dissipation. In that case, two Lindblad
  operators should be included in the simulations: ${\protect \bf L}_{\protect
  \mathrm {loss}} = \protect \sqrt {\kappa (1+n_{\protect \mathrm {th}})}
  \protect \, {\protect \bf a}$ and ${\protect \bf L}_{\protect \mathrm {gain}}
  = \protect \sqrt {\kappa n_{\protect \mathrm {th}}} \protect \, {\protect \bf
  a}^\dagger $, with $n_{\protect \mathrm {th}} = \left ( \exp (\protect \frac
  {\hbar \omega _a}{k_B T}) -1 \right )^{-1}$ the average thermal photon
  number, $T$ the temperature and $k_B$ the Boltzmann constant. This setting
  can be intuitively understood as interpolating between quadrature noise and
  pure photon loss. Indeed, at the level of dissipators, we have $\protect
  \mathcal D[{\protect \bf q}] + \protect \mathcal D[{\protect \bf p}] =
  \protect \mathcal D[{\protect \bf a}] + \protect \mathcal D[{\protect \bf
  a}^\dagger ]$ so that the dissipators associated to one-photon loss and gain
  satisfy: $\protect \mathcal D[{\protect \bf L}_\protect \mathrm {loss}] +
  \protect \mathcal D[{\protect \bf L}_\protect \mathrm {gain}] = \kappa _-
  \protect \, \left ( s ( \protect \mathcal D[{\protect \bf q}] + \protect
  \mathcal D[{\protect \bf p}]) + (1-s) \protect \mathcal D[{\protect \bf
  a}]\right )$ for $\kappa _- = \kappa (1+n_{\protect \mathrm {th}})$ and $s =
  n_{\protect \mathrm {th}}/ (1+n_{\protect \mathrm {th}})$.}\BibitemShut
  {Stop}%
\bibitem [{\citenamefont {Kolesnikow}\ \emph {et~al.}(2024)\citenamefont
  {Kolesnikow}, \citenamefont {Bomantara}, \citenamefont {Doherty},\ and\
  \citenamefont {Grimsmo}}]{kolesnikow2023}%
  \BibitemOpen
  \bibfield  {author} {\bibinfo {author} {\bibfnamefont {X.~C.}\ \bibnamefont
  {Kolesnikow}}, \bibinfo {author} {\bibfnamefont {R.~W.}\ \bibnamefont
  {Bomantara}}, \bibinfo {author} {\bibfnamefont {A.~C.}\ \bibnamefont
  {Doherty}},\ and\ \bibinfo {author} {\bibfnamefont {A.~L.}\ \bibnamefont
  {Grimsmo}},\ }\bibfield  {title} {\bibinfo {title}
  {Gottesman-{{Kitaev-Preskill}} state preparation using periodic driving},\
  }\href {https://doi.org/10.1103/PhysRevLett.132.130605} {\bibfield  {journal}
  {\bibinfo  {journal} {Phys. Rev. Lett.}\ }\textbf {\bibinfo {volume} {132}},\
  \bibinfo {pages} {130605} (\bibinfo {year} {2024})}\BibitemShut {NoStop}%
\bibitem [{\citenamefont {Putterman}\ \emph {et~al.}(2022)\citenamefont
  {Putterman}, \citenamefont {Iverson}, \citenamefont {Xu}, \citenamefont
  {Jiang}, \citenamefont {Painter}, \citenamefont {Brand{\~a}o},\ and\
  \citenamefont {Noh}}]{putterman2022stabilizing}%
  \BibitemOpen
  \bibfield  {author} {\bibinfo {author} {\bibfnamefont {H.}~\bibnamefont
  {Putterman}}, \bibinfo {author} {\bibfnamefont {J.}~\bibnamefont {Iverson}},
  \bibinfo {author} {\bibfnamefont {Q.}~\bibnamefont {Xu}}, \bibinfo {author}
  {\bibfnamefont {L.}~\bibnamefont {Jiang}}, \bibinfo {author} {\bibfnamefont
  {O.}~\bibnamefont {Painter}}, \bibinfo {author} {\bibfnamefont {F.~G.}\
  \bibnamefont {Brand{\~a}o}},\ and\ \bibinfo {author} {\bibfnamefont
  {K.}~\bibnamefont {Noh}},\ }\bibfield  {title} {\bibinfo {title} {Stabilizing
  a bosonic qubit using colored dissipation},\ }\href@noop {} {\bibfield
  {journal} {\bibinfo  {journal} {Physical Review Letters}\ }\textbf {\bibinfo
  {volume} {128}},\ \bibinfo {pages} {110502} (\bibinfo {year}
  {2022})}\BibitemShut {NoStop}%
\bibitem [{Note2()}]{Note2}%
  \BibitemOpen
  \bibinfo {note} {The idea of colored bath engineering is to induce relaxation
  between two energy levels $i$ and $j$ of ${\protect \bf H}_{\protect \mathrm
  {GKP}}$ verifying $|\langle j |{\protect \bf a}| i\rangle |>0$ by coupling
  parametrically the target mode to an ancillary dissipative mode via an
  interaction of the form $g_{ij} e^{i(\omega _{ij}+\omega _a-\omega _b)t}
  {\protect \bf a}{\protect \bf b}^{\dagger }+h.c.$. The coupling strength
  should be chosen such that $g_{ij}\ll |\omega _{ij}-\omega _{ik}|$ for any
  other level $k$ in order not to induce spurious transitions to $|k\rangle $.
  Other types of couplings, in particular of the form $g_{ij} e^{i(\omega
  _{ij}-\omega _a-\omega _b)t} {\protect \bf a}^{\dagger } {\protect \bf
  b}^{\dagger }+h.c.$ may be needed to induce transitions for which ${\protect
  \bf a}$ has a negligible matrix element \cite {sivak2022real}}\BibitemShut
  {NoStop}%
\bibitem [{Note3()}]{Note3}%
  \BibitemOpen
  \bibinfo {note} {An arbitrarily accurate approximation is obtained by
  considering quadrature operators rotated by arbitrarily small angles $\theta
  _a=\protect \frac { \epsilon \epsilon ' }{2\eta }$ and $\theta _b=\protect
  \frac {\pi }{2}-\protect \frac {\eta \epsilon \epsilon '}{4}$ with $\epsilon
  ' \rightarrow 0$, and scaling the terms associated to $j=\pm 1$ in the sum by
  $2/\epsilon '$ \label {footnote:finite_differences}}\BibitemShut {NoStop}%
\bibitem [{\citenamefont {Foster}(1924)}]{foster1924reactance}%
  \BibitemOpen
  \bibfield  {author} {\bibinfo {author} {\bibfnamefont {R.~M.}\ \bibnamefont
  {Foster}},\ }\bibfield  {title} {\bibinfo {title} {A reactance theorem},\
  }\href@noop {} {\bibfield  {journal} {\bibinfo  {journal} {Bell System
  technical journal}\ }\textbf {\bibinfo {volume} {3}},\ \bibinfo {pages} {259}
  (\bibinfo {year} {1924})}\BibitemShut {NoStop}%
\bibitem [{\citenamefont {Nigg}\ \emph {et~al.}(2012)\citenamefont {Nigg},
  \citenamefont {Paik}, \citenamefont {Vlastakis}, \citenamefont {Kirchmair},
  \citenamefont {Shankar}, \citenamefont {Frunzio}, \citenamefont {Devoret},
  \citenamefont {Schoelkopf},\ and\ \citenamefont {Girvin}}]{nigg2012black}%
  \BibitemOpen
  \bibfield  {author} {\bibinfo {author} {\bibfnamefont {S.~E.}\ \bibnamefont
  {Nigg}}, \bibinfo {author} {\bibfnamefont {H.}~\bibnamefont {Paik}}, \bibinfo
  {author} {\bibfnamefont {B.}~\bibnamefont {Vlastakis}}, \bibinfo {author}
  {\bibfnamefont {G.}~\bibnamefont {Kirchmair}}, \bibinfo {author}
  {\bibfnamefont {S.}~\bibnamefont {Shankar}}, \bibinfo {author} {\bibfnamefont
  {L.}~\bibnamefont {Frunzio}}, \bibinfo {author} {\bibfnamefont
  {M.}~\bibnamefont {Devoret}}, \bibinfo {author} {\bibfnamefont
  {R.}~\bibnamefont {Schoelkopf}},\ and\ \bibinfo {author} {\bibfnamefont
  {S.}~\bibnamefont {Girvin}},\ }\bibfield  {title} {\bibinfo {title}
  {Black-box superconducting circuit quantization},\ }\href@noop {} {\bibfield
  {journal} {\bibinfo  {journal} {Phys. Rev. Lett.}\ }\textbf {\bibinfo
  {volume} {108}},\ \bibinfo {pages} {240502} (\bibinfo {year}
  {2012})}\BibitemShut {NoStop}%
\bibitem [{\citenamefont {Smith}\ \emph {et~al.}(2016)\citenamefont {Smith},
  \citenamefont {Kou}, \citenamefont {Vool}, \citenamefont {Pop}, \citenamefont
  {Frunzio}, \citenamefont {Schoelkopf},\ and\ \citenamefont
  {Devoret}}]{smith2016quantization}%
  \BibitemOpen
  \bibfield  {author} {\bibinfo {author} {\bibfnamefont {W.}~\bibnamefont
  {Smith}}, \bibinfo {author} {\bibfnamefont {A.}~\bibnamefont {Kou}}, \bibinfo
  {author} {\bibfnamefont {U.}~\bibnamefont {Vool}}, \bibinfo {author}
  {\bibfnamefont {I.}~\bibnamefont {Pop}}, \bibinfo {author} {\bibfnamefont
  {L.}~\bibnamefont {Frunzio}}, \bibinfo {author} {\bibfnamefont
  {R.}~\bibnamefont {Schoelkopf}},\ and\ \bibinfo {author} {\bibfnamefont
  {M.}~\bibnamefont {Devoret}},\ }\bibfield  {title} {\bibinfo {title}
  {Quantization of inductively shunted superconducting circuits},\ }\href@noop
  {} {\bibfield  {journal} {\bibinfo  {journal} {Physical Review B}\ }\textbf
  {\bibinfo {volume} {94}},\ \bibinfo {pages} {144507} (\bibinfo {year}
  {2016})}\BibitemShut {NoStop}%
\bibitem [{Note4()}]{Note4}%
  \BibitemOpen
  \bibinfo {note} {Here, we have assumed that $\omega _a$ and $\omega _b$ are
  not commensurable and neglected terms in $({\protect \bf b^{\dagger
  }}{\protect \bf b})^k {\protect \bf r}_b$ with $k>0$, whose only impact is to
  renormalize the modular interaction strength $g\rightarrow e^{-\eta _b^2/4}g$
  as detailed in \protect \cref {sm:sec__rwa}.}\BibitemShut {Stop}%
\bibitem [{Note5()}]{Note5}%
  \BibitemOpen
  \bibinfo {note} {This qualitative difference is easily understood by noting
  that dephasing does not change the steady state of the ancilla but kills the
  coherences created by the interaction Hamiltonian ${\protect \bf H}^{\protect
  \mathrm {int}}$, while the thermal population of the buffer changes its
  steady state to a thermal state $\protect \bm {\normalrho }_b^{\protect
  \mathrm {th}} := \protect \frac {1}{1+n_{\protect \mathrm {th}}} \left (
  \protect \frac {n_{\protect \mathrm {th}}}{1+n_{\protect \mathrm {th}}}\right
  )^{{\protect \bf b}^\dagger {\protect \bf b}}$.}\BibitemShut {Stop}%
\bibitem [{Note6()}]{Note6}%
  \BibitemOpen
  \bibinfo {note} {The modular interaction strength $g$ is proportional to the
  bias pulses integrated amplitude $\xi _1$, which decreases with $N$ following
  $\xi _1=2\pi \xi _{\protect \mathrm {max}}/((2N+1)\omega _a)$}\BibitemShut
  {NoStop}%
\bibitem [{Note7()}]{Note7}%
  \BibitemOpen
  \bibinfo {note} {Josephson junctions feature an intrinsic capacitance omitted
  in Fig.~\ref {fig:dissipCircuit}a, which, combined with their kinetic
  inductance, form an oscillator typically resonating around 10---50~GHz for
  standard microfabrication techniques.}\BibitemShut {Stop}%
\bibitem [{\citenamefont {Solinas}\ \emph
  {et~al.}(2015{\natexlab{a}})\citenamefont {Solinas}, \citenamefont
  {Bosisio},\ and\ \citenamefont {Giazotto}}]{solinas2015radiation}%
  \BibitemOpen
  \bibfield  {author} {\bibinfo {author} {\bibfnamefont {P.}~\bibnamefont
  {Solinas}}, \bibinfo {author} {\bibfnamefont {R.}~\bibnamefont {Bosisio}},\
  and\ \bibinfo {author} {\bibfnamefont {F.}~\bibnamefont {Giazotto}},\
  }\bibfield  {title} {\bibinfo {title} {Radiation comb generation with
  extended {Josephson} junctions},\ }\href@noop {} {\bibfield  {journal}
  {\bibinfo  {journal} {Journal of Applied Physics}\ }\textbf {\bibinfo
  {volume} {118}},\ \bibinfo {pages} {113901} (\bibinfo {year}
  {2015}{\natexlab{a}})}\BibitemShut {NoStop}%
\bibitem [{\citenamefont {Solinas}\ \emph
  {et~al.}(2015{\natexlab{b}})\citenamefont {Solinas}, \citenamefont
  {Gasparinetti}, \citenamefont {Golubev},\ and\ \citenamefont
  {Giazotto}}]{solinas2015josephson}%
  \BibitemOpen
  \bibfield  {author} {\bibinfo {author} {\bibfnamefont {P.}~\bibnamefont
  {Solinas}}, \bibinfo {author} {\bibfnamefont {S.}~\bibnamefont
  {Gasparinetti}}, \bibinfo {author} {\bibfnamefont {D.}~\bibnamefont
  {Golubev}},\ and\ \bibinfo {author} {\bibfnamefont {F.}~\bibnamefont
  {Giazotto}},\ }\bibfield  {title} {\bibinfo {title} {A {Josephson} radiation
  comb generator},\ }\href@noop {} {\bibfield  {journal} {\bibinfo  {journal}
  {Scientific reports}\ }\textbf {\bibinfo {volume} {5}},\ \bibinfo {pages} {1}
  (\bibinfo {year} {2015}{\natexlab{b}})}\BibitemShut {NoStop}%
\bibitem [{\citenamefont {Masluk}\ \emph {et~al.}(2012)\citenamefont {Masluk},
  \citenamefont {Pop}, \citenamefont {Kamal}, \citenamefont {Minev},\ and\
  \citenamefont {Devoret}}]{masluk2012microwave}%
  \BibitemOpen
  \bibfield  {author} {\bibinfo {author} {\bibfnamefont {N.~A.}\ \bibnamefont
  {Masluk}}, \bibinfo {author} {\bibfnamefont {I.~M.}\ \bibnamefont {Pop}},
  \bibinfo {author} {\bibfnamefont {A.}~\bibnamefont {Kamal}}, \bibinfo
  {author} {\bibfnamefont {Z.~K.}\ \bibnamefont {Minev}},\ and\ \bibinfo
  {author} {\bibfnamefont {M.~H.}\ \bibnamefont {Devoret}},\ }\bibfield
  {title} {\bibinfo {title} {Microwave characterization of {Josephson} junction
  arrays: Implementing a low loss superinductance},\ }\href@noop {} {\bibfield
  {journal} {\bibinfo  {journal} {Physical review letters}\ }\textbf {\bibinfo
  {volume} {109}},\ \bibinfo {pages} {137002} (\bibinfo {year}
  {2012})}\BibitemShut {NoStop}%
\bibitem [{\citenamefont {Paladino}\ \emph {et~al.}(2014)\citenamefont
  {Paladino}, \citenamefont {Galperin}, \citenamefont {Falci},\ and\
  \citenamefont {Altshuler}}]{paladino20141}%
  \BibitemOpen
  \bibfield  {author} {\bibinfo {author} {\bibfnamefont {E.}~\bibnamefont
  {Paladino}}, \bibinfo {author} {\bibfnamefont {Y.}~\bibnamefont {Galperin}},
  \bibinfo {author} {\bibfnamefont {G.}~\bibnamefont {Falci}},\ and\ \bibinfo
  {author} {\bibfnamefont {B.}~\bibnamefont {Altshuler}},\ }\bibfield  {title}
  {\bibinfo {title} {1/f noise: Implications for solid-state quantum
  information},\ }\href@noop {} {\bibfield  {journal} {\bibinfo  {journal}
  {Reviews of Modern Physics}\ }\textbf {\bibinfo {volume} {86}},\ \bibinfo
  {pages} {361} (\bibinfo {year} {2014})}\BibitemShut {NoStop}%
\bibitem [{\citenamefont {Glazman}\ and\ \citenamefont
  {Catelani}(2021)}]{glazman2021bogoliubov}%
  \BibitemOpen
  \bibfield  {author} {\bibinfo {author} {\bibfnamefont {L.}~\bibnamefont
  {Glazman}}\ and\ \bibinfo {author} {\bibfnamefont {G.}~\bibnamefont
  {Catelani}},\ }\bibfield  {title} {\bibinfo {title} {Bogoliubov
  quasiparticles in superconducting qubits},\ }\href@noop {} {\bibfield
  {journal} {\bibinfo  {journal} {SciPost Physics Lecture Notes}\ ,\ \bibinfo
  {pages} {031}} (\bibinfo {year} {2021})}\BibitemShut {NoStop}%
\bibitem [{\citenamefont {Cardani}\ \emph {et~al.}(2021)\citenamefont
  {Cardani}, \citenamefont {Valenti}, \citenamefont {Casali}, \citenamefont
  {Catelani}, \citenamefont {Charpentier}, \citenamefont {Clemenza},
  \citenamefont {Colantoni}, \citenamefont {Cruciani}, \citenamefont
  {D’Imperio}, \citenamefont {Gironi} \emph {et~al.}}]{cardani2021reducing}%
  \BibitemOpen
  \bibfield  {author} {\bibinfo {author} {\bibfnamefont {L.}~\bibnamefont
  {Cardani}}, \bibinfo {author} {\bibfnamefont {F.}~\bibnamefont {Valenti}},
  \bibinfo {author} {\bibfnamefont {N.}~\bibnamefont {Casali}}, \bibinfo
  {author} {\bibfnamefont {G.}~\bibnamefont {Catelani}}, \bibinfo {author}
  {\bibfnamefont {T.}~\bibnamefont {Charpentier}}, \bibinfo {author}
  {\bibfnamefont {M.}~\bibnamefont {Clemenza}}, \bibinfo {author}
  {\bibfnamefont {I.}~\bibnamefont {Colantoni}}, \bibinfo {author}
  {\bibfnamefont {A.}~\bibnamefont {Cruciani}}, \bibinfo {author}
  {\bibfnamefont {G.}~\bibnamefont {D’Imperio}}, \bibinfo {author}
  {\bibfnamefont {L.}~\bibnamefont {Gironi}}, \emph {et~al.},\ }\bibfield
  {title} {\bibinfo {title} {Reducing the impact of radioactivity on quantum
  circuits in a deep-underground facility},\ }\href@noop {} {\bibfield
  {journal} {\bibinfo  {journal} {Nature communications}\ }\textbf {\bibinfo
  {volume} {12}},\ \bibinfo {pages} {1} (\bibinfo {year} {2021})}\BibitemShut
  {NoStop}%
\bibitem [{\citenamefont {Mannila}\ \emph {et~al.}(2022)\citenamefont
  {Mannila}, \citenamefont {Samuelsson}, \citenamefont {Simbierowicz},
  \citenamefont {Peltonen}, \citenamefont {Vesterinen}, \citenamefont
  {Gr{\"o}nberg}, \citenamefont {Hassel}, \citenamefont {Maisi},\ and\
  \citenamefont {Pekola}}]{mannila2022superconductor}%
  \BibitemOpen
  \bibfield  {author} {\bibinfo {author} {\bibfnamefont {E.~T.}\ \bibnamefont
  {Mannila}}, \bibinfo {author} {\bibfnamefont {P.}~\bibnamefont {Samuelsson}},
  \bibinfo {author} {\bibfnamefont {S.}~\bibnamefont {Simbierowicz}}, \bibinfo
  {author} {\bibfnamefont {J.}~\bibnamefont {Peltonen}}, \bibinfo {author}
  {\bibfnamefont {V.}~\bibnamefont {Vesterinen}}, \bibinfo {author}
  {\bibfnamefont {L.}~\bibnamefont {Gr{\"o}nberg}}, \bibinfo {author}
  {\bibfnamefont {J.}~\bibnamefont {Hassel}}, \bibinfo {author} {\bibfnamefont
  {V.~F.}\ \bibnamefont {Maisi}},\ and\ \bibinfo {author} {\bibfnamefont
  {J.}~\bibnamefont {Pekola}},\ }\bibfield  {title} {\bibinfo {title} {A
  superconductor free of quasiparticles for seconds},\ }\href@noop {}
  {\bibfield  {journal} {\bibinfo  {journal} {Nature Physics}\ }\textbf
  {\bibinfo {volume} {18}},\ \bibinfo {pages} {145} (\bibinfo {year}
  {2022})}\BibitemShut {NoStop}%
\bibitem [{\citenamefont {Anthony-Petersen}\ \emph {et~al.}(2022)\citenamefont
  {Anthony-Petersen}, \citenamefont {Biekert}, \citenamefont {Bunker},
  \citenamefont {Chang}, \citenamefont {Chang}, \citenamefont {Chaplinsky},
  \citenamefont {Fascione}, \citenamefont {Fink}, \citenamefont
  {Garcia-Sciveres}, \citenamefont {Germond} \emph
  {et~al.}}]{anthony2022stress}%
  \BibitemOpen
  \bibfield  {author} {\bibinfo {author} {\bibfnamefont {R.}~\bibnamefont
  {Anthony-Petersen}}, \bibinfo {author} {\bibfnamefont {A.}~\bibnamefont
  {Biekert}}, \bibinfo {author} {\bibfnamefont {R.}~\bibnamefont {Bunker}},
  \bibinfo {author} {\bibfnamefont {C.~L.}\ \bibnamefont {Chang}}, \bibinfo
  {author} {\bibfnamefont {Y.-Y.}\ \bibnamefont {Chang}}, \bibinfo {author}
  {\bibfnamefont {L.}~\bibnamefont {Chaplinsky}}, \bibinfo {author}
  {\bibfnamefont {E.}~\bibnamefont {Fascione}}, \bibinfo {author}
  {\bibfnamefont {C.~W.}\ \bibnamefont {Fink}}, \bibinfo {author}
  {\bibfnamefont {M.}~\bibnamefont {Garcia-Sciveres}}, \bibinfo {author}
  {\bibfnamefont {R.}~\bibnamefont {Germond}}, \emph {et~al.},\ }\bibfield
  {title} {\bibinfo {title} {A stress induced source of phonon bursts and
  quasiparticle poisoning},\ }\href@noop {} {\bibfield  {journal} {\bibinfo
  {journal} {arXiv preprint arXiv:2208.02790}\ } (\bibinfo {year}
  {2022})}\BibitemShut {NoStop}%
\bibitem [{\citenamefont {Bertoldo}\ \emph {et~al.}(2023)\citenamefont
  {Bertoldo}, \citenamefont {Mart{\'\i}nez}, \citenamefont {Nedyalkov},\ and\
  \citenamefont {Forn-D{\'\i}az}}]{bertoldo2023cosmic}%
  \BibitemOpen
  \bibfield  {author} {\bibinfo {author} {\bibfnamefont {E.}~\bibnamefont
  {Bertoldo}}, \bibinfo {author} {\bibfnamefont {M.}~\bibnamefont
  {Mart{\'\i}nez}}, \bibinfo {author} {\bibfnamefont {B.}~\bibnamefont
  {Nedyalkov}},\ and\ \bibinfo {author} {\bibfnamefont {P.}~\bibnamefont
  {Forn-D{\'\i}az}},\ }\bibfield  {title} {\bibinfo {title} {Cosmic muon flux
  attenuation methods for superconducting qubit experiments},\ }\href@noop {}
  {\bibfield  {journal} {\bibinfo  {journal} {arXiv preprint arXiv:2303.04938}\
  } (\bibinfo {year} {2023})}\BibitemShut {NoStop}%
\bibitem [{\citenamefont {Nsanzineza}\ and\ \citenamefont
  {Plourde}(2014)}]{nsanzineza2014trapping}%
  \BibitemOpen
  \bibfield  {author} {\bibinfo {author} {\bibfnamefont {I.}~\bibnamefont
  {Nsanzineza}}\ and\ \bibinfo {author} {\bibfnamefont {B.}~\bibnamefont
  {Plourde}},\ }\bibfield  {title} {\bibinfo {title} {Trapping a single vortex
  and reducing quasiparticles in a superconducting resonator},\ }\href@noop {}
  {\bibfield  {journal} {\bibinfo  {journal} {Physical review letters}\
  }\textbf {\bibinfo {volume} {113}},\ \bibinfo {pages} {117002} (\bibinfo
  {year} {2014})}\BibitemShut {NoStop}%
\bibitem [{\citenamefont {Wang}\ \emph {et~al.}(2014)\citenamefont {Wang},
  \citenamefont {Gao}, \citenamefont {Pop}, \citenamefont {Vool}, \citenamefont
  {Axline}, \citenamefont {Brecht}, \citenamefont {Heeres}, \citenamefont
  {Frunzio}, \citenamefont {Devoret}, \citenamefont {Catelani} \emph
  {et~al.}}]{wang2014measurement}%
  \BibitemOpen
  \bibfield  {author} {\bibinfo {author} {\bibfnamefont {C.}~\bibnamefont
  {Wang}}, \bibinfo {author} {\bibfnamefont {Y.~Y.}\ \bibnamefont {Gao}},
  \bibinfo {author} {\bibfnamefont {I.~M.}\ \bibnamefont {Pop}}, \bibinfo
  {author} {\bibfnamefont {U.}~\bibnamefont {Vool}}, \bibinfo {author}
  {\bibfnamefont {C.}~\bibnamefont {Axline}}, \bibinfo {author} {\bibfnamefont
  {T.}~\bibnamefont {Brecht}}, \bibinfo {author} {\bibfnamefont {R.~W.}\
  \bibnamefont {Heeres}}, \bibinfo {author} {\bibfnamefont {L.}~\bibnamefont
  {Frunzio}}, \bibinfo {author} {\bibfnamefont {M.~H.}\ \bibnamefont
  {Devoret}}, \bibinfo {author} {\bibfnamefont {G.}~\bibnamefont {Catelani}},
  \emph {et~al.},\ }\bibfield  {title} {\bibinfo {title} {Measurement and
  control of quasiparticle dynamics in a superconducting qubit},\ }\href@noop
  {} {\bibfield  {journal} {\bibinfo  {journal} {Nature communications}\
  }\textbf {\bibinfo {volume} {5}},\ \bibinfo {pages} {1} (\bibinfo {year}
  {2014})}\BibitemShut {NoStop}%
\bibitem [{\citenamefont {Gustavsson}\ \emph {et~al.}(2016)\citenamefont
  {Gustavsson}, \citenamefont {Yan}, \citenamefont {Catelani}, \citenamefont
  {Bylander}, \citenamefont {Kamal}, \citenamefont {Birenbaum}, \citenamefont
  {Hover}, \citenamefont {Rosenberg}, \citenamefont {Samach}, \citenamefont
  {Sears} \emph {et~al.}}]{gustavsson2016suppressing}%
  \BibitemOpen
  \bibfield  {author} {\bibinfo {author} {\bibfnamefont {S.}~\bibnamefont
  {Gustavsson}}, \bibinfo {author} {\bibfnamefont {F.}~\bibnamefont {Yan}},
  \bibinfo {author} {\bibfnamefont {G.}~\bibnamefont {Catelani}}, \bibinfo
  {author} {\bibfnamefont {J.}~\bibnamefont {Bylander}}, \bibinfo {author}
  {\bibfnamefont {A.}~\bibnamefont {Kamal}}, \bibinfo {author} {\bibfnamefont
  {J.}~\bibnamefont {Birenbaum}}, \bibinfo {author} {\bibfnamefont
  {D.}~\bibnamefont {Hover}}, \bibinfo {author} {\bibfnamefont
  {D.}~\bibnamefont {Rosenberg}}, \bibinfo {author} {\bibfnamefont
  {G.}~\bibnamefont {Samach}}, \bibinfo {author} {\bibfnamefont {A.~P.}\
  \bibnamefont {Sears}}, \emph {et~al.},\ }\bibfield  {title} {\bibinfo {title}
  {Suppressing relaxation in superconducting qubits by quasiparticle pumping},\
  }\href@noop {} {\bibfield  {journal} {\bibinfo  {journal} {Science}\ }\textbf
  {\bibinfo {volume} {354}},\ \bibinfo {pages} {1573} (\bibinfo {year}
  {2016})}\BibitemShut {NoStop}%
\bibitem [{\citenamefont {Patel}\ \emph {et~al.}(2017)\citenamefont {Patel},
  \citenamefont {Pechenezhskiy}, \citenamefont {Plourde}, \citenamefont
  {Vavilov},\ and\ \citenamefont {McDermott}}]{patel2017phonon}%
  \BibitemOpen
  \bibfield  {author} {\bibinfo {author} {\bibfnamefont {U.}~\bibnamefont
  {Patel}}, \bibinfo {author} {\bibfnamefont {I.~V.}\ \bibnamefont
  {Pechenezhskiy}}, \bibinfo {author} {\bibfnamefont {B.}~\bibnamefont
  {Plourde}}, \bibinfo {author} {\bibfnamefont {M.}~\bibnamefont {Vavilov}},\
  and\ \bibinfo {author} {\bibfnamefont {R.}~\bibnamefont {McDermott}},\
  }\bibfield  {title} {\bibinfo {title} {Phonon-mediated quasiparticle
  poisoning of superconducting microwave resonators},\ }\href@noop {}
  {\bibfield  {journal} {\bibinfo  {journal} {Physical Review B}\ }\textbf
  {\bibinfo {volume} {96}},\ \bibinfo {pages} {220501} (\bibinfo {year}
  {2017})}\BibitemShut {NoStop}%
\bibitem [{\citenamefont {Henriques}\ \emph {et~al.}(2019)\citenamefont
  {Henriques}, \citenamefont {Valenti}, \citenamefont {Charpentier},
  \citenamefont {Lagoin}, \citenamefont {Gouriou}, \citenamefont
  {Mart{\'\i}nez}, \citenamefont {Cardani}, \citenamefont {Vignati},
  \citenamefont {Gr{\"u}nhaupt}, \citenamefont {Gusenkova} \emph
  {et~al.}}]{henriques2019phonon}%
  \BibitemOpen
  \bibfield  {author} {\bibinfo {author} {\bibfnamefont {F.}~\bibnamefont
  {Henriques}}, \bibinfo {author} {\bibfnamefont {F.}~\bibnamefont {Valenti}},
  \bibinfo {author} {\bibfnamefont {T.}~\bibnamefont {Charpentier}}, \bibinfo
  {author} {\bibfnamefont {M.}~\bibnamefont {Lagoin}}, \bibinfo {author}
  {\bibfnamefont {C.}~\bibnamefont {Gouriou}}, \bibinfo {author} {\bibfnamefont
  {M.}~\bibnamefont {Mart{\'\i}nez}}, \bibinfo {author} {\bibfnamefont
  {L.}~\bibnamefont {Cardani}}, \bibinfo {author} {\bibfnamefont
  {M.}~\bibnamefont {Vignati}}, \bibinfo {author} {\bibfnamefont
  {L.}~\bibnamefont {Gr{\"u}nhaupt}}, \bibinfo {author} {\bibfnamefont
  {D.}~\bibnamefont {Gusenkova}}, \emph {et~al.},\ }\bibfield  {title}
  {\bibinfo {title} {Phonon traps reduce the quasiparticle density in
  superconducting circuits},\ }\href@noop {} {\bibfield  {journal} {\bibinfo
  {journal} {Applied physics letters}\ }\textbf {\bibinfo {volume} {115}},\
  \bibinfo {pages} {212601} (\bibinfo {year} {2019})}\BibitemShut {NoStop}%
\bibitem [{\citenamefont {Martinis}(2021)}]{martinis2021saving}%
  \BibitemOpen
  \bibfield  {author} {\bibinfo {author} {\bibfnamefont {J.~M.}\ \bibnamefont
  {Martinis}},\ }\bibfield  {title} {\bibinfo {title} {Saving superconducting
  quantum processors from decay and correlated errors generated by gamma and
  cosmic rays},\ }\href@noop {} {\bibfield  {journal} {\bibinfo  {journal} {npj
  Quantum Information}\ }\textbf {\bibinfo {volume} {7}},\ \bibinfo {pages} {1}
  (\bibinfo {year} {2021})}\BibitemShut {NoStop}%
\bibitem [{\citenamefont {Marchegiani}\ \emph {et~al.}(2022)\citenamefont
  {Marchegiani}, \citenamefont {Amico},\ and\ \citenamefont
  {Catelani}}]{marchegiani2022quasiparticles}%
  \BibitemOpen
  \bibfield  {author} {\bibinfo {author} {\bibfnamefont {G.}~\bibnamefont
  {Marchegiani}}, \bibinfo {author} {\bibfnamefont {L.}~\bibnamefont {Amico}},\
  and\ \bibinfo {author} {\bibfnamefont {G.}~\bibnamefont {Catelani}},\
  }\bibfield  {title} {\bibinfo {title} {Quasiparticles in superconducting
  qubits with asymmetric junctions},\ }\href
  {https://doi.org/10.1103/PRXQuantum.3.040338} {\bibfield  {journal} {\bibinfo
   {journal} {PRX Quantum}\ }\textbf {\bibinfo {volume} {3}},\ \bibinfo {pages}
  {040338} (\bibinfo {year} {2022})}\BibitemShut {NoStop}%
\bibitem [{\citenamefont {Guillaud}\ and\ \citenamefont
  {Mirrahimi}(2019)}]{guillaud2019repetition}%
  \BibitemOpen
  \bibfield  {author} {\bibinfo {author} {\bibfnamefont {J.}~\bibnamefont
  {Guillaud}}\ and\ \bibinfo {author} {\bibfnamefont {M.}~\bibnamefont
  {Mirrahimi}},\ }\bibfield  {title} {\bibinfo {title} {Repetition cat qubits
  for fault-tolerant quantum computation},\ }\href@noop {} {\bibfield
  {journal} {\bibinfo  {journal} {Physical Review X}\ }\textbf {\bibinfo
  {volume} {9}},\ \bibinfo {pages} {041053} (\bibinfo {year}
  {2019})}\BibitemShut {NoStop}%
\bibitem [{Note8()}]{Note8}%
  \BibitemOpen
  \bibinfo {note} {In fact, although the Hadamard gate provides a first simple
  example of our implementation of Clifford gates, it could be implemented even
  more easily in practice: the required $\pi /2$ rotation in phase-space can be
  performed by simply adjusting the clock of the rotating frame in
  software.}\BibitemShut {Stop}%
\bibitem [{\citenamefont {Fukui}\ \emph {et~al.}(2018)\citenamefont {Fukui},
  \citenamefont {Tomita}, \citenamefont {Okamoto},\ and\ \citenamefont
  {Fujii}}]{fukui2018high}%
  \BibitemOpen
  \bibfield  {author} {\bibinfo {author} {\bibfnamefont {K.}~\bibnamefont
  {Fukui}}, \bibinfo {author} {\bibfnamefont {A.}~\bibnamefont {Tomita}},
  \bibinfo {author} {\bibfnamefont {A.}~\bibnamefont {Okamoto}},\ and\ \bibinfo
  {author} {\bibfnamefont {K.}~\bibnamefont {Fujii}},\ }\bibfield  {title}
  {\bibinfo {title} {High-threshold fault-tolerant quantum computation with
  analog quantum error correction},\ }\href@noop {} {\bibfield  {journal}
  {\bibinfo  {journal} {Physical Review X}\ }\textbf {\bibinfo {volume} {8}},\
  \bibinfo {pages} {021054} (\bibinfo {year} {2018})}\BibitemShut {NoStop}%
\bibitem [{\citenamefont {Vuillot}\ \emph {et~al.}(2019)\citenamefont
  {Vuillot}, \citenamefont {Asasi}, \citenamefont {Wang}, \citenamefont
  {Pryadko},\ and\ \citenamefont {Terhal}}]{vuillot2019quantum}%
  \BibitemOpen
  \bibfield  {author} {\bibinfo {author} {\bibfnamefont {C.}~\bibnamefont
  {Vuillot}}, \bibinfo {author} {\bibfnamefont {H.}~\bibnamefont {Asasi}},
  \bibinfo {author} {\bibfnamefont {Y.}~\bibnamefont {Wang}}, \bibinfo {author}
  {\bibfnamefont {L.~P.}\ \bibnamefont {Pryadko}},\ and\ \bibinfo {author}
  {\bibfnamefont {B.~M.}\ \bibnamefont {Terhal}},\ }\bibfield  {title}
  {\bibinfo {title} {Quantum error correction with the toric
  {Gottesman-Kitaev-Preskill} code},\ }\href@noop {} {\bibfield  {journal}
  {\bibinfo  {journal} {Physical Review A}\ }\textbf {\bibinfo {volume} {99}},\
  \bibinfo {pages} {032344} (\bibinfo {year} {2019})}\BibitemShut {NoStop}%
\bibitem [{\citenamefont {Terhal}\ \emph {et~al.}(2020)\citenamefont {Terhal},
  \citenamefont {Conrad},\ and\ \citenamefont {Vuillot}}]{terhal2020towards}%
  \BibitemOpen
  \bibfield  {author} {\bibinfo {author} {\bibfnamefont {B.~M.}\ \bibnamefont
  {Terhal}}, \bibinfo {author} {\bibfnamefont {J.}~\bibnamefont {Conrad}},\
  and\ \bibinfo {author} {\bibfnamefont {C.}~\bibnamefont {Vuillot}},\
  }\bibfield  {title} {\bibinfo {title} {Towards scalable bosonic quantum error
  correction},\ }\href@noop {} {\bibfield  {journal} {\bibinfo  {journal}
  {Quantum Science and Technology}\ }\textbf {\bibinfo {volume} {5}},\ \bibinfo
  {pages} {043001} (\bibinfo {year} {2020})}\BibitemShut {NoStop}%
\bibitem [{\citenamefont {Noh}\ and\ \citenamefont
  {Chamberland}(2020)}]{noh2020fault}%
  \BibitemOpen
  \bibfield  {author} {\bibinfo {author} {\bibfnamefont {K.}~\bibnamefont
  {Noh}}\ and\ \bibinfo {author} {\bibfnamefont {C.}~\bibnamefont
  {Chamberland}},\ }\bibfield  {title} {\bibinfo {title} {Fault-tolerant
  bosonic quantum error correction with the
  surface--{Gottesman-Kitaev-Preskill} code},\ }\href@noop {} {\bibfield
  {journal} {\bibinfo  {journal} {Physical Review A}\ }\textbf {\bibinfo
  {volume} {101}},\ \bibinfo {pages} {012316} (\bibinfo {year}
  {2020})}\BibitemShut {NoStop}%
\bibitem [{\citenamefont {Noh}\ \emph {et~al.}(2022)\citenamefont {Noh},
  \citenamefont {Chamberland},\ and\ \citenamefont {Brand{\~a}o}}]{noh2022low}%
  \BibitemOpen
  \bibfield  {author} {\bibinfo {author} {\bibfnamefont {K.}~\bibnamefont
  {Noh}}, \bibinfo {author} {\bibfnamefont {C.}~\bibnamefont {Chamberland}},\
  and\ \bibinfo {author} {\bibfnamefont {F.~G.}\ \bibnamefont {Brand{\~a}o}},\
  }\bibfield  {title} {\bibinfo {title} {Low-overhead fault-tolerant quantum
  error correction with the surface-{GKP} code},\ }\href@noop {} {\bibfield
  {journal} {\bibinfo  {journal} {PRX Quantum}\ }\textbf {\bibinfo {volume}
  {3}},\ \bibinfo {pages} {010315} (\bibinfo {year} {2022})}\BibitemShut
  {NoStop}%
\bibitem [{\citenamefont {Nathan}\ \emph {et~al.}(2024)\citenamefont {Nathan},
  \citenamefont {O'Brien}, \citenamefont {Noh}, \citenamefont {Matheny},
  \citenamefont {Grimsmo}, \citenamefont {Jiang},\ and\ \citenamefont
  {Refael}}]{nathan2024selfcorrecting}%
  \BibitemOpen
  \bibfield  {author} {\bibinfo {author} {\bibfnamefont {F.}~\bibnamefont
  {Nathan}}, \bibinfo {author} {\bibfnamefont {L.}~\bibnamefont {O'Brien}},
  \bibinfo {author} {\bibfnamefont {K.}~\bibnamefont {Noh}}, \bibinfo {author}
  {\bibfnamefont {M.~H.}\ \bibnamefont {Matheny}}, \bibinfo {author}
  {\bibfnamefont {A.~L.}\ \bibnamefont {Grimsmo}}, \bibinfo {author}
  {\bibfnamefont {L.}~\bibnamefont {Jiang}},\ and\ \bibinfo {author}
  {\bibfnamefont {G.}~\bibnamefont {Refael}},\ }\href@noop {} {\bibinfo {title}
  {Self-correcting {{GKP}} qubit and gates in a driven-dissipative circuit}}
  (\bibinfo {year} {2024}),\ \Eprint {https://arxiv.org/abs/2405.05671}
  {arXiv:2405.05671} \BibitemShut {NoStop}%
\bibitem [{\citenamefont {Sellem}\ \emph {et~al.}(2023)\citenamefont {Sellem},
  \citenamefont {Robin}, \citenamefont {Campagne-Ibarcq},\ and\ \citenamefont
  {Rouchon}}]{sellem2023ifac}%
  \BibitemOpen
  \bibfield  {author} {\bibinfo {author} {\bibfnamefont {L.-A.}\ \bibnamefont
  {Sellem}}, \bibinfo {author} {\bibfnamefont {R.}~\bibnamefont {Robin}},
  \bibinfo {author} {\bibfnamefont {P.}~\bibnamefont {Campagne-Ibarcq}},\ and\
  \bibinfo {author} {\bibfnamefont {P.}~\bibnamefont {Rouchon}},\ }\bibfield
  {title} {\bibinfo {title} {Stability and decoherence rates of a {{GKP}} qubit
  protected by dissipation},\ }in\ \href
  {https://doi.org/10.1016/j.ifacol.2023.10.1776} {\emph {\bibinfo {booktitle}
  {{{IFAC-PapersOnLine}}}}},\ \bibinfo {series} {22nd {{IFAC World Congress}}},
  Vol.~\bibinfo {volume} {56}\ (\bibinfo {year} {2023})\ pp.\ \bibinfo {pages}
  {1325--1332}\BibitemShut {NoStop}%
\bibitem [{\citenamefont {Zettl}(2005)}]{ZettlBook2005}%
  \BibitemOpen
  \bibfield  {author} {\bibinfo {author} {\bibfnamefont {A.}~\bibnamefont
  {Zettl}},\ }\href@noop {} {\emph {\bibinfo {title} {Sturm Liouville
  Theory}}}\ (\bibinfo  {publisher} {Mathematical Surveys and Monographs, vol.
  121, Amer. Math. Soc.},\ \bibinfo {year} {2005})\BibitemShut {NoStop}%
\bibitem [{\citenamefont {Michel}(2019)}]{michelSmallEigenvalues2019a}%
  \BibitemOpen
  \bibfield  {author} {\bibinfo {author} {\bibfnamefont {L.}~\bibnamefont
  {Michel}},\ }\bibfield  {title} {\bibinfo {title} {About small eigenvalues of
  the {{Witten Laplacian}}},\ }\href {https://doi.org/10.2140/paa.2019.1.149}
  {\bibfield  {journal} {\bibinfo  {journal} {Pure and Applied Analysis}\
  }\textbf {\bibinfo {volume} {1}},\ \bibinfo {pages} {149} (\bibinfo {year}
  {2019})}\BibitemShut {NoStop}%
\bibitem [{Note9()}]{Note9}%
  \BibitemOpen
  \bibinfo {note} {We emphasize that, crucially, the Fourier representation of
  $\protect \mathcal L_\sigma $ is sparse, allowing us to numerically
  diagonalize the corresponding matrix of size $(2K_{\protect \mathrm
  {max}}+1)^2$.}\BibitemShut {Stop}%
\bibitem [{\citenamefont {Rymarz}\ \emph
  {et~al.}(2021{\natexlab{b}})\citenamefont {Rymarz}, \citenamefont {Bosco},
  \citenamefont {Ciani},\ and\ \citenamefont
  {DiVincenzo}}]{rymarzHardwareEncodingGrid2021}%
  \BibitemOpen
  \bibfield  {author} {\bibinfo {author} {\bibfnamefont {M.}~\bibnamefont
  {Rymarz}}, \bibinfo {author} {\bibfnamefont {S.}~\bibnamefont {Bosco}},
  \bibinfo {author} {\bibfnamefont {A.}~\bibnamefont {Ciani}},\ and\ \bibinfo
  {author} {\bibfnamefont {D.~P.}\ \bibnamefont {DiVincenzo}},\ }\bibfield
  {title} {\bibinfo {title} {Hardware-{{Encoding Grid States}} in a
  {{Nonreciprocal Superconducting Circuit}}},\ }\href
  {https://doi.org/10.1103/PhysRevX.11.011032} {\bibfield  {journal} {\bibinfo
  {journal} {Physical Review X}\ }\textbf {\bibinfo {volume} {11}},\ \bibinfo
  {pages} {011032} (\bibinfo {year} {2021}{\natexlab{b}})}\BibitemShut
  {NoStop}%
\bibitem [{\citenamefont {Mirrahimi}\ and\ \citenamefont
  {Rouchon}(2015)}]{lecturenotesmirrahimirouchon}%
  \BibitemOpen
  \bibfield  {author} {\bibinfo {author} {\bibfnamefont {M.}~\bibnamefont
  {Mirrahimi}}\ and\ \bibinfo {author} {\bibfnamefont {P.}~\bibnamefont
  {Rouchon}},\ }\href
  {https://cas.mines-paristech.fr/~rouchon/MasterUPMC/LectureNotes-03-18.pdf}
  {\bibinfo {title} {Dynamics and control of open quantum systems}},\ \bibinfo
  {howpublished} {Lecture Notes} (\bibinfo {year} {2015})\BibitemShut {NoStop}%
\bibitem [{\citenamefont {Azouit}\ \emph {et~al.}(2017)\citenamefont {Azouit},
  \citenamefont {Chittaro}, \citenamefont {Sarlette},\ and\ \citenamefont
  {Rouchon}}]{azouitGenericAdiabatic2017}%
  \BibitemOpen
  \bibfield  {author} {\bibinfo {author} {\bibfnamefont {R.}~\bibnamefont
  {Azouit}}, \bibinfo {author} {\bibfnamefont {F.}~\bibnamefont {Chittaro}},
  \bibinfo {author} {\bibfnamefont {A.}~\bibnamefont {Sarlette}},\ and\
  \bibinfo {author} {\bibfnamefont {P.}~\bibnamefont {Rouchon}},\ }\bibfield
  {title} {\bibinfo {title} {Towards generic adiabatic elimination for
  bipartite open quantum systems},\ }\href
  {https://doi.org/10.1088/2058-9565/aa7f3f} {\bibfield  {journal} {\bibinfo
  {journal} {Quantum Science and Technology}\ }\textbf {\bibinfo {volume}
  {2}},\ \bibinfo {pages} {044011} (\bibinfo {year} {2017})}\BibitemShut
  {NoStop}%
\bibitem [{\citenamefont {Azouit}(2017)}]{azouitAdiabaticElimination2017}%
  \BibitemOpen
  \bibfield  {author} {\bibinfo {author} {\bibfnamefont {R.}~\bibnamefont
  {Azouit}},\ }\emph {\bibinfo {title} {Adiabatic Elimination for Open Quantum
  Systems}},\ \href {https://pastel.archives-ouvertes.fr/tel-01743808} {Ph.D.
  thesis} (\bibinfo {year} {2017})\BibitemShut {NoStop}%
\bibitem [{\citenamefont {Grifoni}\ and\ \citenamefont
  {Hänggi}()}]{grifoniDrivenQuantumTunneling1998}%
  \BibitemOpen
  \bibfield  {author} {\bibinfo {author} {\bibfnamefont {M.}~\bibnamefont
  {Grifoni}}\ and\ \bibinfo {author} {\bibfnamefont {P.}~\bibnamefont
  {Hänggi}},\ }\bibfield  {title} {\bibinfo {title} {Driven quantum
  tunneling},\ }\href {https://doi.org/10.1016/S0370-1573(98)00022-2}
  {\bibfield  {journal} {\bibinfo  {journal} {Physics Reports}\ }\textbf
  {\bibinfo {volume} {304}},\ \bibinfo {pages} {229}}\BibitemShut {NoStop}%
\bibitem [{\citenamefont {Verney}\ \emph {et~al.}()\citenamefont {Verney},
  \citenamefont {Lescanne}, \citenamefont {Devoret}, \citenamefont {Leghtas},\
  and\ \citenamefont {Mirrahimi}}]{verneyStructuralInstabilityDriven2019}%
  \BibitemOpen
  \bibfield  {author} {\bibinfo {author} {\bibfnamefont {L.}~\bibnamefont
  {Verney}}, \bibinfo {author} {\bibfnamefont {R.}~\bibnamefont {Lescanne}},
  \bibinfo {author} {\bibfnamefont {M.~H.}\ \bibnamefont {Devoret}}, \bibinfo
  {author} {\bibfnamefont {Z.}~\bibnamefont {Leghtas}},\ and\ \bibinfo {author}
  {\bibfnamefont {M.}~\bibnamefont {Mirrahimi}},\ }\bibfield  {title} {\bibinfo
  {title} {Structural instability of driven {{Josephson}} circuits prevented by
  an inductive shunt},\ }\href
  {https://doi.org/10.1103/PhysRevApplied.11.024003} {\bibfield  {journal}
  {\bibinfo  {journal} {Physical Review Applied}\ }\textbf {\bibinfo {volume}
  {11}},\ \bibinfo {pages} {024003}},\ \Eprint
  {https://arxiv.org/abs/1805.07542} {1805.07542} \BibitemShut {NoStop}%
\bibitem [{\citenamefont {Cohen}\ \emph {et~al.}()\citenamefont {Cohen},
  \citenamefont {Petrescu}, \citenamefont {Shillito},\ and\ \citenamefont
  {Blais}}]{cohenReminiscenceClassicalChaos2023}%
  \BibitemOpen
  \bibfield  {author} {\bibinfo {author} {\bibfnamefont {J.}~\bibnamefont
  {Cohen}}, \bibinfo {author} {\bibfnamefont {A.}~\bibnamefont {Petrescu}},
  \bibinfo {author} {\bibfnamefont {R.}~\bibnamefont {Shillito}},\ and\
  \bibinfo {author} {\bibfnamefont {A.}~\bibnamefont {Blais}},\ }\bibfield
  {title} {\bibinfo {title} {Reminiscence of {{Classical Chaos}} in {{Driven
  Transmons}}},\ }\href {https://doi.org/10.1103/PRXQuantum.4.020312}
  {\bibfield  {journal} {\bibinfo  {journal} {PRX Quantum}\ }\textbf {\bibinfo
  {volume} {4}},\ \bibinfo {pages} {020312}}\BibitemShut {NoStop}%
\bibitem [{\citenamefont {Burgelman}\ \emph {et~al.}()\citenamefont
  {Burgelman}, \citenamefont {Rouchon}, \citenamefont {Sarlette},\ and\
  \citenamefont {Mirrahimi}}]{burgelmanStructurallyStableSubharmonic2022}%
  \BibitemOpen
  \bibfield  {author} {\bibinfo {author} {\bibfnamefont {M.}~\bibnamefont
  {Burgelman}}, \bibinfo {author} {\bibfnamefont {P.}~\bibnamefont {Rouchon}},
  \bibinfo {author} {\bibfnamefont {A.}~\bibnamefont {Sarlette}},\ and\
  \bibinfo {author} {\bibfnamefont {M.}~\bibnamefont {Mirrahimi}},\ }\bibfield
  {title} {\bibinfo {title} {Structurally {{Stable Subharmonic Regime}} of a
  {{Driven Quantum Josephson Circuit}}},\ }\href
  {https://doi.org/10.1103/PhysRevApplied.18.064044} {\bibfield  {journal}
  {\bibinfo  {journal} {Physical Review Applied}\ }\textbf {\bibinfo {volume}
  {18}},\ \bibinfo {pages} {064044}}\BibitemShut {NoStop}%
\bibitem [{\citenamefont {Petrescu}\ \emph {et~al.}()\citenamefont {Petrescu},
  \citenamefont {Le~Calonnec}, \citenamefont {Leroux}, \citenamefont
  {Di~Paolo}, \citenamefont {Mundada}, \citenamefont {Sussman}, \citenamefont
  {Vrajitoarea}, \citenamefont {Houck},\ and\ \citenamefont
  {Blais}}]{petrescuAccurateMethodsAnalysis2023}%
  \BibitemOpen
  \bibfield  {author} {\bibinfo {author} {\bibfnamefont {A.}~\bibnamefont
  {Petrescu}}, \bibinfo {author} {\bibfnamefont {C.}~\bibnamefont
  {Le~Calonnec}}, \bibinfo {author} {\bibfnamefont {C.}~\bibnamefont {Leroux}},
  \bibinfo {author} {\bibfnamefont {A.}~\bibnamefont {Di~Paolo}}, \bibinfo
  {author} {\bibfnamefont {P.}~\bibnamefont {Mundada}}, \bibinfo {author}
  {\bibfnamefont {S.}~\bibnamefont {Sussman}}, \bibinfo {author} {\bibfnamefont
  {A.}~\bibnamefont {Vrajitoarea}}, \bibinfo {author} {\bibfnamefont {A.~A.}\
  \bibnamefont {Houck}},\ and\ \bibinfo {author} {\bibfnamefont
  {A.}~\bibnamefont {Blais}},\ }\bibfield  {title} {\bibinfo {title} {Accurate
  {{Methods}} for the {{Analysis}} of {{Strong-Drive Effects}} in {{Parametric
  Gates}}},\ }\href {https://doi.org/10.1103/PhysRevApplied.19.044003}
  {\bibfield  {journal} {\bibinfo  {journal} {Physical Review Applied}\
  }\textbf {\bibinfo {volume} {19}},\ \bibinfo {pages} {044003}}\BibitemShut
  {NoStop}%
\bibitem [{\citenamefont {Venkatraman}\ \emph {et~al.}()\citenamefont
  {Venkatraman}, \citenamefont {Xiao}, \citenamefont {Cortiñas}, \citenamefont
  {Eickbusch},\ and\ \citenamefont
  {Devoret}}]{venkatramanStaticEffectiveHamiltonian2022}%
  \BibitemOpen
  \bibfield  {author} {\bibinfo {author} {\bibfnamefont {J.}~\bibnamefont
  {Venkatraman}}, \bibinfo {author} {\bibfnamefont {X.}~\bibnamefont {Xiao}},
  \bibinfo {author} {\bibfnamefont {R.~G.}\ \bibnamefont {Cortiñas}}, \bibinfo
  {author} {\bibfnamefont {A.}~\bibnamefont {Eickbusch}},\ and\ \bibinfo
  {author} {\bibfnamefont {M.~H.}\ \bibnamefont {Devoret}},\ }\bibfield
  {title} {\bibinfo {title} {Static {{Effective Hamiltonian}} of a {{Rapidly
  Driven Nonlinear System}}},\ }\href
  {https://doi.org/10.1103/PhysRevLett.129.100601} {\bibfield  {journal}
  {\bibinfo  {journal} {Physical Review Letters}\ }\textbf {\bibinfo {volume}
  {129}},\ \bibinfo {pages} {100601}}\BibitemShut {NoStop}%
\bibitem [{\citenamefont {Xiao}\ \emph {et~al.}()\citenamefont {Xiao},
  \citenamefont {Venkatraman}, \citenamefont {Cortiñas}, \citenamefont
  {Chowdhury},\ and\ \citenamefont
  {Devoret}}]{xiaoDiagrammaticMethodCompute2023}%
  \BibitemOpen
  \bibfield  {author} {\bibinfo {author} {\bibfnamefont {X.}~\bibnamefont
  {Xiao}}, \bibinfo {author} {\bibfnamefont {J.}~\bibnamefont {Venkatraman}},
  \bibinfo {author} {\bibfnamefont {R.~G.}\ \bibnamefont {Cortiñas}}, \bibinfo
  {author} {\bibfnamefont {S.}~\bibnamefont {Chowdhury}},\ and\ \bibinfo
  {author} {\bibfnamefont {M.~H.}\ \bibnamefont {Devoret}},\ }\href
  {https://doi.org/10.48550/arXiv.2304.13656} {\bibinfo {title} {A diagrammatic
  method to compute the effective {{Hamiltonian}} of driven nonlinear
  oscillators}},\ \Eprint {https://arxiv.org/abs/2304.13656} {2304.13656}
  \BibitemShut {NoStop}%
\bibitem [{\citenamefont {Sarlette}\ \emph {et~al.}()\citenamefont {Sarlette},
  \citenamefont {Rouchon}, \citenamefont {Essig}, \citenamefont {Ficheux},\
  and\ \citenamefont {Huard}}]{sarletteQuantumAdiabaticElimination2020}%
  \BibitemOpen
  \bibfield  {author} {\bibinfo {author} {\bibfnamefont {A.}~\bibnamefont
  {Sarlette}}, \bibinfo {author} {\bibfnamefont {P.}~\bibnamefont {Rouchon}},
  \bibinfo {author} {\bibfnamefont {A.}~\bibnamefont {Essig}}, \bibinfo
  {author} {\bibfnamefont {Q.}~\bibnamefont {Ficheux}},\ and\ \bibinfo {author}
  {\bibfnamefont {B.}~\bibnamefont {Huard}},\ }\bibfield  {title} {\bibinfo
  {title} {Quantum adiabatic elimination at arbitrary order for photon number
  measurement},\ }\href {https://doi.org/10.1016/j.ifacol.2020.12.131}
  {\bibfield  {journal} {\bibinfo  {journal} {IFAC-PapersOnLine}\ }\bibinfo
  {series} {21st {{IFAC World Congress}}},\ \textbf {\bibinfo {volume} {53}},\
  \bibinfo {pages} {250}},\ \Eprint {https://arxiv.org/abs/2001.02550}
  {2001.02550} \BibitemShut {NoStop}%
\bibitem [{\citenamefont {Tokieda}\ \emph {et~al.}()\citenamefont {Tokieda},
  \citenamefont {Elouard}, \citenamefont {Sarlette},\ and\ \citenamefont
  {Rouchon}}]{tokiedaCompletePositivityViolation2023}%
  \BibitemOpen
  \bibfield  {author} {\bibinfo {author} {\bibfnamefont {M.}~\bibnamefont
  {Tokieda}}, \bibinfo {author} {\bibfnamefont {C.}~\bibnamefont {Elouard}},
  \bibinfo {author} {\bibfnamefont {A.}~\bibnamefont {Sarlette}},\ and\
  \bibinfo {author} {\bibfnamefont {P.}~\bibnamefont {Rouchon}},\ }\bibfield
  {title} {\bibinfo {title} {Complete {{Positivity Violation}} in
  {{Higher-order Quantum Adiabatic Elimination}}},\ }\href
  {https://doi.org/10.1016/j.ifacol.2023.10.1779} {\bibfield  {journal}
  {\bibinfo  {journal} {IFAC-PapersOnLine}\ }\bibinfo {series} {22nd {{IFAC
  World Congress}}},\ \textbf {\bibinfo {volume} {56}},\ \bibinfo {pages}
  {1333}}\BibitemShut {NoStop}%
\bibitem [{Note10()}]{Note10}%
  \BibitemOpen
  \bibinfo {note} {Note that $\Sha _T^{(N)}$ has $2N+1$ non-zero exponential
  Fourier coefficients, or $N+1$ trigonometric Fourier coefficients as $\Sha
  _T^{(N)}(t) := \protect \frac 1T + \protect \frac 2T \DOTSB \sum@ \slimits@
  _{k=1}^N \cos (k\omega t)$. To study the effect of the control bandwidth, our
  convention is to simply use N, giving the maximum frequency, as the figure of
  merit.}\BibitemShut {Stop}%
\bibitem [{\citenamefont {Gr{\"u}nhaupt}\ \emph {et~al.}(2019)\citenamefont
  {Gr{\"u}nhaupt}, \citenamefont {Spiecker}, \citenamefont {Gusenkova},
  \citenamefont {Maleeva}, \citenamefont {Skacel}, \citenamefont {Takmakov},
  \citenamefont {Valenti}, \citenamefont {Winkel}, \citenamefont {Rotzinger},
  \citenamefont {Wernsdorfer} \emph {et~al.}}]{grunhaupt2019granular}%
  \BibitemOpen
  \bibfield  {author} {\bibinfo {author} {\bibfnamefont {L.}~\bibnamefont
  {Gr{\"u}nhaupt}}, \bibinfo {author} {\bibfnamefont {M.}~\bibnamefont
  {Spiecker}}, \bibinfo {author} {\bibfnamefont {D.}~\bibnamefont {Gusenkova}},
  \bibinfo {author} {\bibfnamefont {N.}~\bibnamefont {Maleeva}}, \bibinfo
  {author} {\bibfnamefont {S.~T.}\ \bibnamefont {Skacel}}, \bibinfo {author}
  {\bibfnamefont {I.}~\bibnamefont {Takmakov}}, \bibinfo {author}
  {\bibfnamefont {F.}~\bibnamefont {Valenti}}, \bibinfo {author} {\bibfnamefont
  {P.}~\bibnamefont {Winkel}}, \bibinfo {author} {\bibfnamefont
  {H.}~\bibnamefont {Rotzinger}}, \bibinfo {author} {\bibfnamefont
  {W.}~\bibnamefont {Wernsdorfer}}, \emph {et~al.},\ }\bibfield  {title}
  {\bibinfo {title} {Granular aluminium as a superconducting material for
  high-impedance quantum circuits},\ }\href@noop {} {\bibfield  {journal}
  {\bibinfo  {journal} {Nature materials}\ }\textbf {\bibinfo {volume} {18}},\
  \bibinfo {pages} {816} (\bibinfo {year} {2019})}\BibitemShut {NoStop}%
\bibitem [{\citenamefont {Peruzzo}\ \emph {et~al.}(2020)\citenamefont
  {Peruzzo}, \citenamefont {Trioni}, \citenamefont {Hassani}, \citenamefont
  {Zemlicka},\ and\ \citenamefont {Fink}}]{peruzzo2020surpassing}%
  \BibitemOpen
  \bibfield  {author} {\bibinfo {author} {\bibfnamefont {M.}~\bibnamefont
  {Peruzzo}}, \bibinfo {author} {\bibfnamefont {A.}~\bibnamefont {Trioni}},
  \bibinfo {author} {\bibfnamefont {F.}~\bibnamefont {Hassani}}, \bibinfo
  {author} {\bibfnamefont {M.}~\bibnamefont {Zemlicka}},\ and\ \bibinfo
  {author} {\bibfnamefont {J.~M.}\ \bibnamefont {Fink}},\ }\bibfield  {title}
  {\bibinfo {title} {Surpassing the resistance quantum with a geometric
  superinductor},\ }\href@noop {} {\bibfield  {journal} {\bibinfo  {journal}
  {Physical Review Applied}\ }\textbf {\bibinfo {volume} {14}},\ \bibinfo
  {pages} {044055} (\bibinfo {year} {2020})}\BibitemShut {NoStop}%
\bibitem [{\citenamefont {Viola}\ and\ \citenamefont
  {Catelani}(2015)}]{viola2015collective}%
  \BibitemOpen
  \bibfield  {author} {\bibinfo {author} {\bibfnamefont {G.}~\bibnamefont
  {Viola}}\ and\ \bibinfo {author} {\bibfnamefont {G.}~\bibnamefont
  {Catelani}},\ }\bibfield  {title} {\bibinfo {title} {Collective modes in the
  fluxonium qubit},\ }\href@noop {} {\bibfield  {journal} {\bibinfo  {journal}
  {Physical Review B}\ }\textbf {\bibinfo {volume} {92}},\ \bibinfo {pages}
  {224511} (\bibinfo {year} {2015})}\BibitemShut {NoStop}%
\bibitem [{\citenamefont {Minev}\ \emph {et~al.}(2021)\citenamefont {Minev},
  \citenamefont {Leghtas}, \citenamefont {Mundhada}, \citenamefont
  {Christakis}, \citenamefont {Pop},\ and\ \citenamefont
  {Devoret}}]{minev2021energy}%
  \BibitemOpen
  \bibfield  {author} {\bibinfo {author} {\bibfnamefont {Z.~K.}\ \bibnamefont
  {Minev}}, \bibinfo {author} {\bibfnamefont {Z.}~\bibnamefont {Leghtas}},
  \bibinfo {author} {\bibfnamefont {S.~O.}\ \bibnamefont {Mundhada}}, \bibinfo
  {author} {\bibfnamefont {L.}~\bibnamefont {Christakis}}, \bibinfo {author}
  {\bibfnamefont {I.~M.}\ \bibnamefont {Pop}},\ and\ \bibinfo {author}
  {\bibfnamefont {M.~H.}\ \bibnamefont {Devoret}},\ }\bibfield  {title}
  {\bibinfo {title} {Energy-participation quantization of {Josephson}
  circuits},\ }\href@noop {} {\bibfield  {journal} {\bibinfo  {journal} {npj
  Quantum Information}\ }\textbf {\bibinfo {volume} {7}},\ \bibinfo {pages}
  {131} (\bibinfo {year} {2021})}\BibitemShut {NoStop}%
\bibitem [{\citenamefont {Manucharyan}(2012)}]{manucharyan2012superinductance}%
  \BibitemOpen
  \bibfield  {author} {\bibinfo {author} {\bibfnamefont {V.}~\bibnamefont
  {Manucharyan}},\ }\emph {\bibinfo {title} {Superinductance}},\ \href@noop {}
  {Ph.D. thesis} (\bibinfo {year} {2012})\BibitemShut {NoStop}%
\bibitem [{\citenamefont {Maleeva}\ \emph {et~al.}(2018)\citenamefont
  {Maleeva}, \citenamefont {Gr{\"u}nhaupt}, \citenamefont {Klein},
  \citenamefont {Levy-Bertrand}, \citenamefont {Dupre}, \citenamefont {Calvo},
  \citenamefont {Valenti}, \citenamefont {Winkel}, \citenamefont {Friedrich},
  \citenamefont {Wernsdorfer} \emph {et~al.}}]{maleeva2018circuit}%
  \BibitemOpen
  \bibfield  {author} {\bibinfo {author} {\bibfnamefont {N.}~\bibnamefont
  {Maleeva}}, \bibinfo {author} {\bibfnamefont {L.}~\bibnamefont
  {Gr{\"u}nhaupt}}, \bibinfo {author} {\bibfnamefont {T.}~\bibnamefont
  {Klein}}, \bibinfo {author} {\bibfnamefont {F.}~\bibnamefont
  {Levy-Bertrand}}, \bibinfo {author} {\bibfnamefont {O.}~\bibnamefont
  {Dupre}}, \bibinfo {author} {\bibfnamefont {M.}~\bibnamefont {Calvo}},
  \bibinfo {author} {\bibfnamefont {F.}~\bibnamefont {Valenti}}, \bibinfo
  {author} {\bibfnamefont {P.}~\bibnamefont {Winkel}}, \bibinfo {author}
  {\bibfnamefont {F.}~\bibnamefont {Friedrich}}, \bibinfo {author}
  {\bibfnamefont {W.}~\bibnamefont {Wernsdorfer}}, \emph {et~al.},\ }\bibfield
  {title} {\bibinfo {title} {Circuit quantum electrodynamics of granular
  aluminum resonators},\ }\href@noop {} {\bibfield  {journal} {\bibinfo
  {journal} {Nature communications}\ }\textbf {\bibinfo {volume} {9}},\
  \bibinfo {pages} {3889} (\bibinfo {year} {2018})}\BibitemShut {NoStop}%
\bibitem [{\citenamefont {Kamenov}\ \emph {et~al.}(2020)\citenamefont
  {Kamenov}, \citenamefont {Lu}, \citenamefont {Kalashnikov}, \citenamefont
  {DiNapoli}, \citenamefont {Bell},\ and\ \citenamefont
  {Gershenson}}]{kamenov2020granular}%
  \BibitemOpen
  \bibfield  {author} {\bibinfo {author} {\bibfnamefont {P.}~\bibnamefont
  {Kamenov}}, \bibinfo {author} {\bibfnamefont {W.-S.}\ \bibnamefont {Lu}},
  \bibinfo {author} {\bibfnamefont {K.}~\bibnamefont {Kalashnikov}}, \bibinfo
  {author} {\bibfnamefont {T.}~\bibnamefont {DiNapoli}}, \bibinfo {author}
  {\bibfnamefont {M.~T.}\ \bibnamefont {Bell}},\ and\ \bibinfo {author}
  {\bibfnamefont {M.~E.}\ \bibnamefont {Gershenson}},\ }\bibfield  {title}
  {\bibinfo {title} {Granular aluminum meandered superinductors for quantum
  circuits},\ }\href@noop {} {\bibfield  {journal} {\bibinfo  {journal}
  {Physical Review Applied}\ }\textbf {\bibinfo {volume} {13}},\ \bibinfo
  {pages} {054051} (\bibinfo {year} {2020})}\BibitemShut {NoStop}%
\bibitem [{\citenamefont {Johansson}\ \emph {et~al.}(2012)\citenamefont
  {Johansson}, \citenamefont {Nation},\ and\ \citenamefont
  {Nori}}]{johansson2012qutip}%
  \BibitemOpen
  \bibfield  {author} {\bibinfo {author} {\bibfnamefont {J.~R.}\ \bibnamefont
  {Johansson}}, \bibinfo {author} {\bibfnamefont {P.~D.}\ \bibnamefont
  {Nation}},\ and\ \bibinfo {author} {\bibfnamefont {F.}~\bibnamefont {Nori}},\
  }\bibfield  {title} {\bibinfo {title} {{{QuTiP}}: {{An}} open-source
  {{Python}} framework for the dynamics of open quantum systems},\ }\href
  {https://doi.org/10.1016/j.cpc.2012.02.021} {\bibfield  {journal} {\bibinfo
  {journal} {Comp. Phys. Com.}\ }\textbf {\bibinfo {volume} {183}},\ \bibinfo
  {pages} {1760} (\bibinfo {year} {2012})}\BibitemShut {NoStop}%
\bibitem [{\citenamefont {Johansson}\ \emph {et~al.}(2013)\citenamefont
  {Johansson}, \citenamefont {Nation},\ and\ \citenamefont
  {Nori}}]{johansson2013qutip}%
  \BibitemOpen
  \bibfield  {author} {\bibinfo {author} {\bibfnamefont {J.~R.}\ \bibnamefont
  {Johansson}}, \bibinfo {author} {\bibfnamefont {P.~D.}\ \bibnamefont
  {Nation}},\ and\ \bibinfo {author} {\bibfnamefont {F.}~\bibnamefont {Nori}},\
  }\bibfield  {title} {\bibinfo {title} {{{QuTiP}} 2: {{A Python}} framework
  for the dynamics of open quantum systems},\ }\href
  {https://doi.org/10.1016/j.cpc.2012.11.019} {\bibfield  {journal} {\bibinfo
  {journal} {Comp. Phys. Com.}\ }\textbf {\bibinfo {volume} {184}},\ \bibinfo
  {pages} {1234} (\bibinfo {year} {2013})}\BibitemShut {NoStop}%
\bibitem [{\citenamefont {Krämer}\ \emph {et~al.}(2018)\citenamefont
  {Krämer}, \citenamefont {Plankensteiner}, \citenamefont {Ostermann},\ and\
  \citenamefont {Ritsch}}]{quantumopticsjl2018}%
  \BibitemOpen
  \bibfield  {author} {\bibinfo {author} {\bibfnamefont {S.}~\bibnamefont
  {Krämer}}, \bibinfo {author} {\bibfnamefont {D.}~\bibnamefont
  {Plankensteiner}}, \bibinfo {author} {\bibfnamefont {L.}~\bibnamefont
  {Ostermann}},\ and\ \bibinfo {author} {\bibfnamefont {H.}~\bibnamefont
  {Ritsch}},\ }\bibfield  {title} {\bibinfo {title} {{{QuantumOptics.jl}}: A
  {{Julia}} framework for simulating open quantum systems},\ }\href
  {https://doi.org/https://doi.org/10.1016/j.cpc.2018.02.004} {\bibfield
  {journal} {\bibinfo  {journal} {Comp. Phys. Com.}\ }\textbf {\bibinfo
  {volume} {227}},\ \bibinfo {pages} {109} (\bibinfo {year}
  {2018})}\BibitemShut {NoStop}%
\bibitem [{\citenamefont {Egger}\ \emph {et~al.}(2021)\citenamefont {Egger},
  \citenamefont {Landa}, \citenamefont {Parr}, \citenamefont {Puzzuoli},
  \citenamefont {Rosand}, \citenamefont {Rupesh}, \citenamefont {Treinish},\
  and\ \citenamefont {Wood}}]{qiskitdynamics2021}%
  \BibitemOpen
  \bibfield  {author} {\bibinfo {author} {\bibfnamefont {D.~J.}\ \bibnamefont
  {Egger}}, \bibinfo {author} {\bibfnamefont {H.}~\bibnamefont {Landa}},
  \bibinfo {author} {\bibfnamefont {A.}~\bibnamefont {Parr}}, \bibinfo {author}
  {\bibfnamefont {D.}~\bibnamefont {Puzzuoli}}, \bibinfo {author}
  {\bibfnamefont {B.}~\bibnamefont {Rosand}}, \bibinfo {author} {\bibfnamefont
  {R.~K.}\ \bibnamefont {Rupesh}}, \bibinfo {author} {\bibfnamefont
  {M.}~\bibnamefont {Treinish}},\ and\ \bibinfo {author} {\bibfnamefont
  {C.~J.}\ \bibnamefont {Wood}},\ }\href
  {https://github.com/Qiskit/qiskit-dynamics} {\bibinfo {title} {Qiskit
  dynamics}} (\bibinfo {year} {2021})\BibitemShut {NoStop}%
\bibitem [{\citenamefont {Riesch}\ and\ \citenamefont
  {Jirauschek}(2019)}]{RIESCH2019290}%
  \BibitemOpen
  \bibfield  {author} {\bibinfo {author} {\bibfnamefont {M.}~\bibnamefont
  {Riesch}}\ and\ \bibinfo {author} {\bibfnamefont {C.}~\bibnamefont
  {Jirauschek}},\ }\bibfield  {title} {\bibinfo {title} {Analyzing the
  positivity preservation of numerical methods for the {Liouville-Von Neumann}
  equation},\ }\href
  {https://doi.org/https://doi.org/10.1016/j.jcp.2019.04.006} {\bibfield
  {journal} {\bibinfo  {journal} {Journal of Computational Physics}\ }\textbf
  {\bibinfo {volume} {390}},\ \bibinfo {pages} {290} (\bibinfo {year}
  {2019})}\BibitemShut {NoStop}%
\bibitem [{\citenamefont {Steinbach}\ \emph {et~al.}(1995)\citenamefont
  {Steinbach}, \citenamefont {Garraway},\ and\ \citenamefont
  {Knight}}]{steinbachgarrawayknight1995}%
  \BibitemOpen
  \bibfield  {author} {\bibinfo {author} {\bibfnamefont {J.}~\bibnamefont
  {Steinbach}}, \bibinfo {author} {\bibfnamefont {B.~M.}\ \bibnamefont
  {Garraway}},\ and\ \bibinfo {author} {\bibfnamefont {P.~L.}\ \bibnamefont
  {Knight}},\ }\bibfield  {title} {\bibinfo {title} {High-order unraveling of
  master equations for dissipative evolution},\ }\href
  {https://doi.org/10.1103/PhysRevA.51.3302} {\bibfield  {journal} {\bibinfo
  {journal} {Phys. Rev. A}\ }\textbf {\bibinfo {volume} {51}},\ \bibinfo
  {pages} {3302} (\bibinfo {year} {1995})}\BibitemShut {NoStop}%
\bibitem [{\citenamefont {Rouchon}\ and\ \citenamefont
  {Ralph}(2015)}]{rouchonralph2015}%
  \BibitemOpen
  \bibfield  {author} {\bibinfo {author} {\bibfnamefont {P.}~\bibnamefont
  {Rouchon}}\ and\ \bibinfo {author} {\bibfnamefont {J.~F.}\ \bibnamefont
  {Ralph}},\ }\bibfield  {title} {\bibinfo {title} {Efficient quantum filtering
  for quantum feedback control},\ }\href
  {https://doi.org/10.1103/PhysRevA.91.012118} {\bibfield  {journal} {\bibinfo
  {journal} {Phys. Rev. A}\ }\textbf {\bibinfo {volume} {91}},\ \bibinfo
  {pages} {012118} (\bibinfo {year} {2015})}\BibitemShut {NoStop}%
\bibitem [{\citenamefont {Cao}\ and\ \citenamefont {Lu}(2021)}]{lucao2021}%
  \BibitemOpen
  \bibfield  {author} {\bibinfo {author} {\bibfnamefont {Y.}~\bibnamefont
  {Cao}}\ and\ \bibinfo {author} {\bibfnamefont {J.}~\bibnamefont {Lu}},\
  }\href {https://doi.org/10.48550/ARXIV.2103.01194} {\bibinfo {title}
  {Structure-preserving numerical schemes for lindblad equations}} (\bibinfo
  {year} {2021}),\ \Eprint {https://arxiv.org/abs/2103.01194} {arXiv:2103.01194
  [math.NA]} \BibitemShut {NoStop}%
\bibitem [{\citenamefont {Jordan}\ \emph {et~al.}(2016)\citenamefont {Jordan},
  \citenamefont {Chantasri}, \citenamefont {Rouchon},\ and\ \citenamefont
  {Huard}}]{jordanAnatomyFluorescence2016}%
  \BibitemOpen
  \bibfield  {author} {\bibinfo {author} {\bibfnamefont {A.~N.}\ \bibnamefont
  {Jordan}}, \bibinfo {author} {\bibfnamefont {A.}~\bibnamefont {Chantasri}},
  \bibinfo {author} {\bibfnamefont {P.}~\bibnamefont {Rouchon}},\ and\ \bibinfo
  {author} {\bibfnamefont {B.}~\bibnamefont {Huard}},\ }\bibfield  {title}
  {\bibinfo {title} {Anatomy of {{Fluorescence}}: {{Quantum}} trajectory
  statistics from continuously measuring spontaneous emission},\ }\href
  {https://doi.org/10.1007/s40509-016-0075-9} {\bibfield  {journal} {\bibinfo
  {journal} {Quantum Stud.: Math. Found.}\ }\textbf {\bibinfo {volume} {3}},\
  \bibinfo {pages} {237} (\bibinfo {year} {2016})}\BibitemShut {NoStop}%
\bibitem [{Note11()}]{Note11}%
  \BibitemOpen
  \bibinfo {note} {Parameter sweeps where performed on a cluster to parallelize
  the simulations, but each simulation could run on a laptop.}\BibitemShut
  {Stop}%
\bibitem [{Note12()}]{Note12}%
  \BibitemOpen
  \bibinfo {note} {In practice, the expectation values defining logical
  coordinates are computed at each time-step rather than after computing the
  whole trajectory, to avoid storage of the full history of $\protect \bm
  {\normalrho }_t$.}\BibitemShut {Stop}%
\end{thebibliography}%
